\documentclass[a4paper,twoside,12pt]{book}
\usepackage{amsmath}
\usepackage{amsfonts}
\usepackage[english]{babel}
\usepackage[ansinew]{inputenc}
\usepackage[final]{graphicx}
\usepackage{varioref}
\usepackage{graphpap}
\usepackage{mathrsfs}
\usepackage{appendix}
\usepackage{booktabs}
\usepackage{psfrag}
\usepackage[backref,linktocpage]{hyperref}
\usepackage[usenames]{color}
\usepackage{amssymb}

\title{\huge \textbf{Trapped surfaces in spacetimes with symmetries and applications to uniqueness theorems}}
\author{\Large \textit{Alberto Carrasco Ferreira}}
\date{\Large April, 2011}

\usepackage{titlesec}
\newcommand{\bigrule}{\titlerule[0.5mm]}
\titleformat{\chapter}[display]
{\bfseries\Huge}
{
 \titlerule
 \filleft 
 \Large\chaptertitlename\
 \Large\thechapter}
{0mm}
{\filleft} 
[\vspace{0.5mm} \bigrule]

\usepackage{fancyhdr}
\renewcommand{\chaptermark}[1]{\markboth{\textbf{\thechapter. #1}}{}}

\def\SS{{\mathfrak S}}

\begin{document}

\newcommand{\bm}[1]{\mbox{\boldmath $#1$}}
\renewcommand{\thechapter}{\arabic{chapter}}
\renewcommand{\thesection}{\arabic{chapter}.\arabic{section}}
\renewcommand{\theequation}{\arabic{chapter}.\arabic{equation}}
\renewcommand{\thefigure}{\arabic{chapter}.\arabic{figure}}
\renewcommand{\thetable}{\Roman{table}}


\newtheorem{thr}{Theorem}[section]
\newtheorem{defi}[thr]{Definition}
\newtheorem{lema}[thr]{Lemma}
\newtheorem{corollary}[thr]{Corollary}
\newtheorem{proposition}[thr]{Proposition}


\def\Journal#1#2#3#4#5#6{#1, ``#2'', {\em #3} {\bf #4}, #5 (#6).}

\def\JournalPrep#1#2#3{#1, ``#2'', #3.}

\def\JGP{J. Geom. Phys.}
\def\JDG{J. Diff. Geom.}
\def\CQG{Class. Quantum Grav.}
\def\JPA{J. Phys. A: Math. Gen.}
\def\PRD{{Phys. Rev.} \bm{D}}
\def\PR{{Phys. Rev.} }
\def\GRG{Gen. Rel. Grav.}
\def\IJT{Int. J. Theor. Phys.}
\def\CMM{Commun. Math. Phys.}
\def\PR{Phys. Rev.}
\def\RMP{Rev. Mod. Phys.}
\def\MNRAS{Mon. Not. Roy. Astr. Soc.}
\def\JMP{J. Math. Phys.}
\def\DG{Diff. Geom.}
\def\CMP{Comm. Math. Phys.}
\def\PRL{Phys. Rev. Lett.}
\def\ARAA{Ann. Rev. Astron. Astroph.}
\def\ANP{Annals Phys.}
\def\AIF{Ann. Inst. Fourier Grenoble}
\def\AP{Ap. J.}
\def\APJL{Ap. J. Lett.}
\def\MPL{Mod. Phys. Lett.}
\def\PREP{Phys. Rep.}
\def\AASF{Ann. Acad. Sci. Fennicae}
\def\ZP{Z. Phys.}
\def\PNAS{Proc. Natl. Acad. Sci. USA}
\def\PLMS{Proc. London Math. Soth.}
\def\AIHP{Ann. Inst. H. Poincar\'e}
\def\AIHPC{Ann. Inst. H. Poincar\'e, Sect C}
\def\ANYAS{Ann. N. Y. Acad. Sci.}
\def\SPJ{Sov. Phys. JETP}
\def\PAWBS{Preuss. Akad. Wiss. Berlin, Sitzber.}
\def\PPLL{Phys. Lett. A }
\def\QJRAS{Q. Jl. R. Astr. Soc.}
\def\CR{C.R. Acad. Sci. (Paris)}
\def\CP{Cahiers de Physique}
\def\NC{Nuovo Cimento}
\def\AM{Ann. of Math.}
\def\APP{Acta Phys. Pol.}
\def\BAMS{Bulletin Amer. Math. Soc}
\def\CPAM{Commun. Pure Appl. Math.}
\def\PJM{Pacific J. Math.}
\def\AGAG{Ann. Global Anal. Geom.}
\def\AHP{Annales Henri Poincar\'e}
\def\PRSL{Proc. Roy. Soc. London}
\def\ATMP{Adv. Theor. Math. Phys.}
\def\CRASP{C.R Acad. Sci. Paris S\'er. A-B}
\def\JGM{Journal of Geom. and Phys.}
\def\CAG{Commun. Anal. Geom.}
\def\IMRN{Int. Math. Res. Not.}
\def\AMM{Amer. Math. Month.}
\def\PAMS{Proc. Amer. Math.  Soc.}
\def\PLB{Physics Letters B}
\def\AnP{Annals Phys.}
\def\ASL{Adv. Sci. Lett.}
\def\NAR{New Astron. Rev}
\def\CJP{Can. J. Phys.}
\def\NAMS{Not. Amer. Math. Soc.}


\def\Omegab{\Omega_{b}}
\def\N{\cal N}
\def\id{(\Sigma,g,K)}
\def\gM{g^{(4)}}
\def\gN{g^{(n)}}
\def\GM{G^{(4)}}
\def\GN{G^{(n)}}
\def\RM{R^{(4)}}
\def\RN{R^{(n)}}
\def\Nu{\nu}
\def\gRW{g_{RWW}^{(4)}}
\def\Sigmatilde{\tilde{\Sigma}}
\def\bd{\partial}
\def\tbd{\partial^{top}}
\def\idfull{(\Sigma,g,K;\rho,{\bf J})}
\def\id{(\Sigma,g,K)}
\def\tr{\mbox{tr}}
\def\nablaM{\nabla}
\def\nablaSigma{\nabla^{\Sigma}}
\def\nablaSigmatilde{\nabla^{\tilde{\Sigma}}}
\def\nablaS{\nabla^{S}}
\def\L{\mathcal{L}}
\def\Db{{\mathfrak D}_{b}}
\def\D{\mathfrak D}
\def\Sb{S_{b}}
\def\S{S}
\def\U{U}
\def\Sm{S_{2}}
\def\Sf{S_{f}}
\def\SN{\Sigma\cap\mathcal{N}}
\def\idfull{(\Sigma,g,K;\rho,{\bf J})}
\def\mots{S}
\def\KVmotsM{\vec{\Pi}}
\def\MCVmotsM{\vec{H}}
\def\MCmotsM{H}
\def\gmots{\gamma}
\def\MCVmotsS{\vec{p}}
\def\MCmotsS{p}
\def\KVmotsS{\vec{\kappa}}
\def\E{\mathcal{S}}
\def\gE{\gamma}
\def\MCVES{\vec{p}_{B}}
\def\MCES{p_{B}}
\def\gl{{g_{0}}}
\def\etal{{\eta_{0}}}
\def\kid{(\Sigma,g,K;N,\vec{Y},\tau)}
\def\kidtilde{(\tilde{\Sigma},g,K;N,\vec{Y},\tau)}
\def\R{R}
\def\RSigma{{R^{\Sigma}}}
\def\C{C}
\def\KES{\kappa}
\def\hE{\gamma}
\def\MCES{p}
\def\kidfull{(\Sigma,g,K;\rho,{\bf J};N,\vec{Y},\tau)}
\def\ext{\{ \lambda>0 \}^{ext}}
\def\kidprima{( \Sigma ',g,K;N,\vec{Y})}
\def\L{\mathcal{L}}
\def\Db{{\mathfrak D}_{b}}
\def\D{\mathfrak D}\def\Sb{S_{b}}
\def\S{S}
\def\U{U}
\def\Sm{S_{2}}
\def\Sf{S_{f}}
\def\ycoor{\hat{y}}
\def\y{y}
\def\Y{Y}
\def\P{P}
\def\Q{Q}
\def\Pcal{\mathcal{P}}
\def\Qcal{\mathcal{Q}}
\def\f{f}
\def\F{F}
\def\Ysol{Y_1}
\def\B{B_{Y,\epsilon}}
\def\Bn{B_{Y_{n},\epsilon_{n}}}
\def\T{T}
\def\p{\mathfrak{p}}
\def\q{\mathfrak{q}}
\def\r{\mathfrak{r}}
\def\auto{\varrho}

\def\JGP{\em J. Geom. Phys.}
\def\JDG{\em J. Diff. Geom.}
\def\CQG{\em Class. Quantum Grav.}
\def\JPA{\em J. Phys. A: Math. Gen.}
\def\PRD{{\em Phys. Rev.} \bm{D}}
\def\GRG{\em Gen. Rel. Grav.}
\def\IJT{\em Int. J. Theor. Phys.}
\def\PR{\em Phys. Rev.}
\def\RMP{\em Rev. Mod. Phys.}
\def\MNRAS{\em Mon. Not. Roy. Astr. Soc.}
\def\JMP{\em J. Math. Phys.}
\def\DG{\em Diff. Geom.}
\def\CMP{\em Commun. Math. Phys.}
\def\APP{\em Acta Phys. Polon.}
\def\PRL{\em Phys. Rev. Lett.}
\def\ARAA{\em Ann. Rev. Astron. Astroph.}
\def\ANP{\em Annals Phys.}
\def\AP{\em Ap. J.}
\def\APJL{\em Ap. J. Lett.}
\def\MPL{\em Mod. Phys. Lett.}
\def\PREP{\em Phys. Rep.}
\def\AASF{\em Ann. Acad. Sci. Fennicae}
\def\ZP{\em Z. Phys.}
\def\PNAS{\em Proc. Natl. Acad. Sci. USA}
\def\PLMS{\em Proc. London Math. Soth.}
\def\AIHP{\em Ann. Inst. H. Poincar\'e}
\def\ANYAS{\em Ann. N. Y. Acad. Sci.}
\def\SPJ{\em Sov. Phys. JETP}
\def\PAWBS{\em Preuss. Akad. Wiss. Berlin, Sitzber.}
\def\PPLL{\em Phys. Lett. A }
\def\QJRAS{\em Q. Jl. R. Astr. Soc.}
\def\CR{\em C.R. Acad. Sci. (Paris)}
\def\CP{\em Cahiers de Physique}
\def\NC{\em Nuovo Cimento}
\def\AM{\em Ann. Math.}
\def\APP{\em Acta Physica Polonica}
\def\BAMS{\em Bulletin Amer. Math. Soc}
\def\CPAM{\em Commun. Pure Appl. Math.}
\def\PJM{\em Pacific J. Math.}
\def\ATMP{\em Adv. Theor. Math. Phys.}
\def\PRSLA{\em Proc. Roy. Soc. London A.}
\def\APPT{\em Ann. Poincar\'e Phys. Theory}

\frontmatter

\maketitle

\newpage
\thispagestyle{empty}
\mbox{}

\newpage
\thispagestyle{empty}

\vspace*{8cm}
\begin{flushright}
A mis padres.\\
\vspace*{1mm}
A mi t\'ia Nena.
\end{flushright}
\newpage
\thispagestyle{empty}
\mbox{}


\newpage
\thispagestyle{empty}

\vspace*{8cm}
\begin{center}
``The transition is a keen one, I assure you, \\
from a schoolmaster to a sailor, and requires \\
a strong decoction of Seneca and the Stoics \\
to enable you to grin and bear it. \\
But even this wears off in time."\\
\vspace{2mm}
{\it Herman Melville, Moby Dick}
\end{center}
\newpage
\thispagestyle{empty}
\mbox{}

\newpage

\chapter{Agradecimientos}

Quiero empezar expresando mi m\'as sincero agradecimiento al profesor Marc Mars Lloret, director de esta tesis doctoral,
por la atenci\'on que me ha prestado durante todos estos a\~nos en los que he tenido la suerte de trabajar a su lado;
por haber compartido conmigo sus ideas y haberme mostrado las l\'ineas de investigaci\'on a seguir en este trabajo;
por tratar siempre de animar mi curiosidad y mi car\'acter cr\'itico por encima de todo;
por su confianza, su paciencia y su apoyo, sin los cuales nunca hubiera podido terminar este trabajo;
por su incansable dedicaci\'on, su total disponibilidad, su trato siempre amable y su sana amistad.

Quiero agradecer a todos los miembros del Departamento de F\'isica Fundamental de la Universidad de Salamanca, por su trato cordial
y por ayudarme siempre que lo he necesitado. Doy las gracias al profesor Walter Simon por su amistad y por compartir su saber conmigo.
Tambi\'en agradezco al profesor Miguel S\'anchez Caja del
Departamento de Geometr\'ia y Topolog\'ia de la Universidad de Granada, que fue mi tutor en los cursos de doctorado, por su hospitalidad, 
su apoyo y su sano inter\'es por mi trabajo durante todos estos a\~nos.

Agradezco a mis compa\~neros de doctorado (Marsopas y otras especies): Cuchi, Jorge, \'Alvaro Due\~nas, \'Alvaro Hern\'andez, Diego, Alberto Soria, Edu y Cristina
por su ayuda, por las risas y por aguantarme todos los d\'ias.
A Toni, el Ave F\'enix de las Marsopas, por haber compartido conmigo tantas inquietudes y tanta magia.
Sin duda, esto hubiera sido mucho m\'as aburrido sin ellos.
Agradezco tambi\'en a la Escuela Kodokai, que me ha llenado de inspiraci\'on y me ha ayudado a estar en forma durante estos a\~nos.

Estoy profundamente agradecido a mis padres, por darme una vida feliz y una buena educaci\'on,
y por la confianza que siempre han depositado en m\'i.
A Marta, por hacer que la vida sea m\'as divertida, y a mi t\'ia Nena, que siempre estuvo a nuestro lado.

Y, por supuesto, a Raquel y a mi pijama azul.
Gracias por existir y por quererme tanto.

Finalmente, agradezco al MICINN por el apoyo econ\'omico prestado.

\newpage
\thispagestyle{empty}
\mbox{}
\newpage

\setcounter{tocdepth}{4}

\setcounter{secnumdepth}{5}
\tableofcontents 

\mainmatter


\chapter{Introduction}
\label{ch:Introduction}

General Relativity, formulated by Einstein in 1915 \cite{Einstein}, is up to the present
date the most accurate theory
to describe gravitational physics. Roughly speaking, this theory establishes that space, time and gravitation
are all of them aspects of a unique structure: the spacetime, a four dimensional manifold
whose geometry is closely related
to its matter contents via the Einstein field equations. 
One of the most striking consequences of General Relativity is the existence of {\it black holes}, that is,
spacetime regions from which no signal can be seen by an observer located
infinitely far from the matter sources.
Black holes in the universe are expected to arise as the final state of
gravitational collapse of sufficiently massive objects, such as massive stars, as the
works by Chandrasekhar, Landau and Oppenheimer and Volkoff \cite{Chandra}
already suggested in the decade of the 1930's.
Despite the fact that many astronomical observations give strong indication that black holes really exist in nature,
a definitive experimental proof of their existence is still lacking.

Although black holes arose first as theoretical predictions of
General Relativity, its modern theory was developed
in the mid-sixties largely in response to the
astronomical discovery of highly energetic and compact objects.
During these years the works of Hawking and Penrose \cite{singularitythrs}
showed that singularities (i.e. ``points''
where the fundamental geometrical quantities are not well-defined)
are commonplace in General Relativity, in particular in the
interior  of black holes.  Singularities have the potential danger of breaking the
predictability power of a theory  because basically anything can happen once
a singularity is visible. However, for the singularities inside black holes
the situation is not nearly as bad, because, in this case, the singularity
is not visible from infinity and hence
the predictability capacity
of the observers lying outside the black hole region remains unaffected. This fact led Penrose to
conjecture that naked singularities (i.e. singularities which do not lie inside a black hole)
cannot occur in any reasonable physical situation \cite{CosmicCensor}. This conjecture, known as the
{\it cosmic censorship hypothesis}, protects the distant observers from the
lack of predictability that occurs in the presence of singularities.
Whether this conjecture is true or not is at present largely unknown (see \cite{Wald1997} for an account of
the situation in the late 90's). Rigorous results are known only in spherical symmetry,
where the conjecture has been proven for several matter models \cite{Christodoulou1999, Dafermos2005}.
In any case, the validity of (some form) of cosmic censorship implies that black holes are the generic end state
of gravitational collapse, and hence fundamental objects in the universe.

Of particular importance is the understanding of equilibrium configurations of black holes.
The {\it uniqueness theorems for static and stationary black holes}, which are considered one of the
cornerstones of the theory of black holes,
also appeared during the sixties mainly motivated by the early work of Israel
\cite{Israel}. These theorems assert that, given a matter model (for example vacuum),
a static or a stationary black hole spacetime belongs necessarily to a specific class of spacetimes
(in the vacuum case, they are Schwarzschild in the static regime and Kerr for the stationary case)
which are univocally
characterized by a few parameters that describe the fundamental properties of the black hole (for
vacuum these parameters are the mass
and the angular momentum of the black hole).
Since, from physical principles, it is expected that astronomical objects
which collapse into a black hole will eventually settle down to a stationary state, the
black hole uniqueness theorems imply that the final state of a generic gravitational collapse (assuming that
cosmic censorship holds) can be described by a very simple spacetime geometry
characterized by a few parameters like the total mass, the electric charge or the angular momentum of the collapsing astronomical object
(or, more precisely, the amount of these physical quantities which is kept by the collapsing object and does not get radiated
away during the process).  The resulting spacetime is therefore
independent of any other of the properties  of the collapsing system (like shape, composition, etc.).
This type of result was, somewhat pompously,  named
{\it ``no hair"} theorems for black holes by Wheeler \cite{Wheeler}.
In 1973 Penrose \cite{Penrose1973} invoked cosmic censorship and the no hair theorems
to deduce
an inequality which imposes a lower bound for the total mass of a spacetime in terms of the area of the
{\it event horizon} (i.e. the boundary) of the black hole which forms during the gravitational collapse.
This conjecture is known as the {\it Penrose inequality}.

The Penrose inequality, like the cosmic censorship conjecture on which it is based, has been proven only in a few particular
cases. Both conjectures therefore remain, up to now, wide open.
One of the intrinsic difficulties for their proof is that
black holes  impose, by its very definition (see e.g. Chapter 12 of \cite{Wald}),
very strong global conditions on a spacetime. From
an evolutive point if view, these objects are of teleological nature because a complete knowledge of the future
is needed to even know if a black hole forms. Determining the future of an initial configuration
(i.e. the metric and its first time derivative on a spacelike hypersurface)
requires
solving the spacetime field equations (either analytical or numerically) with such initial data. The Einstein field
equations are non-linear partial differential equations, so determining the long time behavior of its solutions is
an extremely difficult problem.
In general, the results that can be obtained from present day technology do not give information on the global
structure of the solutions and, therefore, they do not allow to study black holes in an evolutive setting.
As a consequence, the concept of black hole is not very useful in this situation because,
what does it mean that an initial data set represents a black hole? Since the concept of black hole
is central in gravitation, it has turned out to be necessary to replace this global notion by a more local one
that, on the one hand, can be studied in an evolutionary setting and, on the other, hopefully has something to
do with the global concept of black hole. The objects that serve this purpose are
the so-called {\it trapped surfaces}, which are, roughly speaking, compact surfaces without boundary for which the
emanating null rays do not diverge (all the
precise definitions will be given in Chapter \ref{ch:Preliminaires}). The reason for this bending of light ``inwards'' is
the gravitational field and, therefore, these surfaces reveal the presence of an intense gravitational
field. This is expected to indicate that a black hole will in fact form upon evolution. More precisely, under suitable
energy conditions, the maximal Cauchy development of this initial data is known to be causal geodesically incomplete (this is
the content of one of the versions of the singularity theorems, see \cite{S97} for a review). {\it If
cosmic censorship holds}, then a black hole will form. Moreover, it is known that in any black hole
spacetime
the subclass of trapped surfaces called
{\it weakly trapped surfaces} and {\it weakly outer trapped surface} lie
inside the  black hole (see e.g. chapter 9.2 of \cite{HE} and chapter 12.2 of \cite{Wald}),
and so they give an indication of where the back hole event horizon should be in the initial data  (if it forms
at all). In fact, the substitution of the concept of black hole by the concept of trapped surface is so common that
one terminology has replaced the other,
and scientists talk about black hole collision, of black hole-neutron star mergers to refer to evolutions
involving trapped surfaces. However, it should be kept in mind that both concepts
are completely different a priori.

In the context of the Penrose inequality,
the fact that, under cosmic censorship, weakly outer trapped surfaces lie inside the black hole
was used by Penrose to replace the area of the event horizon by the area of weakly outer trapped surfaces
to produce inequalities which, although motivated by the expected global
 structure of the spacetime that forms, can be formulated  directly on the given initial data in a manner
completely independent of its evolution.
A particular case of weakly outer trapped surfaces, the so-called
{\it marginally outer trapped surfaces (MOTS)} (defined as compact surfaces without boundary with
vanishing outer null expansion
$\theta^+$),
are widely considered as the best quasi-local replacements for the event horizon.
From what it has been said, it is clear that proving that these surfaces
can replace black holes is basically the same as proving the validity of cosmic censorship, which is beyond present day knowledge.
The advantage of seeing the problem
from this perspective is that  it allows for simpler questions that can perhaps be solved. One such question is the
Penrose inequality already mentioned. Another one has to do with static and stationary situations. One might think
that, involving no evolution at all, it should be clear that black holes, event horizons and marginally outer trapped
surfaces are essentially the same in an equilibrium configuration. However, although certainly plausible, very little is known about the validity
of this expectation.

The aim of this thesis is precisely to study the properties of
trapped surfaces in spacetimes with symmetries and their possible relation
with the theory of black holes. Even this more modest goal
is vast. We will concentrate
on one aspect of this possible equivalence,
namely {\it whether the static
black hole uniqueness theorems extend to static spacetimes containing MOTS}.
The main result of this thesis states that this question has an affirmative answer, under suitable
conditions on the spacetime.
To solve this question we will have to analyze in depth the properties
of MOTS and weakly outer trapped surfaces
in spacetimes with symmetries, and this will produce a number of results which are,
hopefully, of independent interest.
This study will naturally lead us to
consider a second question,  namely to study the Penrose inequality in
static initial data sets which are not time-symmetric.  Our main result here is
the discovery of a counterexample
of a version of the Penrose inequality that was proposed by Bray and Khuri \cite{BK} not long ago.
It is worth to mention
that most of the results we will obtain in this thesis do not use the
Einstein field equations and, consequently,
they are also valid in any gravitational theory of gravitation in four dimensions.

In the investigations on stationary and static spacetimes there has been
a tendency over the years of reducing the amount of global assumptions in time to
a minimum. This is in agreement with the idea behind cosmic censorship of understanding the
global properties as a consequence of the evolution.
This trend has been particularly noticeable in black hole uniqueness theorems, where
several conditions can be used to capture the notion of black hole (see e.g.
Theorem \ref{thr:electrovacuniqueness0} in Chapter \ref{ch:Preliminaires}).
In this thesis, we will follow this general tendency  and  work
directly on slabs of spacetimes containing suitable spacelike hypersurfaces or, whenever possible,
directly at the initial data level, without assuming the existence of a spacetime where it is embedded.
It should be remarked that the second setting is more  general than the former one.  Indeed,
in some circumstances
the existence of such a spacetime can be proven, for example by using the notion of
Killing development (see \cite{BC} and Chapter \ref{ch:Article1}) or by using well-posedness of the Cauchy problem
and suitable evolution equations for the Killing vector \cite{Coll}. The former,
however, fails at fixed points of the static isometry and the second requires specific matter models,
not just energy
inequalities as we will assume.
Nevertheless, although most of the results of this thesis
will be obtained at the initial data level, we will need to invoke the existence of a spacetime
to complete the proof of the uniqueness result (we emphasize however, that no global
assumption in time is made in that case either). We will also try to make clear which is the difficulty that arises when one attempts
to prove this result directly at the initial data level.

The results obtained in this thesis  constitute, in our opinion, a
step forward in our understanding of how black holes evolve. Regarding the problem
of  establishing a rigorous relationship between black holes and trapped surfaces,
the main result of this thesis (Theorem \ref{uniquenessthr}) shows that,
at least as far as uniqueness of static black holes is concerned, event horizons and MOTS do coincide.
Our uniqueness result for static spacetimes containing MOTS is interesting also independently
of its relationship with black holes. It proves that static configurations are indeed very rigid. This type
of result has several implications. For instance, in any evolution of a collapsing system, it is expected
that an equilibrium configuration is eventually reached. The uniqueness theorems of black holes
are usually invoked to conclude that the spacetime is one of the stationary black holes compatible with the uniqueness
theorem. However, this argument assumes implicitly that one has sufficient information on the spacetime to be able to apply
the uniqueness theorems, which is far from obvious since the spacetime is being constructed during the evolution. In our
setting, as long as the evolution has a MOTS on each time slice, if the spacetime reaches a static configuration, then
it is unique. Related to this issue, it would be very interesting to know if these types of uniqueness results also hold
in an approximate sense, i.e. if a spacetime is {\it nearly} static and contains a MOTS, then the
spacetime is {\it nearly} unique. This problem is, of course, very difficult because it needs a suitable concept
of ``being close to''. In the particular case of the Kerr metric, there exists a notion of an initial
data being close to Kerr \cite{Kroom} which  is based on a suitable characterization of this spacetime \cite{Mars3}.
This closeness notion is defined for initial data sets without boundary and has been extended
to manifolds with boundary under certain circumstances \cite{Kroom2}. It would be of interest
to extend it to the case with a non-empty boundary which is a MOTS.

The static uniqueness result for MOTS is only a first step in this subject. Future work should try to extend this result
to the stationary setting. The problem is, however, considerably more difficult because the techniques known
at present to prove uniqueness of stationary black holes are much less developed than those for
proving uniqueness of static black holes. Assuming however, that the spacetime is axially symmetric (besides
being stationary) simplifies the  black hole uniqueness proof considerably (the problem becomes essentially
a uniqueness proof for a boundary value problem of a non-linear elliptic system on a domain in the
Euclidean plane, see \cite{Heuslerlibro}). The next natural step would be to try and extend this uniqueness result to a setting where
the black hole is replaced by a MOTS. The only result we prove in this thesis in the stationary (non-static) setting
involves MOTS lying in the closure of the exterior region where the Killing
is timelike. We show that in this case
the MOTS cannot penetrate into  the timelike exterior domain (see Theorem   \ref{theorem1}).


In the remaining of this Introduction, we will try to give a general idea of the structure of the thesis and
to discuss its main results.\\

In rough terms, the typical structure of static black holes uniqueness theorems
is the following:\\

{\it Let $(M,\gM)$ be a static solution of the Einstein equations
for a given matter model (for example vacuum) which describes a black hole. Then
$(M,\gM)$ belongs necessarily to a specific class of spacetimes which are univocally
characterized by a number of parameters that can be measured at infinity (in the case of vacuum, the spacetime is necessarily
Schwarzschild and the corresponding parameter is the total mass of the black hole).}\\

There exist static black hole uniqueness theorems for several matter models, such as
vacuum (\cite{Israel}, \cite{MzH}, \cite{Robinson}, \cite{BMuA}, \cite{C}), electro-vacuum (\cite{Israel2}, \cite{MzH2}, \cite{Simon},
\cite{Ruback}, \cite{Simon2}, \cite{MuA}, \cite{C2}, \cite{CT}) and Einstein-Maxwell dilaton (\cite{MuA2}, \cite{MSimon}).
As we will describe in more detail in Chapter \ref{ch:Preliminaires} the most powerful
method for proving these results is the
so called {\it doubling method}, invented by Bunting and Masood-ul-Alam \cite{BMuA} to
show uniqueness in the vacuum case.
This method requires the existence of a complete spacelike hypersurface $\Sigma$ containing
an exterior, asymptotically flat, region $\Sigma^{{ext}}$ such that the Killing
is timelike on $\Sigma^{{ext}}$ and the topological boundary
$\tbd \Sigma^{{ext}}$ is an embedded, compact and non-empty topological manifold.
In static spacetimes, the condition that $(M,\gM)$ is a black hole can be translated
into the existence of such a hypersurface $\Sigma$.
In this setting, the topological boundary $\tbd \Sigma^{{ext}}$ corresponds to the
intersection of the boundary of the domain of outer communications (i.e. the region outside both the black hole
and the white hole) and $\Sigma$. This equivalence, however, is not strict due to
the potential presence of non-embedded Killing prehorizons, which would give rise to
boundaries $\tbd \Sigma^{{ext}}$ which are non-embedded. This issue is important and will be
discussed in detail below. We can however, ignore this subtlety for the purpose of this Introduction.


The type of uniqueness result we are interested in this thesis is of the form:\\

{\it Let $(M,\gM)$ be a static solution of the Einstein equations for a given matter model. Suppose that
$M$ possesses a spacelike hypersurface $\Sigma$ which contains a MOTS.
Then, $(M,\gM)$
belongs to the class of spacetimes established by the uniqueness theorem for static black holes for the
corresponding matter model.}\\

The first result in this direction was given by Miao in 2005 \cite{Miao}, who
extended the uniqueness theorems for vacuum static black holes to the case
of asymptotically flat and time-symmetric slices $\Sigma$ which contain a minimal compact boundary
(it is important to
note that for time-symmetric initial data, a surface is
a MOTS if and only if it is a compact minimal surface).
In this way, Miao was able to
relax the condition of a time-symmetric slice $\Sigma$ having a compact topological
boundary $\tbd \Sigma$ where
the Killing vector vanishes  to simply containing a compact minimal boundary.
Miao's uniqueness result is indeed a generalization of the static uniqueness theorem
of Bunting and Masood-ul-Alam because the
static vacuum field equations imply in the time-symmetric case that the boundary $\tbd \Sigma^{{ext}}$  is necessarily a
totally geodesic surface, which is more restrictive than being a minimal surface.

Miao's result is fundamentally a uniqueness result. However, one of the key ingredients in its
proof consists in showing that no minimal surface can penetrate into the exterior timelike region $\Sigma^{{ext}}$.
As a consequence, Miao's theorem can also be viewed as
a confinement result for minimal surfaces. As a consequence, one can think of extending Miao's result
in three different directions:
Firstly, to allow for other matter models. Secondly, to work with
arbitrary slices and not just time-symmetric ones.  This is important in order to be able to
incorporate so-called degenerate Killing horizons into the problem. Obviously, in the general case
minimal surfaces are no longer suitable and MOTS should be considered.
And finally, try to make the confinement part of the statement as local as possible and
relax the condition of asymptotic flatness to
the existence of suitable exterior barrier.
To that aim it is necessary a proper understanding of the
properties of MOTS and weakly outer trapped surfaces in static spacetimes (or more general, if possible).

For simplicity, let us restrict to the asymptotically flat case for the purpose of the Introduction.
Consider a spacelike hypersurface $\Sigma$ containing
an asymptotically flat end $\Sigma_{0}^{\infty}$.
In what follows, let $\lambda$ be minus the squared norm of the static Killing $\vec{\xi}$. So,
$\lambda>0$ means that $\vec{\xi}$ is timelike. Staticity and asymptotic flatness mean that this Killing vector is timelike
at infinity. Thus, it makes sense to define $\ext$ as the connected component
of $\{ \lambda > 0 \}$ which contains the asymptotically flat end $\Sigma_{0}^{\infty}$ (the
set $\Sigma^{{ext}}$ in the Masood-ul-Alam doubling method is precisely $\ext$).
Since we want to prove the expectation that
MOTS and spacelike sections of the event horizon coincide in static spacetimes,
we will firstly try to ensure that
no MOTS can penetrate into $\ext$.
This result will generalize Miao's theorem as a confinement result and
will extend the well-known confinement result
of MOTS inside the black hole region
(c.f. Proposition 12.2.4 in \cite{Wald})) to the initial data level.
The main tool which will allow us to prove this
result is a recent theorem by Andersson and Metzger \cite{AM} on the existence, uniqueness and regularity
of the outermost MOTS on a given spacelike hypersurface.
This theorem, which will be essential in many places in this thesis,
requires working with trapped surfaces which are {\it bounding}, in the sense that they
are boundaries of suitable regions (see Definition \ref{defi:boundingAM}). Another
important ingredient for our confinement result will be a thorough study of the
causal character that the Killing vector is allowed to have on the outermost MOTS (or, more, generally on stable or strictly
stable MOTS -- all these concepts will be defined below --).
For the case of {\it weakly trapped surfaces} (which are defined by a more restrictive condition than
weakly outer trapped surfaces),
it was
proven in \cite{MS} that no weakly trapped surface can lie
in the region where the Killing vector is timelike provided its mean curvature vector does not vanish identically.
Furthermore, similar restrictions were also obtained for other types of symmetries, such as conformal Killing vectors
(see also \cite{S3} for analogous results in spacetimes with vanishing curvature invariants).

Our main idea to obtain restrictions on the Killing vector on an outermost MOTS $S$
consists on a geometrical
construction \cite{CM1} whereby $S$ is moved first to the past along the
integral lines of the Killing vector  and then back to $\Sigma$ along the outer null geodesics orthogonal to this
newly constructed surface,
producing a new weakly outer trapped surface $S'$, provided the null energy condition (NEC) is satisfied in the spacetime.
If the Killing field $\vec{\xi}$ is timelike anywhere on $S$ then we show that $S'$ lies partially
outside $S$, which is a contradiction with the outermost property of $S$. This simple idea will be central in this thesis
and will be extended in several directions. In particular,
we will generalize the geometric construction to the case
of general vector fields $\vec{\xi}$, not just Killing vectors.
To ensure that $S'$ is weakly outer trapped in this setting we will need to
obtain an explicit expression for the first variation of the outer null expansion $\theta^+$ along $\vec{\xi}$
in terms of the so called {\it deformation tensor} of the metric along $\vec{\xi}$ (Proposition \ref{propositionxitheta}).
This will allow us to obtain results for other types of symmetries,
such as homotheties and conformal Killing vectors,
which are relevant in  many physical situations of interest
(e.g. the Friedmann-Lema\^itre-Robertson-Walker cosmological models).
Another relevant generalization involves analyzing
the infinitesimal
version of the geometric construction. As we will see, the infinitesimal construction
is closely related to the stability properties of the
the first variation of $\theta^{+}$ along $\Sigma$ on a MOTS $S$. This first variation
defines a linear elliptic second order differential operator \cite{AMS} for which
elliptic theory results can be applied.
It turns out that exploiting such results (in particular, the maximum principle for elliptic operators)
the conclusions of the geometric construction  can be sharpened considerably and also extended
to more general MOTS such as stable and strictly stable ones.
(Theorem \ref{TrhAnyXi} and Corollaries \ref{thrstable} and \ref{shear}).

As an explicit application of these results,
we will show that stable MOTS cannot exist in any slice of a large class of Friedmann-Lema\^itre-Robertson-Walker
cosmological models. This class includes
all classic models of matter and radiation dominated eras and also those models with accelerated
expansion which satisfy the NEC (Theorem \ref{thrFRW}).
Remarkably, the geometric construction is more powerful than the elliptic methods in some specific cases.
We will find an interesting situation where this is the case when dealing
with homotheties (including Killing vectors) on outermost MOTS (Theorem \ref{thrkilling}). This will allow
us to prove a result (Theorem \ref{theorem1}) which asserts that, as long as the
spacetime satisfies the NEC,
a Killing vector or homothety cannot be timelike anywhere on a bounding
weakly outer trapped surface
whose exterior lies in a region where the Killing vector is timelike.

Another case when the elliptic theory cannot be applied and we resort to the
geometric procedure deals with
situations when one cannot ensure that the newly constructed surface $S'$ is weakly outer trapped.
However, it can still occur that the portion of $S'$ which lies in the exterior of $S$ has
$\theta^{+}\leq 0$. In this case, we can exploit a result by
Kriele and Hayward \cite{KH97}
in order to construct a weakly outer trapped surface $S''$ outside both $S$ and $S'$ by smoothing
outwards the corner where they intersect.
This will provide us with additional results of interest (Theorems \ref{thrnonelliptic} and \ref{shear2}).
All these results have been published in \cite{CM2} and \cite{CMere2}
and will be presented in Chapter \ref{ch:Article2}.

From then on, we will concentrate exclusively on {\it static} spacetimes.
Chapter \ref{ch:Article1} is devoted to extending Miao's result as a confinement result.
Since in this chapter we will work exclusively at the initial data level,
we will begin by recalling
the concept of a {\it static Killing initial data (static KID)},
(which corresponds to the data and equations one induces on any spacelike hypersurface embedded on a static spacetime, but
viewed as an abstract object on its own, independently of the existence of any embedding into a spacetime).
It will be useful to introduce two scalars $I_{1}, I_{2}$ which correspond to the
invariants of the {\it Killing form} (or Papapetrou field) of the static Killing vector $\vec{\xi}$.
It turns out that $I_{2}$ always vanishes due to staticity and that $I_{1}$
is constant on arc-connected components of $\tbd \{\lambda > 0 \}$ and negative on the
arc-connected components which contains at least a fixed point (Lemma \ref{I1<0}).
Fixed points are initial data translations of
spacetime points where the Killing vector vanishes and, since $I_1$ turns out to
be closely related to the surface gravity of the Killing horizons, this result
extends a well-known result by Boyer
\cite{Boyer} on the structure of Killing horizons to the initial data level.

The general strategy to prove our confinement result for MOTS is to use a contradiction argument. We will
assume that a MOTS can penetrate in the exterior timelike region. By passing to the
outermost MOTS $S$ we will find that the topological boundary of $\tbd \ext$ must intersect both the interior
and the exterior of $S$. It we knew that $\tbd \ext$ is a bounding MOTS, then we could get a contradiction
essentially by smoothing outwards (via the Kriele and Hayward method) these two surfaces. However, it is not true
that $\tbd \ext$ is a bounding MOTS in general. There are simple examples even in Kruskal where this property fails.
The problem lies in the fact that $\tbd \ext$ can intersect both the black hole and the white hole event horizons
(think of the Kruskal spacetime for definiteness) and then the boundary $\tbd \ext$ is, in general, not smooth on the
bifurcation surface. To avoid this situations we need to assume a condition which essentially imposes
that $\tbd \ext$ intersects only the black hole or only the white hole region.
Furthermore, the possibility of $\tbd \ext$ intersecting the white hole region must be removed to ensure that this smooth
surface is in fact a MOTS and not a {\it past} MOTS.
The precise statement of this final condition
is given in points (i) and (ii) of Proposition \ref{is_a_MOTS}, but the more intuitive idea above is sufficient for this
Introduction. Since we will need to mention this condition below, we refer to it as ($\star$).
In this way, in Proposition \ref{is_a_MOTS}, we prove that
every arc-connected component of
$\tbd \{\lambda>0\}$ is an injectively immersed submanifold with $\theta^{+}=0$. 
However, injectively immersed submanifolds may well not be embedded.
Since, in order to find a contradiction
we need to construct a bounding weakly outer trapped surface, and these are necessarily embedded,
we need to care about proving that the injective immersion is an embedding
(i.e. an homeomorphism with the induced topology in the image). In the case with $I_1 \neq 0$
this is easy. In the case of components with $I_1 =0$ (so-called {\it degenerate} components), the problem is
difficult and open. This issue is very closely related to the possibility that there may
exist non-embedded Killing prehorizons in a static spacetime which has already been mentioned before. This problem,
which has remained largely overlooked in the black hole uniqueness theory until very recently \cite{Cc},
is important and very interesting. However, it is beyond the scope of this thesis.
For our purposes it is sufficient to assume an extra condition on degenerate components of $\tbd \ext$
which easily implies that they are embedded submanifolds. This condition is that every arc-connected
component of $\tbd \ext$ with $I_1=0$ is topologically closed. This requirement will appear in all
the main results in this thesis precisely in order to avoid dealing with the possibility of non-embedded
Killing prehorizons. If one can eventually prove that such objects simply do not exist (as we
expect), then this condition can simply be dropped in all the results below.
Our main confinement result is given in Theorem \ref{theorem2}.
The results of Chapter \ref{ch:Article1} have been published in \cite{CM1} and \cite{CMere1}.



Theorem \ref{theorem2} leads directly to a uniqueness result (Theorem \ref{uniquenessthr0})
which already generalizes Miao's result as a uniqueness statement. The idea of the uniqueness
proof is to show that the presence of a MOTS boundary in an initial data set implies, under
suitable conditions, that $\tbd\ext$ is a compact embedded surface {\it without boundary}.
This is precisely the main hypothesis that is made in order to apply
the doubling method of Bunting and Masood-ul-Alam. Thus, assuming that the matter
model is such that static black hole uniqueness holds, then we can conclude uniqueness in the case with MOTS.
The strategy is therefore to reduce the uniqueness theorem for MOTS to the uniqueness theorem for black holes.
This idea is in full agreement with our main theme of showing that MOTS and black holes
are the same in a static situation.

Theorem \ref{uniquenessthr0} is, however, not fully satisfactory because
it still requires condition ($\star$) on $\tbd \ext$.
Since $\tbd\ext$ is a fundamental object in the doubling method, it would be preferable if no conditions
are a priori imposed on it. Chapter \ref{ch:Article4} is devoted to
obtaining a uniqueness result for static spacetimes containing weakly outer trapped surfaces with
no a priori restrictions on $\tbd \ext$ (besides the condition on components with $I_1=0$ which
we have already mentioned). In Chapter \ref{ch:Article1} the fact that
$\tbd \ext$ is closed (i.e. compact and without boundary) is proven as a consequence of its
smoothness. However, when condition ($\star$) is dropped,
we know that $\tbd \ext$ is not smooth in general, and in principle, it may have a non-empty manifold boundary.
Therefore, we will
need a better understanding of the structure of the set $\tbd \{\lambda>0\}$ when
($\star$) is not assumed. In this case, our methods of Chapter \ref{ch:Article1} do not work and we will be forced
to invoke the existence of a spacetime where the initial data set is embedded. By exploiting a
construction by R\'acz and Wald in \cite{RW} we show that, in an embedded static KID,
the set $\tbd \{\lambda>0\}$ 
is a finite union of smooth, compact and embedded surfaces, possibly with boundary.
Moreover,
at least one of the two null expansions $\theta^+$ or $\theta^-$ vanishes identically on each one of these surfaces
(Proposition \ref{proposition1}).
With this result at hand we then prove that
the set  $\tbd \ext$ coincides with the outermost bounding MOTS (Theorem \ref{mainthr})
provided the spacetime satisfies the NEC and that
the
{\it past weakly outer trapped region} $T^{-}$ is included in the {\it weakly outer trapped region} $T^{+}$.
It may seem that the condition $T^{-}\subset T^+$ is very
similar to ($\star$): In some sense, both try to avoid that the
slice intersects first the white hole horizon when moving from the outside.
However, it is important to
remark that $T^+$ and $T^-$ have a priori nothing to do with Killing horizons and that
the condition $T^{-}\subset T^{+}$ is not a condition directly on $\tbd \ext$.
Our main uniqueness theorem is hence Theorem \ref{uniquenessthr}, which states that, under
reasonable hypotheses, MOTS and spacelike sections of Killing horizons do coincide in
static spacetimes. If the static spacetime is a black hole (in the global sense) then
the event horizon is a Killing horizon. This shows the equivalence between MOTS
and (spacelike sections of) the event horizon in the static setting.

The last part of this thesis is devoted to the study of the Penrose
inequality in initial data sets which are not time-symmetric.
The standard version of the Penrose inequality bounds the ADM mass of the spacetime in terms of the
smallest area of all surfaces which enclose the outermost MOTS. 
The huge problem in proving this inequality has led several authors to propose more general and simpler looking versions
of the Penrose inequality (see \cite{Mars2} for a review).
In particular, in a recent proposal by Bray and Khuri \cite{BK}, a Penrose inequality
has been conjectured in terms of
the area of  so-called outermost {\it generalized apparent horizon}
in a given asymptotically flat initial data set. Generalized apparent horizons are more general
than weakly outer trapped surfaces and have interesting analytic and geometric properties.
The Penrose inequality conjectured by Bray and Khuri reads
\begin{equation}\label{ineq}
M_{\scriptscriptstyle ADM}\geq \sqrt{\frac{|S_{out}|}{16\pi}},
\end{equation}
where $M_{\scriptscriptstyle ADM}$ is the total ADM mass of a given slice and
$|S_{out}|$ is the area of the outermost generalized apparent horizon $S_{out}$.
This new inequality has several appealing properties, like being invariant under time reversals,
the fact that no minimal area enclosures are involved and that it implies the standard
Penrose inequality. On the other hand, this version
is not directly supported by any heuristic argument based on cosmic censorship, as the standard Penrose inequality.
In fact, as a consequence of a theorem by Eichmair \cite{Eichmair} on the
existence, uniqueness and regularity of the outermost generalized apparent horizon, there exist slices
in the Kruskal spacetimes (for which $\tbd \ext$ intersects both the black hole and the white hole event horizons),
with the property that its outermost generalized apparent horizon lies, at least partially, inside the domain of outer communications.
In Chapter \ref{ch:Article3} we present a counterexample of (\ref{ineq}) precisely by studying this type of slices in the
Kruskal spacetime.

The equations that define a generalized apparent horizon are non-linear elliptic PDE. Thus, we intend to determine
properties of the solutions of these equations for slices sufficiently close to the time-symmetric
slice of the Kruskal spacetime. Since the outermost generalized apparent horizon in the time-symmetric
slice is the well-known bifurcation surface, we can exploit the
implicit function theorem to show that
any solution of the linearized equation for the generalized apparent horizon corresponds to the linearization
of a solution of the non-linear problem (Proposition \ref{proposition}). With this existence result at hand,
we find a generalized apparent horizon $\hat{S}$
which turns out to be located entirely inside the domain of outer communications and which has
area larger than $16 \pi M^2_{\scriptscriptstyle ADM}$, this violating (\ref{ineq}). This would give
a counterexample to the Bray and Khuri conjecture provided $\hat{S}$ is
either the outermost generalized apparent horizon $S_{out}$ or else, the latter has not smaller
area than the former one. Finally, we will prove that the area of $S_{out}$ is, indeed, at least as large
as the area of $\hat{S}$, which gives a counterexample to (\ref{ineq}) (Theorem \ref{theorem}).
It is important to remark that the existence of this counterexample
does not invalidate the approach given by Bray and Khuri in \cite{BK} to prove the standard Penrose inequality but it does
indicate that the emphasis
must not be on generalized apparent horizons.
This result has been  published in \cite{CM3} and \cite{CMere3}.

Before going into our new results, we start with a
preliminary chapter where the fundamental definitions and results required to
understand this thesis are stated and briefly discussed. This chapter contains
in particular, a detailed sketch of the Bunting and Masood-ul-Alam method
to prove uniqueness of electro-vacuum static black holes. We have
preferred to collect all the preliminary material in one chapter
to facilitate the reading of the thesis.
We have also
found it convenient to include
two mathematical appendices. One where some well-known definitions of manifolds
with boundary
and topology are included (Appendix \ref{ch:appendix1}) and
another one that collects a number of theorems in mathematical analysis (Appendix \ref{ch:appendix2})
which are used as tools in the main text.

\renewcommand{\theequation}{\arabic{chapter}.\arabic{section}.\arabic{equation}}

\chapter{Preliminaries}
\label{ch:Preliminaires}

\section{Basic elements in a geometric theory of gravity}

The fundamental concept in any geometric theory of gravity is that of spacetime.
A {\bf spacetime} is a connected $n$-dimensional smooth differentiable manifold $M$ without boundary
endowed with a Lorentzian metric $\gN$. All manifolds considered in this thesis will be Hausdorff.
(see Appendix \ref{ch:appendix1} for the definition).
A Lorentzian metric is 
a metric with signature $(-,+,+,...,+)$. The covariant derivative associated with the Levi-Civita connection of $\gN$
will be denoted by $\nabla^{(n)}$ and the corresponding Riemann, Ricci and scalar curvature
tensors 
will be denoted by $\RN_{\mu\nu\alpha\beta}$, $\RN_{\mu\nu}$ and $\RN$, respectively
(where $\mu,\nu,\alpha,\beta=0,...,n-1$). We follow the sign conventions of \cite{Wald}.
We will denote by $T_{\p} M$ the tangent space to $M$ at
a point $\p\in M$, by $TM$ the tangent bundle to $M$
(i.e. the collection of the tangent spaces at every point of $M$)
and by $\mathfrak{X}(M)$ the set of smooth sections of $TM$ (i.e. vector fields on $M$).
\begin{defi}
According to the sign of its squared norm,
a vector $\vec{v}\in T_{\p}M$ is:
\begin{itemize}
\item Spacelike, if $\left.\gN_{\mu\nu}v^{\mu}v^{\nu}\right|_{\p}>0$.
\item Timelike, if $\left.\gN_{\mu\nu}v^{\mu}v^{\nu}\right|_{\p}<0$.
\item Null, if $\left.\gN_{\mu\nu}v^{\mu}v^{\nu}\right|_{\p}=0$.
\item Causal, if $\left.\gN_{\mu\nu}v^{\mu}v^{\nu}\right|_{\p}\leq 0$.
\end{itemize}
\end{defi}

\begin{defi}
A spacetime $(M,\gN)$ is {\bf time orientable} if and only if there exists a vector field $\vec{u}\in \mathfrak{X}(M)$ which is timelike everywhere on $M$.\\
Consider a time orientable spacetime $(M,\gN)$.
A {\bf time orientation} is a selection of a timelike vector field $\vec{u}$ which is declared to be future directed.\\
A {\bf time oriented} spacetime is a time orientable spacetime after a time orientation has been selected.
\end{defi}

In a time oriented manifold, causal vectors can be classified in two types: future directed or past directed.

\begin{defi}
Let $(M,\gN)$ be a spacetime with time orientation $\vec{u}$. Then, a causal vector $\vec{v}\in T_{\p}M$ is
	\begin{itemize}
	\item future directed if $\left. \gN_{\mu\nu}u^{\mu}v^{\nu} \right|_{\p}\leq 0$.
	\item past directed if $\left. \gN_{\mu\nu}u^{\mu}v^{\nu} \right|_{\p}\geq 0$.
	\end{itemize}
\end{defi}


Throughout this thesis all spacetimes 
are oriented (see Definition
\ref{defi:orientablemanifold} in Appendix \ref{ch:appendix1})
and time oriented.

General Relativity is a geometric theory of gravity in four dimensions in which
the spacetime metric $\gM$ satisfies the Einstein field equations,
which in geometrized units, $G=c=1$ (where $G$ is the Newton gravitational constant and $c$ is the speed of light in vacuum), takes the form:
\begin{equation}\label{Einsteinequation}
\GM_{\mu\nu}+\Lambda\gM_{\mu\nu}=8\pi T_{\mu\nu},
\end{equation}
where $\GM_{\mu\nu}$ is the so-called Einstein tensor, $\GN_{\mu\nu}\equiv\RN_{\mu\nu}-\frac12 \RN\gN_{\mu\nu}$ (in $n$ dimensions),
$\Lambda$ is the so-called cosmological constant and $T_{\mu\nu}$ is the stress-energy tensor which describes the matter contents of the spacetime.
In such a framework, freely falling test bodies are assumed to travel along the causal (timelike for massive particles and null for massless
particles) geodesics of the spacetime $(M,\gM)$.



Due to general physical principles, it is expected that many dynamical processes
tend to a stationary final state.
Studying these stationary configurations is therefore an essential step for understanding any physical theory.
This is the case, for example, in gravitational collapse processes in General Relativity
which are expected to
settle down to a stationary system. 
Since the fundamental object in gravity is the spacetime metric $\gM$, the existence of
symmetries in the spacetime is expressed in terms of a group of isometries, that is, diffeomorphisms of the spacetime
manifold $M$
which leave the metric unchanged. The infinitesimal generator of the isometry group defines a so-called
{\it Killing vector field}. Conversely, a Killing vector field defines a local
isometry, i.e. a local group of  diffeomorphisms, each of which is an isometry of $(M,\gM)$. If the Killing
vector field is complete then the local group is, in fact, a global group of isometries (or, simply, an isometry).
Throughout this thesis, we will mainly work at the local level without assuming that the Killing vector fields are
complete, unless otherwise stated.
More precisely, consider a spacetime $(M,\gM)$ and a
vector field $\vec{\xi}\in \mathfrak{X}(M)$. The Lie derivative
$\mathcal{L}_{\vec{\xi}}\,\gM_{\mu\nu}$ describes how
the metric is deformed along the local group of diffeomorphisms
generated by $\vec{\xi}$. We thus define the
{\bf metric deformation tensor} associated to
$\vec{\xi}$, or simply
deformation tensor, as
\begin{equation}\label{mdt}
a_{\mu\nu}(\vec{\xi}\,)\equiv \mathcal{L}_{\vec{\xi}}\,\gM_{\mu\nu}=
\nablaM_{\mu}\xi_{\nu}+\nablaM_{\nu}\xi_{\mu},
\end{equation}
where, throughout this thesis, $\nabla$ will denote the covariant derivative of $\gM$.
If $a_{\mu\nu}(\vec{\xi}\,)=0$, then the vector field $\vec{\xi}$ is a {\bf Killing vector field} or simply a Killing vector.\\
If the Killing field is timelike on some non-empty set, then the spacetime is called {\bf stationary}. If, furthermore, the
Killing field is integrable, i.e.
\begin{equation}\label{integrablekilling}
\xi_{[\mu}\nablaM_{\nu}\xi_{\alpha]}=0
\end{equation}
where the square brackets denote anti-symmetrization, then the spacetime
is called {\bf static}.\\
Other important types of isometries are the following.
If the Killing field is spacelike and
the isometry group generated is $U(1)$, then the spacetime has a {\bf cyclic symmetry}.
If, furthermore, there exists a regular axis of symmetry, then the spacetime is {\bf axisymmetric}. 
If the isometry group is $SO(3)$ with orbits being spacelike 2-spheres (or points), then the spacetime is {\bf spherically symmetric}.\\
Other special forms of $a_{\mu\nu}(\vec{\xi}\,)$ define special types of vectors which are also interesting. In particular,
$a_{\mu\nu}(\vec{\xi}\,)=2\phi \gM_{\mu\nu}$ (with $\phi$ being a scalar function)
defines a {\bf conformal Killing vector} and
$a_{\mu\nu}(\vec{\xi}\,)=2 C \gM_{\mu\nu}$ (with $C$ being a constant) corresponds to
a {\bf homothety}.\\

Regarding the matter contents of the spacetime, represented by $T_{\mu\nu}$,
we will not assume a priori any specific matter model, such as vacuum, electro-vacuum, perfect fluid, etc.
However, we will often restrict the class of models in such a way that various types of so-called
energy conditions are satisfied (c.f. Chapter 9.2 in \cite{Wald}).
These are inequalities involving $T_{\mu\nu}$ acting on certain causal vectors and are satisfied by most physically
reasonable matter models. In fact, since in General Relativity without cosmological constant, the Einstein equations impose $\GM_{\mu\nu}=8\pi T_{\mu\nu}$, these conditions can be
stated directly in terms of the Einstein tensor. We choose to define the energy conditions directly in terms of $\GM_{\mu\nu}$. This is preferable
because then all our results hold in any geometric theory of gravity independently of whether the Einstein field equations hold or not.
Obviously, these inequalities are truly energy conditions only in specific theories as, for instance, General Relativity with $\Lambda=0$.
Throughout this thesis, we will often need to impose the
so-called {\it null energy condition (NEC)}.

\begin{defi}
A spacetime $(M,\gM)$ satisfies the {\bf null energy condition (NEC)} if the Einstein tensor $\GM_{\mu\nu}$
satisfies $\GM_{\mu\nu}  k^{\mu}k^{\nu} |_\p \geq 0$ for any null vector $\vec{k}\in T_{\p} M$ and all $\p\in M$.
\end{defi}

Other usual energy conditions are the {\it weak energy condition} and the {\it dominant energy condition (DEC)}.

\begin{defi}
A spacetime $(M,\gM)$ satisfies the {\bf weak energy condition} if the Einstein tensor $\GM_{\mu\nu}$
satisfies that $\GM_{\mu\nu} t^{\mu}t^{\nu}|_\p\geq 0$ for any timelike vector $\vec{t}\in T_{\p}M$ and all $\p\in M$.
\end{defi}

\begin{defi}
A spacetime $(M,\gM)$ satisfies the {\bf dominant energy condition (DEC)} if the Einstein tensor $\GM_{\mu\nu}$
satisfies that $-{\GM}{}^{\nu}_{\mu} t^{\mu}|_\p$ is a future directed causal vector for any future directed timelike vector $\vec{t}\in T_{\p} M$ and all $\p\in M$.
\end{defi}

{\bf Remark.} Obviously, the DEC implies the NEC. $\hfill \square$

\setcounter{equation}{0}
\section{Geometry of surfaces in Lorentzian spaces}
\label{sc:GeometryOfSurfaces}

\subsection{Definitions}
\label{ssc:GeometryOfSurfacesDefinitions}

In this subsection we will motivate and introduce several types of surfaces, such as trapped surfaces and marginally outer trapped surfaces,
that will play an important role in this thesis. We will also discuss several relevant known results concerning them.
For an extensive classification of surfaces in Lorentzian spaces, see \cite{S1}.
Let us begin with some previous definitions and notation.

In what follows, $M$ and $\Sigma$ are two smooth differentiable manifolds, $\Sigma$ possibly with boundary,
with dimensions $n$ and $s$, respectively, satisfying $n\geq s$.

\begin{defi}
Let $\Phi: \Sigma\rightarrow M$ be a smooth map between $\Sigma$ and $M$. Then $\Phi$ is an {\bf immersion}
if its differential has maximum rank (i.e. $rank(\Phi)=s$) at every point.
\end{defi}

The set $\Phi(\Sigma)$ is then said to be {\it immersed} in $M$. However $\Phi(\Sigma)$ can fail to be a
manifold because it can intersect itself. \\
To avoid self-intersections, one has to consider {\it injective immersions}.
In fact, we will say that $\Phi(\Sigma)$ is a {\bf submanifold} of $M$ if
$\Sigma$ is injectively immersed in $M$. All immersions considered in this thesis will be submanifolds.
For simplicity, and since no confusion usually arises,
we will frequently denote by the same symbol ($\Sigma$ in this case) both the manifold $\Sigma$
(as an abstract manifold) and
$\Phi(\Sigma)$ (as a submanifold).
Similarly, and unless otherwise stated,
we will use the same convention for contravariant tensors. More specifically, a contravariant tensor defined on $\Sigma$
and pushed-forward to $\Phi(\Sigma)$ will be usually denoted by the same symbol. Notice however
that $\Phi(\Sigma)$ admits two topologies which
are in general different: the induced topology as a subset of $M$ and the manifold topology defined by $\Phi$ from
$\Sigma$. When referring to topological concepts in injectively immersed submanifolds we will always use the subset
topology unless otherwise stated.

Next, we will define the first and the second fundamental forms of a submanifold.

\begin{defi}
Consider a smooth manifold $M$ endowed with a metric $\gN$ and let $\Sigma$ be a submanifold of $M$.
Then, the {\bf first fundamental form} of $\Sigma$ is the tensor field $g$ on $\Sigma$ defined as
\[
g=\Phi^{*}\left( \gN \right),
\]
where $\Phi^{*}$ denotes the pull-back of the injective immersion $\Phi: \Sigma\rightarrow M$.
\end{defi}

According to the algebraic properties of its first fundamental form, a submanifold can be classified as follows.
\begin{defi}
A submanifold $\Sigma$ of a spacetime $M$ is:
\begin{itemize}
\item {\bf Spacelike} if $g$ is non-degenerate and positive definite.
\item {\bf Timelike} if $g$ is non-degenerate and non-positive definite.
\item {\bf Null} if $g$ is degenerate.
\end{itemize}
\end{defi}
The following result is straightforward and well-known (see e.g. \cite{citesubmanifold})

\begin{proposition}
Let $\Sigma$ be a submanifold of $M$. Then, the first fundamental form $g$ of $\Sigma$ is non-degenerate (and, therefore, a {\it metric})
at a point $\p\in \Sigma$ if and only if
\begin{equation}\label{nnsubmanifold}
T_{\p} M=T_{\p}\Sigma \oplus (T_{\p} \Sigma)^{\perp},
\end{equation}
where $(T_{\p} \Sigma)^{\perp}$ denotes the set of normal vectors to $\Sigma$ at $\p$.
\end{proposition}
We will denote $(T_{\p}M)^{\perp}$ by $N_{\p}M$ and we will call this set the normal space to $\Sigma$ at $\p$. The collection of all normal spaces
forms a vector bundle over $\Sigma$ which is called the normal bundle and is denoted by $N\Sigma$.
From now on, unless otherwise stated, we will only consider submanifolds satisfying (\ref{nnsubmanifold})
at every point.
Let us denote by $\nablaSigma$ the covariant derivative associated with $g$.

Next, consider two arbitrary vectors $\vec{X},\vec{Y}\in\mathfrak{X}(\Sigma)$. According to
(\ref{nnsubmanifold}), the derivative
$\nabla^{(n)}_{\vec{X}}\vec{Y}$, as a vector on ${T M}$, can be split according to
\[
\nabla^{(n)}_{\vec{X}}\vec{Y}=\left( \nabla^{(n)}_{\vec{X}}\vec{Y} \right){}^{T} + \left( \nabla^{(n)}_{\vec{X}}\vec{Y}
\right){}^{\perp},
\]
where the superindices $T$ and $\perp$ denote the tangential and normal parts with respect to $\Sigma$.
The following is an important result in the theory of submanifolds \cite{citesubmanifold}.
\begin{thr}
With the notation above, we have
\[
\left( \nabla^{(n)}_{\vec{X}}\vec{Y} \right){}^{T}=\nablaSigma_{\vec{X}}\vec{Y}.
\]
\end{thr}
The extrinsic geometry of the submanifold is encoded in its second fundamental form.
\begin{defi}
The {\bf second fundamental form vector} $\vec{K}$ of $\Sigma$ in $M$ is a symmetric linear map
$\vec{K}:\mathfrak{X}(\Sigma)\times \mathfrak{X}(\Sigma)\rightarrow  N\Sigma$ defined by
\[
\vec{K}(\vec{X},\vec{Y})= - \left( \nabla^{(n)}_{\vec{X}}\vec{Y} \right){}^{\perp},
\]
for all $\vec{X},\vec{Y}\in\mathfrak{X}(\Sigma)$.
\end{defi}

{\bf Remark.} Our sign convention is such that the second fundamental form vector of a 2-sphere in the
Euclidean 3-space points outwards. $\hfill \square$


\begin{defi}
The {\bf mean curvature vector} of $\Sigma$ in $M$ is defined as $\vec{H}\equiv \tr_{\,\Sigma}\vec{K}$ (where $\tr_{\,\Sigma}$ denotes the
trace with the induced metric $g$ on $T_{\p}\Sigma$ for any $\p\in \Sigma$).
\end{defi}


\begin{defi}
We will define an {\bf embedding} $\Phi$ as an injective immersion such that $\Phi: \Sigma\rightarrow \Phi(\Sigma)$ is
an homeomorphism with the topology on
$\Phi(\Sigma)$ induced from $M$. The image $\Phi(\Sigma)$ will be called an embedded submanifold.
\end{defi}

\begin{defi}
A {\bf surface} $S$ is a smooth, orientable, codimension two, embedded submanifold of $M$
with positive definite first fundamental form $\gamma$.
\end{defi}

From now on we will focus on 4-dimensional spacetimes $(M,\gM)$. For a surface $S\subset M$
we have the following result.
\begin{lema}\label{thr:lk}
The normal bundle of $S$ admits two vector fields $\left\{ \vec{l}_{+},\vec{l}_{-} \right\}$ which are null and future directed everywhere, and
which form a basis of $NS$ in $TM$ at every point $\p\in S$.
\end{lema}
{\bf Proof.}
Let $\p\in S$ and $(U_{\alpha},\varphi_{\alpha})$ be any chart at $\p$ belonging to the positively oriented atlas of $M$.
Let us define $\{\vec{l}_{+}^{\,\, U_{\alpha}},\vec{l}_{-}^{\,\, U_{\alpha}}\}$ as the solution of the set of equations
\begin{eqnarray}\label{lk}
\left.\gM(\vec{l}_{\pm}^{\,\,U_{\alpha}},\vec{e}_{A})\right|_{\p}=0,\qquad \quad
\left.\gM(\vec{l}_{\pm}^{\,\,U_{\alpha}},\vec{l}_{\pm}^{\,\,U_{\alpha}})\right|_{\p}=0,\nonumber\\
\left.\gM(\vec{l}_{+}^{\,\,U_{\alpha}},\vec{l}_{-}^{\,\,U_{\alpha}})\right|_{\p}=-2,\qquad\quad
\left.\gM(\vec{l}_{+}^{\,\,U_{\alpha}},\vec{u})\right|_{\p}=-1,\\
\hspace{-3cm}\left. {\bf \eta}^{(4)}(\vec{l}_{-}^{\,\,U_{\alpha}},\vec{l}_{+}^{\,\,U_{\alpha}},\vec{e}_{1},\vec{e}_{2})\right|_{\p}>0.
\nonumber
\end{eqnarray}
where the vectors $\{\vec{e}_{A}\}$ ($A=1,2$) are the coordinate basis in $U_{\alpha}$, $\vec{u}$ is the timelike
vector which defines the time-orientation for the spacetime and ${\bf \eta}^{(4)}$ is the volume form of $(M,\gM)$.
It is immediate to check that $\{\vec{l}_{+}^{\,\,U_{\alpha}},\vec{l}_{-}^{\,\,U_{\alpha}}\}$ exists and is
unique. 
The last equation is necessary in order to avoid the ambiguity
$\vec{l}_{+}^{\,\,U_{\alpha}}\leftrightarrow\vec{l}_{-}^{\,\,U_{\alpha}}$ allowed by the previous four equations.

The set $\{\vec{l}_{+}^{\,\,U_{\alpha}},\vec{l}_{-}^{\,\,U_{\alpha}}\}$ defines two vector fields if and only if
this definition is
independent of the chart. Select any other positively oriented chart $(U_{\beta},\varphi_{\beta})$ at $\p$. Let
$\{\vec{e'}_{1},\vec{e'}_{2}\}$ be the corresponding coordinate basis, which is related with
$\{\vec{e}_{1},\vec{e}_{2}\}$ by
${{e'_{A}}}^{\mu}=A_{\nu}^{\mu} e_{A}^{\nu}$ ($A,B=1,2$), where $A_{\nu}^{\mu}$ denotes the Jacobian. Since $U_{\alpha}$ and
$U_{\beta}$ belong to the positively oriented atlas, we have that $\text{det} A >0$ everywhere.

The first four equations in (\ref{lk}) force that either
$\vec{l}_{\pm}^{\,\,U_{\beta}}=\vec{l}_{\pm}^{\,\,U_{\alpha}}$ or
$\vec{l}_{\pm}^{\,\,U_{\beta}}=\vec{l}_{\mp}^{\,\,U_{\alpha}}$.
However, the second possibility would imply
\[
\left. {\bf \eta}^{(4)}(\vec{l}_{-}^{\,\,U_{\beta}},\vec{l}_{+}^{\,\,U_{\beta}},\vec{e'}_{1},\vec{e'}_{2})\right|_{\p}=
\left. (\text{det} A)\, {\bf \eta}^{(4)}(\vec{l}_{+}^{\,\,U_{\alpha}},\vec{l}_{-}^{\,\,U_{\alpha}},\vec{e}_{1},\vec{e}_{2})\right|_{\p}<0,
\]
which contradicts the fifth equation in (\ref{lk}) for $U_{\beta}$. Consequently $\{\vec{l}_{+},\vec{l}_{-}\}$ does not depend on the chart, which proves the result.
$\hfill \blacksquare$ \\

{\bf Remark.} From now on we will take the vector fields $\vec{l}_{+}$, $\vec{l}_{-}$ to be partially normalized
to satisfy ${l_{+}}_{\mu}l_{-}^{\mu}=-2$, as in the proof of the lemma.
Note that these vector fields are then defined modulo a transformation
$\vec{l}_{+}\rightarrow F\vec{l}_{+}$, $\vec{l}_{-}\rightarrow \frac 1F \vec{l}_{-}$,
where $F$ is a positive function on $S$. $\hfill \square$ \\

For a surface $S$, $\nablaS$ will denote the covariant derivative associated with $\gamma$ and
$\vec{\Pi}$ and $\vec{H}$ will denote the second fundamental form vector and the mean curvature of $S$ in $M$.
The physical meaning of the causal character of $\vec{H}$ is closely related to the first variation of area,
which we briefly discuss next.
Let $\vec{
\nu}$ be a normal variation vector on $S$, i.e. a vector defined in a neighbourhood of $S$ in $M$ which, on $S$, is orthogonal to $S$. 
Choose $\vec{\nu}$ to be compactly supported on $S$ (which obviously places no restrictions when $S$ itself is compact).
The vector $\vec{
\nu}$ generates a one-parameter local group $\{ \varphi_{\tau} \}_{\tau\in I}$ of transformations
where $\tau$ is the canonical parameter and $I\subset \mathbb{R}$ is an interval containing $\tau=0$. We then define a one parameter
family of surfaces $S_{\tau}\equiv \varphi_{\tau}(S)$, which obviously satisfies $S_{\tau=0}=S$.
Let $|S_{\tau}|$ denote the area of the surface $S_{\tau}$. The formula of the first variation of area states (see
e.g. \cite{Chavel})
\begin{equation}\label{firstvariation}
\delta_{\vec{\nu}}|S|\equiv\left.\frac{d|S_{\tau}|}{d\tau}\right|_{\tau=0}=\int _{S} H_{\mu}\nu^{\mu}\eta_{S}.
\end{equation}
{\bf Remark.} It is important to indicate that, when $S$ is boundaryless, expression (\ref{firstvariation})
holds regardless of
whether the variation $\vec{\nu}$ is normal or not.
This formula is valid for any dimensions of $M$ and $S$, provided $\text{dim} M>\text{dim} S$. $\hfill \square$ \\

The first variation of area justifies the definition of a {\it minimal surface} as follows.
\begin{defi}
A surface $S$ is {\bf minimal} if and only if $\vec{H}=0$.
\end{defi}

According to (\ref{firstvariation}), if $\vec{H}$ is timelike and future directed
(resp. past directed) everywhere on $S$, then the
area of $S$ will decrease along any non-zero causal future (resp. past) direction.
If a surface is such that its area does not increase for any future variation, one may say that the surface
is, in some sense, trapped.
Thus, according to the previous discussion, we find that the {\it trappedness} of a surface is intimately
related with the causal character and time orientation of its mean curvature vector $\vec{H}$.
In what follows, we will introduce various notions of trapped surface. For that,
it will be useful to consider a null basis $\{\vec{l}_{+},\vec{l}_{-}\}$ for the normal bundle of $S$ in $M$, as before.
Then, the mean curvature vector decomposes as
\begin{equation}\label{MeanCurvature}
\vec{H}=-\frac 12 \left( \theta^{-}\vec{l}_{+} + \theta^{+}\vec{l}_{-} \right),
\end{equation}
where $\theta^{+}\equiv {l_{+}}^{\mu}H_{\mu}$ and $\theta^{-}\equiv {l_{-}}^{\mu}H_{\mu}$ are the null expansions
of $S$ along $\vec{l}_{+}$ and $\vec{l}_{-}$, respectively.
It is worth to remark that these null expansions $\theta^{\pm}$ are equal to the divergence on $S$
of light rays (i.e. null geodesics) emerging orthogonally from $S$ along $\vec{l}_{\pm}$. Thus,
the negativity of both $\theta^{+}$ and $\theta^{-}$ indicates the presence of strong gravitational fields which bend the light
rays sufficiently so that both are contracting.
Thus, this leads to various concepts of trapped surfaces, as follows.
\begin{defi}\label{defi:MTS}
A closed (i.e. compact and without boundary) surface is a:
\begin{itemize}
\item {\bf Trapped surface} if $\theta^{+}< 0$ and $\theta^{-}< 0$. Or equivalently, if $\vec{H}$ is timelike and future directed.
\item {\bf Weakly trapped surface} if $\theta^{+}\leq 0$ and $\theta^{-}\leq 0$. Or equivalently, if $\vec{H}$ is causal and future directed.
\item {\bf Marginally trapped surface} if either, $\theta^{+}=0$ and $\theta^{-}\leq 0$ everywhere, or, $\theta^{+}\leq 0$ and $\theta^{-}= 0$ everywhere. Equivalently, if $\vec{H}$ is future directed and either proportional to $\vec{l}_{+}$ or proportional to $\vec{l}_-$ everywhere.
\end{itemize}
\end{defi}

If the signs of the inequalities are reversed then we have trappedness along the past directed
causal vectors orthogonal to $S$. 
Thus,
\begin{defi}
A closed surface is a:
\begin{itemize}
\item {\bf Past trapped surface} if $\theta^{+}> 0$ and $\theta^{-}>0 $. Or equivalently if $\vec{H}$ is timelike and past directed.
\item {\bf Past weakly trapped surface} if $\theta^{+}\geq 0$ and $\theta^{-}\geq 0$. Or equivalently if $\vec{H}$ is causal and past directed.
\item {\bf Past marginally trapped surface} if either, $\theta^{+}=0$ and $\theta^{-}\geq 0$ everywhere, or $\theta^{+}\geq 0$ and $\theta^{-}=0$ everywhere.
Equivalently, $\vec{H}$ is past directed and either proportional to $\vec{l}_{+}$ or proportional to $\vec{l}_-$
everywhere.
\end{itemize}
\end{defi}


We also define ``untrapped" surface as a kind of strong complementary of the above.

\begin{defi}
A closed surface is
{\bf untrapped} if $\theta^{+}\theta^{-}<0$, or equivalently if $\vec{H}$ is spacelike everywhere.
\end{defi}

Notice that, according to these definitions, a closed {\it minimal} surface is both weakly trapped
and marginally trapped, as well as past weakly trapped and past marginally trapped.

Because of their physical meaning as indicators of strong gravitational fields,
trapped surfaces are widely considered as good natural quasi-local replacements for black holes.
Let us briefly recall the definition of a black hole which, as already mentioned in the Introduction,
involves global hypotheses in the spacetime. First, it requires
a proper definition of asymptotic flatness in terms of the conformal compactification of the spacetime
(see e.g. Chapter 11 of \cite{Wald}).
Besides, it also requires that the spacetime is {\it strongly asymptotically predictable},
(see Chapter 12 of \cite{Wald} for a precise definition). A strongly asymptotically predictable spacetime $(M,\gM)$ is
then said to contain a black hole if
$M$ is not contained in the causal past of future null infinity $J^{-}(\mathscr{I}^{+})$. The {\bf black hole region} ${\mathcal{B}}$ is defined as
$\mathcal{B}=M\setminus J^{-}(\mathscr{I}^{+})$. The topological boundary $\mathcal{H}_{\mathcal{B}}$ of $\mathcal{B}$ in $M$ is called the {\bf event horizon}. Similarly,
we can define the 
{\bf white hole region} $\mathcal{W}$ as the complementary of the causal future of past null infinity, i.e. $M\setminus J^{+}(\mathscr{I}^{-})$,
and the {\bf white hole event horizon} $\mathcal{H}_{\mathcal{W}}$ as its topological boundary. Finally,
the {\bf domain of outer communications} is defined as $M_{DOC}\equiv J^{-}(\mathscr{I}^{+})\cap J^{+}(\mathscr{I}^{-})$.
Hawking and Ellis show (see Chapter 9.2 in \cite{HE}) that weakly trapped surfaces lie inside the black hole region
in a spacetime provided this spacetime is future asymptotically predictable.
However, as we already pointed out in the Introduction, the study of trapped surfaces is specially interesting when
no global assumptions are imposed on the spacetime and
the concept of black hole is not available.
It is worth to remark that trapped surfaces are also fundamental ingredients in several versions of singularity theorems of General Relativity (see e.g. Chapter 9 in \cite{Wald}).

Note that all the surfaces introduced above are defined by restricting both null expansions $\theta^+$ and $\theta^-$.
When only one of the null expansions is restricted, other interesting types of surfaces are obtained:
the {\it outer} trapped surfaces, which will be the fundamental objects of this thesis.

\begin{figure}[h]
\begin{center}
\psfrag{p}{$\p$}
\psfrag{l+}{$\vec{l}_{+}$}
\psfrag{l-}{$\vec{l}_{-}$}
\psfrag{m}{{$\vec{m}$}}
\psfrag{n}{{$\vec{n}$}}
\includegraphics[width=6cm]{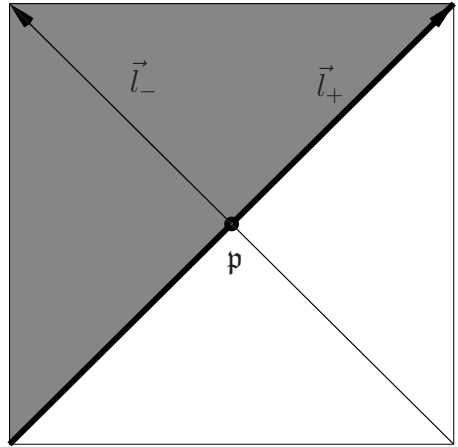}
\caption{
This figure represents the normal space to $S$ in $M$ at a point $\p\in S$. 
If $S$ is outer trapped, the mean curvature vector $\vec{H}$ points into the shaded region. If $S$ is a MOTS, $\vec{H}$ points into the direction
of the bold line.
}
\label{fig:MOTS}
\end{center}
\end{figure}

Again, consider a surface $S$.
Suppose that for some reason one of the future null directions can be geometrically selected so that it points
into the ``outer" direction of $S$ (shortly, we will find a specific setting where this selection is meaningful).
In that situation we will always denote by $\vec{l}_{+}$ the vector pointing along this outer null direction.
We will say that $\vec{l}_{+}$ is the future outer null direction, and similarly, $\vec{l}_{-}$ will be the
future inner null direction. We define the following types of surfaces (c.f. Figure \ref{fig:MOTS}).
\begin{defi}\label{defi:MOTS}
A closed surface is:
\begin{itemize}
\item {\bf Outer trapped} if $\theta^+ < 0$.
\item {\bf Weakly outer trapped} if $\theta^{+}\leq 0$.
\item {\bf Marginally outer trapped (MOTS)} if $\theta^{+}=0$.
\item {\bf Outer untrapped} if $\theta^{+}>0$.
\end{itemize}
\end{defi}

\begin{figure}[h]
\begin{center}
\psfrag{p}{$\p$}
\psfrag{l+}{$\vec{l}_{+}$}
\psfrag{l-}{$\vec{l}_{-}$}
\psfrag{m}{{$\vec{m}$}}
\psfrag{n}{{$\vec{n}$}}
\includegraphics[width=6cm]{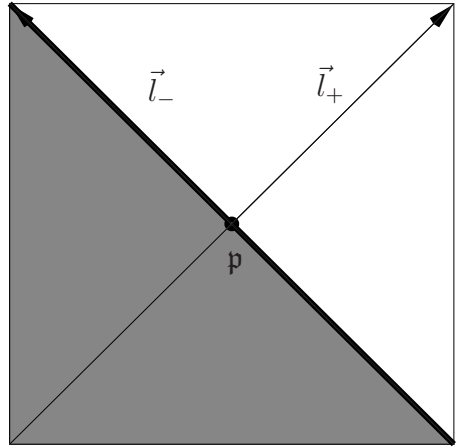}
\caption{
On the normal space $N_{\p}S$ for any point $\p\in S$, the mean curvature vector $\vec{H}$ points into the shaded region if $S$ is past outer trapped, and into the direction
of the bold line if $S$ is a past MOTS.
}
\label{fig:pastMOTS}
\end{center}
\end{figure}

As before, these definitions depend on the time orientation
of the spacetime. If the time orientation is reversed but the notion of {\it outer} is unambiguous, then
$-\vec{l}_{-}$ becomes the new future outer null direction.
Since the null expansion of $-\vec{l}_{-}$ is $-\theta^{-}$, the following definitions become natural (c.f. Figure \ref{fig:pastMOTS}).

\begin{defi}\label{defi:PastMOTS}
A closed surface is:
\begin{itemize}
\item {\bf Past outer trapped} if $\theta^- > 0$.
\item {\bf Past weakly outer trapped} if $\theta^{-}\geq 0$.
\item {\bf Past marginally outer trapped (past MOTS)} if $\theta^{-}=0$.
\item {\bf Past outer untrapped} if $\theta^{-}<0$.
\end{itemize}
\end{defi}

As for weakly trapped surfaces,
weakly outer trapped surfaces are always inside the black hole region
provided the spacetime is strongly asymptotically predictable.
In fact, in one of the simplest dynamical situations, namely the Vaidya spacetime,
Ben-Dov has proved \cite{BenDov} that
the event horizon is the boundary of the
spacetime region containing weakly outer trapped surfaces, proving in this particular case a previous conjecture by Eardley \cite{Eardley}. On the other hand, Bengtsson and Senovilla
have shown \cite{SB} that the spacetime region containing weakly trapped surfaces does
not extend to the event horizon.
This result suggests that the concept of
weakly outer trapped surface does capture the essence
of a black hole better than that of weakly trapped surface.

Two other interesting classes of surfaces that also depend on a choice of outer direction are the so-called
{\it generalized trapped surfaces} and its marginal case,
{\it generalized apparent horizons}.
They were specifically introduced by Bray and Khuri while studying a new approach to
prove the Penrose inequality \cite{BK}.
\begin{defi}\label{defi:GAH}
A closed surface is a:
\begin{itemize}
\item {\bf Generalized trapped surface} if $\left.\theta^{+}\right|_{\p}\leq 0$ or $\left.\theta^{-}\right|_{\p}\geq 0$ at each point $\p\in S$.
\item {\bf Generalized apparent horizon} if either $\left.\theta^{+}\right|_{\p}=0$ with $\left.\theta^{-}\right|_{\p}\leq 0$ or $\left.\theta^{-}\right|_{\p}=0$ with $\left.\theta^{+}\right|_{\p}\geq 0$ at each point $\p\in S$.
\end{itemize}
\end{defi}

\begin{figure}[h]
\begin{center}
\psfrag{p}{$\p$}
\psfrag{l+}{$\vec{l}_{+}$}
\psfrag{l-}{$\vec{l}_{-}$}
\psfrag{m}{{$\vec{m}$}}
\psfrag{n}{{$\vec{n}$}}
\includegraphics[width=6cm]{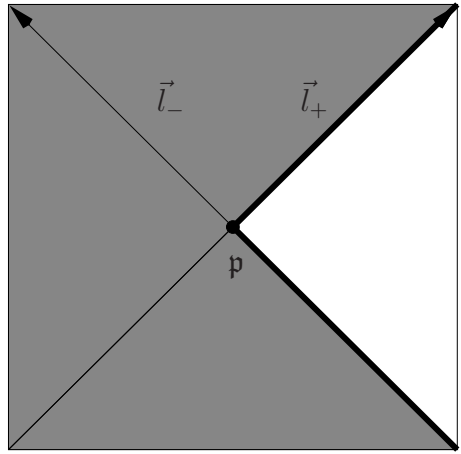}
\caption{This figure represents the normal space of a surface $S$ in $M$ at a point $\p\in S$. 
For generalized trapped surfaces, the mean curvature vector $\vec{H}$ points into the shaded region. For generalized apparent horizons, $\vec{H}$ points into the direction
of the bold line.}
\label{fig:GAH}
\end{center}
\end{figure}

It is clear from Figures \ref{fig:MOTS}, \ref{fig:pastMOTS} and \ref{fig:GAH} that the
set of generalized trapped surfaces includes both the set of weakly outer trapped surfaces and the set of
past weakly outer trapped surfaces as particular cases. 

In this thesis we will often consider surfaces embedded in a spacelike hypersurface $\Sigma\subset M$.
For this reason, it will be useful to give a (3+1) decomposition of the null expansions and to reformulate
the previous definitions in terms of objects defined directly on $\Sigma$.

\begin{defi}
A {\bf hypersurface} $\Sigma$ of $M$ is an embedded, connected 
spacelike submanifold, possibly with boundary, of codimension 1.
\end{defi}
Let us consider a hypersurface $\Sigma$ of $M$ and denote by $g$ its induced metric, by $\vec{K}$ its second fundamental form vector and by $K$ the second fundamental form, defined as $K(\vec{X},\vec{Y})=-{\bf n}(\vec{K}(\vec{X},\vec{Y}))$,
where $\bf n$ is the unit,
future directed, normal 1-form to $\Sigma$ and $\vec{X}, \vec{Y}\in \mathfrak{X}(\Sigma)$.


Consider a surface $S$ embedded in $\id$ 
As before, we denote by $\gamma$, $\vec{\Pi}$ and $\vec{H}$ the induced metric, the second fundamental form vector and the mean curvature
vector of $S$ as a submanifold of $(M,\gM)$, respectively.
As a submanifold of $\Sigma$, $S$ will also have a second fundamental form vector $\vec{\kappa}$ and a
mean curvature vector $\vec{p}$. 
From their definitions, we immediately have
\[
\vec{\Pi}(\vec{X},\vec{Y})=\vec{K}(\vec{X},\vec{Y})+\vec{\kappa}(\vec{X},\vec{Y}),
\]
where $\vec{X},\vec{Y}\in\mathfrak{X}(S)$. Taking trace on $S$ we find
\begin{equation}\label{Hpvec}
\vec{H}=\vec{p}+\gamma^{AB}\vec{K}_{AB},
\end{equation}
where $\vec{K}_{AB}$ is the pull-back of $\vec{K}_{ij}$ ($i,j=1,2,3$) onto $S$. 
Assume that an outer null direction $\vec{l}_{+}$ can be selected on $S$. Then, after a suitable rescaling of $\vec{l}_{+}$ and $\vec{l}_{-}$,
we can define $\vec{m}$ univocally on $S$ as the unit
vector tangent to $\Sigma$ which satisfies
\begin{eqnarray}
\vec{l}_{+}&=&\vec{n}+\vec{m},\label{l+mn} \\
\vec{l}_{-}&=& \vec{n}-\vec{m}.\label{l-mn}
\end{eqnarray}
By construction, $\vec{m}$ is normal to $S$ in $\Sigma$ and will be denoted as the {\it outer normal}.



Multiplying (\ref{Hpvec}) by $\vec{l}_+$ and by $\vec{l}_-$ we find
\begin{equation}\label{Hp}
\theta^{\pm}=\pm p + q,
\end{equation}
where $p\equiv p_{i}m^{i}$ and $q\equiv \gamma^{AB}K_{AB}$.
All objects in (\ref{Hp}) are intrinsic to $\Sigma$.
This allows us to
reformulate the definitions above in terms of $p$ and $q$.
The following table summarizes the types of surfaces mostly
used in this thesis.

\begin{table}[h]
\begin{center}
\begin{tabular}{|l|l|}
\hline
{Outer trapped surface} & $p < -q$\\
\hline
{Weakly outer trapped surface} & $p \leq -q$\\
\hline
Marginally outer trapped surface (MOTS) & $p=-q$\\
\hline
Outer untrapped surface & $p>-q$\\
\hline
Past outer trapped surface & $p<q$\\
\hline
Past weakly outer trapped surface & $p\leq q$\\
\hline
Past marginally outer trapped surface (past MOTS) & $p=q$\\
\hline
Past outer untrapped surface & $p>q$\\
\hline
Generalized trapped surface & $p\leq |q|$\\
\hline
Generalized apparent horizon & $p=|q|$\\
\hline
\end{tabular}
\caption{Definitions of various types of trapped surfaces in terms of the mean curvature $p$ of $S\subset \Sigma$ and
the trace $q$ on $S$ of the second
fundamental form of $\Sigma$ in $M$.}
\end{center}
\end{table}





Having defined the main types of surfaces used in this thesis, let us next consider
the important concept of stability of a MOTS.

\subsection{Stability of marginally outer trapped surfaces (MOTS)}
\label{ssc:GeometryOfSurfacesStability}

Let us first recall the concept of stability for minimal surfaces.
Let $S$ be a closed minimal surface embedded in a Riemannian $3$-dimensional manifold $(\Sigma,g)$.
From (\ref{firstvariation}), $S$ is an extremal of area for all variations (normal or not).
In order to study whether this extremum is a minimum, a maximum or a saddle point,
it is necessary to analyze the second variation of area.
A minimal surface is called {\it stable} if
the second variation of area is non-negative for all smooth variations.
This definition becomes operative once an explicit form for the second variation is obtained.
For closed minimal surfaces the crucial object is the so-called {\it stability operator}, defined as follows.
Consider a variation vector $\psi \vec{m}$ normal to $S$ within $\Sigma$.
Let us denote by a sub-index $\tau$
the magnitudes which correspond to the surfaces $S_{\tau}=\varphi_{\tau}(S)$ (where, as before, $\{\varphi_{\tau}\}_{\tau\in I\subset\mathbb{R}}$ denotes the one-parameter local group of transformations generated by any vector $\vec{\nu}$ satisfying $\left.\vec{\nu} \right|_{S}=\psi\vec{m}$).
For any covariant tensor $\Gamma$ defined on $S$, let us define the variation of $\Gamma$ along $\psi{\vec{m}}$
as $\delta_{\psi\vec{m}} \Gamma \equiv \left.\frac{d}{d\tau} \left[ \varphi_{\tau}^{*}(\Gamma_{\tau})
\right]\right|_{\tau=0}$,
where $\varphi_{\tau}^{*}$ denotes the pull-back of $\varphi_{\tau}$ (this definition does not depend on
the extension of the
vector $\psi\vec{m}$ outside $S$).
The stability operator $L^{min}_{\vec{m}}$ is then defined as
\begin{equation}\label{stabilityoperatorforminimalsurfaces}
L^{min}_{\vec m} \psi\equiv\delta_{\psi\vec{m}}p=-\Delta_{S}\psi - (\RSigma_{ij}{m}^{i}{m}^{j}+{\kappa}_{ij}\kappa^{ij})\psi,
\end{equation}
where
$\Delta_{S}=\nablaS_{A}{\nablaS}^{A}$ is the Laplacian on $S$
and $\RSigma_{ij}$ denotes the Ricci tensor of $(\Sigma,g)$. The second equality follows from a direct computation
(see e.g. \cite{Chavel}).

In terms of the stability operator, the formula for the second variation of area of a
closed minimal surface is given by
\begin{equation}\label{secondvariationforminimalsurfaces0}
{\delta^{2}_{\psi\vec{m}}} |S|
=\int_{S} \psi L^{min}_{\vec{m}}\psi \eta_{S}.\nonumber
\end{equation}

The operator $L^{min}_{\vec{m}}$ is linear, elliptic and formally self-adjoint (see Appendix \ref{ch:appendix2}
for the definitions).
Being self-adjoint implies that the principal eigenvalue $\auto$
can be represented by the Rayleigh-Ritz formula (\ref{Rayleigh-Ritz}),
and therefore the second variation of area 
can be bounded according to
\begin{equation}\label{secondvariationforminimalsurfaces}
{\delta^{2}_{\psi\vec{
m}}} |S|\geq \auto\int_{S}\psi^{2}\eta_S ,\nonumber
\end{equation}
where equality holds when $\psi$ is a principal eigenfunction (i.e. an eigenfunction corresponding to $\auto$).
This implies that ${\delta^{2}_{\psi\vec{m}}} |S|\geq 0$ for all smooth variations is equivalent to $\auto\geq 0$.
Thus, {\it a minimal surface is stable if and only if $\auto\geq 0$}.

A related construction can be performed for MOTS.
Consider a MOTS $S$ embedded in a spacelike hypersurface
$\Sigma$ of a spacetime $M$.
As embedded submanifolds of $\Sigma$, MOTS are not minimal surfaces in general. 
Consequently, any connection
between stability and the second variation of area is lost.
However, the stability for
minimal surfaces involves the sign of the variation $\delta_{\psi\vec{
m}}p$ (see (\ref{stabilityoperatorforminimalsurfaces})), so
it is appropriate to define stability of MOTS in terms of the sign of first
variations of $\theta^{+}$.

A formula for the first variation of $\theta^{+}$ was derived by Newman in \cite{Newman} for arbitrary immersed
spacelike submanifolds. The derivation was simplified by Andersson, Mars and Simon in \cite{AMS}.

\begin{lema}\label{lema:variationtheta+general0}
Consider a surface $S$ embedded in a spacetime $(M,\gM)$. Let $\{\vec{l}_{+},\vec{l}_{-}\}$ be a future directed null basis in the normal bundle of $S$ in $M$, 
partially
normalized to satisfy ${l_{+}}_{\mu}l_{-}^{\mu}=-2$.
Any variation vector $\vec{\nu}$ can be decomposed on $S$ as $\vec{
\nu}=\vec{
\nu}^{\,\parallel}+b\vec{l}_{+}-\frac u2 \vec{l}_{-}$, where $\vec{
\nu}^{\,\parallel}$ is tangent to $S$ and
$b$ and $u$ are functions on $S$. Then,
\begin{eqnarray}\label{variationtheta+general0}
\delta_{\vec{
\nu}}\theta^{+}&=&-\frac{\theta^{+}}{2}{l_{-}}^{\mu}\delta_{\vec{
\nu}}{{l_{+}}}_{\mu} +\vec{
\nu}^{\,\parallel}(\theta^{+}) -
b\left( {\Pi^{\mu}}_{AB}{\Pi^{\nu}}^{AB}{l_{+}}_{\mu}{l_{+}}_{\nu}+G_{\mu\nu}{l_{+}}^{\mu}{l_{+}}^{\nu} \right) - \Delta_{S}u \nonumber\\
&& + 2s^{A}\nablaS_{A}u+ \frac{u}{2}\left( R_{S}-H^{2}-G_{\mu\nu}{l_{+}}^{\mu}{l_{-}}^{\nu}-2s_{A}s^{A}+2\nablaS_{A}s^{A} \right),
\end{eqnarray}
where $R_{S}$ denotes the scalar curvature of $S$, $H^{2}=H_{\mu}H^{\mu}$ 
and $s_{A}=-\frac12 {l_{-}}_{\mu}\nablaM_{\vec{e}_{A}}{l_{+}}^{\mu}$, with $\{\vec{e}_{A}\}$ being a local basis for $TS$.
\end{lema}

Expression (\ref{variationtheta+general0}) can be particularized when the variation is restricted to $\Sigma$,
i.e. when $\vec{\nu}=\psi\vec{m}$ for an arbitrary function $\psi$. Writing $\vec{l}_{\pm}=\vec{n}\pm\vec{m}$ as before, we have
$\vec{\nu}=\frac{\psi}{2}(\vec{l}_{+}-\vec{l}_{-})$ and hence $\vec{\nu}^{\,\parallel}=0$, $b=\frac{\psi}{2}$,
$u=\psi$. As a consequence of Lemma \ref{lema:variationtheta+general0} we have the following \cite{AMS}.


\begin{defi}\label{defi:stabilityoperator}
The {\bf stability operator} $L_{\vec{m}}$ for a MOTS $S$ is defined by
\begin{equation}\label{stabilityoperator}
L_{\vec{m}}\psi\equiv \delta_{\psi\vec{m}}\theta^{+}=-\Delta_{S}\psi + 2s^{A}\nablaS_{A}\psi + \left(\frac 12 R_{S}- Y- s_{A}s^{A}+\nablaS_{A}s^{A} \right)\psi,
\end{equation}
where
\begin{equation}\label{Y}
Y\equiv \frac 12 \Pi _{AB}^{\mu}{\Pi^{\nu}}^{AB}{l_{+}}_{\mu}{l_{+}}_{\nu}+G_{\mu\nu}l_{+}^{\mu}n^{\nu}.
\end{equation}
\end{defi}

{\bf Remark.} In terms of objects on $\Sigma$, a simple computation using $\vec{l}_{\pm}=\vec{n}\pm\vec{m}$ shows that $s_{A}=m^{i}e_{A}^{j}K_{ij}$. $\hfill \square$ \\

If we consider a variation along $\vec{l}_{+}$, then (\ref{variationtheta+general0}) implies that, on a MOTS,
\begin{equation}
\delta_{\psi\vec{l}_{+}}\theta^{+}=-\psi W,\label{raychaudhuri}
\end{equation}
where
\begin{equation}\label{W}
W= \Pi _{AB}^{\mu}{\Pi^{\nu}}^{AB}{l_{+}}_{\mu}{l_{+}}_{\nu}+G_{\mu\nu}l_{+}^{\mu}l_{+}^{\nu}.
\end{equation}
This is the well-known Raychaudhuri equation for a MOTS (see e.g. \cite{Wald}).

Note that $W$ is non-negative provided the NEC holds and
$Y$ is non-negative if the DEC holds (recall that $\vec{n}$ is timelike).



The operator $L_{\vec{m}}$ is linear and elliptic
which implies that it has a discrete spectrum.
However, due to the presence of a first order term, it is not formally self-adjoint (see Appendix \ref{ch:appendix2}) in general.
Nevertheless, it is still true (c.f. Lemma (\ref{PrincipleEigenvalue} in Appendix \ref{ch:appendix2}))
that there exists an eigenvalue $\auto$ with smallest real part. This
eigenvalue is called the {\it principal eigenvalue} and it has the following properties:
\begin{enumerate}
\item It is real.
\item Its eigenspace (the set of smooth real functions $\psi$ on $S$ satisfying $L_{\vec{m}}\psi=\auto\psi$) is one-dimensional.
\item An eigenfunction $\psi$ of $\auto$ vanishes at one point $\p\in S$ if and only if it vanishes everywhere on $S$ (i.e. the principal eigenfunctions do not change sign).
\end{enumerate}
The stability of minimal surfaces could be rewritten in terms of the sign of the principal eigenvalue of
its stability operator. In \cite{AMS2}, \cite{AMS} the following definition of stability of MOTS is put forward.

\begin{defi}\label{defi:stableMOTS}
A MOTS $S\subset\Sigma$ is {\bf stable in $\Sigma$} if the principal eigenvalue $\auto$ of the stability operator
$L_{\vec{m}}$ is non-negative.
$S$ is {\bf strictly stable in $\Sigma$} if $\auto>0$.
\end{defi}

For simplicity, since no confusion will arise, 
we will refer to {\it stability
in $\Sigma$} simply as {\it stability}. 

For stable MOTS, there is no scalar quantity which is non-decreasing for arbitrary variations, like the area for stable minimal
surfaces. However, in the minimal surface case, the formula
\[
<\phi,\psi>_{L^2}\auto=<L_{\vec{m}}^{min}\phi,\psi>_{L^2}=<\phi,L_{\vec{m}}^{min}\psi>_{L^2},
\]
where $\phi$ is a principal eigenfunction of $L_{\vec{m}}^{min}$,
implies that if there exists a positive variation $\psi\vec{m}$ for which
$\delta_{\psi\vec{m}}p\geq 0$, then $\auto\geq 0$ and the minimal surface is stable. A similar result can be proven for MOTS \cite{AMS}:

\begin{proposition}\label{prop:stablevariation}
Let $S\subset \Sigma$ be a MOTS. Then $S$ is stable if and only if there exists a function $\psi\geq 0$, $\psi\not\equiv 0$ on
$S$ such that $\delta_{\psi\vec{m}}\theta^{+}\geq 0$. Furthermore, $S$ is strictly stable if and only if, in addition,
$\delta_{\psi\vec{m}}\theta^{+}\not\equiv 0$.
\end{proposition}


{\bf Remark.} For the case of {\it past} MOTS simply change
$\vec{n}\rightarrow -\vec{n}$, $\vec{l}_{+}\rightarrow -\vec{l}_{-}$, $\vec{l}_{-}\rightarrow -\vec{l}_{+}$,
$s_{A}\rightarrow -s_{A}$ and $\theta^{+}\rightarrow -\theta^{-}$
in equations (\ref{stabilityoperator}), (\ref{Y}), (\ref{raychaudhuri}), (\ref{W}) and, also, in Proposition \ref{prop:stablevariation}. $\hfill \square$ \\

Thus, Proposition \ref{prop:stablevariation} tells us that a (resp. past) MOTS
$S$ is strictly stable
if and only if there exists an outer variation with strictly increasing (resp. decreasing) $\theta^{+}$ (resp. $\theta^{-}$).
This suggests that the presence of
surfaces with negative $\theta^{+}$ (resp. positive $\theta^{-}$) outside $S$ may be related with the stability
property of $S$. This can be made precise by introducing the following notion.

\begin{defi}\label{defi:locallyoutermostMOTS}
A (resp. past) MOTS $S\subset\Sigma$ is {\bf locally outermost} if there exists a two-sided neighbourhood of $S$ on $\Sigma$ whose exterior part
does not contain any (resp. past) weakly outer trapped surface.
\end{defi}

The following proposition gives the relation between these concepts \cite{AMS2}.
\begin{proposition}\mbox{}
\begin{enumerate}
\item A {\it strictly stable} MOTS (or past MOTS) is necessarily {\it locally outermost}.
\item A {\it locally outermost} MOTS (or past MOTS) is necessarily {\it stable}.
\item None of the converses is true in general.
\end{enumerate}
\end{proposition}



\subsection{The trapped region}
\label{ssc:GeometryOfSurfacesTrappedRegion}

In this section we will extend the notion of locally outermost to a {\it global} concept and
state a theorem by Andersson and Metzger \cite{AM}
on the existence, uniqueness and regularity of the outermost MOTS on a spacelike hypersurface $\Sigma$.
We will also see that an analogous result holds
for the outermost generalized apparent horizon (Eichmair, \cite{Eichmair}).
Both results will play a fundamental role throughout this thesis. 




The result by Andersson and Metzger is local in the sense that it works for any
{\it compact} spacelike hypersurface $\Sigma$ with boundary $\bd \Sigma$ as long as the boundary $\bd \Sigma$
splits in two disjoint non-empty components $\bd \Sigma= \bd^{-}\Sigma \cup \bd^{+}\Sigma$.
Neither of these components is assumed to be connected a priori.
Andersson and Metzger deal with surfaces which are {\it bounding with respect to} the boundary
$\bd^{+}\Sigma$ which plays the role of outer untrapped {\it barrier}. 
Both concepts are defined as follows.

\begin{defi}\label{defi:barrier}
Consider a spacelike hypersurface $\Sigma$ possibly with boundary. 
A closed surface $\Sb\subset\Sigma$ is a {\bf barrier with interior
$\Omegab$} if there exists a manifold with boundary $\Omegab$ which is topologically closed and such that
$\bd \Omegab=\Sb\bigcup\underset{a}{\cup}(\bd \Sigma)_{a}$, where $\underset{a}{\cup}(\bd \Sigma)_{a}$ is
a union (possibly empty) of connected components of $\bd \Sigma$.

\end{defi}

{\bf Remark.} For simplicity, when no confusion arises,
we will often refer to a barrier $\Sb$ with interior $\Omegab$ simply as a
{\it barrier} $\Sb$. $\hfill \square$\\

The concept of a barrier will give us a criterion to
define the exterior and the interior of a special type of surfaces called {\it bounding}. More precisely,

\begin{defi}\label{defi:boundingAM}
Consider a spacelike hypersurface $\Sigma$ possibly with boundary with a barrier
$\Sb$ with interior $\Omegab$. 
A surface $S\subset \Omegab\setminus \Sb$ is {\bf bounding with respect to the barrier} $\Sb$ if
there exists a compact manifold $\Omega\subset{\Omegab}$
with boundary such that $\bd\Omega= S\cup\Sb$.
The set $\Omega\setminus S$ will be called the
{\bf exterior} of $S$ in $\Omegab$ and $(\Omegab\setminus\Omega)\cup S$ the {\bf interior} of $S$ in $\Omegab$.
\end{defi}

\begin{figure}[h]
\begin{center}
\psfrag{Sb}{{\color{red}{\small{$S_b$}}}}
\psfrag{pS}{\small{$\bd \Sigma$}}
\psfrag{S1}{\color{blue}{\small{$S_{1}$}}}
\psfrag{S2}{\color{blue}{\small{$S_{2}$}}}
\psfrag{Ob}{\small{$\Omega_b$}}
\psfrag{O1}{{\small{$\Omega_{1}$}}}
\includegraphics[width=6cm]{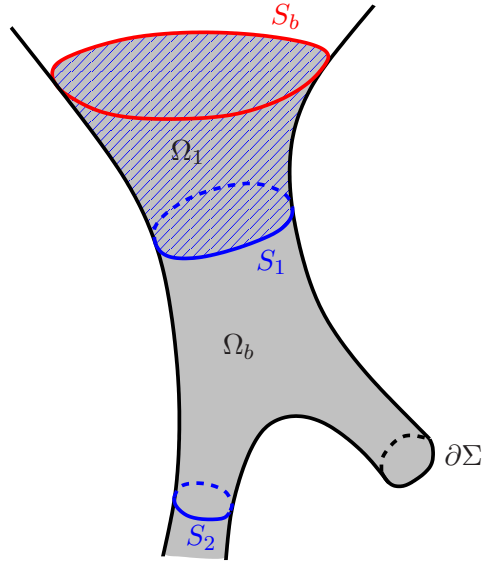}
\caption{In this graphic example, the surface $S_{b}$ (in red) is a barrier
with interior $\Omegab$ (in grey). The surface $S_{1}$ is bounding with respect to
$\Sb$ with $\Omega_{1}$ (the stripped area) being its exterior in $\Omega_{b}$.
The surface $S_{2}$ fails to be bounding with respect to $S_{b}$ because its ``exterior"
would contain $\bd \Sigma$.
}
\label{fig:boundingAM0}
\end{center}
\end{figure}

\begin{figure}[h]
\begin{center}
\psfrag{m}{{{\small{$\vec{m}$}}}}
\psfrag{pS+}{\color{red}{\small{$\bd^+ \Sigma$}}}
\psfrag{pS-}{\small{$\bd^- \Sigma$}}
\psfrag{S1}{\color{blue}{\small{$S_{1}$}}}
\psfrag{S2}{\color{blue}{\small{$S_{2}$}}}
\psfrag{S}{\small{$\Sigma$}}
\psfrag{S3}{\color{blue}{\small{$S_{3}$}}}
\includegraphics[width=10cm]{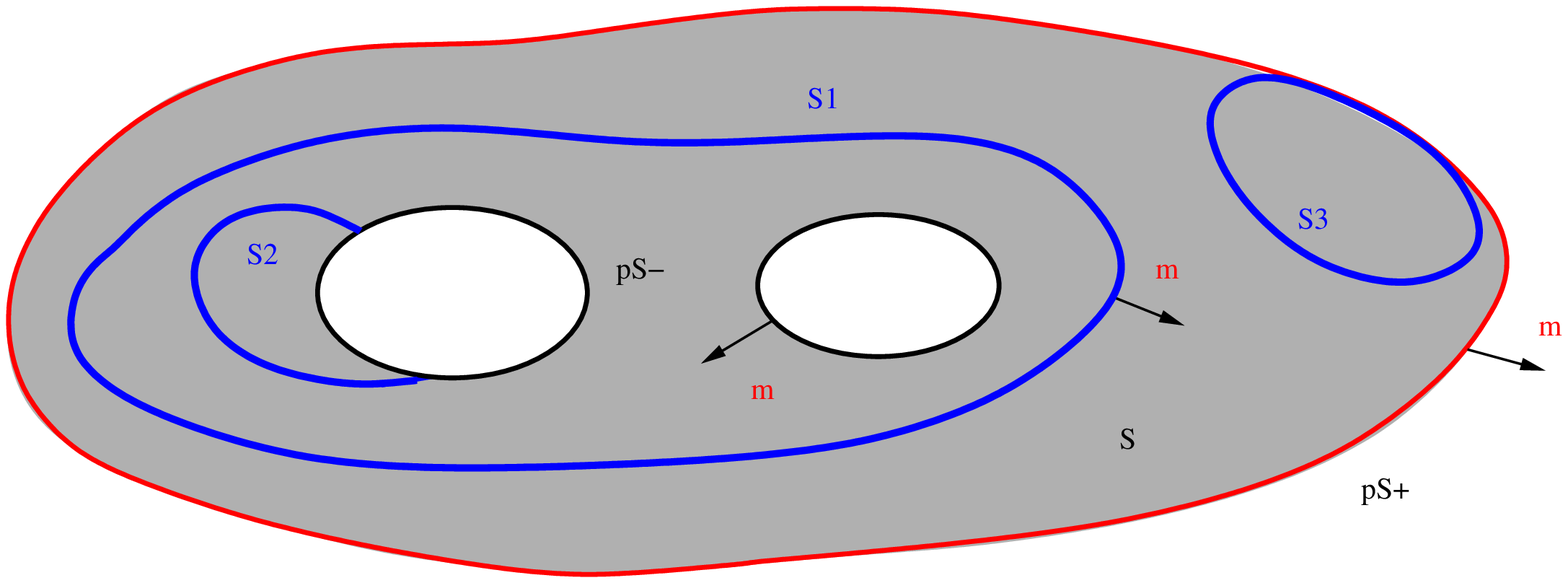}
\caption{A manifold $\Sigma$ with boundary $\bd \Sigma= \bd^{-} \Sigma\cup \bd^{+}\Sigma$.
The boundary $\bd^{+}\Sigma$ is a barrier whose interior coincides with $\Sigma$.
The surface
$S_{1}$ is bounding with respect to $\bd^{+}\Sigma$,
while $S_{2}$ and $S_{3}$ fail to be bounding.
The figure also shows the outer normal $\vec{m}$ as defined in the text.
}
\label{fig:boundingAM}
\end{center}
\end{figure}

{\bf Remark.} Note that a surface $S$ which is bounding with respect to a barrier $S_{b}$ is always disjoint to
$S_b$ and that its
exterior is always not empty.
Again, for simplicity and when no confusion arises, we will often refer to a surface which is
bounding with respect
a barrier simply as a {\it bounding surface}.
Notice that, in the topology of $\Omegab$, the exterior of a bounding surface $S$ in $\Omegab$ is
topologically open (because for every point
$\p\in \bd\Omegab$ there exists an open set $U\subset \Omegab$ such that $\p\in U$), while its interior is topologically closed. For graphic examples of surfaces
which are bounding with respect to a barrier see figures \ref{fig:boundingAM0} and \ref{fig:boundingAM}. $\hfill \square$ \\


The concept of bounding surface allows for a meaningful definition of
{\it outer null direction}. 
For that, define the vector $\vec{m}$ as the unit vector normal to $S$ in $\Sigma$ which points into the
exterior of $S$ in $\Omegab$. For $\Sb$, $\vec{m}$ will be taken to point outside of $\Omegab$.
Then, we will select the outer and the inner null vectors, $\vec{l}_{+}$ and $\vec{l}_{-}$ as those
null vectors orthogonal to $S$ or $\Sb$ which satisfy equations (\ref{l+mn}) and (\ref{l-mn}), respectively.

\begin{defi}\label{defi:encloses}
Given two surfaces $S_1$ and $S_{2}$ which are bounding with respect to a barrier $\Sb$,
we will say that $S_1$ encloses $S_2$ if the exterior of $S_2$ contains the exterior
of $S_1$.
\end{defi}


\begin{defi}\label{defi:outermostsigmatilde}
A (past) MOTS $S\subset \Sigma$ which is bounding with respect to a barrier $\Sb$ is {\bf outermost}
if there is no other (past) weakly outer trapped surface
in $\Sigma$ which is bounding with respect to $\Sb$ and enclosing $S$.
\end{defi}

Since bounding surfaces split $\Omegab$ into an exterior and an interior region,
it is natural to consider the points inside a
bounding weakly outer trapped surface $S$ as ``trapped points".
The region containing trapped points is called {\it weakly outer trapped region}
and will be essential for the formulation of the result by Andersson and Metzger. More precisely,

\begin{defi}\label{defi:trappedregion}
Consider a spacelike hypersurface containing a barrier $\Sb$ with interior $\Omegab$.
The {\bf weakly outer trapped region} $T^{+}$ of $\Omegab$ is the union of the interiors of all bounding weakly outer
trapped surfaces in $\Omegab$.
\end{defi}

Analogously,
\begin{defi}\label{defi:pasttrappedregion}
The {\bf past weakly outer trapped region} $T^{-}$ of $\Omegab$ is the union of the interiors of all bounding past weakly outer
trapped surfaces in $\Omegab$.
\end{defi}

The fundamental result by Andersson and Metzger, which will be an important tool in this thesis, reads as follows.

\begin{thr}[Andersson, Metzger, 2009 \cite{AM}]\label{thr:AM}
Consider a compact spacelike hypersurface $\Sigmatilde$ with boundary $\bd \Sigmatilde$.
Assume that the boundary can be split in
two non-empty disjoint components $\bd \Sigmatilde= \bd^{-}\Sigmatilde \cup \bd^{+}\Sigmatilde$
(neither of which are necessarily connected) and
take $\bd^{+}\Sigmatilde$ as a barrier with interior $\Sigmatilde$.
Suppose that $\theta^{+}[\bd^{-}\Sigmatilde]\leq 0$ and $\theta^{+}[\bd^{+}\Sigmatilde]>0$
(with respect to the outer normals defined above).
Then the topological boundary $\tbd T^{+}$ of the weakly outer trapped region of $\Sigmatilde$
is a smooth MOTS which is bounding with respect to $\bd^{+}\Sigmatilde$ and stable.
\end{thr}

{\bf Remark.}
Since no bounding MOTS can penetrate into the exterior of
$\tbd T^{+}$, by definition, this theorem shows the existence,
uniqueness and smoothness of the outermost bounding MOTS in a compact hypersurface.
Note also that another consequence of this result is the fact that
the set $T^{+}$ is topologically closed (because it is the interior of the bounding surface $\tbd T^{+}$). $\hfill \square$ \\

The proof of this theorem uses the Gauss-Bonnet Theorem in several places and, therefore,
this result is valid only in (3+1) dimensions.

If we reverse the time orientation of the spacetime, an analogous result for the topological
boundary of the past weakly outer trapped region $T^{-}$
follows.
Indeed, if the hypotheses on the sign of the outer null expansion of the components of $\bd \Sigmatilde$
are replaced by
$\theta^{-}[\bd^{-}\Sigmatilde]\geq 0$ and $\theta^{-}[\bd^{+}\Sigmatilde]<0$ then the conclusion is that
$\tbd T^{-}$ is a smooth past MOTS which is bounding with respect to $\bd^{+}\Sigmatilde$ and stable.
\\


As we mentioned before, a similar result for the existence of the outermost generalized apparent horizon also exists.
It has been recently obtained by Eichmair \cite{Eichmair}.
\begin{thr}[Eichmair, 2009 \cite{Eichmair}]\label{thr:Eichmair}
Let $(\Sigmatilde, g, K)$ be a compact n-dimensional
spacelike hypersurface in an (n+1)-dimensional spacetime,
with $3\leq n\leq 7$ and boundary
$\bd \Sigmatilde$.
Assume that the boundary can be split in
two non-empty disjoint components $\bd \Sigmatilde= \bd^{-}\Sigmatilde \cup \bd^{+}\Sigmatilde$
(neither of which are necessarily connected) and take $\bd^{+}\Sigmatilde$ as a barrier with interior
$\Sigmatilde$. Suppose that the inner boundary $\bd^{-}\Sigmatilde$ is a generalized trapped surface,
and the outer boundary
satisfies $p>|q|$ with respect to the outer normals defined above.

Then there exists a unique $C^{2,\alpha}$ (i.e. belonging to the H\"older space $C^{2,\alpha}$,
with $0<\alpha\leq 1$, see Appendix \ref{ch:appendix2}) generalized apparent horizon $S$ which is
bounding with respect to $\bd^{+}\Sigmatilde$ and outermost (i.e. there is
no other bounding generalized trapped surface in $\Sigmatilde$ enclosing $S$). Moreover,
$S$ has smaller area than any other surface enclosing it.
\end{thr}

The proof of this result does not use the
Gauss-Bonnet theorem or any other specific property of $3$-dimensional spaces, so it not restricted to
(3+1) dimensions. However, it is based on regularity of
minimal surfaces, which implies that the dimension of $\Sigmatilde$ must be at most seven
(in higher dimensions minimal hypersurfaces need not be regular everywhere, see e.g. \cite{Giusti}).

The area minimizing property of the outermost bounding generalized apparent horizon makes this type of surfaces
potentially interesting for the Penrose inequality, as we will discuss in the next section. 

\setcounter{equation}{0}
\section{The Penrose inequality}
\label{sc:PenroseInequality}

The Penrose inequality involves the concept of the total ADM mass of a spacetime, so
we start with a brief discussion about mass in
General Relativity.

The notion of {\it energy} in General Relativity is not as clear as in other physical theories. 
The energy-momentum tensor $T_{\mu\nu}$ represents the matter contents of a spacetime
and therefore should contribute to the total energy of a spacetime. 
However, the {\it gravitational field}, represented by the metric tensor $\gM$, must also contribute to
the total energy of the spacetime.
In agreement with the Newtonian limit, a suitable {\it gravitational energy density} should be
an expression quadratic in the first derivatives of the metric $\gM$. However, since at any point we can make
the metric to be Minkowskian and the Christoffel symbols to vanish, there is no non-trivial scalar object
constructed from the metric and its first derivatives alone. Therefore, a natural notion of energy density in
General Relativity does not exist. The same problem is also found in other geometric theories of gravity.
Nevertheless, there does exist a useful notion of the {\it total energy} in 
the so-called asymptotically flat spacetimes.

The term {\it asymptotic flatness} was introduced in General Relativity to express the idea of a spacetime corresponding to an isolated system.
It involves restrictions on the spacetime ``far away" form the sources.
There are several notions of asymptotic flatness according to the type of infinity considered (see e.g. Chapter 11.1 of \cite{Wald}),
namely limits along null directions (null infinity) or limits along spacelike directions (spacelike infinity).
The idea is to define the mass as integrals in the asymptotic region where the gravitational field is sufficiently weak so that
integrals become meaningful (i.e. independent of the coordinate system).
According to the type of infinity considered there are
two different concepts: the {\it Bondi energy-momentum} where
the integral is taken at null infinity and the {\it ADM energy-momentum} where the integral is taken at spatial infinity.
Both are vectors in a suitable four dimensional vector space and
transform as a Lorentz vector under suitable transformations. Moreover, the Lorentz length of this vector is either a conserved quantity upon evolution (ADM) or monotonically decreasing in advanced time (Bondi). 
An interesting and more precise discussion about the definitions of both Bondi and ADM energy-momentum tensors can be found in
Chapter 11.2 of \cite{Wald}.
Because of its relation with the Penrose inequality we are specially interested in the ADM energy-momentum.
To make these concepts precise we need to define first {\it asymptotic flatness} for spacelike hypersurfaces.

\begin{defi}\label{defi:asymptoticallyflatend}
An {\bf asymptotically flat end} of a spacelike hypersurface $\id$ is a subset $\Sigma_{0}^{\infty}\subset\Sigma$ which
is diffeomorphic to $\mathbb{R}^3 \setminus
\overline{B_{R}}$, where $B_{R}$ is an open ball of radius $R$. Moreover, in the Cartesian coordinates $\{ x^i \}$
induced by the diffeomorphism, the following decay holds
\begin{equation}\label{asymptoticallyflatend}
g_{ij}-\delta_{ij}=O^{(2)}(1/r),\quad\quad K_{ij}=O^{(2)}(1/r^{2}),
\end{equation}
where $r=|x|=\sqrt{x^{i}x^{j}\delta_{ij}}$.
\end{defi}
Here, a function $f(x^i)$ is said to be $O^{(k)}(r^n), k\in\mathbb{N}\cup \{0\}$ if $f(x^i)=O(r^n)$, $\partial_{j}f(x^{i})=O(r^{n-1})$ and
so on for all derivatives up to and including the $k$-th ones.

\begin{defi}\label{defi:asymptoticallyflat}
A spacelike hypersurface $\id$, possibly with boundary, is {\bf asymptotically flat} if $\Sigma=\mathcal{K}\cup \Sigma^{\infty}$, where $\mathcal{K}$ is a compact set and
$\Sigma^{\infty}=\underset{a}{\cup}\Sigma_{a}^{\infty}$ is a finite union of asymptotically flat ends $\Sigma_{a}^{\infty}$.
\end{defi}

\begin{defi}
Consider a spacelike hypersurface $\id$ with a selected asymptotically flat end
$\Sigma_{0}^{\infty}$. Then,
the ADM energy-momentum ${\bf P}_{\scriptscriptstyle ADM}$ associated with $\Sigma_{0}^{\infty}$ is
defined as the spacetime vector with components
\begin{eqnarray}
{P_{\scriptscriptstyle ADM}}_0=E_{\scriptscriptstyle ADM}\equiv \underset{r\rightarrow \infty}{lim} \frac{1}{16\pi} \overset{3}{\underset{j=1}{\sum}} \int_{S_{r}}\left( \partial_{j}g_{ij}-\partial_{i}g_{jj} \right)dS^{i},
\label{ADMenergy}\\
{P_{\scriptscriptstyle ADM}}_{i}= {p_{\scriptscriptstyle ADM}}_{i}\equiv \underset{r\rightarrow \infty}{lim} \frac{1}{8\pi}\int_{S_{r}}\left( K_{ij}-g_{ij}\tr{K} \right)dS^{j},\label{ADMmomentum}
\end{eqnarray}
where $\{x^i\}$ are the Cartesian coordinates induced by the diffeomorphism which defines the asymptotically flat end,
$S_r$ is the surface at constant $r$ and $dS^{i}=m^{i}dS$ with $\vec{m}$ being the outward unit normal and $dS$ the area element.

The quantity $E_{\scriptscriptstyle ADM}$ is called the ADM energy while ${\bf p}_{\scriptscriptstyle ADM}$ the ADM spatial momentum.
\end{defi}


\begin{defi}\label{defi:ADMmass}
The ADM mass is defined as
\[
M_{\scriptscriptstyle ADM}=\sqrt{E_{\scriptscriptstyle ADM}^{2}-\delta^{ij}{P_{\scriptscriptstyle ADM}}_{i}{P_{\scriptscriptstyle ADM}}_{j}}.
\]
\end{defi}

A priori, these definitions depend on the choice of the coordinates $\{x^{i}\}$.
However, the decay in $g$ and $K$ at infinity implies that
${\bf P}_{\scriptscriptstyle ADM}$ is indeed a geometric quantity provided $\GM_{\mu\nu}n^{\mu}$ decays as
$1/r^{4}$ at infinity \cite{citeADM}. The notion of ADM mass is in fact independent of the
coordinates as long as the decay
(\ref{asymptoticallyflatend}) is replaced by
\begin{equation}
g_{ij}-\delta_{ij}=O^{(2)}(1/r^{\alpha}),\qquad\quad K_{ij}=O^{(1)}(1/r^{1+\alpha}),
\end{equation}
with $\alpha>\frac{1}{2}$ \cite{Bartnik1}.

A fundamental property of the ADM energy-momentum is its causal character. The
Positive Mass Theorem (PMT) of
Schoen and Yau \cite{SY} (also proven by Witten \cite{Witten} using spinors) establishes that the ADM energy is non-negative and the ADM mass is real
(c.f. Section 8.2 of \cite{Strauman} for further details). More precisely,

\begin{thr}[Positive mass theorem (PMT), Schoen, Yau, 1981]\label{thr:PMT0}
Consider an asymptotically flat spacelike hypersurface $\id$ without boundary
satisfying the DEC. Then the total ADM energy-momentum $\vec{P}_{\scriptscriptstyle ADM}$
is a future directed causal vector. Furthermore, $\vec{P}_{\scriptscriptstyle ADM}=0$ if and only if $(\Sigma,g,K)$
is a slice of the Minkowski spacetime.
\end{thr}

The global conditions required for the PMT were relaxed in \cite{CBartnik} where
$\Sigma$ was allowed to be complete and contain an asymptotically flat end 
instead of being necessarily asymptotically flat (see Theorem \ref{thr:PMT1} below). 
The PMT has also been extended to other situations of interest. Firstly, it holds
for spacelike hypersurfaces admitting corners on a surface,
provided the mean curvatures of the surface from
one side and the other satisfy the right inequality \cite{Miao2}. It has also been proved for spacelike hypersurfaces
{\it with boundary} provided this boundary is
composed by either future or past weakly outer trapped surfaces \cite{GHHP}.
Since future weakly outer trapped surfaces are intimately related
with the existence of black holes (as we have already pointed out above), this
type of PMT is usually referred to as
{\it PMT for black holes}. Having introduced these notions we can now describe the Penrose inequality.

During the seventies, Penrose \cite{Penrose1973} conjectured that the total ADM mass of a spacetime containing a black
hole that settles down to a stationary state must satisfy the inequality
\begin{equation}\label{penrose0}
M_{\scriptscriptstyle ADM}\geq \sqrt{\frac{|{\mathscr{H}}|}{16\pi}},
\end{equation}
where $|{\mathscr{H}}|$ is the area of the event horizon at one instant of time. Moreover, equality happens if and only if
the spacetime is the Schwarzschild spacetime. The plausibility argument by Penrose goes as follows \cite{Penrose1973}. Assume a spacetime $(M,\gM)$ which is globally well-behaved in
the sense of being strongly asymptotically predictable and admitting a
complete future null infinity $\mathscr{I}^{+}$ (see \cite{Wald} for definitions).
Suppose that $M$ contains a non-empty black hole region. 
The black hole event horizon ${\mathcal{H_{B}}}$ 
is a null hypersurface at least Lipschitz continuous. 
Next, consider a spacelike Cauchy hypersurface $\Sigma\subset M$ (see e.g. Chapter 8 of \cite{Wald} for the definition of a Cauchy hypersurface)
with ADM mass $M_{\scriptscriptstyle ADM}$. Clearly
${\mathcal{H_{B}}}$ and $\Sigma$ intersect in a two-dimensional Lipschitz manifold. This represents the event horizon
at one instant of time.
Let us denote by $\mathscr{H}$ this intersection and by $|\mathscr{H}|$ its area (the manifold is almost everywhere $C^{1}$ so the
area makes sense).
Consider now any other cut $\mathscr{H}_1$ lying
in the causal future of $\mathscr{H}$. The black hole area theorem \cite{H1}, \cite{H2}, \cite{CDGH} states that
$|\mathscr{H}_1|\geq |\mathscr{H}|$ provided the NEC holds. Physically, it is reasonable to expect that the spacetime
settles down to some vacuum 
equilibrium configuration (if an electromagnetic field is present, the conclusions would be essentially the same).
Then, the uniqueness theorems for stationary black holes (which hold under suitable assumptions \cite{CL-C}, \cite{L-C})
imply that the
spacetime must approach the Kerr spacetime. In the Kerr spacetime the area of any cut of the event horizon $\mathscr{H}_{Kerr}$
takes the value $|\mathscr{H}_{Kerr}|=8\pi M_{Kerr}\left( M_{Kerr}+\sqrt{{M_{Kerr}}^{2}-L_{Kerr}^{2}/{M_{Kerr}}^{2}} \right)$ where $M_{Kerr}$ and $L_{Kerr}$ are respectively the total mass and the
total angular momentum of the Kerr spacetime (the angular momentum can be defined also as a suitable
integral at infinity). This means that $M_{Kerr}$ is the asymptotic value of the Bondi mass along the future null infinite
$\mathscr{I}^{+}$. 
Assuming that the Bondi mass
tends to the $M_{\scriptscriptstyle ADM}$ of the initial slice, inequality (\ref{penrose0}) follows because
the Bondi mass cannot increase along the evolution. Moreover, equality holds if and only if $\Sigma$ is a slice
of the Kruskal extension of the Schwarzschild spacetime.

It is important to remark than inequality (\ref{penrose0}) is global in the sense that, in order to locate the cut $\mathscr{H}$, it is necessary to know
the global structure of the spacetime.
Penrose proposed to estimate the area $|\mathscr{H}|$ from below in terms of the area of certain surfaces which
can be defined independently of the future evolution of the spacetime.
The validity of these estimates relies on the validity of the cosmic censorship. 
These types of inequalities
are collectively called {\it Penrose inequalities} and they are interesting for several reasons. First of all, they would provide a strengthening of the
PMT. Moreover, 
they would also give indirect support to the validity of cosmic censorship, which is a basic ingredient in their derivation.

There are several versions of the Penrose inequality. Typically one considers closed
surfaces $S$ embedded in a spacelike hypersurface with a selected asymptotically flat end
$\Sigma_{0}^{\infty}$ which are {\it bounding} with respect to a suitable large sphere
in $\Sigma_{0}^{\infty}$. This leads to the following definition:

\begin{defi}\label{defi:bounding}
Consider a spacelike hypersurface $\id$ possibly with boundary
with a selected asymptotically flat end $\Sigma_{0}^{\infty}$.
Take a sphere $\Sb\subset\Sigma_{0}^{\infty}$ with $r=r_{0}=const$ large enough so that the spheres with
$r\geq r_{0}$ are outer untrapped with respect to the direction pointing into the asymptotic region in
$\Sigma_{0}^{\infty}$. Let $\Omegab=\Sigma\setminus\{r>r_{0}\}$, which is obviously topologically
closed and satisfies $\Sb\subset\bd\Omegab$. Then $\Sb$ is a barrier with interior $\Omegab$.
A surface $S\subset  \Sigma$ will be called {\bf bounding} if it is bounding with respect to $\Sb$.
\end{defi}

{\bf Remark 1.} It is well-known that on an asymptotically flat end $\Sigma_{0}^{\infty}$, the surfaces at constant
$r$ are, for large enough $r$, outer untrapped.
Essentially, this definition establishes a specific
form of selecting the barrier in hypersurfaces containing a selected asymptotically flat end. $\hfill \square$ \\

{\bf Remark 2.} Obviously, the definitions of exterior and interior of a bounding surface
(Definition \ref{defi:boundingAM}),
enclosing (Definition \ref{defi:encloses}),
outermost (Definition \ref{defi:outermostsigmatilde}) and $T^{\pm}$
(Definitions \ref{defi:trappedregion} and \ref{defi:pasttrappedregion}), given in the previous section,
are applicable in the asymptotically
flat setting. Moreover, since $r_{0}$ can be taken as large as desired,
the specific choice of $S_b$ and $\Omegab$ is not relevant for the definition of bounding (once the asymptotically flat end has been selected). 
Because of that, when considering asymptotically flat ends, we will refer to the
exterior of $S$ in $\Omegab$ as the {\it exterior of $S$ in $\Sigma$}. $\hfill \square$ \\

\begin{figure}[h]
\begin{center}
\psfrag{Sb}{\color{red}{$S_{b}$}}
\psfrag{Sigma}{$\Sigma$}
\psfrag{S1}{\color{blue}{$S_{1}$}}
\psfrag{Omega}{$\Omega_b$}
\psfrag{Sigma0}{$\Sigma_{0}^{\infty}$}
\psfrag{S2}{\color{blue}{$S_{2}$}}
\includegraphics[width=7cm]{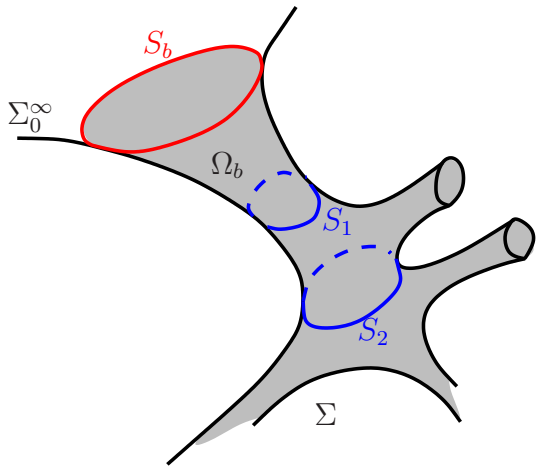}
\caption{The hypersurface $\Sigma$ possesses an asymptotically flat end $\Sigma_{0}^{\infty}$
but also other types of ends and boundaries. The surface $S_b$, which represents a large sphere in $\Sigma_{0}^{\infty}$
and is outer untrapped, is a barrier with interior $\Omega_b$ (in grey). The surface
$S_1$ is bounding with respect to $S_{b}$ (c.f. Definition \ref{defi:boundingAM}) and therefore is bounding.
The surface $S_2$ fails to be bounding (c.f. Figure \ref{fig:boundingAM0}).
}
\label{fig:bounding1}
\end{center}
\end{figure}


The standard version of the Penrose inequality reads
\begin{equation}\label{penrose1}
M_{\scriptscriptstyle ADM}\geq \sqrt{\frac{A_{{min}}(\tbd T^{+})}{16\pi}},
\end{equation}
where $A_{min}(\tbd T^{+})$ is the minimal area necessary to enclose
$\tbd T^{+}$.
This inequality (\ref{penrose1}) is a consequence of the heuristic argument outlined before because 
(under cosmic censorship) 
$\mathscr{H}$ encloses $\tbd T^{+}$ 
The minimal area enclosure of $\tbd T^+$ needs to be taken because
$\mathscr{H}$ could still have less area than $\tbd T^{+}$ \cite{Horowitz}.

By reversing the time orientation,
the same argument yields
(\ref{penrose1}) with $\tbd T^{+}$
replaced by $\tbd T^{-}$. In  general,
neither $\tbd T^{+}$ encloses $\tbd T^{-}$ nor
vice versa. In the case that $K_{ij}=0$, these inequalities simplify because $T^{+} = T^{-}$ and $\tbd T^{+}$
is the outermost minimal surface (i.e. a minimal surface enclosing
any other bounding minimal surface
in $\Sigma$) and, hence, its own minimal area enclosure.
The inequality in this case is called
{\it Riemannian Penrose inequality} and it has been proven for connected
$\tbd T^+$  in \cite{HI} and in the general case in \cite{Bray}
using a different method.
In the non-time-symmetric case, (\ref{penrose1}) is not invariant
under time reversals. Moreover, the minimal area
enclosure of a given surface $S$ can be a rather complicated object
typically consisting of portions of $S$ together with portions of minimal surfaces
outside of $S$. This complicates the problem
substantially. This has led several authors to
propose simpler looking versions of the inequality, even if they
are not directly
supported by cosmic censorship. Two of such extensions are
\begin{eqnarray}
M_{\scriptscriptstyle ADM} \geq \sqrt{\frac{A_{{min}}( \tbd (T^{+} \cup T^{-} ))}{16\pi}},
\quad M_{\scriptscriptstyle ADM} \geq \sqrt{\frac{|\tbd (T^{+} \cup T^{-} )|}{16\pi}},
\label{penrose2}
\end{eqnarray}
(see e.g. \cite{Karkowski-Malec2005}).
These inequalities are immediately stronger
than (\ref{penrose1})
and have the advantage of being invariant under time
reversals. The second inequality avoids even the use of minimal area enclosures. Neither
version  is supported by
cosmic censorship and at present there is little evidence for their validity.
However, both reduce to the standard version in the Riemannian case and
both hold in spherical symmetry. No counterexamples
are known either. It would be interesting to have either stronger
support for them, or else to find a
counterexample.

Recently, Bray and Khuri proposed \cite{BK} a new method to
approach the general (i.e. non time-symmetric) Penrose inequality. The
basic idea was to modify the Jang equation \cite{Jang}, \cite{SY}
so that
the product manifold $\Sigma \times \mathbb{R}$ used to construct the graphs which define
the Jang equation is endowed with a
warped type metric of the form $-\varphi^2 dt^2 + g$ instead of the product metric.
Their aim was to reduce the general Penrose inequality to
the Riemannian Penrose inequality on the graph manifold. A discussion on the
type of divergences that could possibly occur for the generalized
Jang equation led the authors
to consider a new type of trapped surfaces which they called
{\bf generalized trapped surfaces} and {\bf generalized apparent horizons} (defined in Section \ref{ssc:GeometryOfSurfacesDefinitions}).
This type of surfaces have very interesting properties.
The most notable one is given by Theorem \ref{thr:Eichmair} \cite{Eichmair} which
guarantees the existence, uniqueness and $C^{2,\alpha}$-regularity of the outermost generalized apparent horizon
$S_{out}$.
The Penrose inequality proposed by these authors reads
\begin{eqnarray}\label{penroseBK}
M_{\scriptscriptstyle ADM}\geq\sqrt{\frac{|S_{out}|}{16\pi}},
\end{eqnarray}
with equality only if the spacetime is Schwarzschild.
This inequality has several
remarkable properties that makes it very appealing \cite{BK}.
First of all, the definition
of generalized apparent horizon, and hence the corresponding Penrose inequality,
is insensitive to time reversals. Moreover, there is no need of taking
the minimal area enclosure of $S_{out}$, as this surface has less area than any of
its enclosures (c.f. Theorem \ref{thr:Eichmair}). Since MOTS are automatically generalized
trapped surfaces, $S_{out}$ encloses the outermost MOTS $\tbd T^{+}$. Thus, (\ref{penroseBK})
is stronger than (\ref{penrose1}) and its proof would also
establish the standard version of the Penrose inequality. Moreover, Khuri has proven
\cite{Khuri3}
that no generalized trapped surfaces exist in Minkowski, which is a necessary condition
for the validity of (\ref{penroseBK}). Another interesting property of this version,
and one of its motivations discussed in \cite{BK}, is that the
equality case in (\ref{penroseBK})
covers a larger number of slices of Kruskal than the equality case in (\ref{penrose1}).
Recall that the rigidity statement of any
version of the Penrose inequality asserts that equality implies that $(\Sigma,g,K)$ is
a hypersurface of Kruskal. However, {\it which} slices of Kruskal satisfy the equality case may depend
on the version under consideration. The more slices having this property, the more accurate the version can
be considered. For any slice $\Sigma$ of Kruskal we can define $\Sigma^{+}$ as
the intersection of $\Sigma$ with the domain of outer communications. Bray and Khuri noticed that
whenever $\tbd \Sigma^{+}$ intersects both the black hole and the white hole event horizons,
then the standard version (\ref{penrose1}) gives, in fact, a strict inequality. Although (\ref{penroseBK})
does not give equality for all slices of Kruskal, it does so in all cases where the boundary
of $\Sigma^{+}$ is a $C^{2,\alpha}$ surface (provided this boundary is the outermost generalized apparent horizon).
It follows that version (\ref{penroseBK}) contains more
cases of equality than (\ref{penrose1}) and is therefore more accurate. It should be stressed that the second inequality in
(\ref{penrose2}) gives equality for
{\it all} slices of Kruskal, so in this sense it would be optimal.

Despite its appealing properties, (\ref{penroseBK}) is {\it not} directly supported
by cosmic censorship. The reason is that the outermost generalized apparent horizon need not always
lie inside the event horizon. A simple example \cite{Mars2}
is given by a slice $\Sigma$ of Kruskal such that
$\tbd T^{+}$ (which corresponds to the intersection of $\Sigma$ with the black hole event
horizon) and $\tbd T^{-}$ (the intersection $\Sigma$ with the white hole horizon)
meet transversally. Since both surfaces are generalized trapped surfaces,
Theorem \ref{thr:Eichmair}
implies that there must exist a unique $C^{2,\alpha}$ outermost
generalized apparent horizon enclosing
both. This surface must therefore penetrate into the exterior region $\Sigma^+$ somewhere, as claimed.
We will return to the issue of the Penrose inequality in Chapter \ref{ch:Article3}, where
we will find a counterexample of (\ref{penroseBK}) precisely by studying the outermost generalized apparent horizon in this type of slices
in the Kruskal spacetime. 
For further information about the present status of the Penrose inequality, see \cite{Mars2}.


\setcounter{equation}{0}
\section{Uniqueness of Black Holes}
\label{sc:UniquenessOfBlackHoles}

According to cosmic censorship, any gravitational collapse that settles down to a stationary state should approach a stationary
black hole.
The {\it black hole uniqueness theorems} aim to 
classify all the stationary black hole solutions of Einstein equations.
In this section we will first summarize briefly the status of stationary black hole uniqueness theorems.
We will also describe in some detail a powerful method (the so-called {\it doubling method} of
Bunting and Masood-ul-Alam) to prove uniqueness for {\it static} black holes which will be
essential in Chapter \ref{ch:Article4}.

In the late sixties and early seventies the properties of equilibrium states of black holes were extensively studied by many
theoretical physicists interested in the gravitational collapse process.
The first uniqueness theorem for black holes was found by W. Israel in 1967 \cite{Israel}, who found the very surprising result that a
{static}, topologically spherical vacuum black hole is described by the Schwarzschild solution.
In the following years, several works (\cite{MzH}, \cite{Robinson}, \cite{BMuA}) established that the Schwarzschild solution indeed exhausts
the class of static vacuum black holes with non-degenerate horizons. The method of the proofs in \cite{Israel}, \cite{MzH}, \cite{Robinson} consisted in constructing two
integral identities which were used to investigate the geometric properties of the level surfaces of the norm of the static Killing.
This method proved uniqueness under the assumption of connectedness and non-degeneracy of the event horizon. The
hypothesis on the connectedness of the horizon was dropped by Bunting and Masood-ul-Alam \cite{BMuA} who devised a new method
based on finding a suitable conformal rescalling which allowed using the rigidity part of the PMT to conclude uniqueness.
This method, known as the {\it doubling method} is, still nowadays, the most powerful method to prove uniqueness of black holes in
the static case. 
Finally, the hypothesis on the non-degeneracy of the event horizon was dropped by Chru\'sciel \cite{C} in 1999 who
applied the doubling
method across the non-degenerate components and applied the PMT for complete manifolds with one asymptotically flat end
(Theorem \ref{thr:PMT1} below) to conclude uniqueness (the Bunting and Masood-ul-Alam conformal rescalling transforms the degenerate components into cylindrical ends).
The developments in the uniqueness of static electro-vacuum black holes go in parallel to the
developments in the vacuum case. 
Some remarkable works which played an important role in the general proof of the uniqueness of static electro-vacuum black holes are \cite{Israel2}, \cite{MzH2}, \cite{Simon},
\cite{Ruback}, \cite{Simon2}, \cite{MuA}, \cite{C2}, \cite{CT}.
Uniqueness of static black holes using the doubling method has also been proved for other matter models, as for instance
the Einstein-Maxwell-dilaton model \cite{MuA2}, \cite{MSimon}.

During the late sixties, uniqueness of {\it stationary} black holes also started to take shape. 
In fact, the works of Israel, Hawking, Carter and Robinson, between 1967 and 1975,
gave an almost complete proof that the Kerr black hole was the
only possible stationary vacuum black hole. The first step was given by Hawking (see \cite{HE}) who proved that
the intersection
of the event horizon with a Cauchy hypersurface has $\mathbb{S}^2$-topology. The next step, also due to Hawking
\cite{HE} was the demonstration of the so-called
Hawking Rigidity Theorem, which states that
a stationary black hole must be static or axisymmetric. Finally, the work of Carter \cite{Carter} and
Robinson \cite{Robinson2} succeeded in proving that
the Kerr solutions are the only possible stationary axisymmetric black holes.
Nevertheless, due to the fact that the Hawking Rigidity Theorem requires analyticity of all objects involved,
uniqueness was proven only for analytic spacetimes.
The recent work \cite{CL-C} by Chru\'sciel and Lopes Costa
has contributed substantially to reduce the hypotheses and to fill several gaps
present in the previous arguments. Similarly, uniqueness of stationary electro-vacuum black holes has been
proven for analytic spacetimes.
Some remarkable works for the stationary electro-vacuum case are \cite{Carter2}, \cite{Mazur}
and, more recently, \cite{L-C}, where weaker
hypotheses 
are assumed for the proof.
Uniqueness of stationary and axisymmetric black holes has also been proven for non-linear $\sigma$-models in \cite{BMG}.
The Hawking Rigidity Theorem has not been generalized to non-linear $\sigma$-models and, hence, axisymmetry
is required in this case.
It is also worth to remark that, in the case of matter models modeled with Yang-Mills fields, uniqueness of
stationary black holes is not true
in general and counterexamples exist \cite{BartnikMcKinnon}.

In this thesis we will be interested in uniqueness theorems for static {\it quasi-local} black holes and, particularly, in the
doubling method of Bunting and Masood-ul-Alam. In the remainder of this chapter, we will describe this method in some detail by giving a sketch
of the proof of the uniqueness theorem for static electro-vacuum black holes. 

\subsection{Example: Uniqueness for electro-vacuum static black holes}
\label{ssc:electro-vacuum}

Let us start with some definitions.
An electro-vacuum solution of the Einstein field equations is a triad $(M,\gM,{\bf F})$, where ${\bf F}$ is the source-free electromagnetic tensor, i.e. a
2-form satisfying the Maxwell equations which no sources, i.e.
\begin{eqnarray*}
{\nablaM}^{\mu}F_{\mu\nu}=0, \\
\nablaM_{[\alpha}F_{\mu\nu]}=0,
\end{eqnarray*}
and $(M,\gM)$ is the spacetime satisfying the Einstein equations with energy-momentum tensor
\[
T_{\mu\nu}=\frac{1}{4\pi} \left(F_{\mu\alpha}{F_{\nu}}^{\alpha}-\frac14 F_{\alpha\beta}F^{\alpha\beta}\gM_{\mu\nu}\right).
\]

We call a stationary electro-vacuum spacetime an electro-vacuum spacetime admitting a stationary Killing vector field $\vec{\xi}$,
satisfying $\mathcal{L}_{\vec{\xi}} F_{\mu\nu}=0$. Let us define the electric and magnetic fields with respect to $\vec{\xi}$ as
\begin{eqnarray*}
E_{\mu}&=&-F_{\mu\nu}\xi^{\nu},\\
B_{\mu}&=&(*F)_{\mu\nu}\xi^{\nu},
\end{eqnarray*}
respectively. Here, $*{\bf F}$ denotes the Hodge dual of ${\bf F}$ defined as
\[
(*F)_{\mu\nu}=\frac 12 \eta^{(4)}_{\mu\nu\alpha\beta}F^{\alpha\beta}.
\]

From the Maxwell equations and $\mathcal{L}_{\vec{\xi}}F_{\mu\nu}=0$ it follows easily that
$d {\bf E}=0$ and $d {\bf B}=0$ which implies that, at least locally, there exist two functions $\phi$ and $\psi$, called the
{\bf electric} and {\bf magnetic
potentials}, so that ${\bf E}=-d \phi$ and ${\bf B}=-d \psi$, respectively. These potentials are defined up to an additive constant and they satisfy
$\vec{\xi}(\phi)=\vec{\xi}(\psi)=0$.

\begin{defi}
A stationary electro-vacuum spacetime $(M,\gM,{\bf F})$ with Killing field $\vec{\xi}$ is said to be {\bf purely electric} with respect to
$\vec{\xi}$ if and only if ${\bf B}=0$.
\end{defi}
For simplicity, we will restrict ourselves to the purely electric case. In fact, the general case can be reduced to the purely electric case
by a transformation called {\it duality rotation} \cite{Heusler2}.

In the static case there exists an important simplification which allows to reduce the formulation of the uniqueness theorem for black holes
in terms of conditions on a spacelike hypersurface instead of conditions on the spacetime.
The fact is that, under suitable circumstances,
the presence of an event horizon in a  static spacetime implies the existence of an asymptotically flat
hypersurface 
with compact topological boundary such that the static Killing field is causal everywhere and
null precisely on the boundary.
Then, the uniqueness theorem for static electro-vacuum black holes can be stated simply as follows.

\begin{thr}[Chru\'sciel, Tod, 2006 \cite{CT}]\label{thr:electrovacuniqueness0}
Let $(M,\gM,F)$ be a static solution of the Einstein-Maxwell equations. Suppose that $M$ contains
a simply connected asymptotically flat hypersurface $\Sigma$ with non-empty topological boundary such that ${\Sigma}$ is the union
of an asymptotically flat end and a compact set, such that:
\begin{itemize}
\item The topological boundary $\tbd \Sigma$ is a compact, 2-dimensional embedded topological submanifold.
\item The static Killing vector field is causal on $\Sigma$ and null only on $\tbd \Sigma$.
\end{itemize}
Then, after performing a duality rotation of the electromagnetic field if necessary:
\begin{itemize}
\item If $\tbd \Sigma$ is connected, then $\Sigma$ is diffeomorphic to $\mathbb{R}^{3}$ minus a ball. Moreover, there exists
a neighbourhood of $\Sigma$ in $M$ which is isometrically diffeomorphic to an open subset of the
Reissner-Nordstr\"om spacetime.

\item If $\tbd \Sigma$ is not connected, then $\Sigma$ is diffeomorphic to $\mathbb{R}^{3}$ minus a finite
union of disjoint balls and there exists a neighborhood of $\Sigma$ in $M$ which is isometrically
diffeomorphic to an open subset of the standard Majumdar-Papapetrou spacetime.
\end{itemize}
\end{thr}

{\bf Remark.} The standard Majumdar-Papapetrou spacetime
is the manifold $(\mathbb{R}^{3}\setminus \overset{n}{\underset{i=1}{\cup}}\p_{i})
\times \mathbb{R}$ endowed with the metric $ds^2=\frac{-dt^2}{u^2}+u^2(dx^2+dy^2+dz^2)$,
where
$u=1+\overset{n}{\underset{i=1}{\sum}}\frac{q_i}{r_{i}}$ with $q_i$ being a constant and $r_{i}$ the Euclidean distance to $\p_{i}$. $\hfill \square$ \\

In what follows we will give a sketch of the proof of the Theorem \ref{thr:electrovacuniqueness0}.
Firstly, we need some results concerning the boundary of the set
$\{ \p\in M: \left.\lambda\right|_{\p}>0 \}$, where
$\lambda \equiv -\xi_{\mu}\xi^{\mu}$, i.e. minus the squared norm of the stationary Killing field $\vec{\xi}$.

Let us start with some definitions.

\begin{defi}
Let $(M,\gM)$ be a spacetime with a Killing vector $\vec{\xi}$. A {\bf Killing prehorizon} $\mathcal{H}_{\vec{\xi}}$ of $\vec{\xi}$
is a null, 3-dimensional submanifold (not necessarily embedded), at least $C^{1}$, such that
$\vec{\xi}$ is tangent to $\mathcal{H}_{\vec{\xi}}$, null and different from
zero.
\end{defi}

\begin{defi}
A {\bf Killing horizon} is an embedded Killing prehorizon.
\end{defi}

Next, let us introduce a quantity $\kappa$ defined on a Killing prehorizon in any stationary spacetime.
Clearly, on a Killing prehorizon $\mathcal{H}_{\vec{\xi}}$ we have $\lambda=0$.
It implies that $\nablaM_{\mu}\lambda$ is normal to $\mathcal{H}_{\vec{\xi}}$.
Now, since $\vec{\xi}$ is null and tangent to $\mathcal{H}_{\vec{\xi}}$, it is also normal to $\mathcal{H}_{\vec{\xi}}$.
Since, moreover $ \vec{\xi}\,\big|_{\mathcal{H}_{\vec{\xi}}}$ is nowhere zero, it follows that there exists a function $\kappa$ such that
\begin{equation}\label{kappa0}
\nablaM_{\mu}\lambda=2\kappa \xi_{\mu}.
\end{equation}
$\kappa$ is called the {\bf surface gravity} on $\mathcal{H}_{\vec{\xi}}$. The following result states the constancy of $\kappa$ on
a Killing prehorizon in a static spacetime.
\begin{lema}[R\'acz, Wald, 1996 \cite{RW}]\label{lema:RW}
Let $\mathcal{H}_{\vec{\xi}}$ be a Killing prehorizon for an integrable Killing vector $\vec{\xi}$. Then $\kappa$ is constant on each arc-connected
component of $\mathcal{H}_{\vec{\xi}}$.
\end{lema}
{\bf Remark.} This lemma also holds in stationary spacetimes provided the DEC holds. Its proof can be found in Chapter 12 of \cite{Wald}. $\hfill \square$ \\

This lemma allows to classify Killing prehorizons in static spacetimes in two types with very different behavior.
\begin{defi}
An arc-connected Killing prehorizon $\mathcal{H}_{\vec{\xi}}$ is called {\bf degenerate} when $\kappa=0$ and {\bf non-degenerate} when $\kappa\neq 0$.
\end{defi}

Since $\nablaM_{\mu}\lambda\neq 0$ on a non-degenerate Killing prehorizon, the set $\{\lambda=0\}$ defines
an embedded submanifold (c.f. \cite{Cc}).
\begin{lema}
Non-degenerate Killing prehorizons are Killing horizons.
\end{lema}

The next lemma guarantees the existence of a Killing prehorizon in a static spacetime. This lemma will be used several times along this thesis.
For completeness, we find it appropriate to include its proof (we essentially follow \cite{C}).
\begin{lema}[Vishveshwara, 1968 \cite{Vishveshwara}, Carter, 1969 \cite{Carter69}]\label{lema:VC}
Let $(M,\gM)$ be a static spacetime with Killing vector $\vec{\xi}$. Then the set
$\mathcal{N}_{\vec{\xi}}\equiv \tbd \{\lambda>0\}\cap \{\vec{\xi}\neq 0\}$, if non-empty, is a smooth Killing prehorizon.
\end{lema}

{\bf Proof.} Consider a point $\p\in \mathcal{N}_{\vec{\xi}}$. Due to the Fr\"obenius's theorem (see e.g. \cite{Frobenius}), staticity implies that there exists a neighbourhood
$\mathcal{V}_{0}\subset M$ of $\p$, with $\vec{\xi}\,\big|_{\mathcal{V}_{0}}\neq 0$, which
(for $\mathcal{V}_{0}$ small enough)
is foliated by a family of smooth embedded
submanifolds $\Sigma_{t}$
of codimension one and orthogonal to $\vec{\xi}$.
In particular, $\p\in \Sigma_{0}$, where $\Sigma_{0}$ denotes a leaf of this foliation.

Now consider the leaves $\Sigma_\alpha$ of the $\Sigma_{t}$ foliation
such that $\Sigma_{\alpha}\cap \{\lambda\neq0\}\neq \emptyset$.
The staticity
condition (\ref{integrablekilling}) implies
\[
\xi_{[\nu}\nabla_{\mu]}\lambda=\lambda \nabla_{[\mu}\xi_{\nu]},
\]
which on $\mathcal{V}_{0}\cap \{\lambda\neq0\}$ reads
\begin{equation}\label{eqVC}
\xi_{[\nu}\nabla_{\mu]}(\ln{|\lambda|})=\nabla_{[\mu}\xi_{\nu]}.
\end{equation}
Let $\vec{W}$ and $\vec{Z}$ be smooth vector fields on $\mathcal{V}_{0}$
such that
$\vec{W}$ satisfies $\xi_{\mu}W^{\mu}=1$ and $\vec{Z}$ is
tangent to the leaves $\Sigma_{t}$. At points of $\Sigma_{\alpha}$ on which $\lambda\neq0$, the contraction
of equation (\ref{eqVC}) with $Z^{\mu}W^{\nu}$ gives
\[
Z^{\mu}\nabla_{\mu}(\ln{|\lambda|})=2Z^{\mu}W^{\nu}\nabla_{[\mu}\xi_{\nu]}.
\]
The right-hand side of this equation is uniformly bounded on $\Sigma_{\alpha}$,
which implies that $\ln{|\lambda|}$ is uniformly bounded
on $\Sigma_{\alpha} \cap \{\lambda\neq0\}$. This is only possible if
$\Sigma_{\alpha} \cap \{\lambda=0\}= \emptyset$. 
Consequently, $\lambda$ is either positive, or negative, or zero in each leaf of the foliation $\Sigma_{t}$.
In particular, it implies that 
$\{\lambda=0\}\cap\mathcal{V}_{0}$ is a union of leaves of the $\Sigma_{t}$ foliation.

It only remains to prove that each arc-connected component of
$\tbd \{\lambda>0\}\cap \mathcal{V}_{0}$ coincides with one of these leaves.
For that, take coordinates $\{z,x^{A}\}$ in $\mathcal{V}_{0}$ in such a way that
the coordinate $z$ characterizes the leaves of the foliation $\Sigma_t$ and $\p=(z=0,x^A=0)$
(this is possible because each leaf of $\Sigma_{t}$
is an embedded submanifold of $\mathcal{V}_{0}$).
Note that the leaf $\Sigma_0\ni \p$ is then defined by $\{z=0\}$.
In this setting, we just need to prove that $\{z=0\}$ coincides with an arc-connected component of $\tbd\{\lambda>0\}\cap\mathcal{V}_{0}$.
Due to the fact that  $\p\in \tbd \{\lambda>0\}\cap \mathcal{V}_{0}$, there exists a
sequence of points $\p_{i}\in\mathcal{V}_{0}$ with $\lambda>0$
which converge
to $\p$ and have coordinates $(z(\p_{i}),x^{A}(\p_{i}))$. Since the coordinate $z$ characterizes
the leaves and $\lambda$ is either positive, or negative, or zero in each leaf, it follows that
the sequence of points $\p'_{i}$ with coordinates $(z(\p_{i}),0)$ also has $\lambda>0$ and
tends to $\p$. By the same reason, given any point $\q\in \{z=0\}$ with coordinates
$(0,x_{0}^{A})$, the sequence of points
$\q_{i}=(z(\p_{i}),x_{0}^{A})$ tends to $\q$ and lies in  $\{\lambda>0\}$.
Therefore, $\{z=0\}$ is composed precisely by the points of the arc-connected component
of $\tbd\{\lambda>0\}\cap \mathcal{V}_{0}$ which contains $\p$.
This implies that every arc-connected component of $\tbd \{\lambda>0\}\cap \mathcal{V}_{0}$ coincides
with a leaf $\Sigma_{t}$ where $\lambda\equiv 0$ (and $\vec{\xi}\neq 0$).
Finally, this local argument can be extended to the whole set $\mathcal{N}_{\vec{\xi}}$ simply by taking a
covering of $\mathcal{N}_{\vec{\xi}}$ by suitable open neighbourhoods
$\mathcal{V}_{\beta}\subset M$.
$\hfill \blacksquare$ \\

\begin{figure}[h]
\begin{center}
\psfrag{lambda}{{$\mathcal{N}_{\vec{\xi}}$}}
\includegraphics[width=9cm]{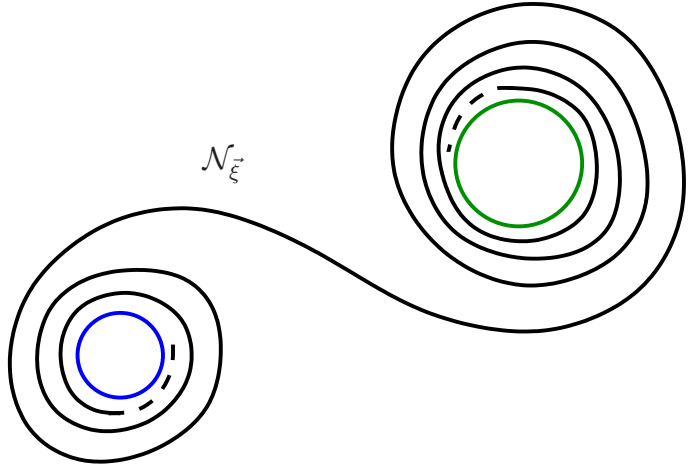}
\caption {The figure illustrates a situation where
$\mathcal{N}_{\vec{\xi}}=\tbd \{\lambda>0\}\cap \{\vec{\xi}\neq 0\}$ fails to be embedded. In this figure, the Killing vector is nowhere zero, causal
everywhere and null precisely on the plotted line. Here,
$\mathcal{N}_{\vec{\xi}}$ has three arc-connected components: two spherical and one with spiral form.
The fact that
the spiral component accumulates around the spheres implies that the whole set $\mathcal{N}_{\vec{\xi}}$
is not embedded. Moreover, the spiral arc-connected component, which is itself embedded, is not compact.
}
\label{fig:spiral}
\end{center}
\end{figure}

{\bf Remark.} Although each arc-connected component of $\tbd \{\lambda>0\}\cap \mathcal{V}_{\beta}$ is an embedded
submanifold of $\mathcal{V}_{\beta}\subset M$, the whole set $\mathcal{N}_{\vec{\xi}}$ may fail to be
embedded
in $M$ (see Figure \ref{fig:spiral}). Thus, a priori, degenerate Killing prehorizons may
fail to be embedded. As mentioned before, this possibility has been overlooked
in the literature until recently \cite{Cc}.  The occurrence of non-embedded Killing prehorizons poses
serious difficulties for the uniqueness proofs. One way to deal with these objects
is to make hypotheses that simply exclude them. In Proposition \ref{prop:Chrusciel} below, the
hypothesis that $\tbd \Sigma$ is a compact and embedded topological manifold is made precisely
for this purpose. Another possibility is to prove that these prehorizons do not exist. At present,
this is only known under strong global hypotheses on the spacetime (c.f. Definition \ref{Iplusregularity} below).
It is an interesting open problem to either find an example of a non-embedded Killing prehorizon
or else to prove that they do not exist. $\hfill \square$ \\

The hypotheses of Theorem \ref{thr:electrovacuniqueness0} require the existence of a hypersurface $\Sigma$
with topological boundary such that $\lambda\geq 0$ everywhere and $\lambda=0$ precisely on $\tbd \Sigma$. It is clear then that $\tbd \Sigma \subset \tbd U$, where $U \equiv \{ \p\in M: \left.\lambda\right|_{\p}>0 \}$,
but, in general, $\tbd \Sigma$
will not lie in a Killing prehorizon because it can still happen that $\vec{\xi}=0$ on a subset of $\tbd U$.
However, the set of points where $\vec{\xi}= 0$ cannot be very ``large" as the next result guarantees.
\begin{thr}[Boyer, 1969 \cite{Boyer}, Chru\'sciel, 1999 \cite{C}]\label{thr:Boyer}
Consider a static spacetime $(M,\gM)$ with Killing vector $\vec{\xi}$.
Let $\p\in\tbd\{\lambda>0\}$ be a fixed point (i.e. $\vec{\xi}\,\big|_{\p}=0$).
Then $\p$ belongs to a connected, spacelike, smooth, totally geodesic, 2-dimensional surface $S_{0}$ which is
composed by fixed points. Furthermore, $S_{0}$ lies in the closure of a non-degenerate
Killing horizon $\mathcal{H}_{\vec{\xi}}$
\end{thr}


Therefore, using Lemma \ref{lema:VC} and Theorem \ref{thr:Boyer}, we can assert that  {\it $\tbd \{\lambda>0\}$ belongs to the closure of a Killing prehorizon}.

The manifold $\mbox{int}(\Sigma)$ admits, besides the induced metric,
a second metric $h$ called {\it orbit space metric} which is a key object in the uniqueness proof. Let us first define
the projector orthogonal to $\vec{\xi}$.

\begin{defi}
On the open set $U\equiv \{ \lambda>0 \}\subset M$, the {\bf projector orthogonal} to $\vec{\xi}$, denoted by $h_{\mu\nu}$, is defined as
\begin{equation}\label{h}
h_{\mu\nu}\equiv \gM_{\mu\nu}+\frac{\xi_{\mu}\xi_{\nu}}{\lambda}.
\end{equation}
This tensor has the following properties:
\begin{itemize}
\item It is symmetric, i.e. $h_{\mu\nu}=h_{\nu\mu}$.
\item It has rank 3.
\item It satisfies $h_{\mu\nu}\xi^{\mu}=0$
\end{itemize}
\end{defi}

On $U$ we can also define the function $V=+\sqrt{\lambda}$. The hypersurface
$\mbox{int}(\Sigma)$ is fully contained in $U$.  Let
$\Phi: \mbox{int}(\Sigma)\rightarrow U\subset M$ denote the embedding of $\mbox{int}(\Sigma)$ in
$U$, then the pull-back of the projector
$\Phi^{*}(h)$ is a Riemannian metric on $\Sigma$. We will denote by the same symbols
$h$, $V$ and $\phi$ both the objects in $U\subset M$ and their corresponding pull-backs in $\mbox{int}(\Sigma)$.

The Einstein-Maxwell field equations for a purely electric stationary electro-vacuum spacetime 
are equivalent to the
following equations on $\mbox{int}(\Sigma)$ see e.g. \cite{Heuslerlibro}.
\begin{eqnarray}
V \Delta_{h}\phi&=&D_{i}V{D}^{i}\phi,\label{EinsteinEV1}\\
V \Delta_{h} V&=&D_{i}\phi {D}^{i}\phi, \label{EinsteinEV2}\\
V R_{ij}(h)&=&D_{i}D_{j}V+\frac1V\left( D_{k}\phi {D}^{k}\phi h_{ij} - 2D_{i}\phi {D}_{j}\phi \right), \label{EinsteinEV3}
\end{eqnarray}
where $D$ and $R_{ij}(h)$ are the covariant derivative and the Ricci tensor of the Riemannian metric $h$, respectively. Indices are raised
and lowered with $h_{ij}$ and its inverse $h^{ij}$.

In the asymptotically flat end $\Sigma^{\infty}_{0}$ of $\mbox{int}(\Sigma)$,
the Einstein equations on $\mbox{int}(\Sigma)$ and
(\ref{asymptoticallyflatend}) that $V$ and $\phi$ decay as
\begin{equation}\label{VQ}
V=1-\frac{M_{\scriptscriptstyle ADM}}{r} + O^{(2)}(1/r^{2}), \qquad \qquad \phi=\frac{Q}{r}+O^{(2)}(1/r^2),
\end{equation}
where $Q$ is a constant (called the {\bf electric charge} associated with $\Sigma^{\infty}_{0}$), and
$M_{\scriptscriptstyle ADM}$ is the corresponding ADM mass.


A crucial step for the uniqueness proof is to understand the behavior of
the Riemannian metric $h$ near the boundary $\tbd \Sigma$.
This is the aim of the following proposition.

\begin{proposition}[Chru\'sciel, 1999 \cite{C}]\label{prop:Chrusciel}
Let $\Sigma$ be a spacelike hypersurface in a static spacetime $(M,\gM)$ with Killing vector $\vec{\xi}$.
Suppose that $\lambda\geq 0$ on $\Sigma$ with $\lambda =0$ precisely on its topological boundary $\tbd \Sigma$
which is assumed to be a compact, 2-dimensional and embedded topological manifold. Then
\begin{enumerate}
\item Every arc-connected component $(\tbd \Sigma)_{d}$ which intersects a $C^2$ degenerate Killing horizon corresponds to a complete
cylindrical asymptotic end of $(\Sigma,h)$.
\item $(\overline{\Sigma},h)$ admits a differentiable structure such that every arc-connected component $(\tbd \Sigma)_{n}$ of $\tbd \Sigma$ which intersects
a non-degenerate Killing horizon is a totally geodesic boundary of $(\Sigma,h)$ with $h$ being smooth up to and including the boundary.
\end{enumerate}
\end{proposition}

This proposition shows that the Riemannian manifold
$(\overline{\Sigma} \setminus \underset{d}{\cup} (\tbd \Sigma)_d,h)$
is the union of asymptotically flat ends,
complete cylindrical asymptotic ends and compact sets with totally geodesic boundaries.
Let us define $\Sigmatilde \equiv \overline{\Sigma} \setminus \underset{d}{\cup}(\tbd \Sigma)_d$.


Now we are ready to explain the doubling method itself. Recall that the final aim is to show
that the spacetime is either Reissner-Nordstr\"om or Majumdar-Papapetrou.
Both have the property that $(\Sigmatilde,h)$ is conformally flat
(i.e. there exists a positive function $\Omega$, called the conformal factor, such that the metric $\Omega^{2}h$
is the flat metric). Moreover, conformal flatness together with sufficient information on the conformal factor would imply,
via the Einstein field equations, that the spacetime is in fact Reissner-Nordstr\"om or Majumdar-Papapetrou.

A powerful method to prove that a given metric is flat is by using the rigidity part of the PMT.
Unfortunately Theorem \ref{thr:PMT0} cannot be applied directly to $(\Sigmatilde,h)$ because,
first, $\Sigmatilde$ is a manifold with boundary, and second, $(\Sigmatilde,h)$ has in general cylindrical asymptotic ends
and therefore it is not
asymptotically flat.


The presence of boundaries was dealt with by Bunting and Masood-ul-Alam who invented a method which constructs a new
manifold without boundary to which the PMT can be applied.

To simplify the presentation, let us assume for a moment that $(\Sigmatilde,h)$ has no cylindrical ends, so this manifold is the union of asymptotically ends and a compact interior with
totally geodesic boundaries (by Proposition \ref{prop:Chrusciel}).
Next, find two conformal factors $\Omega_{+}>0$ and $\Omega_{-}>0$ such that
\begin{itemize}
\item $h_{+}\equiv \Omega_{+}^{2} h$ is asymptotically flat, has vanishing mass and $R(h_{+})\geq 0$,
where $R(h_{+})$ is the scalar curvature of $h_{+}$.
\item $h_{-}\equiv \Omega_{-}^{2}h$ admits a one point (let us denote it by $\Upsilon$) compactification of the asymptotically flat infinity, and $R(h_{-})\geq 0$.
\end{itemize}

Then the 
idea 
is to glue the manifolds $(\Sigmatilde,h_{+})$ and $(\Sigmatilde\cup \Upsilon, h_{-})$ 
across the boundaries to produce a complete, asymptotically
flat manifold $(\hat{\Sigma},\hat{h})$ with no boundaries, vanishing mass and
non-negative scalar curvature $\hat{R}\geq 0$. 
In order to glue the two manifolds with sufficient differentiability, the following two conditions are required:
\begin{itemize}
\item $\left. \Omega_{+}\right|_{\bd \Sigmatilde}=\left. \Omega_{-}\right|_{\bd \Sigmatilde}$,
\item $\left.\vec{m}(\Omega_{+})\right|_{\bd \Sigmatilde}=-\left.\vec{m}(\Omega_{-})\right|_{\bd \Sigmatilde}$.
\end{itemize}
where $\vec{m}$ is the unit normal pointing to the interior $\Sigmatilde$ in each of the copies.

\begin{figure}[h]
\begin{center}
\psfrag{pS}{\small{\color{red}{$\bd \Sigmatilde$}}}
\psfrag{p}{\color{red}{$\Upsilon$}}
\psfrag{S+}{$(\Sigmatilde,h_+)$}
\psfrag{S-}{$(\Sigmatilde,h_-)$}
\includegraphics[width=7cm]{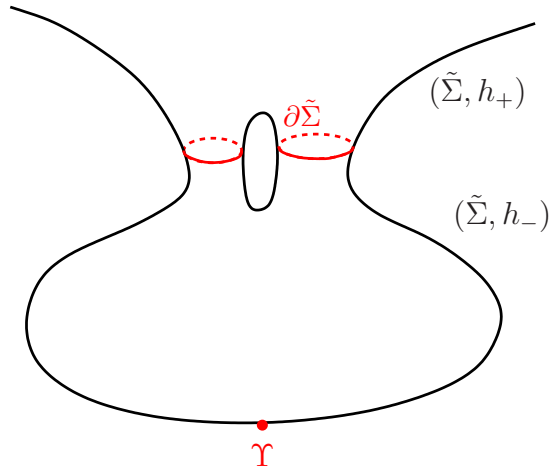}
\caption{The doubled manifold $(\hat{\Sigma},\hat{h})$ resulting from gluing $(\Sigmatilde,h_{+})$ and $(\Sigmatilde\cup \Upsilon, h_{-})$.}
\label{fig:doubling}
\end{center}
\end{figure}

Theorem \ref{thr:PMT0} can be applied to $(\hat{\Sigma},\hat{h})$ to conclude that this space is in fact Euclidean.

When the spacetime also has degenerate horizons the doubling method across non-degenerate components can still
be done. The resulting manifold however is no longer asymptotically flat since it contains asymptotically
cylindrical ends, so Theorem \ref{thr:PMT0} cannot be applied directly.
Fortunately, there exists
a suitable generalization of the PMT that covers this case.
The precise statement is the following.

\begin{thr}[Bartnik, Chru\'sciel, 1998 \cite{CBartnik}]\label{thr:PMT1}
Let $(\hat{\Sigma},\hat{h})$ be a smooth complete Riemannian manifold with an asymptotically flat end $\hat{\Sigma}^{\infty}_{0}$ and with
a smooth one-form $\hat{\bf E}$ satisfying $\hat{D}_{i}\hat{E}^{i}=0$ and $\hat{E}_{i}dx^{i}=\frac{\hat{Q}}{r^2} dr + o(\frac{1}{r^2})$
in $\hat{\Sigma}^{\infty}_{0}$, where $\hat{Q}$ is a constant called electric charge. Suppose that $\hat{h}$ satisfies $R(\hat{h})\geq 2\hat{E}_{i} \hat{E}^{i}$ and that
\[
\int_{\hat{\Sigma}^{\infty}_{0}} \left( R(\hat{h})- 2\hat{E}_{i} \hat{E}^{i} \right) \eta_{\hat{h}}< \infty.
\]
Then the ADM mass $\hat{M}_{\scriptscriptstyle ADM}$ of $\hat{\Sigma}^{\infty}_{0}$ satisfies $\hat{M}_{\scriptscriptstyle ADM}\geq |\hat{Q}|$ and equality holds if and only if
locally $\hat{h}=u^{2}(dx^2+dy^2+dz^2)$, $\hat{\bf E}=\frac{du}{u}$ and $\Delta_{\delta}u=0$.
\end{thr}

{\bf Remark.} 
As a consequence of this result, it is no longer necessary to require that $(\Sigmatilde,h_-)$ admits a one-point compactification. It is only necessary to assume that
$(\overline{\Sigma_{0}^{\infty}},h_-)$ is complete. $\hfill \square$ \\

It is clear from the discussions above that the key to prove Theorem \ref{thr:electrovacuniqueness0}
is to find suitable conformal factors which allow to conclude that
$({\Sigmatilde},{h})$ is conformally flat. 
For the 
static electro-vacuum case, two conformal factors have been considered, one due to Ruback \cite{Ruback},
$\Omega_{\pm}=\frac{1\pm V+\phi}{2}$, and another proposed by Masood-ul-Alam \cite{MuA},
$\Omega_{\pm}=\frac{(1\pm V)^{2}-\phi^2}{4}$.
Recently, Chru\'sciel
has showed \cite{C2} that the Ruback conformal factor is the only one which works when
degenerate Killing horizons are allowed a priori.

We will therefore consider only the Ruback conformal factors
$\Omega_{\pm}=\frac{1\pm V+\phi}{2}$.
The first thing to do is to check that $\Omega_{\pm}$ are strictly positive 
on $\Sigmatilde$. 
This was shown by Ruback \cite{Ruback} and extended by Chru\'sciel \cite{C2} and Chru\'sciel and Tod \cite{CT} when
there are degenerate horizons.
\begin{proposition}[Ruback, 1988, Chru\'sciel, 1998, Chru\'sciel, Tod, 2006]\label{proposition:positivityoftheconformalfactor}
On $\Sigmatilde$ it holds $|\phi|\leq 1-V$. Moreover, equality at one point only occurs when the spacetime
is the standard Majumdar-Papapetrou spacetime.
\end{proposition}
This proposition implies $\Omega_{-}>0$ unless we have Majumdar-Papapetrou. Moreover, since
$V\geq 0$ on $\Sigmatilde$, we have $\Omega_{+}\geq \Omega_{-}>0$ except for the
standard Majumdar-Papapetrou.


The remaining ingredients are as follows:
\begin{itemize}
\item The matching conditions for the gluing procedure 
follow easily from the fact that
$\left.V\right|_{\bd \Sigmatilde}=0$, which immediately implies  $\left.\Omega_{+}\right|_{\bd\Sigmatilde}=\left.\Omega_{-}\right|_{\bd \Sigmatilde}$ and
$\left.\vec{m}(\Omega_{+})\right|_{\bd \Sigmatilde}=-\left.\vec{m}(\Omega_{-})\right|_{\bd \Sigmatilde}$. 
%
\item The asymptotically flat end $(\Sigma_{0}^{\infty})$ becomes a complete end with respect
to the metric $h_{-}$. This follows from the asymptotic form
$\Omega_{-}=\frac{1}{4r}(M_{\scriptscriptstyle ADM}-Q)+O(1/r^{2})$ and the fact that $M_{\scriptscriptstyle ADM}> |Q|$ which
follows from the positivity of $\Omega_{-}$. 
\item The field ${\bf {E}}_{\pm}\equiv\frac{-(1+\phi)d\phi + VdV}{V(1+\phi\pm V)}$
has the following asymptotic behavior
\[
{\bf E}_{+}=\frac12\frac{M_{\scriptscriptstyle ADM}+Q}{r^{2}}dr+o(1/r^{2}),
\]
and satisfies, from the Einstein field equations, that
$D^{\pm}_{i}{E_{\pm}}^{i}=0$ and $R(h_{\pm})=2E^{i}_{\pm}{E_{\pm}}_{i}$, where $R(h_{\pm})$ is the scalar curvature of $h_{\pm}$.
\item A direct computation gives that the ADM mass and the electric charge of $(\hat{\Sigma},\hat{h})$ satisfy,
\[
\hat{M}_{\scriptscriptstyle ADM}=\hat{Q}.
\]
\end{itemize}

Therefore, the rigidity part of Theorem \ref{thr:PMT1} can be applied, to conclude
$\hat{h}=u^{2}g_{E}$, where $u$ is a specific function of $(V,\phi)$ and $g_{E}$ is the Euclidean metric.
Consequently,  $h$ (which was conformally related with $\hat{h}$) is conformally flat. 
The original proof used at this point the explicit form of $u(\phi,V)$ together with the field equations to conclude that
$(\Sigmatilde,h)$ corresponds to the metric of the $\{t=0\}$ slice of Reissner-Nordstr\"om spacetime
with $M>|Q|$.
This last step has been simplified
recently
by Gonz\'alez and Vera in \cite{GV} who show that the Reissner-Nordstr\"om and the
Majumdar-Papapetrou
spacetimes are indeed the only static electro-vacuum
spacetimes for which $(\Sigmatilde,h)$ is asymptotically flat and conformally flat.

Summarizing, we have obtained that, in the case when Theorem \ref{thr:PMT1} can be applied, the spacetime is
Reissner-Nordstr\"om, and in the cases when it cannot be applied the spacetime is
already the standard Majumdar-Papapetrou spacetime. 
We conclude then that a static and electro-vacuum spacetime corresponding to a
black hole must be either the Reissner-Nordstr\"om spacetime (where $\tbd \Sigma$ is connected) or
the standard Majumdar-Papapetrou spacetime (where $\tbd \Sigma$ is non-connected), which proves Theorem \ref{thr:electrovacuniqueness0}.


{\bf Remark.} The compactness assumption for the embedded topological submanifold $\tbd \Sigma$ is used in order to ensure
that $(\hat{\Sigma},\hat{h})$ is complete. It would be interesting to study whether this condition can be relaxed or not. $\hfill \square$ \\

We will finish this chapter by giving a brief discussion about the global approach of Theorem \ref{thr:electrovacuniqueness0}.
In several works (\cite{C}, \cite{Cc} and \cite{CG}) Chr\'usciel and Galloway have
studied sufficient hypotheses which ensure that a black hole spacetime possesses a spacelike
hypersurface $\Sigma$ like the one required in Theorem \ref{thr:electrovacuniqueness0} and, also,
which assumptions are needed to conclude uniqueness for the whole spacetime (or at least for the domain of outer communications)
The first work on the subject, namely \cite{C}, deals with the vacuum case and requires, among other things,
the spacetime to be analytic
(although this hypothesis was not explicitly mentioned in \cite{C} and
it was included only in the correction \cite{Cc}). 
This hypothesis is needed to avoid the existence
of non-embedded degenerate Killing prehorizons, which implies that
$\tbd \Sigma$ may fail to be compact and embedded as required in Theorem \ref{thr:electrovacuniqueness0}. 
In \cite{Cc}, Chru\'sciel was able to drop the analyticity assumption by assuming
a second Killing vector on $M$ generating a $U(1)$ action and a global
hypothesis (named $I^{+}$-regularity in the later paper \cite{CL-C}).
Finally, in \cite{CG} the assumption on the existence of a second
Killing field was removed and the result was explicitly extended to the electro-vacuum case.
Before giving the statement of such a result, let us define the property of $I^{+}$-regularity of a spacetime.

\begin{defi}
\label{Iplusregularity}
Let $(M,\gM)$ be a stationary spacetime containing an asymptotically flat end and let $\vec{\xi}$ be the stationary Killing vector field on $M$.
$(M,\gM)$ is {\bf $I^{+}$-regular} if $\vec{\xi}$ is complete, if the domain of outer communications $M_{DOC}$ is globally hyperbolic, and
if $M_{DOC}$ contains a spacelike, connected, acausal hypersurface $\Sigma$ containing an asymptotically flat end, the closure $\overline{\Sigma}$
of which is a $C^0$ manifold with boundary, consisting of the union of a compact set and a finite number of asymptotically flat ends, such that
$\tbd \Sigma$ is an embedded surface satisfying
\[
\tbd \Sigma\subset {\mathcal{E}^{+}}\equiv \tbd M_{DOC}\cap I^{+}(M_{DOC}),
\]
with $\tbd \Sigma$ intersecting every generator of $\mathcal{E}^{+}$ just once.
\end{defi}


Then the result by Chru\'sciel and Galloway states the following.

\begin{thr}[Chru\'sciel and Galloway, 2010 \cite{CG}]\label{thr:electrovacuniqueness1}
Let $(M,\gM)$ be a static solution of the electro-vacuum Einstein equations. Assume that $(M,\gM)$ is $I^{+}$-regular.
Then the conclusions of Theorem \ref{thr:electrovacuniqueness0} hold.
Moreover, $M_{DOC}$ is isometrically diffeomorphic to the domain of outer communications of either the
Reissner-Nordstr\"om spacetime or the standard Majumdar-Papapetrou spacetime.
\end{thr}

\chapter{Stability of marginally outer trapped surfaces and symmetries}
\label{ch:Article2}

\setcounter{equation}{0}
\section{Introduction}\label{sc:A2sectionintroduction}


As we have already mentioned in Chapter \ref{ch:Introduction}, although the main aim of this thesis is to study properties of
certain types of trapped surfaces, specially weakly outer trapped surfaces and MOTS, in stationary and static configurations, isometries are not the only type of symmetries which can be involved in physical situations of interest.
For instance, many relevant spacetimes
admit other types of symmetries, such as
conformal symmetries, e.g. in Friedmann-Lema\^itre-Robertson-Walker (FLRW) cosmological models.
Another interesting example appears when studying the critical collapse, which is a universal
feature of many matter models. Indeed, the critical solution, which
separates those configurations that disperse from those that form black holes,
are known to admit either a continuous
or a discrete self-similarity.
Therefore, it is interesting to understand the relationship between
trapped surfaces and several special types of symmetries. This is precisely the aim of this
chapter.

A recent interesting example of this interplay has been given in
\cite{BenDov}, \cite{SB}, \cite{SB2} where the location of the boundaries of
the spacetime set containing weakly trapped surfaces and weakly outer trapped surfaces 
was analyzed, firstly, in the Vaidya spacetime \cite{BenDov}, \cite{SB} (which is one of the simplest dynamical situations) and, later,
in spherically symmetric spacetimes in general \cite{SB2}.
In these analyses the presence of
symmetries turned out to be fundamental.
In the important case of isometries, general results on
the relationship between weakly trapped surfaces and
Killing vectors were discussed in \cite{MS}, where the first variation of area was used to
obtain several restrictions on the existence of weakly
trapped surfaces in spacetime regions possessing a causal Killing vector. More specifically,
weakly trapped surfaces
can exist in the region where the Killing vector is timelike only if their mean curvature vanishes identically.
By obtaining a general identity for the first variation of area in terms of the
deformation tensor of an arbitrary vector (defined in equation (\ref{mdt})),
similar restrictions were obtained for spacetimes admitting other types of symmetries, such as
conformal Killing vectors or Kerr-Schild vectors (see \cite{CHS} for its definition). The same idea was also applied in \cite{S3}
to obtain analogous results in spacetimes with
vanishing curvature invariants.
The interplay between isometries and
dynamical horizons (which are spacelike hypersurfaces foliated
by marginally trapped surfaces) was considered in \cite{AG05}
where it was proven that dynamical horizons cannot exist
in spacetime regions containing a nowhere vanishing causal Killing vector, provided the
spacetime satisfies the NEC.
Regarding MOTS, the relation between stable MOTS and isometries was
considered in \cite{AMS}, where
it was shown that, given a strictly stable MOTS $S$
in a hypersurface $\Sigma$ (not necessarily spacelike),
any Killing vector on $S$ tangent
to $\Sigma$ must in fact be tangent to $S$.

In the present chapter, we will study the interplay between stable
and outermost properties of MOTS in
spacetimes possessing special types of vector fields $\vec{\xi}$, including
isometries, homotheties and conformal Killing vectors. In fact, we
will find results involving completely general vector
fields $\vec{\xi}$ and then, we will particularize them to the different types of symmetries.
More precisely, we will find
restrictions on $\vec{\xi}$ on stable, strictly stable and locally outermost MOTS $S$ in a given
spacelike hypersurface $\Sigma$, or alternatively, forbid the
existence of a MOTS in certain regions where $\vec{\xi}$ fails to
satisfy those restrictions.
In what follows, we give a brief summary of the present chapter.

The fundamental idea which will allow us to
obtain the results of this chapter will be introduced in Section \ref{sc:A2sectionbasics}.
As we will see, it will consist in a geometrical construction which can potentially restrict
a vector field $\vec{\xi}$ on the outermost MOTS $S$. The geometrical procedure
will involve the analysis of the stability operator $L_{\vec{m}}$ of a MOTS acting on a certain function $Q$.
It will turn out that the results obtained by
the geometric construction can, in most cases, be sharpened
considerably by using the maximum principle of elliptic operators.
This will also allow us to extend the validity of the results from
the outermost case to the case of stable and strictly stable MOTS.
However, the defining expression
(\ref{stabilityoperator}) for the stability operator $L_{\vec{m}}Q$ has a priori nothing to do with
the properties of the vector field $\vec{\xi}$, which makes the method of little use. Our
first task will be therefore to obtain an alternative (and completely general) expression for $L_{\vec{m}}Q$
in terms of $\vec{\xi}$,
or more specifically, in terms of its deformation tensor $a_{\mu\nu}(\vec{\xi}\,)$. We will devote
Section \ref{sc:A2variation} to doing this. The result, given in Proposition \ref{propositionxitheta}, is
thoroughly used in this chapter and also has independent interest.

With this expression at hand, we will be able to analyze under which
conditions our geometrical procedure gives restrictions on $\vec{\xi}$.
In Section \ref{sc:A2LmQ} we will concentrate on the case where $L_{\vec{m}}Q$ has a
sign everywhere on $S$.
The main result of Section \ref{sc:A2LmQ} will be given in Theorem
\ref{TrhAnyXi}, which holds for any vector field $\vec{\xi}$. This
result will be then particularized to conformal Killing vectors
(including homotheties and Killing vectors) in Corollary \ref{thrstable}. Under the additional
restriction that the homothety or the Killing vector is everywhere
causal and future (or past) directed, strong restrictions on the
geometry of the MOTS will be derived (Corollary \ref{shear}). As a
consequence, we will prove that in a plane wave spacetime any stable
MOTS must be orthogonal to the direction of propagation of the
wave. Marginally trapped surfaces will be also discussed in this
section.

As an explicit application of the results on conformal Killing
vectors, we will show, in Subsection \ref{ssc:A2sectionFLRW}, that stable
MOTS cannot exist in any spacelike hypersurface in FLRW
cosmological models provided the density $\mu$ and pressure $p$
satisfy the inequalities $\mu \geq 0$, $\mu \geq 3 p $ and $\mu
+ p \geq 0$. This includes, for instance, all classic models of
matter and radiation dominated eras and also those models with
accelerated expansion which satisfy the NEC. Subsection
\ref{ssc:A2sectiongeometric} will deal with one case where, in contrast with
the standard situation, the geometric construction does in fact
give sharper results than the elliptic theory. One of these results, together
with Theorem \ref{thr:AM} by Andersson and Metzger, will imply an interesting result (Theorem \ref{theorem1}) for
weakly outer trapped surfaces in stationary spacetimes.

In the case when $L_{\vec{m}} Q$ is not assumed to have a definite sign,
the maximum principle loses its power. However, as we will discuss in Section \ref{sc:A2sectionnonelliptic},
a result
by Kriele and Hayward \cite{KH97} will allow us to exploit our geometric
construction again to obtain additional results.
This will produce a theorem (Theorem \ref{thrnonelliptic}) which holds for
general vector fields $\vec{\xi}$ on any locally outermost MOTS.
As in the previous section, we will particularize the result to conformal
Killing vectors, and then to causal Killing vectors and
homotheties which, in this case, will be allowed to change their time
orientation on $S$

The results presented in this chapter have been published mainly in the papers \cite{CM2}, \cite{CMere2} and partly in \cite{CM1} and
\cite{CMere1}.

\setcounter{equation}{0}
\section{Geometric procedure}
\label{sc:A2sectionbasics}

Consider a spacelike hypersurface $(\Sigma,g,K)$ which is embedded in a
spacetime $(M,\gM)$ with a vector field $\vec{\xi}$ defined on a neighbourhood of $\Sigma$.
Assume that $\Sigma$ possesses a barrier $\Sb$ with interior $\Omegab$ and let $S\subset\Sigma$ be a bounding MOTS
with respect to $\Sb$ (and therefore an {\it exterior region} of $S$ in $\Omegab$ can be properly defined).
The idea we want to exploit
consists in constructing under certain circumstances a new weakly outer trapped surface $\S_{\tau}\subset\Omegab$
which lies, at least partially, outside $\S$.
This fact will provide a contradiction in the case when $\S$ is the outermost bounding MOTS
and will allow us
to obtain restrictions on the vector $\vec{\xi}$ on $\S$. As we will see below,
this simple idea will allow us to
obtain results also for stable, strictly stable and locally outermost MOTS, irrespectively of whether they are bounding or not,
by using the
theory of elliptic second order operators. 

\begin{figure}
\begin{center}
\psfrag{t}{$\tau$}
\includegraphics[width=9cm]{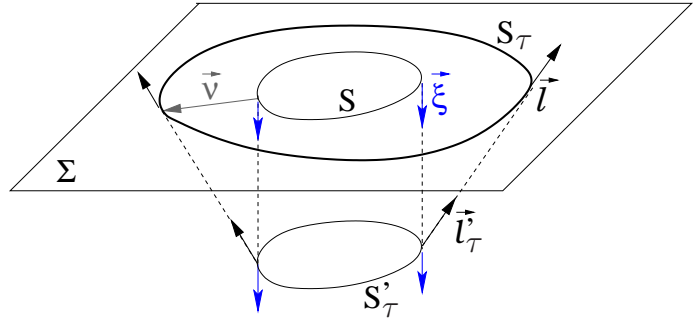}
\caption {The figure represents how the new surface $S_t$ is
constructed from the original surface $S$. The intermediate
surface $S^{\prime}_{\tau}$ is obtained from $S$ by dragging along
$\vec{\xi}$ a parametric amount $\tau$. Although $\vec{\xi}$ has been
depicted as timelike here, this vector can be in fact of any
causal character.}
\label{fig:construction}
\end{center}
\end{figure}

The geometric procedure to construct the new surface $\S_{\tau}$ consists in
moving $\S$ first along the integral lines of
$\vec{\xi}$ a parametric amount $\tau$. This gives a new surface $S^{\prime}_{\tau}$.
Next, take the
null normal ${{\vec{l}_{+}}^{\prime}}(\tau)$ on this surface
which coincides with the continuous deformation of
the outer null normal $\vec{l}_{+}$ on $S$ normalized to satisfy $l_{+}^{\mu}n_{\mu}=-1$ (where $\vec{n}$ denotes the unit vector normal to
$\Sigma$ and future directed) and consider the null hypersurface generated by null geodesics
with tangent vector ${\vec{l}_{+}^{\prime}}(\tau)$. This hypersurface is smooth close
enough to $\S^{\prime}_{\tau}$. Being null, its intersection with the spacelike
hypersurface $\Sigma$ is transversal and hence defines a smooth surface
$S_{\tau}$ (for $\tau$ sufficiently small). By this construction, a point $\p$ on $S$
describes a curve in $\Sigma$ when $\tau$ is varied. The tangent vector of
this curve on $\S$, denoted by $\vec{\nu}$,
will define the variation vector generating the
one-parameter family $\{ \S_{\tau} \}_{\tau\in I\subset\mathbb{R}}$ on a neighbourhood of $S$ in $\Sigma$.
Figure \ref{fig:construction} gives a graphic representation of this
construction.

\begin{figure}
\begin{center}
\psfrag{xi}{$\vec{\xi}$}
\psfrag{N}{$N\vec{n}$}
\psfrag{Y}{$\vec{Y}$}
\psfrag{Sigma}{$\Sigma$}
\includegraphics[width=9cm]{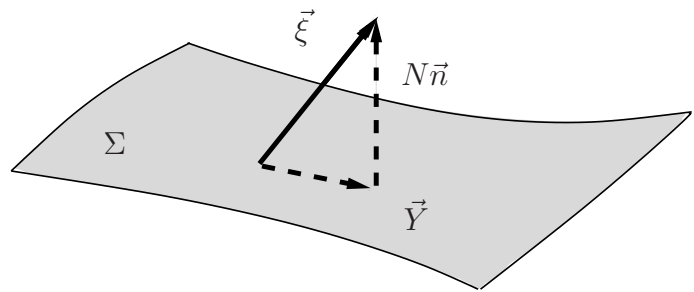}
\caption {The vector $\vec{\xi}$ decomposed into normal $N\vec{n}$ and tangential $\vec{Y}$ components.}
\label{fig:XiNY}
\end{center}
\end{figure}

Let us decompose the vector $\vec{\xi}$ into normal and
tangential components with respect to $\Sigma$, as
$\vec{\xi}=N\vec{n}+\vec{Y}$ (see Figure \ref{fig:XiNY}).
On $\S$ we will further decompose
$\vec{Y}$ in terms of a tangential component
$\vec{Y}^{\parallel}$, and a normal component $(Y_{i}m^{i})\vec{m}$,
where $\vec{m}$ is the unit vector normal to $S$ in $\Sigma$ which points to the exterior
of $S$ in $\Sigma$. Therefore, $\vec{\xi} |_S =N_S \vec{n}+ (Y_{i} m^{i}) \vec{m}  + \vec{Y}^{\parallel}$, where $N_S$ is
the value of $N$ on the surface. In order to study the variation
vector $\vec{\nu}$, let us expand the embedding functions $\left\{
x^\mu\left( y^A,\tau \right) \right\}$ of the surface $\S_{\tau}$ (where
$\left\{ y^A \right\}$ are intrinsic coordinates of $\S$) as
\begin{equation}\label{embedding2}
x^{\mu}\left( y^A,\tau \right)=x^{\mu}\left( y^A,0 \right) +
\xi^\mu\left( y^A,0 \right)\tau + F(y^A){l'_{+}}(\tau)^{\mu}\left(
y^A\right)\tau + O(\tau^2),
\end{equation}
where $F(y^A)$ is a function to be adjusted. Since $\vec{\nu}$
defines the variation of $\S$ to first order, equation
(\ref{embedding2}) implies that we only need to evaluate the vector
${\vec{l}_{+}^{\prime}}(\tau)$ to zero order in $\tau$, which obviously
coincides with $\vec{l}_{+}$. It follows then that $\vec{\nu}$ is a
linear combination (with functions) of $\vec{\xi}$ and
$\vec{l}_{+}$. The amount we need to move $S^{\prime}_{\tau}$ in order to
go back to $\Sigma$ can be determined by imposing $\vec{\nu}$ to
be tangent to $\Sigma$. Since $\vec{\nu}(y^{A})=\vec{\xi}(y^{A})+F(y^{A})\vec{l}_{+}(y^{A})$, multiplication with $\vec{n}$ gives
$0=N_{S}+F$. Thus, $F=-N_{S}$ and $\vec{\nu}=\vec{\xi}-N_{S}\vec{l}_{+}$. Using the previous decomposition for $\vec{\xi}$ and
$\vec{l}_{+}=\vec{n}+\vec{m}$ we can rewrite $\vec{\nu}=Q\vec{m}+\vec{Y}^{\parallel}$, where
\begin{equation}\label{Q}
Q= (Y_{i}m^{i}) -N_S = {\xi}_{\mu}l_{+}^{\mu}\,
\end{equation}
determines at first order the amount and sense to which a point $\p\in S$ moves along the normal direction.

Let us consider for a moment the simplest
case that $\vec{\xi}$ is a Killing vector. Suppose $\S$ is a MOTS which is bounding with respect to a barrier
$\Sb$ with interior $\Omegab$. 
Since the null expansion does not
change under an isometry, it follows that the surface $S^{\prime}_{\tau}$ is
also a bounding MOTS for the spacelike hypersurface obtained by moving $\Sigma$ along the integral curves of $\vec{\xi}$ an amount $\tau$.
Moving back to $\Sigma$ along the null hypersurface
gives a contribution to $\theta^{+} [S_\tau]$ which is easily computed
to be $\left.\frac{d}{d\tau}\left[\hat{\varphi}_{\tau}^{*}(\theta^{+}[S_{\tau}]) \right]\right|_{\tau=0}=\left. \frac{1}{2}N{\theta^{+}}^{2}[S]+
N W\right|_{{S}}$ which is the well-known Raychaudhuri equation (which has already appeared before in equation (\ref{raychaudhuri}) for
the particular case of MOTS), where
$\hat{\varphi}_{\tau}:S\rightarrow S_{\tau}$ is the diffeomorphism defined by the geometrical construction above
and $W$ was defined in equation (\ref{W}) and is non-negative provided the NEC holds.
It implies that if $N_{S}<0$ and $W\neq 0$ everywhere, then $\theta^{+}[S_{\tau}]< 0$ provided
$\tau$ is positive and sufficiently small 
and the NEC holds. Therefore, $S_{\tau}$ is a bounding (provided $\tau$ is sufficiently small)
weakly outer trapped surface which lies partially outside $S$ if $Q>0$ somewhere.
This is impossible if $S$ is an outermost bounding MOTS by Theorem \ref{thr:AM} of Andersson and Metzger.
Thus, the function $Q$ must be
non-positive everywhere on any outermost bounding MOTS $S$ for which
$N_{S}<0$ and $W\neq 0$ everywhere.

Independently of whether $\vec{\xi}$ is a Killing vector or not, the more favorable case to obtain restrictions on the generator
$\vec{\xi}$ on a given outermost bounding MOTS is when the newly constructed surface $S_{\tau}$ is bounding and weakly outer trapped.
This is guaranteed for small enough $\tau$ when $\delta_{\vec{\nu}}\theta^{+}$ is
strictly negative everywhere, because then this first order terms becomes dominant for small enough $\tau$.
Due to the fact that the tangential part of $\vec{\nu}$ does not
affect the variation of $\theta^{+}$ along $\vec{\nu}$ for a MOTS (c.f. (\ref{variationtheta+general0})),
it follows that $\delta_{\vec
{\nu}}\theta^{+}=L_{\vec{m}}Q$, where $L_{\vec{m}}$ is the stability operator for MOTS defined in (\ref{stabilityoperator}).
Since the vector $\vec{\nu}=Q\vec{m}+\vec{Y}^{\parallel}$ determines to first order the direction to
which a point $\p\in S$ moves, it is clear that $L_{\vec{m}}Q<0$ everywhere and $Q>0$ somewhere is impossible
for an outermost bounding MOTS. This is precisely the argument we have used above and is intuitively very clear.
However, this geometric method does not provide the
most powerful way of finding this type of restriction. Indeed,
when the first order term $L_{\vec{m}}Q$ vanishes at some points, then
higher order coefficients come necessarily into play, which makes
the geometric argument of little use. It is remarkable that using the
elliptic results described in Appendix \ref{ch:appendix2}, most of
these situations can be treated in a satisfactory way.
Furthermore, since the elliptic methods only use infinitesimal
information, there is no need to restrict oneself to outermost
bounding MOTS, and the more general case of stable or strictly stable
MOTS (not necessarily bounding) can be considered.

Unfortunately, the general expression of $L_{\vec{m}}Q$ given in equation (\ref{stabilityoperator})
is not directly linked to the vector $\vec{\xi}$, which is clearly unsuitable for our aims.
In the case of
Killing vectors, the point of view of moving $\S$ along $\vec{\xi}$ and then back to
$\Sigma$ gives a simple method of calculating
$L_{\vec{m}} Q$. For more general vectors, however, the motion along $\vec{\xi}$
will give a non-zero contribution to $\theta^+$ which needs
to be computed (for Killing vectors this term was known to be zero
via a symmetry argument, not from
a direct computation). In order to do this, it becomes necessary
to have an alternative, and completely general, expression for
$\delta_{\vec{\xi}}\, \theta^+$ directly in terms of the deformation tensor $a_{\mu\nu}(\vec{\xi}\,)$
associated with $\vec{\xi}$. This is the aim of the following section.

\setcounter{equation}{0}
\section{Variation of the expansion and the metric deformation tensor}
\label{sc:A2variation}

Let us derive an identity for $\delta_{\vec{\xi}}\, \theta^+$
in terms of $a_{\mu\nu}(\vec{\xi}\,)$. This result will be important later on in this chapter,
and may also be of
independent interest. We derive this expression
in full generality, without assuming $\S$ to be a MOTS and for the expansion
$\theta_{\vec{\eta}}$ along any normal vector $\vec{\eta}$ of $\S$ (not necessarily a null normal) i.e.
\[
\theta_{\vec{\eta}}\equiv H_{\mu}\eta^{\mu},
\]
where $\vec{H}$ denotes the mean curvature of $S$ in $M$.

To do this calculation, we need to take derivatives of tensorial objects defined
on each one of $\S^{\prime}_{\tau}$. For a given point $\p \in \S$, these tensors live on different spaces,
namely the tangent spaces of $\varphi_{\tau} (\p)$, where $\varphi_{\tau}$ is the one-parameter local
group of diffeomorphisms generated by $\vec{\xi}$.
In order to define the variation, we need to pull-back all these tensors to the point $\p$ before doing the derivative.
We will denote the resulting derivative by $\mathscr{L}_{{\vec{\xi}}}$. In general, this operation is not the standard
Lie derivative $\mathcal{L}_{\vec{\xi}}$ on tensors because it is applied to tensorial objects on each $S^{\prime}_{\tau}$
which may not define tensor fields on $M$ (e.g. when these surfaces intersect each other).
Nevertheless, both derivatives do coincide when acting on spacetime tensor fields
(e.g. the metric $\gM$) 
which will simplify the calculation considerably.

Notice in particular that the definition of $\theta_{\vec{\eta}}$ depends on the choice of $\vec{\eta}$
on each of the surfaces $S^{\prime}_{\tau}$. Thus $\delta_{\vec{\xi}}\, \theta_{\vec{\eta}}
\equiv \left. \mathscr{L}_{\vec{\xi\,}} \theta_{\vec{\eta}}\right|_{S}$ will necessarily
include a term of the form $\mathscr{L}_{\vec{\xi}}\, \eta_{\alpha}$ which is not uniquely defined
(unless $\vec{\eta}$ can be uniquely defined on each $S^{\prime}_{\tau}$, which is usually not
the case). Nevertheless, for the
case of MOTS and when $\vec{\eta} = \vec{l}_{+}$ this a priori ambiguous term
becomes determined, as we will see. The general expression
for $\delta_{\vec{\xi}}\, \theta_{\vec{\eta}}$ is given in the following proposition.

\begin{proposition}\label{propositionxitheta} Let $\S$ be a surface on a spacetime $(M,\gM)$,
$\vec{\xi}$ a vector field defined on $M$ with deformation tensor $a_{\mu\nu}(\vec{\xi}\,)$ and
$\vec{\eta}$ a vector field normal to $\S$
and extend $\vec{\eta}$ to a smooth map $\vec{\eta}:(-\epsilon,\epsilon)\times S \rightarrow TM$
satisfying $\vec{\eta}(0,\p)=\vec{\eta}(\p)$ and $\vec{\eta}(\tau,\p)\in (T_{\varphi_{\tau}(\p)}S'_{\tau})^{\perp}$ where
$\varphi_{\tau}$ is the local group of diffeomorphisms generated by $\vec{\xi}$ and $S'_{\tau}=\varphi_{\tau}(S)$.
Then, the variation along $\vec{\xi}$
of the expansion $\theta_{\vec{\eta}}$ on $\S$ reads
\begin{eqnarray}\label{xithetau}
\delta_{\vec{\xi}}\,\theta_{\vec{\eta}}&=&
H^{\mu}\mathscr{L}_{\vec{\xi}}\,\eta_{\mu}-a_{AB}(\vec{\xi}\,)
\Pi^{AB}_{\mu}\eta^{\mu}\nonumber \\
&&\qquad\left .
+ \gamma^{AB}e_{A}^{\alpha}e_{B}^{\rho}\eta^{\nu}
    \left[ \frac{1}{2}\nabla_{\nu}a_{\alpha\rho}(\vec{\xi}\,) - \nabla_{\alpha}a_{\nu\rho}(\vec{\xi}\,)\right]
\right|_{\S},
\end{eqnarray}
where $\vec{\Pi}_{AB}$ denotes the second fundamental form vector of $S$ in $M$,
and $a_{AB}(\vec{\xi}\,)\equiv e_{A}^{\alpha}e_{B}^{\beta}a_{\alpha\beta}(\vec{\xi}\,)$, with $\{\vec{e}_{A}\}$ being
a local basis for $TS$.
\end{proposition}

{\bf Proof.} Since $\theta_{\vec{\eta}}=H_{\mu}\eta^{\mu}=\gamma^{AB}\Pi_{AB}^{\mu}\eta_{\mu}$, the
variation we need to calculate involves three terms
\begin{equation}\label{Lietheta}
\mathscr{L}_{\vec{\xi}}\,\theta_{\vec{\eta}}=\left(\mathscr{L}_{\vec{\xi}}\,\gamma^{AB}\right)
\Pi_{AB}^{\mu}\eta_{\mu} + \gamma^{AB}\left(\mathscr{L}_{\vec{\xi}}\,\Pi_{AB}^{\mu}\right)+ H^{\mu}\left(\mathscr{L}_{\vec{\xi}}\,\eta_{\mu}\right).
\end{equation}
In order to do the calculation, we will choose ${\varphi_{\tau}}_{\star}
(\vec{e}_A)$ as the basis of tangent vectors at $\varphi_{\tau}(\p) \in
S^{\prime}_{\tau}$ (we refer to ${\varphi_{\tau}}_{\star} (\vec{e}_A)$ merely
as $\vec{e}_A$ in the following to simplify the notation). This
entails no loss of generality and implies $\mathscr{L}_{\vec{\xi}}\,\vec{e}_{A}=0$,
which makes the calculation simpler. Our aim is to express each
term of (\ref{Lietheta}) in terms of $a_{\mu\nu}(\vec{\xi}\,)$. For the first
term, we need to calculate $\mathscr{L}_{\vec{\xi}}\,\gamma^{AB}$. We start with
$\mathscr{L}_{\vec{\xi}}\,\gamma_{AB}=\mathscr{L}_{\vec{\xi}} \left ( \gM (\vec{e}_A, \vec{e}_B )
\right ) = ( \mathscr{L}_{\vec{\xi}\,} g ) \left (\vec{e}_A,\vec{e}_B \right ) =
( \mathcal{L}_{\vec{\xi}\,} g ) \left (\vec{e}_A,\vec{e}_B \right ) =
a_{\mu\nu}(\vec{\xi}\,) e_{A}^{\mu}e_{B}^{\nu}\equiv a_{AB}(\vec{\xi}\,)$, which immediately
implies $\mathscr{L}_{\vec{\xi}\,}\gamma^{AB}=-a_{CD}(\vec{\xi}\,)\gamma^{AC}\gamma^{BD}$, so
that the first term in (\ref{Lietheta}) becomes
\begin{equation}\label{firstterm}
\mathscr{L}_{\vec{\xi}}\,\gamma^{AB} \Pi_{AB}^{\mu}\eta_{\mu}=
- a_{AB}(\vec{\xi}\,)\Pi^{AB}_ {\mu}\eta^{\mu}.
\end{equation}

The second term $\gamma^{AB}(\mathscr{L}_{\vec{\xi}}\Pi_{AB}^{\mu}) \eta_{\mu}$ is more complicated.
It is useful to introduce the projector to the normal space of $S$,
$h_{\nu}^{\mu}\equiv \delta^{\mu}_{\nu} - \gM_{\nu\beta}e_{A}^{\mu}e_{B}^{\beta}\gamma^{AB}$.
From the previous considerations, it follows that
$\mathscr{L}_{\vec{\xi}}\, h_{\nu}^{\mu}=e_{A}^{\mu}e_{B}^{\beta}
    (  a^{AB}(\vec{\xi}\,) \gM_{\nu\beta} - \gamma^{AB}a_{\nu\beta}(\vec{\xi}\,))$, which implies
$\left( \mathscr{L}_{\vec{\xi}\,} h_{\nu}^{\mu}  \right)\eta_{\mu}=0$ and hence
\begin{eqnarray}
\mathscr{L}_{\vec{\xi}}\, (\Pi^{\mu}_{AB} ) \eta_{\mu} = -
\mathscr{L}_{\vec{\xi}} \left ( h^{\mu}_{\nu} e_{A}^{\alpha}\nabla_{\alpha}e_{B}^{\nu} \right)
\eta_{\mu} = -
\eta_{\nu}\mathscr{L}_{\vec{\xi}}\left( e_{A}^{\alpha}\nabla_{\alpha}e_{B}^{\nu}\right).
\label{LiePi}
\end{eqnarray}

Therefore we only need to evaluate
$\mathscr{L}_{\vec{\xi}} \left( e_{A}^{\alpha}\nabla_{\alpha}e_{B}^{\nu}\right)$. It is well-known
that for an arbitrary
vector field $\vec{v}$, $\mathcal{L}_{\vec{\xi}}\nabla_{\alpha}v^{\nu}-\nabla_{\alpha}\mathcal{L}_{\vec{\xi}}\,v^{\nu} =
v^{\rho}\nabla_{\alpha}\nabla_{\rho}\xi^{\nu}+{\RM}^{\nu}_{\,\,\,\rho\sigma\alpha}v^{\rho}\xi^{\sigma}$.
However, this expression is not directly applicable to the
variational derivative we are calculating and we need the  following closely related lemma.
\begin{lema}\label{lema:Liee}
\begin{eqnarray}
\mathscr{L}_{\vec{\xi}}\left( {e}_A^{\alpha}\nabla_{\alpha}{e}_{B}^{\nu}\right)
={e}_A^{\alpha}{e}_{B}^{\rho}\nabla_{\alpha}\nabla_{\rho}\xi^{\nu}
+{\RM}^{\nu}_{\,\,\,\rho\sigma\alpha}{e}_{A}^{\alpha}{e}_{B}^{\rho}\xi^{\sigma}.
\label{Liee}
\end{eqnarray}
\end{lema}
{\bf Proof of Lemma \ref{lema:Liee}.}
Choose coordinates $y^A$ on $S$ and extend them as constants along $\vec{\xi}$.
This gives coordinates on each one of $S'_{\tau}$.
Define $e_{A}^{\alpha}=\frac{\partial x^{\alpha}}{\partial y^{A}}$, where $x^{\mu}(y^{A},\tau)$ are the embedding functions of
$S'_{\tau}$ in $M$ in spacetime coordinates $x^{\mu}$.
The map $\varphi_{-\tau}: M\rightarrow M$ relates every point $\p\in S_{\tau}$ with
coordinates $\{x^{\alpha}\}$ to a point $\varphi_{-\tau}(\p)\in S$ with coordinates
$\{\varphi_{-\tau}^{\,\alpha}(x^{\beta})\}$.
By definition,
$\mathscr{L}_{\vec{\xi}}(e^{\mu}_{A}\nabla_{\mu}e_{B}^{\nu})\equiv
\frac{d}{d\tau}\left( (\varphi_{-\tau})_{*}
(e_{A}^{\mu}\nabla_{\mu}e_{B}^{\nu}) \right)$. 
Using that $\frac{\partial \varphi_{-\tau}^{\,\alpha}(x^{\beta})}{\partial \tau}=-\xi^{\alpha}$, it is immediate to obtain
\begin{eqnarray*}
&&\left.\frac{d}{d\tau}\left( (\varphi_{-\tau})_{*}
(e_{A}^{\mu}\nabla_{\mu}e_{B}^{\nu}) \right)\right|_{\tau=0}=
\left.\frac{d}{d\tau}\left[ (e_{A}^{\mu}\nabla_{\mu}e_{B}^{\alpha}) \frac{\partial \varphi_{-\tau}^{\,\nu}}{\partial x^{\alpha}}
\right]\right|_{\tau=0}\\
&&\qquad =\frac{\partial}{\partial \tau} (e_{A}^{\mu}\nabla_{\mu}e_{B}^{\nu}(y^{C},\tau)) -\partial _{\alpha}\xi^{\nu}e_{A}^{\mu}\nabla_{\mu}e_{B}^{\alpha}\\
&&\qquad
=\frac{\partial}{\partial \tau}\left[ \frac{\partial^{2}x^{\nu}}{\partial y^{A}\partial y^{B}} +\Gamma_{\alpha\rho}^{\nu}\frac{\partial x^{\alpha}}{\partial y^{A}}\frac{\partial x^{\rho}}{\partial y^{B}}\right]
-\partial_{\mu}\xi^{\nu}\left[ \frac{\partial^{2}x^{\mu}}{\partial y^{A}\partial y^{B}} +\Gamma_{\alpha\rho}^{\mu}\frac{\partial x^{\alpha}}{\partial y^{A}}\frac{\partial x^{\rho}}{\partial y^{B}}\right].
\end{eqnarray*}
On the other hand,
\begin{eqnarray*}
&&{e}_A^{\alpha}{e}_{B}^{\rho}\nabla_{\alpha}\nabla_{\rho}\xi^{\nu}
    +{\RM}^{\nu}_{\,\,\,\rho\sigma\alpha}{e}_{A}^{\alpha}{e}_{B}^{\rho}\xi^{\sigma}\\
&&\qquad= \frac{\partial x^{\alpha}}{\partial y^{A}}\frac{\partial x^{\rho}}{\partial y^{B}}\left[
        \partial_{\alpha}
        \partial_{\rho}\xi^{\nu}+
        \Gamma_{\mu\rho}^{\nu}
        \partial_{\alpha}\xi^{\mu}+
        \Gamma_{\mu\alpha}^{\nu}
        \partial_{\rho}\xi^{\mu}
        -\Gamma_{\alpha\rho}
        ^{\mu}\partial_{\mu}\xi^{\nu}+  \xi^{\sigma}\partial_{\sigma}\Gamma_{\alpha\rho}^{\nu}
       \right]\\
&&\qquad= \frac{\partial^{3}x^{\nu}}{\partial \tau\partial y^{A} \partial y^{B}}-
    \frac{\partial^{2}x^{\rho}}{\partial y^{A}\partial y^{B}}\partial_{\rho}\xi^{\nu}+
    \frac{\partial x^{\rho}}{\partial y^{B}}\Gamma_{\mu\rho}^{\nu}\partial_{\tau}
        \left( \frac{\partial x^{\mu}}{\partial y^{A}} \right)+
    \frac{\partial x^{\alpha}}{\partial y^{A}}\Gamma_{\mu\alpha}^{\nu}\partial_{\tau}
        \left( \frac{\partial x^{\mu}}{\partial y^{B}} \right)\\
&&\qquad\qquad + \frac{\partial x^{\alpha}}{\partial y^{A}}\frac{\partial x^{\rho}}{\partial y^B}\left[
        \partial_{\tau}\Gamma_{\alpha\rho}^{\nu}-\Gamma_{\alpha\rho}^{\mu}\partial_{\mu}\xi^{\nu}\right]\\
&&\qquad =
\frac{\partial}{\partial \tau}\left[ \frac{\partial^{2}x^{\nu}}{\partial y^{A}\partial y^{B}} +\Gamma_{\alpha\rho}^{\nu}\frac{\partial x^{\alpha}}{\partial y^{A}}\frac{\partial x^{\rho}}{\partial y^{B}}\right]
-\partial_{\mu}\xi^{\nu}\left[ \frac{\partial^{2}x^{\mu}}{\partial y^{A}\partial y^{B}} +\Gamma_{\alpha\rho}^{\mu}\frac{\partial x^{\alpha}}{\partial y^{A}}\frac{\partial x^{\rho}}{\partial y^{B}}\right],\\
\end{eqnarray*}
where we have used
\[
{{\RM}^{\nu}}_{\rho\sigma\alpha}=\partial_{\sigma}\Gamma_{\rho\alpha}^{\nu}-\partial_{\alpha}\Gamma_{\rho\sigma}^{\nu} +\Gamma_{\gamma\sigma}^{\nu}\Gamma_{\rho\alpha}^{\gamma}-\Gamma_{\gamma\alpha}^{\nu}\Gamma_{\rho\sigma}^{\gamma},
\]
in the first equality and
$\xi^{\mu}=\frac{\partial x^{\mu}(y^{A},\tau)}{\partial \tau}$ in the second one.
This proves the lemma.
$\hfill \blacksquare$ \\

We can now continue with the proof of Proposition \ref{propositionxitheta}. It only remains to express the quantity
$\nabla_{\alpha}\nabla_{\rho}\xi^{\nu}
+{\RM}^{\nu}_{\,\,\,\rho\sigma\alpha}\xi^{\sigma}$ in terms of $a_{\mu\nu}(\vec{\xi}\,)$.
To that end, we take a derivative of $\nabla_{\nu}\xi_{\rho}+\nabla_{\rho}\xi_{\nu}=a_{\nu\rho}(\vec{\xi}\,)$ to get
\[
\nabla_{\alpha}\nabla_{\nu}\xi_{\rho}+\nabla_{\alpha}\nabla_{\rho}\xi_{\nu}=\nabla_{\alpha}a_{\nu\rho}(\vec{\xi}),
\]
and use
the Ricci identity $\nabla_{\alpha}\nabla_{\nu}\xi_{\rho}-\nabla_{\nu}\nabla_{\alpha}\xi_{\rho}=-{\RM}_{\sigma\rho\alpha\nu}\xi^{\sigma}$ to obtain
\[
\nabla_{\nu}\nabla_{\alpha}\xi_{\rho} + \nabla_{\alpha}\nabla_{\rho}\xi_{\nu}=
{\RM}_{\sigma\rho\alpha\nu}\xi^{\sigma} + \nabla_{\alpha}a_{\nu\rho}(\vec{\xi}\,).
\]
Now, write the three equations obtained from this one by cyclic
permutation of the three indices.
Adding two of them and subtracting the third one we find
\begin{eqnarray*}
\nabla_{\alpha}\nabla_{\rho}\xi_{\nu} &=&
\frac12({\RM}_{\sigma\rho\alpha\nu}+{\RM}_{\sigma\nu\rho\alpha}-{\RM}_{\sigma\alpha\nu\rho})\xi^{\sigma}\\
&&\qquad +
\frac12\left[ \nabla_{\alpha}a_{\nu\rho}(\vec{\xi}\,)+\nabla_{\rho}a_{\alpha\nu}(\vec{\xi}\,)-\nabla_{\nu} a_{\alpha\rho}(\vec{\xi}\,) \right].
\end{eqnarray*}
which, after using the
first Bianchi identity $\RM_{\sigma\rho\alpha\nu}+\RM_{\sigma\nu\rho\alpha}+\RM_{\sigma\alpha\nu\rho}=0$, leads to
\[
\nabla_{\alpha}\nabla_{\rho}\xi_{\nu} =
{\RM}_{\sigma\alpha\rho\nu}\xi^{\sigma}
+
\frac12\left[ \nabla_{\alpha}a_{\nu\rho}(\vec{\xi}\,)+\nabla_{\rho}a_{\alpha\nu}(\vec{\xi}\,)-\nabla_{\nu} a_{\alpha\rho}(\vec{\xi}\,) \right].
\]
Substituting (\ref{Liee}) and this expression into (\ref{LiePi}) yields
\begin{equation}\label{secondterm}
\gamma^{AB}\mathscr{L}_{\vec{\xi}}\,\Pi_{AB}^{\mu}\eta_{\mu}=
     \gamma^{AB}e_{A}^{\alpha}e_{B}^{\rho}\eta^{\nu}
    \left[ \frac{1}{2}\nabla_{\nu}a_{\alpha\rho}(\vec{\xi}\,) - \nabla_{\alpha}a_{\nu\rho}(\vec{\xi}\,)\right].
\end{equation}
Inserting (\ref{firstterm}) and (\ref{secondterm})
into equation (\ref{Lietheta})  proves the proposition.
$\hfill \blacksquare$ \\

We can now particularize
to the outer null expansion in a MOTS.
\begin{corollary}
\label{corollaryxitheta}
If $S$ is a MOTS then
\begin{eqnarray}\label{xitheta} \delta_{\vec{\xi}}\,\theta^{+}
&=&-\frac14\theta^{-}a_{\mu\nu}(\vec{\xi}\,)l_{+}^{\mu}l_{+}^{\nu}-a_{AB}(\vec{\xi}\,)\Pi^{AB}_{\mu}l_{+}^{\mu}\nonumber \\
&& \qquad
\left .
+ \gamma^{AB}e_{A}^{\alpha}e_{B}^{\rho} l_{+}^{\nu}
    \left[ \frac{1}{2}\nabla_{\nu}a_{\alpha\rho}(\vec{\xi}\,) - \nabla_{\alpha}a_{\nu\rho}(\vec{\xi}\,)\right]
\right|_{\S}.
\end{eqnarray}
\end{corollary}

{\bf Proof.}  The normal vector $\vec{l}_+^{\prime}(\tau)$ defined on each of the surfaces $S^{\prime}_{\tau}$ is null.
Therefore, using $\mathscr{L}_{\vec{\xi}} \, {\gM}^{\mu\nu}=\mathcal{L}_{\vec{\xi}} \, {\gM}^{\mu\nu} = - a^{\mu\nu}(\vec{\xi}\,)$,
\begin{eqnarray}
0 = \mathscr{L}_{\vec{\xi}} \left ( {l_{+}^{\prime}}_{\mu}(\tau) {l_{+}^{\prime}}_{\nu}(\tau) {\gM}^{\mu\nu} \right ) = 2 l_{+}^{\mu}
\mathscr{L}_{\vec{\xi}}
\,{l_{+}^{\prime}}_{\mu}(\tau)
- a_{\mu\nu}(\vec{\xi}\,) l_{+}^{\mu} l_{+}^{\nu}.
\label{null}
\end{eqnarray}
Since, on a MOTS  $\vec{H}=-\frac12 \theta^{-}\vec{l}_{+}$, it follows
$H^{\mu} \mathscr{L}_{\vec{\xi}}\, {l_{+}^{\prime}}_{\mu}(\tau) = -\frac{1}{2} \theta^{-} l_{+}^{\mu} \mathscr{L}_{\vec{\xi}} \, {l_{+}^{\prime}}_{\mu}(\tau) =
- \frac{1}{4} \theta^{-} a_{\mu\nu}(\vec{\xi}\,) l_{+}^{\mu} l_{+}^{\nu}$, and the corollary follows
from (\ref{xithetau}).  $\hfill \blacksquare$ \\

{\bf Remark.} 
Formula (\ref{xitheta}) holds in general for arbitrary surfaces $\S$
at any point where $\theta^{+}=0$. $\hfill \square$

\setcounter{equation}{0}
\section{Results provided $L_{\vec{m}} Q$ has a sign on $\S$}
\label{sc:A2LmQ}

In this section we will give several
results provided $L_{\vec{m}}Q$ has a definite sign on $S$. In this case,
a
direct application of
Lemma \ref{lemmaelliptic} for a MOTS $\S$ with stability operator
$L_{\vec{m}}$ leads to the following result.
\begin{lema}
\label{lemmaelliptic2}
Let $\S$ be a stable MOTS on a spacelike hypersurface $\Sigma$.
If $\left. L_{\vec{m}}Q\right|_{\S}\leq 0$ (resp. $\left.L_{\vec{m}}Q\right|_{\S}\geq 0$) and not identically zero,
then $\left. Q\right|_{\S}<0$ (resp. $\left. Q\right|_{\S}>0$).

Furthermore, if $\S$ is strictly stable and
$\left. L_{\vec{m}}Q\right|_{\S}\leq 0$ (resp. $\left.L_{\vec{m}}Q\right|_{\S}\geq 0$) then
$\left. Q\right|_{\S}\leq 0$ (resp. $\left. Q\right|_{\S}\geq 0$) and it vanishes at one
point only if it vanishes everywhere on $\S$.
\end{lema}

The general idea then is to combine Lemma
\ref{lemmaelliptic2} with the general calculation for the
variation of $\theta^+$ obtained in the previous section to get
restrictions on special types of generators $\vec{\xi}$ on a
stable or strictly stable MOTS.
Our first result is fully general in the sense that it is valid for any generator $\vec{\xi}$.
\begin{thr}
\label{TrhAnyXi}
Let $S$ be a stable MOTS on a spacelike hypersurface $\Sigma$ and $\vec{\xi}$ a vector field
on $S$ with deformation tensor $a_{\mu\nu}(\vec{\xi}\,)$. With the notation above, define
\begin{eqnarray}
Z &=& -\frac14\theta^{-}a_{\mu\nu}(\vec{\xi}\,)l_{+}^{\mu}l_{+}^{\nu}-a_{AB}(\vec{\xi}\,)\Pi^{AB}_{\mu}l_{+}^{\mu}\nonumber\\
&&
\qquad \left .
+ \gamma^{AB}e_{A}^{\alpha}e_{B}^{\rho} l_{+}^{\nu}
    \left[ \frac{1}{2}\nabla_{\nu}a_{\alpha\rho}(\vec{\xi}\,) - \nabla_{\alpha}a_{\nu\rho}(\vec{\xi}\,)\right]
+ N W \right|_{\S},
\label{Z}
\end{eqnarray}
where $W= \Pi _{AB}^{\mu}{\Pi^{\nu}}^{AB}{l_{+}}_{\mu}{l_{+}}_{\nu}+G_{\mu\nu}l_{+}^{\mu}l_{+}^{\nu}$, and assume $Z \leq 0$ everywhere on $S$.
\begin{itemize}
\item[(i)] If $Z \neq 0$ somewhere, then $\xi_{\mu}l_{+}^{\mu}  < 0$ everywhere.
\item[(ii)] If $S$ is strictly stable, then $\xi_{\mu}l_{+}^{\mu} \leq 0$
everywhere and vanishes at one point only if it vanishes everywhere.
\end{itemize}
\end{thr}

{\bf Proof.}
Consider the
first variation of $S$ defined by the vector $\vec{\nu} = \vec{\xi} - N_S \vec{l}_{+} =
Q \vec{m} + \vec{Y}^{\parallel}$. From equation (\ref{variationtheta+general0}) and Definition \ref{defi:stabilityoperator}
we have
$\delta_{\vec{\nu}}\, \theta^{+} = L_{\vec{m}} Q$. On the other hand, linearity of this variation under addition gives
$\delta_{\vec{\nu}}\, \theta^{+} = \delta_{\vec{\xi}}\, \theta^{+} - \delta_{N_S \vec{l}_{+}} \theta^{+}$.
The Raychaudhuri equation for MOTS establishes that $\delta_{N_S \vec{l}_{+}} \theta^{+} = - N_S W$ (see (\ref{raychaudhuri}) and (\ref{W})) and the
identity (\ref{xitheta}) gives $L_{\vec{m}} Q = Z$. Since $Q = \xi_{\mu}l_{+}^{\mu}$,
the result follows directly from Lemma \ref{lemmaelliptic2}. $\hfill \blacksquare$ \\

{\bf Remark.} The theorem also holds if all the inequalities are reversed. This follows
directly by replacing $\vec{\xi} \rightarrow - \vec{\xi}$. $\hfill \square$ \\

This theorem gives information about the relative position between the generator
$\vec{\xi}$ and the outer null normal $\vec{l}_{+}$ and has, in principle, many potential consequences.
Specific applications require considering spacetimes
having special vector fields for which sufficient information about its deformation tensor is available.
Once such a vector is known to exist, the result above can be used either to restrict the
form of $\vec{\xi}$ in stable or strictly stable MOTS or, alternatively, to
restrict the regions of the spacetime where such MOTS are allowed to be present.

Since conformal vector fields (and homotheties and isometries as particular cases) have very special deformation
tensors, the theorem above gives interesting information for spacetimes admitting such symmetries.

\begin{corollary}\label{thrstable} Let $\S$ be a
stable MOTS in a hypersurface $\Sigma$ of a spacetime $(M,\gM)$ which admits
a
conformal Killing vector $\vec{\xi}$,  $\mathcal{L}_{\vec{\xi}} \gM_{\mu\nu} = 2 \phi \gM_{\mu\nu}$ (including
homotheties $\phi=C$, and isometries $\phi=0$).
\begin{itemize}
\item[(i)] If
$2 \vec{l}_{+} (\phi) + N W |_{S} \leq 0$ and
not identically zero, then $\xi_{\mu}l_{+}^{\mu} |_S <0$.
\item[(ii)] If $S$ is strictly stable and
$2 \vec{l}_{+} (\phi) + N W |_{S} \leq 0$ then
$\xi_{\mu}l_{+}^{\mu} |_S \leq 0$ and vanishes at one point only if it vanishes everywhere
\end{itemize}
\end{corollary}

{\bf Remark 1.} As before, the theorem is still true if all inequalities are reversed.
$\hspace*{1cm} \hfill \square$ \\

\begin{figure}
\begin{center}
\psfrag{p}{$\p$}
\includegraphics[width=9cm]{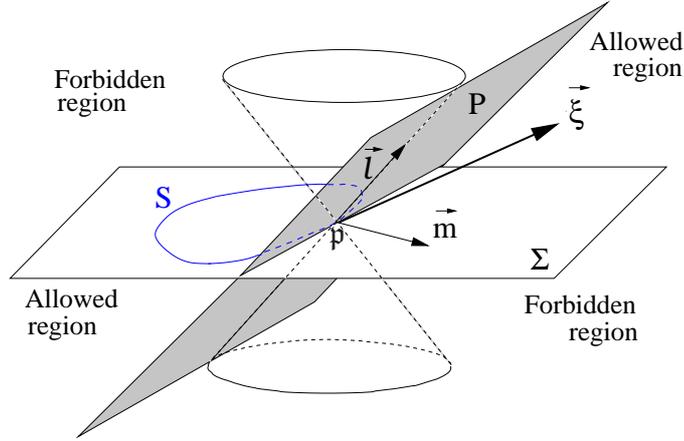}
\caption {The planes $T_{\p} \Sigma$ and $P\equiv T_{\p} S\oplus \mbox{span}\{
\vec{l}_{+}\, |_{\p} \}$ divide the tangent space $T_{\p}M$
in four regions. By Corollary \ref{thrstable}, if $S$ is strictly
stable and $\vec{\xi}$ is a Killing vector or a homothety in a
spacetime satisfying the NEC which points above $\Sigma$
everywhere, then $\vec{\xi}$ cannot enter into the forbidden
region at any point (and similarly, if $\vec{\xi}$ points below
$\Sigma$ everywhere). The allowed region includes the plane $P$.
However, if there is a point with $W \neq 0$ where $\vec{\xi}$ is
not tangent to $\Sigma$, then the result is also valid for stable
MOTS with $P$ belonging to the forbidden region.}
\label{fig:fig1}
\end{center}
\end{figure}

{\bf Remark 2.} In the case of homotheties and Killing vectors, the
condition of the theorem demands that $N_S W \leq 0$. Under the
NEC, this holds provided $N_S \leq 0$, i.e. when $\vec{\xi}$
points below $\Sigma$ everywhere  on $S$ (where the term ``below''
includes also the tangential directions). For strictly stable $S$,
the conclusion of the theorem is that the homothety or the Killing
vector must lie above the null hyperplane defined by the tangent
space of $\S$ and the outer null normal $\vec{l}_{+}$ at each point $\p
\in S$. If the MOTS is only assumed to be stable, then the theorem
requires the extra condition that $\vec{\xi}$ points strictly
below $\Sigma$ at some point with $W \neq 0$. In this case, the
conclusion is stronger and forces $\vec{\xi}$ to lie strictly
above the null hyperplane everywhere. By changing the orientation
of $\vec{\xi}$, it is clear that similar restrictions arise when
$\vec{\xi}$ is assumed to point {\it above} $\Sigma$. Figure \ref{fig:fig1}
summarizes the allowed and forbidden regions for $\vec{\xi}$ in
this case. $\hfill \square$ \\

{\bf Proof.}
We only need to show that
$Z = 2 \vec{l}_{+} (\phi) + N W |_{S}$ for
conformal Killing vectors. This follows at once from (\ref{Z}) and
$a_{\mu\nu}(\vec{\xi}\,) = 2 \phi \gM_{\mu\nu}$ after using orthogonality of $\vec{e}_A$ and
$\vec{l}_{+}$. Notice in particular that $Z$ is the same for isometries and for
homotheties.
$\hfill \blacksquare$ \\

This corollary has an interesting consequence in spacetime regions where there exists
a Killing vector or a homothety $\vec{\xi}$ which is causal everywhere.

\begin{corollary}
\label{shear} Let a  spacetime $(M,\gM)$ satisfying the NEC admit a
causal Killing vector or homothety $\vec{\xi}$ which is future
(or past) directed everywhere on a stable MOTS $\S \subset \Sigma$.
Then,
\begin{itemize}
\item[(i)] The second fundamental form $\Pi^{+}_{AB}$ along $\vec{l}_{+}$ (i.e. $\Pi^{+}_{AB}\equiv \Pi_{AB}^{\mu}{l_{+}}_{\mu}$)
and $\GM_{\mu\nu}l_{+}^{\mu}l_{+}^{\nu}$ vanish identically on every point $\p \in S$
where $\vec{\xi}|_{\p} \neq 0$.
\item[(ii)] If $S$ is strictly stable, then $\vec{\xi} \propto \vec{l}_{+}$
everywhere.
\end{itemize}

\end{corollary}

{\bf Remark.} If we assume that there exists an open neighbourhood of $\S$ in $M$
where the Killing vector or homothety $\vec{\xi}$ is causal and future (or past)
directed everywhere then the conclusion (i) can be generalized to say that
$\Pi^{+}_{AB}$ and $G_{\mu\nu}l_{+}^{\mu}l_{+}^{\nu}$ vanish identically on $S$.
The reason  is that such a $\vec{\xi}$ cannot vanish anywhere in this neighbourhood  (and
consequently neither on $S$). For Killing vectors this result is proven
in Lemma 3.2 in \cite{BEM}\footnote{We thank Miguel S\'anchez Caja for pointing this out.}.
A simple generalization shows that
the same holds for homotheties, as follows.
Suppose that $\vec{\xi}\,|_{\p\in S}=0$. Take a timelike
affine-parametrized geodesic $\gamma$ passing through $\p$ with
future directed unit tangent vector $\vec{v}$.
A simple computation gives that, if $\vec{\xi}$ is a homothety with constant $C$,
$v^{\mu}\nabla_{\mu}(\xi_{\nu}v^{\nu})=-C$. Supposing $C>0$, this
implies that the causal vector $\vec{\xi}$ is future directed on the future of $\p$
and past directed on the past of $\p$ contradicting the fact that $\vec{\xi}$ is future (past)
directed everywhere on a neighbourhood of $\S$ in $M$. A similar argument works if $C<0$.

Point (ii) can be generalized to locally outermost MOTS using a finite construction. We will
prove this in Theorem \ref{corollaryextended} below. $\hfill \square$ \\

{\bf Proof.}
We can assume, after reversing the sign of $\vec{\xi}$ if necessary,
that $\vec{\xi}$ is past directed, i.e. $N_S  \leq 0$.

Under the NEC, $W$ is the sum of two non-negative terms, so in
order to prove (i) we only need to show that $W =0$ on points
where $\vec{\xi} \neq 0$, i.e. at points where $N_S <0$. Assume,
on the contrary, that $W \neq 0$ and $N_S <0$ happen
simultaneously at a point $\p \in S$. It follows that $N_S W \leq
0$ everywhere and non-zero at $\p$. Thus, we can apply statement
(i) of Corollary \ref{thrstable} to conclude $Q <0$ everywhere.
Hence $N_S Q \geq 0$ and not identically zero on $S$. Recalling
the decomposition $\vec{\xi} = N_S \vec{l}_{+} + Q \vec{m} +
\vec{Y}^{\parallel}$, the squared  norm of this vector is
\begin{eqnarray}
\label{normsquare}
\xi_{\mu}\xi^{\mu} = 2N_{\S}Q+Q^2+
{{Y}^{\parallel}}_{\mu}{{Y}^{\parallel}}^{\mu}.
\end{eqnarray}
This is the sum of non-negative terms, the
first one not identically zero. This contradicts the condition of $\vec{\xi}$ being causal.

To prove the second statement, we notice that point (ii) in
Corollary \ref{thrstable} implies $Q \leq 0$, and hence $N_S Q
\geq 0$. The only way (\ref{normsquare}) can be negative or zero
is if $Q=0$ and  $\vec{Y}^{\parallel} =0$, i.e. $\vec{\xi} \propto
\vec{l}_{+}$.
$\hfill \blacksquare$ \\

This corollary extends Theorem 2 in \cite{MS} to the case of
stable MOTS and implies, for instance, that any
strictly stable MOTS in a plane wave spacetime (which by definition
admits a null and nowhere zero Killing vector field $\vec{\xi}\,$)
must be aligned with the direction of propagation of the wave (in the
sense that $\vec{\xi}$ must be one of the null normals to the surface).
It also implies that any spacetime admitting a nowhere zero and causal Killing vector (or homothety) whose energy-momentum tensor
satisfies the DEC and does not admit a null eigenvector cannot contain
any stable MOTS. This is because $\GM_{\mu\nu}l_{+}^{\mu}l_{+}^{\nu}=0$ and the DEC implies
$\GM_{\mu\nu}l_{+}^{\mu}\propto l_{\nu}$ and $\GM_{\mu\nu}$ would have a null eigenvector.
For perfect fluids this result holds even without the DEC provided $\mu+ p\neq 0$ (this is because in this case
$\GM_{\mu\nu}l_{+}^{\mu}l_{+}^{\nu}=(\mu+p)(l_{+}^{\mu}u_{\mu})^{2}\neq 0$ -- where $\mu$
is the density, $p$ the pressure and $\vec{u}$ is the 4-velocity of the fluid--).

The results above hold for stable or strictly stable MOTS. Among such surfaces,
marginally trapped surfaces are of special interest. Our next result restricts (and in some cases
forbids) the existence of such surfaces in spacetimes admitting Killing vectors, homotheties
or conformal Killings.

\begin{thr}
\label{thrfirstvariation} Let $\S$ be a stable MOTS in a spacelike
hypersurface $\Sigma$ of a spacetime $(M,\gM)$ which satisfies the
NEC and admits a conformal Killing vector $\vec{\xi}$ with
conformal factor $\phi \geq 0$ (including homotheties with
$C\geq0$ and Killing vectors). Suppose furthermore that either (i)
$( 2\vec{l}_{+}(\phi)+NW  ) |_S \not \equiv 0$ or (ii) $S$ is strictly
stable and $\xi_{\mu}l_{+}^{\mu}|_S \not \equiv 0$. Then
the following holds.
\begin{itemize}
\item[(a)]
If $ 2\vec{l}_{+}(\phi)+NW  |_S \leq 0$  then
$S$ cannot be a marginally trapped surface, unless $\vec{H} \equiv 0$. The latter
case is excluded if $\phi|_S \not \equiv 0$.
\item[(b)] If $ 2\vec{l}_{+}(\phi)+NW |_S \geq 0$ then
$S$ cannot be a past marginally trapped surface, unless $\vec{H} \equiv 0$. The latter
case is excluded if $\phi|_S \not \equiv 0$.
\end{itemize}
\end{thr}

{\bf Remark.} The statement obtained from this one by reversing all the inequalities is also true.
This is a direct consequence of the freedom in changing
$\vec{\xi} \rightarrow - \vec{\xi}$. $\hfill \square$ \\

{\bf Proof.} We will only prove case (a). The argument for case
(b) is similar. The idea is taken from \cite{MS} and consists of
performing a variation of $S$ along the conformal Killing vector
and evaluating the change of area in order to get a contradiction
if $S$ is marginally trapped. The difference is that here
we do not make any a priori assumption on the causal character for
$\vec{\xi}$. Corollary \ref{thrstable} provides us with sufficient
information for the argument to go through.

The first variation of area (\ref{firstvariation}) gives
\begin{equation}\label{variationofareaMOTS}
\delta_{\vec{\xi}\,} |S|=-\frac12\int_{S}\theta^{-} \xi_{\mu}l_{+}^{\mu}
\eta_{\S},
\end{equation}
where we have used $\vec{H} = -
\frac{1}{2} \theta^{-} \vec{l}_{+}$.  Now, since $2\vec{l}_{+}(\phi)+NW  |_S \leq 0$, and
furthermore either hypothesis (i) or (ii) holds,  Corollary
\ref{thrstable} implies that $\xi_{\mu}{l}_{+}^{\mu} |_S < 0$.

On the other hand, $\vec{\xi}$ being a conformal Killing vector,
the induced metric on $S^{\prime}_{\tau}$ is related to the metric on
$S$ by conformal rescaling.
A simple computation gives $\delta_{\vec{\xi}\,}{\eta_{S}}= \frac12 \gamma^{AB}(\mathcal{L}_{\vec{\xi}\,}g)(\vec{e}_{A},\vec{e}_{B})\eta_{S}$
(see e.g. \cite{MS}), which for the particular case of conformal Killing vectors gives the following.
\begin{equation}\label{variationofareaconformalKilling}
\delta_{\vec{\xi}\,} |S|=2\int_S \phi\eta_S,
\end{equation}
This quantity is non-negative due to $\phi\geq 0$ and not identically zero if $\phi \neq 0 $ somewhere.
Combining (\ref{variationofareaMOTS}) and (\ref{variationofareaconformalKilling}) we conclude that
if $\theta^{-} \leq 0 $ (i.e. $S$ is marginally trapped) then necessarily $\theta^{-}$ vanishes identically
(and so does $\vec{H}$). Furthermore, if $\phi |_S$ is non-zero somewhere, then $\theta^{-}$ must necessarily
be positive somewhere, and $S$ cannot be marginally trapped. $\hfill \blacksquare$ \\

\subsection{An application: No stable MOTS in Friedmann-Lema\^itre-Robertson-Walker spacetimes}
\label{ssc:A2sectionFLRW}

In this subsection we apply Corollary \ref{thrstable} to show that
a large subclass of Friedmann-Lema\^itre-Robertson-Walker (FLRW)
spacetimes do not admit stable MOTS on {\it any} spacelike
hypersurface. Obtaining this type of results for metric
spheres only requires a straightforward calculation, and is
therefore simple. The power of the method is that it provides a
general result involving no assumption on the geometry of the MOTS
or on the spacelike hypersurface where it is embedded. The only
requirement is that the scale factor and its time derivative
satisfy certain inequalities. This includes, for instance all FLRW
cosmological models satisfying the NEC with accelerated expansion, as
we shall see in Corollary \ref{corollaryFLRW} below.

Recall that the FLRW metric is
\begin{eqnarray*}
\gM_{FLRW}=-dt^2+a^{2}(t)\left[ dr^2+\chi^{2}(r;k)d\Omega^{2} \right],
\end{eqnarray*}
where $a(t)>0$ is the scale factor
and $\chi(r;k)= \{\sin{r}, r, \sinh r \}$ for $k=\{1,0,-1\}$, respectively.
The Einstein tensor of this metric is of perfect fluid type
and reads
\begin{eqnarray}
&&\GM_{\mu\nu} = (\mu + p) u_{\mu} u_{\nu} + p \gM_{\mu\nu}, \quad \vec{u} = \partial_t, \quad
\mu = \frac{3(\dot{a}^2(t)+k)}{a^2(t)}, \label{EM_FLRW0}\\
&&\qquad\qquad \mu + p =  2\left ( \frac{\dot{a}^2(t)+k}{a^2(t)}-
\frac{\ddot{a}(t)}{a(t)} \right)
\label{EM_FLRW}
\end{eqnarray}
where dot stands for derivative with respect to $t$.
\begin{thr}\label{thrFRW}
There exists no stable MOTS in any spacelike
hypersurface of a FLRW spacetime $(M,\gM_{FLRW})$ satisfying
\begin{equation}\label{conditionFLRW}
\frac{{\dot{a}}^{2}(t)+k}{a(t)} \geq 0 , \quad
-\frac{\dot{a}^{2}(t)+k}{a(t)}\leq
\ddot{a}(t)\leq \frac{{\dot{a}}^{2}(t)+k}{a(t)}.
\end{equation}
\end{thr}

{\bf Remark.} In terms of the energy-momentum contents of the
spacetime, these three conditions read, respectively, $\mu \geq
0$, $\mu \geq 3 p$ and $\mu + p \geq 0$. As an example, in the
absence of a cosmological constant they are satisfied as soon as
the weak energy condition is imposed and the pressure is not too large
(e.g. for the matter and radiation dominated eras). The class of
FLRW satisfying (\ref{conditionFLRW}) is clearly very large (c.f.
Corollary \ref{corollaryFLRW} below). We also remark that Theorem
\ref{thrFRW} agrees with the fact that the causal
character of the hypersurface which separates the trapped from the
non-trapped {\it spheres} in FLRW spacetimes depends precisely on
the quantity $\mu^2(\mu+p)(\mu-3p)$ (c.f. \cite{S97}). $\hfill \square$ \\

{\bf Proof.} The FLRW spacetime admits a conformal Killing vector
$\vec{\xi}=a(t) \vec{u}$ with conformal factor $\phi=\dot{a}(t)$.
Since this vector is timelike and future directed, it follows
that  $\xi_{\mu}l_{+}^{\mu}|_S<0$
for any spacelike surface $S$ embedded in a spacelike hypersurface $\Sigma$.
If we can show that
$2\vec{l}_{+}(\phi)+NW|_S \geq 0$,
and non-identically zero for any $S$, then the sign reversed of point (i) in Corollary \ref{thrstable}
implies that $S$ cannot be a stable MOTS, thus proving the
result. The proof therefore relies on finding
conditions on the scale factor which imply the validity of this inequality on any $S$.
First of all, we notice that the second fundamental form
$\Pi^+_{AB}$ can be made as small as desired on
a suitably chosen $S$. Thus, recalling that $W={\Pi^{+}}_{AB}{\Pi^{+}}^{AB}+\GM_{\mu\nu}l_{+}^{\mu}l_{+}^{\nu}$, it is clear that
the inequality that needs to be satisfied is
\begin{equation}\label{conditioncorollary2}
\left. 2\vec{l}_{+}(\phi)+N \GM_{\mu\nu}l_{+}^{\mu}l_{+}^{\nu}\right|_S \geq 0,
\end{equation}
and positive somewhere. In order to evaluate this expression
recall that $\vec{u} = a^{-1} \vec{\xi} = a(t)^{-1} N \vec{n} + a(t)^{-1} \vec{Y}$.
Let us write
$\vec{Y} = Y \vec{e}$, where $\vec{e}$ is unit and let $\alpha$ be the hyperbolic angle of $\vec{u}$ in the
basis $\{\vec{n}, \vec{e}\, \}$, i.e. $\vec{u} = \cosh \alpha \, \vec{n} + \sinh \alpha \,\vec{e}$. It follows immediately
that $N = a(t) \cosh \alpha$ and $Y = a(t) \sinh \alpha$. Furthermore, multiplying $\vec{u}$ by the normal
vector to the surface $S$ in $\Sigma$ we find $u_{\mu}m^{\mu}= \cos \varphi \sinh \alpha$, where
$\varphi$ is the angle between $\vec{m}$ and $\vec{e}$. With this notation, let us calculate
the null vector $\vec{l}_{+}$. Writing $\vec{l}_{+} = A \vec{u}  + \vec{b}$, with
$\vec{b}$ orthogonal to $\vec{u}$, it follows $b_{\mu}b^{\mu} = A^2$ from the condition of
$\vec{l}_{+}$ being  null. On the other hand we have the decomposition
$A \vec{u} + \vec{b} = \vec{l}_{+} = \vec{n} + \vec{m}$. Multiplying by $\vec{u}$ we immediately get
$A =  \cosh \alpha - \cos \varphi \sinh \alpha$, and, since $\phi = \dot{a}(t)$ only depends on $t$,
\begin{equation}\label{l(phi)FRW}
\vec{l}_{+}(\phi)=\left( \cosh{\alpha}-\cos{\varphi}\sinh{\alpha} \right)\ddot{a}(t).
\end{equation}
The following expression for  $\GM_{\mu\nu}l_{+}^{\mu}l_{+}^{\nu}$ follows directly
from $\vec{l}_{+} = A \vec{u} + \vec{b}$ and (\ref{EM_FLRW0}),
(\ref{EM_FLRW}),
\begin{eqnarray}
\GM_{\mu\nu}l_{+}^{\mu}l_{+}^{\nu}&=&A^{2}(\mu+ p )\nonumber \\
\label{gll} &=& 2\left(
\cosh{\alpha}-\cos{\varphi}\sinh{\alpha} \right)^2 \left (
\frac{\dot{a}^2(t)+k}{a^2(t)}-\frac{\ddot{a}(t)}{a(t)} \right).
\end{eqnarray}
Inserting  (\ref{l(phi)FRW}) and (\ref{gll}) into
(\ref{conditioncorollary2}) and dividing by $2 A^2 \cosh \alpha$
(which is positive) we find the equivalent condition
\begin{equation}\label{conditioncorollary2c}
\left (
\frac{1}{\cosh{\alpha}\left(\cosh{\alpha}-\cos{\varphi}\sinh{\alpha}\right)}-1
\right )\ddot{a}(t)+\frac{\dot{a}^{2}(t)+k}{a(t)} \geq 0,
\end{equation}
and non-zero somewhere. The dependence on $S$ only arises
through the function
$f(\alpha,\varphi) = \cosh \alpha ( \cosh \alpha - \cos \varphi \sinh \alpha)$.
Rewriting this as $f = 1/2 ( 1 + \cosh (2 \alpha) - \cos \varphi \sinh (2\alpha) )$
it is immediate to show that $f$ takes all values in $(1/2, +\infty)$. Hence,
$\left[ \cosh{\alpha}\left(\cosh{\alpha}-\cos{\varphi}\sinh{\alpha}\right)\right]^{-1} -1 $ takes all values between
$-1$ and $1$.
In order to satisfy (\ref{conditioncorollary2c}) on all this
range, it is necessary and sufficient that the three inequalities in
(\ref{conditionFLRW}) are satisfied. $\hfill \blacksquare$ \\

The following corollary gives a particularly interesting case
where all the conditions of Theorem \ref{thrFRW}
are satisfied.

\begin{corollary}
\label{corollaryFLRW} Consider a FLRW spacetime $(M,\gM_{FLRW})$
satisfying the NEC. If $\ddot{a}(t)>0$, then there exists no
stable MOTS in any spacelike hypersurface of $(M,\gM_{FLRW})$
\end{corollary}

{\bf Proof.}
The null energy condition gives  $0 \leq \mu+p =2\left (
\frac{\dot{a}^2(t)+k}{a^2(t)} -\frac{\ddot{a}(t)}{a(t)} \right)$. This implies the first and third
inequalities in (\ref{conditionFLRW}) if $\ddot{a}>0$. The remaining condition
$- \frac{\dot{a}^2(t)+k}{a(t)} \leq \ddot{a}$ is also obviously satisfied provided $\ddot{a} >0$.
$\hfill \blacksquare$ \\

\subsection{A consequence of the geometric construction of $S_{\tau}$}
\label{ssc:A2sectiongeometric}

We have emphasized at the beginning of this section that the restrictions obtained directly
by the geometric procedure
of moving $S$ along $\vec{\xi}$ and then back to $\Sigma$  are intuitively
clear but typically weaker than those obtained by using elliptic theory results.
There are some cases, however, where
the reverse  actually holds, and the geometric construction provides stronger results.
We will present
one of these cases in this subsection.

Corollary \ref{thrstable} gives restrictions on ${\xi}_{\mu}{l}_{+}^{\mu} |_S$ for Killing vectors and homotheties in spacetimes
satisfying the NEC, provided $\vec{\xi}$ is future or past
directed everywhere. However, when $W$ vanishes identically, the
result only gives useful information in the strictly stable case.
The reason is that $W \equiv 0$ implies $L_{\vec{m}} Q \equiv 0$ and, for
marginally stable MOTS (i.e. when the principal eigenvalue of $L_{\vec{m}}$ vanishes), the maximum
principle is not strong enough to conclude that $Q$ must have a
sign. There is at least one case where marginally stable MOTS
play an important role, namely after a jump in the outermost MOTS
in a (3+1) foliation of the spacetime (see \cite{AMMS} for
details). As we will see next, the geometric construction does
give restrictions in this case even when $W$ vanishes identically.



\begin{thr}\label{thrkilling} Consider a spacetime $(M,\gM)$ possessing a
Killing vector or a homothety $\vec{\xi}$ and satisfying the NEC.
Suppose $M$ contains a compact spacelike hypersurface $\Sigmatilde$
with boundary consisting in the disjoint union of a weakly outer trapped surface $\bd^{-}
\Sigmatilde$ and an outer untrapped
surface $\bd^{+} \Sigmatilde$ (neither of which are necessarily connected) and take $\bd^{+}\Sigmatilde$ as a barrier with interior
$\Sigmatilde$. Without loss of generality, assume that $\Sigmatilde$ is
defined locally by a level function $T=0$ with $T>0$ to the future
of $\Sigmatilde$
and let $S$ be the outermost
MOTS which is bounding with respect to $\bd^{+}\Sigmatilde$.
If $\vec{\xi}(T) \leq 0 $ on some spacetime
neighbourhood of $S$, then $\xi^{\mu}{l}_{\mu}^{+} \leq 0$
everywhere on $\S$.
\end{thr}

{\bf Remark 1.} As usual, the theorem still holds if all the inequalities involving
$\vec{\xi}$ are reversed. $\hfill \square$ \\


{\bf Remark 2.} The simplest way to ensure that $\vec{\xi}(T)\leq 0 $ on some neighbourhood
of $S$ is by imposing a condition merely on $\S$, namely ${\xi}_{\mu}{n}^{\mu} |_S > 0 $,
because then $\vec{\xi}$ lies
strictly below $\Sigmatilde$  on $S$ and this property is obviously preserved sufficiently near $S$ (i.e.
$\vec{\xi}$ points strictly below the level set of $T$ on a sufficiently small spacetime neighbourhood of $S$).
We prefer imposing directly the condition $\vec{\xi}(T)\leq 0$ on a spacetime neighbourhood of $S$ because this allows
$\vec{\xi}\,|_{S}$ to be tangent to $\Sigma$. $\hfill \square$ \\

{\bf Proof.} First note that the hypersurface $\Sigmatilde$ satisfies the assumptions of Theorem \ref{thr:AM} which
ensures that an outermost MOTS $S$ which is bounding with respect to $\bd^{+}\Sigmatilde$
does exist and, therefore, no weakly outer trapped surface can penetrate in its
exterior region.
Then, the idea is precisely to use the geometric procedure described above to construct $S_{\tau}$
and use the fact that $S$ is the outermost bounding MOTS to conclude that $S_{\tau}$ (${\tau}>0$) cannot have points
outside $S$. Here we move $S$ a small but finite amount $\tau$, in contrast
to the elliptic results before, which only involved infinitesimal displacements. We want to have information
on the sign of the outer expansion of $S_{\tau}$ in order to make sure that a weakly outer trapped surface forms.
The first part of the displacement is along $\vec{\xi}$ and gives $S^{\prime}_{\tau}$.
Let us first see that all these surfaces are MOTS. For Killing vectors,
this follows at once from symmetry arguments. For homotheties 
we have the identity
\begin{equation}\label{noindependentterm}
\delta_{\vec{\xi}}\,\theta^{+}= \left ( -\frac{1}{2}
l_{-}^{\alpha}\mathscr{L}_{\vec{\xi}}\,
{l_{+}^{\prime}}_{\alpha}(\tau)
-2C
\right )\theta^{+},
\end{equation}
which follows directly from (\ref{xithetau}) with $\vec{\eta} = \vec{l}_{+}$ after using
$l_{+}^{\mu} \mathscr{L}_{\vec{\xi}} \, {l_{+}^{\prime}}_{\mu}(\tau)
 = \frac{1}{2} a_{\mu\nu}(\vec{\xi}\,) l_{+}^{\mu} l_{+}^{\nu} = 0$, see
(\ref{null}). Expression (\ref{noindependentterm}) holds for each
one of the surfaces $\{S'_{\tau}\}$, independently of them being MOTS
or not. Since this variation vanishes on MOTS and the starting
surface $S$ has this property, it follows that each surface
$S^{\prime}_{\tau}$ ($\tau>0$) is also a MOTS. Moving back to $\Sigmatilde$
along the null hypersurface introduces, via the Raychaudhuri
equation (\ref{raychaudhuri}), a non-positive term $N_S W$ in the outer null expansion,
provided the motion is to the future. Hence, $S_{\tau}$ for small but
finite ${\tau}>0$ is a weakly outer trapped surface provided
$\vec{\xi}$ moves to the past of $\Sigmatilde$. This is ensured if
$\vec{\xi} (T) \leq 0$ near $S$, because $T$ cannot become
positive for small enough $\tau$. On the other hand, since a point $\p
\in S$ moves initially along the vector field $\nu = \vec{\xi} -
N_S \vec{l}_{+} = Q \vec{m} + \vec{Y}^{\parallel}$, where
$Q={\xi}_{\mu}l_{+}^{\mu}$ as usual, it follows that $Q>0$
somewhere  implies (for small enough $\tau$) that the bounding weakly outer
trapped surface $S_{\tau}$ has a portion lying strictly to the outside
of $S$ which, due to Theorem \ref{thr:AM} by Andersson and Metzger, is a contradiction to $S$ being the outermost
bounding MOTS. Hence
$Q\leq 0$ everywhere and the theorem is proven. $\hspace*{1cm} \hfill \blacksquare$ \\

It should be remarked that the assumption of $\vec{\xi}$ being a Killing vector or a homothety is important
for this result. Trying to generalize it for instance to conformal Killings fails in general because
then the right hand side of
equation (\ref{noindependentterm}) has an additional term $2\vec{l}_{+}(\phi)$, not proportional to
$\theta^{+}$. This means that moving a MOTS along
a conformal Killing does not lead to another MOTS in general. The method can however, still give useful information
if $\vec{l}_{+} (\phi)$ has the appropriate sign, so that $S^{\prime}_{\tau}$ is in fact weakly
outer trapped. We omit the details.

An immediate consequence of the finite construction of $S_{\tau}$ 
is the extension of point (ii) of Corollary \ref{shear} to locally outermost MOTS.

\begin{thr}\label{corollaryextended}
Let $(M,\gM)$ be a spacetime satisfying NEC and admitting a causal Killing vector or homothety $\vec{\xi}$
which is future (past) directed on a locally outermost MOTS $\S \subset \Sigma$. Then $\vec{\xi}\propto
\vec{l}_{+}$ everywhere on $\S$.
\end{thr}

{\bf Proof.} As before, let
$\Sigma$ be defined locally by a level function $T=0$ with $T>0$ to the future of
$\Sigma$. Assume that $\vec{\xi}$ is past directed (the future directed case is similar). Then, the
assumption $\vec{\xi}(T) \leq 0 $
on some spacetime neighbourhood
of $S$ of Theorem \ref{thrkilling}
is automatically satisfied. Then we can use the finite construction therein
to find a weakly outer trapped surface which, due to the fact that $\vec{\xi}$ is causal (and
past directed), does not penetrate in the interior part of the two-sided neighbourhood of $\S$.
In fact, this new trapped surface will have points strictly outside $\S$ if on some point of $\S$
$\vec{\xi}\not\propto \vec{l}_+$ which proves the result. $\hfill \blacksquare$ \\

Finally, Theorem \ref{corollaryextended} together with Theorem \ref{thr:AM} lead to the following result.

\begin{thr}\label{theorem1}
Consider a spacelike hypersurface $\id$ possibly with boundary in a spacetime satisfying
the NEC and possessing a Killing vector or a homothety $\vec{\xi}$ with squared norm $\xi_{\mu}\xi^{\mu}=-\lambda$. 
Assume that $\Sigma$ possesses a barrier $\Sb$ with interior $\Omegab$ 
which is outer untrapped with respect to the direction pointing outside of $\Omegab$.

Consider any surface $S$ which is bounding with respect to $\Sb$.
Let us denote by $\Omega$ the exterior of $S$ in $\Omegab$.
If $S$ is weakly outer trapped and 
${\Omega}\subset \{\lambda>0\}$,
then $\lambda$ cannot be strictly positive on any point $\p\in S$.
\end{thr}

{\bf Remark.} 
When {\it weakly outer trapped surface} is replaced by the stronger condition
of being a {\it weakly trapped surface with non-vanishing mean curvature}, then
this theorem can be proven by a simple argument based on the first variation of area \cite{MS}.
In that case, the assumption of $S$ being bounding becomes unnecessary. It would be
interesting to know if Theorem \ref{theorem1} holds for arbitrary weakly outer trapped surfaces,
not necessarily bounding. $\hfill \square$ \\

{\bf Proof.}
We argue by contradiction. Suppose a weakly outer trapped surface $S$ satisfying the assumptions of the theorem and
with $\lambda>0$ at some point.
Theorem \ref{thr:AM} implies that an outermost MOTS $\tbd T^{+}$ which is bounding
with respect to $\Sb$ exists in the closure of the exterior $\Omega$ of $S$ in $\Omegab$.
In particular, $\tbd T^{+}$ is a locally outermost MOTS.
The hypothesis $\Omega\subset\{\lambda>0\}$ implies that the vector
$\vec{\xi}$ is causal everywhere on $\tbd T^{+}$, either future or
past directed. Moreover, the fact that $\lambda>0$ on some point of $S$
implies that the Killing vector is timelike in
some non-empty set of $\tbd T^{+}$, which contradicts Theorem \ref{corollaryextended}.
$\hfill \blacksquare$ \\

\begin{figure}
\begin{center}
\psfrag{S}{\color{blue}{$S$}}
\psfrag{Sb}{{$S_{b}$}}
\psfrag{lambda}{$\{\lambda>0\}$}
\psfrag{Sigma}{$\Sigma$}
\psfrag{Ob}{$\Omega$}
\psfrag{O}{\color{blue}{$\Omega$}}
\includegraphics[width=10cm]{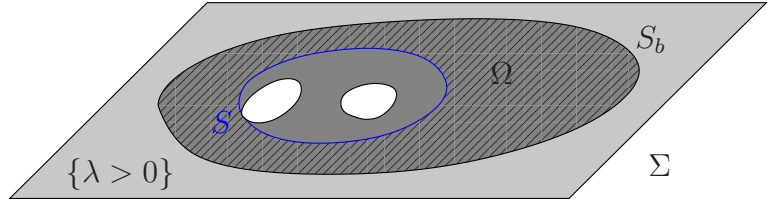}
\caption {Theorem \ref{theorem1} excludes the possibility pictured in this figure, where $S$ (in blue) is a weakly outer trapped
surface which is bounding with respect to the outer trapped barrier $\Sb$. The grey (both light and dark) regions represent
the region where $\lambda>0$. The
dark grey region represents the interior of $S_{b}$,
while the striped area corresponds to $\Omega$, which is the exterior of $S$ in $\Omega_{b}$.}
\label{fig:theorem1}
\end{center}
\end{figure}

The following result is a particularization of Theorem \ref{theorem1} to the case when the hypersurface
$\Sigma$ possesses an asymptotically flat end.

\begin{thr}
Let $\id$ be a spacelike hypersurface in a spacetime satisfying the NEC and possessing a
Killing vector or homothety $\vec{\xi}$. Suppose that $\Sigma$ possesses an asymptotically flat
end $\Sigma_{0}^{\infty}$.

Consider any bounding surface $S$ (see Definition \ref{defi:bounding}).
Let us denote by $\Omega$ the exterior of $S$ in $\Sigma$.
If $S$ is weakly outer trapped and
${\Omega}\subset{\{\lambda>0\}}$,
then $\lambda$ cannot be strictly positive on any point $\p\in S$.
\end{thr}

{\bf Proof.} The result is a direct consequence of Theorem \ref{theorem1}.
$\hfill \blacksquare$ \\


Two immediate corollaries follow.

\begin{corollary}\label{corollary}
Consider a spacelike hypersurface $\id$ in a spacetime satisfying the NEC and
possessing a Killing vector or a homothety $\vec{\xi}$.
Assume that $\Sigma$ has a selected asymptotically flat end $\Sigma_{0}^{\infty}$ and $\lambda>0$
everywhere on $\Sigma$.
Then
there exists no bounding weakly outer trapped surface in $\Sigma$.
\end{corollary}

\begin{corollary}\label{corollary2}
Let $\id$ be a spacelike hypersurface of the Minkowski spacetime.
Then there exists no bounding weakly outer trapped surface in $\Sigma$.
\end{corollary}

The second Corollary is obviously a particular case
of the first one because the vector $\partial_t$
in Minkowskian coordinates is strictly stationary everywhere, in particular on $\Sigma$.
The non-existence result of a bounding weakly outer trapped surface in a Cauchy surface
of Minkowski spacetime is however, well-known
as this spacetime is obviously regular predictable (see \cite{HE} for definition) and then the
proof of Proposition $9.2.8$ in \cite{HE} gives the result.

So far, all the results we have obtained require that the quantity $L_{m}Q$ does not change
sign on the MOTS $\S$. In the next section we will relax this condition.

\setcounter{equation}{0}
\section{Results regardless of the sign of $L_{\vec{m}} Q$}
\label{sc:A2sectionnonelliptic}

When $L_{\vec{m}} Q$ changes sign on $\S$, the elliptic methods exploited
in the previous section lose their power. Moreover, for
sufficiently small $\tau$, the surface $\S_{\tau}$ defined by the
geometric construction above necessarily fails to be weakly outer
trapped. Thus, obtaining restrictions in this case becomes a much
harder problem.

However, for locally outermost MOTS $S$,
an interesting situation arises when $\S_{\tau}$ lies partially outside $S$ and
happens to be weakly outer trapped in that exterior region.
More precisely, if a connected component of the subset of $S_{\tau}$ which lies outside
$S$ turns out to have non-positive outer null expansion, then using a smoothing result
by Kriele and Hayward \cite{KH97}, we will be able to construct a new weakly outer trapped surface outside
$S$, thus leading to a contradiction with the fact that $S$ is locally
outermost (or else giving restrictions on the
generator $\vec{\xi}\, $).

The result by Kriele and Hayward states, in rough terms,
that given two surfaces which intersect on a curve, a new smooth surface
can be constructed lying outside the previous ones in such a way that the
outer null expansion does not increase in the process. The precise statement is as follows.

\begin{lema}[Kriele, Hayward, 1997 \cite{KH97}]
\label{lemasmoothness}
Let $S_{1},S_{2}\subset \Sigma$ be
smooth two-sided surfaces which intersect transversely on a smooth
curve $\gamma$. Suppose that the exterior regions of $S_1$ and $S_2$ are properly defined in $\Sigma$ and let
$U_{1}$ and $U_{2}$ be respectively tubular neighbourhoods of $S_{1}$ and $S_{2}$ and $U^{-}_{1}$ and $U^{-}_{2}$ their interior parts.
Assume it is possible to choose one connected
component of each set $S_{1}\setminus
\gamma$ and $S_{2}\setminus\gamma$, say $S_{1}^{+}$ and $S_{2}^{+}$ respectively,
such that
$S_{1}^{+}\cap U^{-}_{2}=\emptyset$ and $S^{+}_{2}\cap U^{-}_{1} =\emptyset$.
Then, for any neighbourhood $V$ of $\gamma$ in $\Sigma$
there
exists a smooth surface $\tilde{S}$ and a continuous and piecewise
smooth bijection $\Phi\colon S_{1}^{+}\cup S_{2}^{+}\cup \gamma\rightarrow
\tilde{S}$ such that
\begin{enumerate}
\item $\Phi(\p)=\p$,  $\forall \p\in\left( S_{1}^{+} \cup S_{2}^{+}\right)\setminus
V$
\item $\left.{\theta}^{+}[\tilde{S}]\right|_{\Phi(\p)}\leq
\left.{\theta}^{+}[S_{A}^{+}]\right|_{\p}$ $\forall \p\in
S_{A}^{+}$ ($A=1,2$). 
\end{enumerate}
Moreover $\tilde{S}$ lies in the connected component
of $V\setminus\left( S^{+}_1\cup S^{+}_2\cup\gamma \right)$
lying in the exterior regions of both $S_1$ and $S_2$.
\end{lema}

\begin{figure}
\begin{center}
\psfrag{g}{$\gamma$}
\psfrag{S1}{{$S_1$}}
\psfrag{S2}{$S_2$}
\psfrag{S1+}{\color{blue}{$S_{2}^{+}$}}
\psfrag{S2+}{\color{blue}{$S_{1}^{+}$}}
\psfrag{U1}{{$U_{1}$}}
\psfrag{U2}{$U_{2}$}
\psfrag{U1+}{$U_{1}^{-}$}
\psfrag{U2+}{{$U_{2}^{-}$}}
\psfrag{S}{\color{red}{$\tilde{S}$}}
\includegraphics[width=10cm]{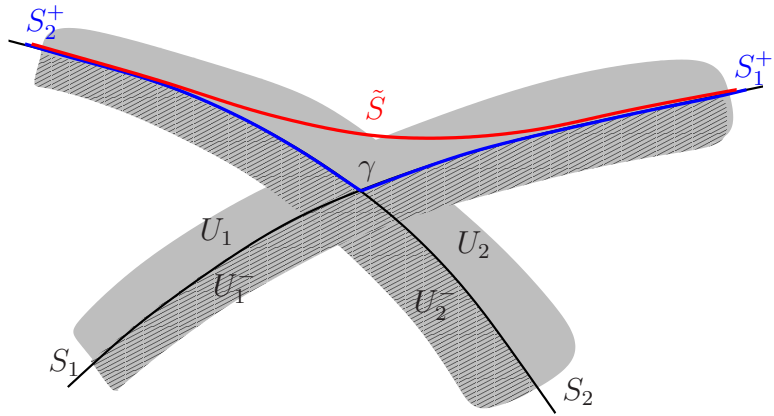}
\caption {The figure represents the two surfaces $S_{1}$ and $S_{2}$ which intersects in a curve $\gamma$,
(where one dimension has been suppressed). The two intersecting
grey regions are the tubular neighbourhoods $U_1$ and $U_2$ and, inside them,
the stripped regions represents their interior parts, $U_{1}^{-}$ and $U_{2}^{-}$. The sets $S_{1}^{+}$ and $S_{2}^{+}$, in blue color,
are then taken to
be the connected components of $S_{1}\setminus \gamma$ and $S_{2}\setminus \gamma$ which do not intersect
$U_{2}^{-}$ and $U_{1}^{-}$, respectively. Finally, the red line represents the smooth surface $\tilde{S}$ which has smaller $\theta^{+}$ than
$S_{1}$ and $S_{2}$..
}
\label{fig:KH}
\end{center}
\end{figure}

{\bf Remark.} It is important to emphasize that the statement of this result is slightly different from the
one appearing in the original paper \cite{KH97} by Kriele and Hayward.
Indeed, the assumptions made in \cite{KH97} are
rather ambiguous and restrictive in the sense that
the outer normals of $S_{1}$ and $S_{2}$ are required to form an angle (defined only by a figure),
not smaller than 90 degrees. This condition is not necessary for the lemma to work.
This result also appears quoted in \cite{AM} where
the assumptions are wrongly formulated (although the result is properly used throughout the paper).
In our paper \cite{CM2}, where Lemma \ref{lemasmoothness} is also formulated,
the hypotheses are incomplete as well. $\hfill \square$ \\ 

This result will allow us to adapt the arguments above without having to assume
that  $L_{\vec{m}} Q$ has a constant sign on $S$. The argument will be again by contradiction,
i.e. we will assume a locally outermost MOTS $S$ and, under suitable
circumstances, we will be able to find a new weakly outer trapped
surface lying outside $S$. Since the conditions are much weaker than in the previous section,
the conclusion is also weaker. It is, however, fully general in the sense that it holds for
any vector field $\vec{\xi}$ on $S$. Recall that $Z$ is defined in equation (\ref{Z}).

\begin{thr}\label{thrnonelliptic}
Let $S$ be a locally outermost MOTS in a
spacelike hypersurface $\Sigma$ of a spacetime $(M,\gM)$. Denote by $\U_0$
a connected component of
the set $\{\p\in\S ; \xi_{\mu}{l}_{+}^{\mu} |_{\p}>0 \}$. Assume
$U_0\neq \emptyset$ and that its boundary $\gamma\equiv\tbd U_0$ is either empty, or it satisfies
that the function $\xi_{\mu}{l}_{+}^{\mu}$ has a non-zero gradient everywhere on $\gamma$, i.e.
$d(\xi_{\mu}{l}_{+}^{\mu}) |_{\gamma} \neq 0$.

Then, there exists a point $\p\in\overline{U_0}$ such that $\left.Z\right|_{\p}\geq 0$.
\end{thr}

{\bf Proof.}  As mentioned, we will use a contradiction argument. Let us therefore assume that
\begin{equation}\label{conditionsubset}
Z|_{\p} <0, \quad \forall {\p} \in \overline{\U_0}.
\end{equation}
The aim is to construct a weakly outer trapped surface near $S$ and outside of it. This will contradict
the condition of $S$ being locally outermost.

First of all we observe that $Z$ cannot be negative everywhere on $S$, because
then Theorem \ref{TrhAnyXi}
(recall that outermost MOTS are
always stable) would imply $Q\equiv(\xi_{\mu}{l}_{+}^{\mu}) <0$
everywhere and $U_0$ would be
empty against hypothesis. Consequently, under (\ref{conditionsubset}),
$U_0$ cannot coincide with $S$ and $\gamma\equiv \tbd \U_0 \neq \emptyset$. Since
$\left.Q\right|_{\gamma}=0$ and, by assumption,
$\left. dQ \right|_{\gamma}\neq 0$ it follows that
$\gamma$ is a smooth embedded curve.
Taking $\mu$ to be a local coordinate on $\gamma$, it is clear that
$\{\mu,Q\}$ are coordinates of a neighbourhood of $\gamma$ in $S$.
We will coordinate a small enough neighbourhood of $\gamma$ in $\Sigma$
by Gaussian coordinates $\{ u,\mu,Q\}$ such that $u=0$ on $S$ and $u>0$
on its exterior.

\begin{figure}
\begin{center}
\includegraphics[width=10cm]{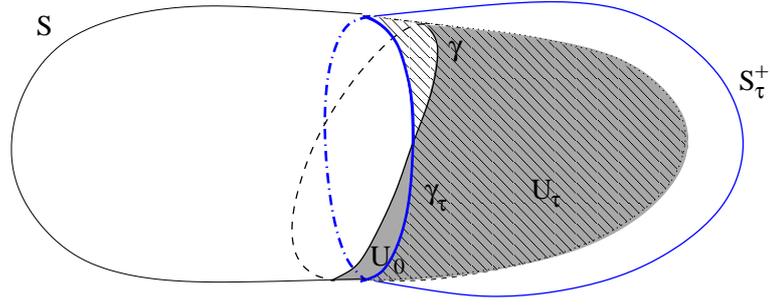}
\caption {The figure represents both intersecting surfaces $S$ and $S^{+}_{\tau}$ together with the curves
$\gamma$ and $\gamma_{\tau}$. The shaded region corresponds to $U_0$ and the stripped region to $U_{\tau}$.}
\label{fig:St}
\end{center}
\end{figure}

By moving $S$ along $\vec{\xi}$ a finite but small parametric amount ${\tau}$
and back to $\Sigma$ with the outer null geodesics, as described in
Section \ref{sc:A2sectionbasics}, we construct a family of surfaces $\{ S_{\tau} \}_{\tau}$.
The curve that each point $\p \in S$ describes via this construction has tangent
vector $\vec{\nu} = Q\vec{m}+\vec{Y}^{\parallel} |_{\S}$
on $S$. In a small neighbourhood of $\gamma$, the normal component of this vector, i.e.
$Q \vec{m}$, is smooth and only vanishes on $\gamma$. This implies that
for small enough $\tau$, $S_{\tau}$ are graphs over $S$ near $\gamma$. We will always
work on this neighbourhood, or suitable restrictions thereof.
In the Gaussian coordinates above, this graph is of the form $\{ u=\hat{u}(\mu,Q,\tau),\mu,Q \}$.
Since the normal unit vector to $S$ is simply
$\vec{m} = \partial_u$ in these coordinates and the normal component of $\vec{\nu}$ is $Q \vec{m}$,
the graph function $\hat{u}$ has the
following Taylor  expansion
\begin{equation}\label{graph}
\hat{u}(\mu,Q,\tau)=Q\tau+O(\tau^2).
\end{equation}
Our next aim is to use this expansion to conclude that
the intersection of $S$ and $S_\tau$ near
$\gamma$ is an embedded curve $\gamma_\tau$ for all small enough $\tau$. To do that
we will apply the implicit function theorem for functions to the equation $\hat{u}=0$. It is useful
to introduce a new function
$v(\mu,Q,\tau)=\frac{\hat{u}(\mu,Q,\tau)}{\tau}$, which is still smooth (thanks to
(\ref{graph})) and vanishes
at $\tau=0$ only on the curve $\gamma$. Moreover, its derivative with respect to
$Q$
is nowhere zero on $\gamma$, in fact
$\left.\frac{\partial
v}{\partial Q}\right|_{(\mu,0,0)}=1$ for all $\mu$.
The implicit function theorem implies that there
exist a unique function $Q=\varphi(\mu,\tau)$ which solves the equation
$v(\mu,Q,\tau)=0$, for small enough $\tau$. Obviously, this function is also the
unique solution near $\gamma$
of $\hat{u}(\mu,Q,\tau)=0$ for $\tau>0$. Consequently,
the intersection of $S$ and $S_{\tau}$ ($\tau>0$)
lying in the
neighbourhood of $\gamma$ where we are working on is an embedded curve $\gamma_{\tau}$.
Since $\gamma$ separates $S$ into two or more connected components, the same is true
for $\gamma_{\tau}$ for small enough $\tau$ (note that $\gamma$ need not be connected and
the number of connected components of $S\setminus \gamma$ may be bigger than two).
Recall that $\gamma$ is the boundary of a connected set $\U_0$. Hence,
by construction, there is only one connected component
of $S_{\tau}\setminus \gamma_{\tau}$ which has
$v (\mu,Q,{\tau}) >0$ near
$\gamma$ (i.e. that lies in the exterior of $S$ near $\gamma$). Let us denote it by $S_{\tau}^{+}$.
$S^{+}_{\tau}$
in fact lies fully outside of $S$, not just in a neighborhood of $\gamma$, as we see next.
First of all, note that $Q>0$ on $\U_{0}$. 
We have just seen that $\gamma_{\tau}$ is a continuous
deformation of $\gamma$. Let us denote by $U_{\tau}$ the domain in $S$
obtained by deforming $U_0$ when the boundary moves from $\gamma$ to $\gamma_{\tau}$ (See Figure \ref{fig:St}).
It is obvious that $S_{\tau}^+$ is obtained by moving $U_{\tau}$ first along $\vec{\xi}$
an amount $\tau$ and then back to $\Sigma$ by null hypersurfaces. The closed subset of $U_{\tau}$
lying outside the tubular neighbourhood
where we applied the implicit function theorem is, by construction
a proper subset of $U _0$.
Consequently, on this closed set $Q$ is uniformly bounded below
by a positive constant. Given that $Q$ is the first order term of the normal variation,
all these points move outside of $S$. This proves that $S^+_{\tau}$ is fully outside $S$ for
sufficiently small $\tau$. Incidentally this also shows that $S^+_{\tau}$ is a graph over $U_{\tau}$.

The next aim is to show that the outer null expansion
of $\S_{\tau}$ is non-positive everywhere on $\S^{+}_{\tau}$.
To that aim, we will prove that, for small enough $\tau$, $Z$ is strictly negative
everywhere on $U_{\tau}$. Since $Z$ is the first order term in the variation of $\theta^{+}$, this
implies that the outer null expansion of $S^+_{\tau}$ satisfies $\theta^+[S^{+}_{\tau}] <0$ for $\tau>0$
small enough.

By assumption (\ref{conditionsubset}), $Z$ is strictly negative on
$U_0$. Therefore, this quantity is automatically negative in the
portion  of $U_{\tau}$ lying in $U_0$ (in particular, outside the
tubular neighbourhood where we applied  the implicit function
theorem). The only difficulty comes from the fact that $\gamma_{\tau}$
may move outside $U_0$ at some points and we only have information
on the sign of $Z$ on $\overline{U_0}$. To address this issue, we
first notice that $Q$ defines a distance function to $\gamma$
(because $Q$ vanishes on $\gamma$ and its gradient is nowhere
zero). Consequently, the fact that $Z$ is strictly negative on
$\gamma$ (by assumption (\ref{conditionsubset})) and that this
curve is compact imply that there exists a $\delta>0$ such that,
inside the tubular neighbourhood of $\gamma$, $|Q|<\delta$ implies
$Z <0$. Moreover, the function $Q = \varphi(\mu,\tau)$, which defines
$\gamma_{\tau}$, is such that it vanishes at $\tau=0$ and depends smoothly
on $\tau$. Since $\mu$ takes values on a compact set, it follows that
for each $\delta^{\prime}>0$, there exists an
$\epsilon(\delta^{\prime})>0$, independent of $\mu$ such that $|\tau|
< \epsilon(\delta^{\prime})$ implies $|Q| = |\varphi(\mu,\tau)| <
\delta^{\prime}$. By taking $\delta^{\prime} = \delta$, it follows
that, for $|\tau| < \epsilon (\delta)$, $U_{\tau}$ is contained in a
$\delta$-neighbourhood of $U_0$ (with respect to the distance
function $Q$) and consequently $Z<0$ on this set, as claimed. We
restrict to $0 < \tau < \epsilon(\delta)$ from now on.

Summarizing, so far we have shown that $\S_{\tau}^{+}$ lies fully
outside $\S$ and has $\theta^{+}[S_{\tau}^{+}]<0$. The final task is to
use Lemma \ref{lemasmoothness} to construct a weakly outer trapped
surface strictly outside $\S$.
Denote by $S_{\tau}^{*}$ the complement of $U_{\tau}$ in $S$, which may
have several connected components. 
For any connected component $\gamma_{\tau}^{i}$ of $\gamma_{\tau}$ there exists
a neighbourhood $W^{*}_{\tau ,i}$ of $\gamma_{\tau}^{i}$ in $S_{\tau}^{*}\subset S$
which lies in the exterior of $S_{\tau}$ (because the intersection between $S$ and $S_{\tau}$ is transverse).
Similarly, there is a connected neighbourhood $W^{+}_{\tau ,i}$ of $\gamma_{\tau}^{i}$
in $S_{\tau}^{+}\subset S_{\tau}$ which lies in the exterior of $S$.
The smoothing argument of Lemma \ref{lemasmoothness} can be therefore applied locally on each
union $W^{*}_{\tau, i}\cup \gamma_{\tau}^{i}\cup W^{+}_{\tau,i}$ to produce a weakly outer trapped surface $\tilde{S}$
which lies outside $S$, leading a contradiction. This surface $\tilde{S}$ is constructed in such a way that
$\tilde{S}= S^{*}_{\tau}$ in $S^{*}_{\tau}\setminus \left(\underset{i}{\cup}W^{*}_{\tau, i}\right)$ and
$\tilde{S}=S^{+}_{\tau}$ in $S^{+}_{\tau}\setminus \left(\underset{i}{\cup}W^{+}_{\tau, i}\right)$.
$\hfill \blacksquare$ \\



{\bf Remark.} As usual, this theorem also holds if all the inequalities are reversed.
Note that in this case $U_0$ is defined to be a connected component of the set
$\{\p\in\S; (\xi_{\mu}{l}_{+}^{\mu}) |_{\p}<0 \}$. For the proof simply take $\tau<0$
instead of $\tau>0$ (or equivalently move along $-\vec{\xi}$ instead of $\vec{ \xi}$). $\hfill \square$ \\


Similarly as in the previous section, this theorem can be particularized to the case
of conformal Killing vectors, as follows (recall that $Z=2\vec{l}_{+}(\phi)+NW$ in the conformal Killing case, see Corollary \ref{thrstable}).

\begin{corollary}\label{corollarynonelliptic}
Under the assumptions of Theorem \ref{thrnonelliptic}, suppose that $\
\vec{\xi}$ is a conformal Killing vector with conformal factor $\phi$
(including homotheties $\phi=C$ and isometries $\phi=0$).

Then, there exists $\p\in\overline{U_0}$ such that
$2\vec{l}_{+}(\phi)+N_{S}({\Pi^{+}_{AB}}{\Pi^{+}}^{AB}+\GM_{\mu\nu}l_{+}^{\mu}l_{+}^{\nu})
|_{\p}\geq 0$.
\end{corollary}

If the conformal Killing is in fact a homothety or a Killing
vector and it is causal everywhere, the result can be strengthened
considerably. The next result extends Corollary \ref{shear} in a
suitable sense to the cases when the generator is not assumed to
be either future or past everywhere. Since its proof requires an
extra ingredient we write it down as a theorem.

\begin{thr}\label{shear2}
In a spacetime $(M,\gM)$ satisfying the NEC and admitting a Killing
vector or homothety  $\vec{\xi}$, consider a locally outermost
MOTS $\S$ in a spacelike hypersurface $\Sigma$. Assume that
$\vec{\xi}$ is causal on $S$ and that
$W={\Pi^{+}_{AB}}{\Pi^{+}}^{AB}+\GM_{\mu\nu}l_{+}^{\mu}l_{+}^{\nu}$ is non-zero everywhere
on $\S$. Define $U\equiv\{\p\in\S ;
(\xi_{\mu}{l}_{+}^{\mu})|_{\p}>0\}$ and assume that this set is
neither empty nor covers all of $S$. Then, on each connected
component $U_{\alpha}$ of $U$ there exist a point $\p\in\tbd U_{\alpha}$
with $d(\xi_{\mu}{l}_{+}^{\mu})|_{\p}=0$.
\end{thr}

{\bf Remark 1.} The same conclusion holds on the boundary of each connected components of the set $\{
\p\in\S ; (\xi_{\mu}{l}_{+}^{\mu})|_{\p}<0\}$. This is obvious since $\vec{\xi}$ can be changed
to $-\vec{\xi}$. $\hfill \square$ \\

{\bf Remark 2.} The case $\tbd U=\emptyset$, excluded by assumption in this theorem, can only occur
if $\vec{\xi}$  is future or past everywhere on $S$. Hence, this case is already
included in  Corollary \ref{shear}. $\hfill \square$ \\

{\bf Proof.} We first show that on any point in $U$ we have $N_S < 0$, which
has as an immediate consequence that $N_S \leq 0$ on any point in $\overline{U}$.
The former statement is a consequence of the decomposition
$\vec{\xi}=N\vec{l}_{+}+Q\vec{m}+\vec{Y}^{\parallel}$, where
$Q=(\xi_{\mu}{l}_{+}^{\mu})$. The condition that $\vec{\xi}$ is causal then implies
$\xi_{\mu}\xi^{\mu}=2N_{\S}Q+Q^{2}+{Y^{\parallel}}^{2} \leq 0$. This can only happen at a point where
$Q>0$ (i.e. on $U$) provided $N_S < 0$ there. Moreover, if at any point $\q$ on the boundary $\tbd U$
we have $N_S |_{\q}=0$, then necessarily the full vector $\vec{\xi}$ vanishes at this point. This implies,
in particular, that the geometric construction of $S_{\tau}$ has the property that $\q$ remains invariant.

Having noticed these facts, we will now argue by contradiction,
i.e. we will assume that there exists a connected component
$U_{0}$ of $U$ such that $d(\xi_{\mu}{l}_{+}^{\mu}) |_{\tbd
U_0}\neq 0$ everywhere. In these circumstances, we can follow the
same steps as in the proof of Theorem \ref{thrnonelliptic} to show
that, for small enough $\tau$ the surface $S_{\tau}$ has a portion
$S^{+}_{\tau}$ lying in the exterior of $S$ and which, in the Gaussian
coordinates above, is a graph over a subset $U_{\tau}$ which is a
continuous deformation of $U_0$. Moreover, the boundary of $U_{\tau}$
is a smooth embedded curve $\gamma_{\tau}$. The only difficulty with
this construction is that we cannot use $N_S W = Z  <0$ everywhere
on $\overline{U}_0$, in order to conclude that $\theta^{+}[S^{+}_{\tau}]
< 0$, as we did before. The reason is that there may be points on
$\tbd U_0$ where $N_S=0$. However, as already noted, these
points have the property that {\it do not move at all} by the
construction of $S_{\tau}$, i.e. the boundary $\gamma_{\tau}$ (which is
the intersection of $S$ and $S^+_{\tau}$) can only move outside of
$U_0$ at points  where $N_S$ is strictly negative. Hence on the
interior points of $U_{\tau}$ we have $N_S <0$ everywhere, for
sufficiently small $\tau$. Consequently the first order terms in the
variation of $\theta^{+}$, namely $Z = N_s W$,  is strictly negative
on all the interior points of $U_{\tau}$. This implies that $S^{+}_{\tau}$
has negative outer null expansion everywhere except possibly on
its boundary $\gamma_{\tau}$. By continuity, we conclude
$\theta^{+}[S_{\tau}^{+}] \leq 0$ everywhere.
The smoothing argument of the proof of Theorem \ref{thrnonelliptic} implies that
a smooth weakly outer trapped surface can be constructed outside the
locally outermost MOTS $S$. This gives a contradiction. Therefore,
there exists $\p\in\tbd U_0$ such that
$d(\xi_{\mu}{l}_{+}^{\mu})|_{\p}=0$, as claimed. $\hfill \blacksquare$ \\

{\bf Remark} The assumption $\left. dQ \right|_{ \gamma}\neq 0$ is a technical requirement for
using the smoothing argument of
Lemma  \ref{lemasmoothness}.  This is why we had to
include an assumption on $d Q |_{\gamma}$ in Theorem \ref{thrnonelliptic}  and also that the conclusion
of Theorem \ref{shear2} is stated in terms of the existence of critical points for $Q$.
If Lemma \ref{lemasmoothness} could be strengthened so as to remove
this requirement, then Theorem \ref{shear2} could be rephrased as stating that any
outermost MOTS in a region where there is a causal Killing vector (irrespective
of its future or past character) must have at least one point where the shear and
$\GM_{\mu\nu}l_{+}^{\mu}l_{+}^{\nu}$ vanish simultaneously.

In any case, the existence of critical points for a function in the boundary of {\it every}
connected component of $\{ Q >0 \}$ and {\it every} connected
component of $\{ Q <0 \}$ is obviously a highly non-generic situation. So, locally outermost
MOTS in regions where there is a causal Killing vector or homothety can at most occur under
very exceptional circumstances. $\hfill \square$ \\

\chapter{Weakly outer trapped surfaces in static spacetimes}
\label{ch:Article1}

\setcounter{equation}{0}
\section{Introduction}\label{introduction}

In the next two chapters we will concentrate on {\it static} spacetimes.
As we have remarked in Chapter
\ref{ch:Introduction}, one of the main aims of this thesis is to extend the
uniqueness theorems for static black holes to static spacetimes containing MOTS.
This chapter is
devoted to obtaining a proper understanding of MOTS in static spacetimes, which will be essential
to prove the uniqueness result in the next chapter.

The first answer to the question of whether the uniqueness theorems for static black holes extend to static
spacetimes containing MOTS was given by Miao in 2005 \cite{Miao}, who proved uniqueness for the particular
case
of time-symmetric, asymptotically flat and vacuum spacelike hypersurfaces possessing a minimal compact boundary (note that in a time-symmetric slice
compact minimal surfaces are MOTS and vice versa).
This result generalized the
classic uniqueness result of Bunting and Masood-ul-Alam \cite{BMuA}
for vacuum static black holes which states the following.
\begin{thr}\label{thr:BMuA}
Consider a vacuum spacetime $(M,\gM)$ with a static Killing vector $\vec{\xi}$.
Assume that $(M,\gM)$ possesses a connected, asymptotically flat
spacelike hypersurface $(\Sigma,g,K)$ which is time-symmetric (i.e. $K=0$, $\vec{\xi}\perp \Sigma$), has
non-empty compact boundary $\bd \Sigma$ and is such that the static Killing vector $\vec{\xi}$ is causal on $\Sigma$ and
null only on $\bd \Sigma$.\\
Then $(\Sigma,g)$ is isometric to $\left(\mathbb{R}^{3}\setminus B_{{M_{Kr}}/2}(0),{(g_{Kr})}_{ij}=\left( 1+\frac{M_{Kr}}{2|x|} \right)^{4}\delta_{ij}\right)$ for some
$M_{Kr}>0$, i.e. the $\{t=0\}$ slice of the Kruskal spacetime
with mass $M_{Kr}$ outside and including the horizon. Moreover, there exists a neighbourhood of $\Sigma$ in $M$ which is isometrically diffeomorphic to the closure of the domain of outer communications
of the Kruskal spacetime.
\end{thr}

In other words, this theorem asserts that a
time-symmetric slice $\Sigma$ of a non-degenerate
static vacuum black hole
must
be a time-symmetric slice of the Kruskal spacetime.
Miao was able to reach the same conclusion under much weaker assumptions, namely
by simply assuming that the boundary of $\Sigma$ is a closed minimal surface.
As in Bunting and Masood-ul-Alam's theorem, Miao's
result deals with time-symmetric and asymptotically flat spacelike hypersurfaces embedded in static vacuum spacetimes.
More precisely,

\begin{thr}\label{thr:Miao}
Consider a vacuum spacetime $(M,\gM)$ with a static Killing vector $\vec{\xi}$.
Assume that $(M,\gM)$ possesses a connected, asymptotically flat spacelike hypersurface
$(\Sigma,g,K)$ which is time-symmetric and such that $\bd\Sigma$ is a (non-empty) compact minimal
surface.\\ 
Then $(\Sigma,g)$ is isometric to $\left(\mathbb{R}^{3}\setminus B_{M_{Kr}/2}(0),{(g_{Kr})}_{ij}=\left( 1+\frac{M_{Kr}}{2|x|} \right)^{4} \delta_{ij}\right)$ for some
$M_{Kr}>0$, i.e. the $\{t=0\}$ slice of the Kruskal spacetime
with mass $M_{Kr}$ outside and including the horizon. Moreover, there exists a neighbourhood of $\Sigma$ in $M$ which is isometrically diffeomorphic to the closure of the domain of outer communications
of the Kruskal spacetime.
\end{thr}

A key ingredient in Miao's proof was to show that 
the existence of a closed minimal surface implies the existence
of an asymptotically flat end $\Sigma^{\infty}$ with smooth topological boundary $\tbd \Sigma^{\infty}$ such that
$\vec{\xi}$ is timelike on $\Sigma^{\infty}$ and vanishes on $\tbd \Sigma^{\infty}$.
Miao then proved that $\tbd \Sigma^{\infty}$ coincides in fact with the minimal boundary $\bd\Sigma$
of the original manifold.
Hence, the strategy was to reduce Theorem \ref{thr:Miao} to the
Bunting and Massod-ul-Alam uniqueness theorem of black holes.

As a consequence of the static vacuum field equations the set of points where the
Killing vector vanishes in a time-symmetric slice is known to be a totally geodesic surface.
Totally geodesic surfaces are of course minimal and in this sense Theorem \ref{thr:Miao}
is a generalization of Theorem \ref{thr:BMuA}. In fact, Theorem \ref{thr:BMuA} allows us to rephrase
Miao's theorem as follows:
{\it No minimal surface can penetrate in the exterior region where the Killing vector is timelike in any
time-symmetric and asymptotically flat slice of a static vacuum spacetime.}
In this sense, Miao's result can be regarded as a confinement result for MOTS
in time-symmetric slices of static vacuum spacetimes.
Here, it is important to remark that a general confinement result of this type was already known when
suitable global hypotheses in time are assumed in the spacetime. In this case,
weakly outer trapped surfaces must lie inside the black hole region (see e.g. Proposition 12.2.4 in \cite{Wald}).
Consequently, Theorem \ref{thr:Miao} can also be viewed as an extension of this
result to the initial data setting (which drops completely all global assumptions in time) for the particular case
of time-symmetric, static vacuum slices.

We aim 
to generalize Miao's theorem
in three different directions. Firstly, we want to
allow for non-vanishing matter as long as the NEC is satisfied. Secondly,
the slices will no longer be required to be time-symmetric. In this situation the natural
replacement for minimal surfaces are MOTS.
And finally, we intend to relax the condition of asymptotic flatness to just assuming the
presence of an outer untrapped surface (of course, this will not
be possible for the uniqueness theorem, but it is possible when viewing Miao's result as
a confinement result).
The proof given by Miao relies strongly on the vacuum field equations, so
we must resort to different methods.
Obviously, a fundamental step for our purposes is a proper understanding of MOTS
in static spacetimes.

In this chapter we
explore the properties of MOTS in static spacetimes. The main result of this chapter is Theorem \ref{theorem2}
which
extends Theorem \ref{thr:Miao} as a confinement result for MOTS by asserting that
no MOTS which are bounding
can {\it penetrate} into the exterior region where the static Killing
is timelike provided some hypotheses hold.
In fact, this result for MOTS also holds for weakly outer trapped surfaces.
It is important to note that Theorem \ref{theorem1} in the previous
chapter already forbids the existence of weakly outer trapped surfaces
whose exterior lies in the region where the Killing vector is timelike,
and which penetrates into the timelike region (recall that the exterior of $S$ does not contain $S$, by definition).
However, this result does not exclude the existence of a weakly outer trapped surface penetrating into the timelike region
but not lying
entirely in the causal region. This is the situation we exclude in Theorem \ref{theorem2}.
The essential ingredients to prove this result will be a combination of the ideas that
allowed us to prove Theorem \ref{theorem1}
together with a detailed study of
the properties of the boundary of the region where the static Killing is timelike.
Besides a confinement result, Miao's theorem is also (and fundamentally) a uniqueness theorem.
The generalization of Miao's result
as a uniqueness result will be studied in the next chapter, where several of the results of the present
chapter will be applied.



As we remarked in the introductory chapter, a general tendency in investigations involving
stationary and static spacetimes over the years has been to relax the global hypotheses in time and
work at the
initial data level as much as possible.
Good examples of this fact are the statements of Theorems \ref{thr:BMuA} and \ref{thr:Miao} above, where
the existence of a spacelike hypersurface with suitable properties is, in fact, sufficient for the proof.
Following this trend, all the results of this chapter will be proved by working
directly on spacelike hypersurfaces,
with no need of invoking a spacetime containing them.
These spacelike hypersurfaces, considered as abstract objects on their own,
will be called {\it initial data sets}. Some of these results generalize known properties
of static spacetimes to the initial data setting and, consequently, can be of independent interest.

We finish this introduction with a brief summary of the chapter.
In Section \ref{preliminaries} we
define initial data set as well as Killing initial data (KID). Then we
introduce the so-called Killing form and give some of its properties.
In Section \ref{staticityKID}
we discuss the implications of imposing staticity
on a Killing initial data set and state a number of useful properties of the boundary of the
set where the static Killing vector is timelike, which will be fundamental to prove Theorem \ref{theorem2}.
Some of the technical work required in this section is
related to the fact that we are not a priori assuming the existence of a spacetime. 
Finally, Section \ref{mainresults} is devoted to stating and proving Theorem \ref{theorem2}.

The results presented in this chapter have been published in \cite{CM1}, \cite{CMere1}.

\setcounter{equation}{0}
\section{Preliminaries}
\label{preliminaries}

\subsection{Killing Initial Data (KID)} \label{sKID}

We start with the standard definition of initial data set \cite{BC}.
\begin{defi}
An \textbf{initial data set} $(\Sigma,g,K;\rho,{\bf J})$ is a
3-dimensional connected manifold $\Sigma$, possibly with boundary,
endowed with a Riemannian metric $g$, a symmetric, rank-two
tensor $K$, a scalar $\rho$ and a one-form $\bf{J}$
satisfying the so-called {\it constraint equations},
\begin{eqnarray*}
2 \rho & = & \RSigma+ (\tr_{\,\Sigma}  K)^{2}-K_{ij}K^{ij},  \\
-  J_{i} & = & {\nablaSigma}_j({K_{i}}^{j}- \tr_{\,\Sigma}  K \delta_{i}^{j}),
\end{eqnarray*}
where $\RSigma$ and $\nablaSigma$ are respectively the scalar curvature and the covariant derivative of $(\Sigma,g)$
and $\tr_{\,\Sigma}  K= g^{ij}K_{ij}$.
\end{defi}

For simplicity, we will often write $\id$ instead of $\idfull$
when no confusion arises.

In the framework of the Cauchy problem for the Einstein field equations,
$\Sigma$ is a spacelike hypersurface of a spacetime
$(M,\gM)$, $g$ is
the induced metric and $K$ is the second fundamental form.
The {\bf initial data energy density} $\rho$ and {\bf energy flux} ${\bf J}$
are defined by $\rho \equiv \GM_{\mu\nu}n^{\mu}n^{\nu}, J_{i} \equiv
-\GM_{\mu\nu}n^{\mu}e_{i}^{\nu}$, where $\GM_{\mu\nu}$ is the Einstein tensor of $\gM$,
$\vec{n}$ is the unit future directed vector normal to $\Sigma$
and $\{ \vec{e}_i \}$ is a local basis for $\mathfrak{X}(\Sigma)$.
When $\rho=0$ and ${\bf J}=0$, the initial data set
is said to be {\bf vacuum}.

As remarked in the previous section,
we will regard initial data sets as abstract objects
on their own, independently of the existence of a spacetime where they may be embedded,
unless explicitly stated.

Consider for a moment a spacetime
$(M,\gM)$ possessing a Killing vector field $\vec{\xi}$ and let $\id$ be an initial data set in this spacetime.
We can decompose $\vec{\xi}$ along $\Sigma$ into a normal and
a tangential component as
\begin{equation}\label{killingdecomposition}
\vec{\xi}=N\vec{n}+Y^{i}\vec{e}_{i}
\end{equation}
(see Figure \ref{fig:XiNY}), where $N = -\xi^{\mu}n_{\mu}$. Note that with this decomposition
\[
\lambda\equiv -\xi_{\mu}\xi^{\mu}=N^{2}-Y^{2}.
\]
Inserting (\ref{killingdecomposition}) into the Killing equations and performing a 3+1 splitting
on $\id$ it follows (see \cite{Coll}, \cite{BC}),
\begin{eqnarray}
2NK_{ij} +  2\nablaSigma_{(i}Y_{j)}&=&0, \hspace{98mm} \label{kid1} \\
\mathcal{L}_{\vec{Y}}K_{ij}  +  \nablaSigma_{i}\nablaSigma_{j}N&=&N\left(
\RSigma_{ij}+ \tr_{\,\Sigma} K K_{ij}-2K_{il}K_{j}^{l}\right.
-\tau_{ij}\nonumber\\
&&\left.\qquad\qquad\qquad\qquad\qquad\qquad + \frac{1}{2}g_{ij}(\tr_{\,\Sigma} \tau-\rho) \right), \label{kid2}
\end{eqnarray}
where 
the parentheses in (\ref{kid1}) denotes symmetrization,
$\tau_{ij} \equiv \GM_{\mu\nu}e_{i}^{\mu}e_{j}^{\nu}$ are the
remaining components of the Einstein tensor and
$\tr_{\,\Sigma} \tau=g^{ij}\tau_{ij}$. Thus, the following definition
of Killing initial data becomes natural \cite{BC}.

\begin{defi} An initial data set $\idfull$ endowed with a scalar $N$, a
vector $\vec{Y}$ and a symmetric tensor
$\tau_{ij}$ satisfying equations
(\ref{kid1}) and (\ref{kid2}) is called a \textbf{Killing initial data}
\textit{(KID)}.
\end{defi}
In particular, if a KID has $\rho=0$, ${\bf J}=0$ and $\tau=0$ then it is said to be a
{\bf vacuum KID}.

A point $\p \in \Sigma$  where $N=0$ and $\vec{Y}=0$ is a {\bf fixed point}.
This name is motivated by the fact that when the KID is embedded into a spacetime with a
local isometry, the corresponding Killing vector $\vec{\xi}$ vanishes at $\p$ and the isometry has
a fixed point there.

A natural question regarding KID is whether they can be embedded into a spacetime
$(M,\gM)$ such that $N$ and $\vec{Y}$ correspond to a Killing vector $\vec{\xi}$.
The simplest case
where existence is guaranteed involves ``transversal'' KID, i.e. when
$N\ne0$ everywhere. Then, the following spacetime, called
\textbf{Killing development} of $(\Sigma,g,K)$,
 can be constructed
\begin{equation}
\label{killingdevelopment}
\left( \Sigma\times\mathbb{R},\quad g^{(4)}=-\hat{\lambda}dt^{2}
+2\hat{Y}_{i}dtdx^{i}+ \hat{g}_{ij}dx^{i}dx^{j} \right)
\end{equation}
where
\begin{equation}\label{killingdevelopment2}
\hat{\lambda}(t,x^{i})\equiv (N^2 - Y^i Y_i )(x^{i}), \quad \hat{g}_{ij}(t,x^{k})\equiv
g_{ij}(x^{k}), \quad \hat{Y}^{i}(t,x^{j})\equiv Y^{i}(x^{j}).
\end{equation}
Notice that $\partial_{t}$
is a complete Killing field with orbits diffeomorphic to $\mathbb{R}$
which, when evaluated on $\Sigma \equiv \{t=0 \}$
decomposes as $\partial_{t}=N\vec{n}+Y^{i}\vec{e}_{i}$, in agreement with
(\ref{killingdecomposition}). The Killing development is the unique
spacetime with these properties.
Further details can be found in \cite{BC}. Notice also that the Killing development
can be constructed for any connected subset of $\Sigma$ where $N \neq 0$ everywhere.

We will finish this subsection by giving the definition of asymptotically flat KID, which is just
the same as for asymptotically flat spacelike hypersurface {\it but adding} the suitable decays for the quantities $N$ and $\vec{Y}$.
\begin{defi}
\label{asymptoticallyflat1}
A KID $\kid$ is {\bf asymptotically flat} if
$\Sigma=\mathcal{K} \cup \Sigma^{\infty}$, where $\mathcal{K}$ is a compact set and
$\Sigma^{\infty}=\underset{a}{\bigcup}\Sigma^{\infty}_{a}$ is a finite union
with each $\Sigma^{\infty}_{a}$,
called an asymptotic end, being diffeomorphic
to $\mathbb{R}^3 \setminus
\overline{B_{R_{a}}}$, where $B_{R_{a}}$ is an open ball of radius $R_{a}$. Moreover, in
the Cartesian coordinates $\{ x^i \}$
induced by the diffeomorphism, the following decay holds
\begin{eqnarray}
    N-A_{a} = O^{(2)}(1/r),\qquad g_{ij}-{\delta}_{ij}&=&O^{(2)}(1/r),\nonumber\\
    Y^{i}-C^{i}_{a}=O^{(2)}(1/r), \qquad\qquad
    K_{ij}&=&O^{(2)}(1/r^{2}).\nonumber
\end{eqnarray}
where $A_{a}$ and $\{C^{i}_{a} \}_{i=1,2,3}$
are constants such that $A^2_{a}-{\delta}_{ij} C^{i}_{a} C^{j}_{a}>0$
for each $a$, and $r=\left(x^{i}x^{j}\delta_{ij} \right)^{1/2}$.
\end{defi}

{\bf Remark.} The condition on the constants $A_a, C^i_a$ is imposed to ensure that
the KID is timelike near infinity on each asymptotic end. $\hfill \square$

\subsection{Killing Form on a KID}\label{subsectionkillingform}

A useful object in spacetimes with a Killing vector $\vec{\xi}$ is
the two-form $\nabla_{\mu}\xi_{\nu}$, usually called {\bf Killing form} or also
Papapetrou field. This tensor will play a relevant role  below. Since
we intend to work directly on the initial data set,
we need to define a suitable tensor on $\id$
which corresponds to the Killing form  whenever a spacetime is present.
Let $\kid$ be a KID in $(M,\gM)$. Clearly we need
to restrict and decompose  $\nabla_{\mu}\xi_{\nu}$ onto $\kid$
and try to get an expression in terms of $N$ and $\vec{Y}$ and its spatial
derivatives. In order to use
(\ref{killingdecomposition}) we first extend
$\vec{n}$ to a neighbourhood of $\Sigma$ as a timelike unit and hypersurface
orthogonal, but otherwise arbitrary, vector field
(the final expression we obtain will be independent of this
extension), and define $N$ and $\vec{Y}$ so that $\vec{Y}$ is
orthogonal to $\vec{n}$ and (\ref{killingdecomposition}) holds.
Taking covariant derivatives we find
\begin{equation}
\label{KF}
\nabla_{\mu}\xi_{\nu}=\nabla_{\mu}Nn_{\nu}+N\nabla_{\mu}n_{\nu}+\nabla_{\mu}Y_{\nu}.
\end{equation}
Notice that, by construction,
$\nabla_{\mu}n_{\nu} |_{\Sigma}=K_{\mu\nu}-n_{\mu}a_{\nu} |_{\Sigma}$
where $a_{\nu}=n^{\alpha}\nabla_{\alpha}n_{\nu}$ is the acceleration of $\vec{n}$.
To elaborate $\nabla_{\mu} Y_{\nu}$ we recall that $\nablaSigma$-covariant derivatives
correspond to spacetime covariant derivatives projected
onto $\Sigma$. Thus, from
$\nablaSigma_{\mu} Y_{\nu} \equiv h^{\alpha}_{\mu} h^{\beta}_{\nu} \nabla_{\alpha} Y_{\beta}$,
where $h^{\mu}_{\nu}=\delta^{\mu}_{\nu}+n^{\mu}n_{\nu}$
is the
projector orthogonal to $\vec{n}$, and expanding we find
\begin{eqnarray*}
\nabla_{\mu}Y_{\nu} |_{\Sigma} &=&
\nablaSigma_{\mu}Y_{\nu}-n_{\mu}\left( n^{\alpha}\nabla_{\alpha}Y_{\beta} \right)
h^{\beta}_{\nu}
-n_{\nu}\left( n^{\beta}\nabla_{\alpha}Y_{\beta} \right)
h^{\alpha}_{\mu}
+n_{\mu}n_{\nu}n^{\alpha}n^{\beta}\nabla_{\alpha}Y_{\beta} |_{\Sigma}\\
&=&
\nablaSigma_{\mu}Y_{\nu}-n_{\mu}\left( n^{\alpha}\nabla_{\alpha}Y_{\beta} \right)
h^{\beta}_{\nu}+
n_{\nu}\left( Y^{\beta}\nabla_{\alpha}n_{\beta} \right)
h^{\alpha}_{\mu}
+n_{\mu}n_{\nu}n^{\alpha}n^{\beta}\nabla_{\alpha}Y_{\beta} |_{\Sigma}\\
&=&
\nablaSigma_{\mu}Y_{\nu}-n_{\mu}\left( n^{\alpha}\nabla_{\alpha}Y_{\beta} \right)
h^{\beta}_{\nu}+
K_{\mu\alpha}Y^{\alpha}n_{\nu}+n_{\mu}n_{\nu}n^{\alpha}n^{\beta}\nabla_{\alpha}Y_{\beta} |_{\Sigma},
\end{eqnarray*}
Substitution into (\ref{KF}), using $\nabla_{\mu}N=\nablaSigma_{\mu}N-n_{\mu}n^{\alpha}\nabla_{\alpha}N$, gives
\begin{eqnarray}
\label{killingform1}
\nabla_{\mu}\xi_{\nu}\big|_{\Sigma}&=&n_{\nu}\left( \nablaSigma_{\mu}N+K_{\mu\alpha}Y^{\alpha} \right) -
n_{\mu}\left( Na_{\nu}+n^{\alpha}h^{\beta}_{\nu} \nabla_{\alpha}Y_{\beta}
\right)  \nonumber\\
&& +
(\nablaSigma_{\mu}Y_{\nu}+NK_{\mu\nu}) + n_{\mu}n_{\nu}\left(
n^{\alpha}n^{\beta}\nabla_{\alpha}Y_{\beta}-
n^{\alpha}\nabla_{\alpha}N \right) |_{\Sigma}.
\end{eqnarray}
The Killing equations then require $
n^{\alpha}n^{\beta}\nabla_{\alpha}Y_{\beta} |_{\Sigma}=
n^{\alpha}\nabla_{\alpha}N |_{\Sigma} $ and
$ \nablaSigma_{\mu}N+K_{\mu\alpha}Y^{\alpha}|_{\Sigma}= Na_{\mu}+n^{\alpha}h^{\beta}_{\mu} \nabla_{\alpha}Y_{\beta} |_{\Sigma}$,
so that (\ref{killingform1}) becomes, after using (\ref{kid1}),
\begin{equation}
\label{killingform2}
\left. \nabla_{\mu}\xi_{\nu} \right |_{\Sigma} =
\left . n_{\nu}\left( \nablaSigma_{\mu}N+K_{\mu\alpha}Y^{\alpha} \right) -
n_{\mu}\left( \nablaSigma_{\nu}N+K_{\nu\alpha}Y^{\alpha} \right) +
\frac12\left( \nablaSigma_{\mu}Y_{\nu}-\nablaSigma_{\nu}Y_{\mu} \right) \right |_{\Sigma}.
\end{equation}
This expression involves solely objects defined on $\Sigma$. However,
it still involves four-di\-men\-sio\-nal objects. In order to work
directly on the KID, we introduce an auxiliary
four-dimensional vector space on each point of $\Sigma$ as follows (we
stress that we are {\it not} constructing a spacetime, only a
Lorentzian vector space attached to each point on the KID).

At every point $\p\in\Sigma$ define the vector space $V_{\p}=T_{\p}\Sigma\oplus\mathbb{R}$,
and endow this space with the Lorentzian metric
$\gl |_{\p} =g |_{\p}\oplus\left( -\delta \right)$, where ${\delta}$ is the
canonical metric on $\mathbb{R}$. Let $\vec{n}$ be the unit
vector tangent to the fiber $\mathbb{R}$.
 Having a metric we
can lower and raise indices of tensors in $T_{\p}\Sigma\oplus\mathbb{R}$.
In particular define ${\bf n}=\gl(\vec{n},\cdot)$. Covariant tensors
$Q$ on $T_{\p} \Sigma$ can be canonically extended to tensors of the same type on
$V_{\p} = T_{\p}\Sigma\oplus\mathbb{R}$ (still denoted with the same symbol) simply by noticing
that any vector in $V_{\p}$ is
of the form $\vec{X} + a \vec{n}$, where $\vec{X} \in T_{\p} \Sigma$ and $a \in \mathbb{R}$. The extension
is defined (for a type $m$ covariant tensor) by $Q (\vec{X_1} + a_1 \vec{n}, \cdots, \vec{X}_m + a_m \vec{n})
\equiv Q(\vec{X_1}, \cdots, \vec{X}_m)$. In index notation, this extension will
be expressed simply by changing Latin to Greek indices. It is clear that
the collection of $\left( T_{\p}\Sigma\oplus\mathbb{R},\gl \right)$ at
every $ \p\in\Sigma$ contains no more information than just $(\Sigma,g)$.
In particular, this construction allows us to redefine the energy conditions
appearing in Chapter \ref{sc:A2sectionbasics} at the initial data level.
Let us give the definition of NEC for an initial data set.

\begin{defi}
An initial data set $\id$ satisfies the {\bf null energy condition} (NEC) if for all
$\p\in\Sigma$ the tensor $\GM_{\mu\nu}\equiv \rho n_{\mu}n_{\nu}+J_{\mu}n_{\nu}+n_{\mu}J_{\nu}+\tau_{\mu\nu}$
on $T_{\p}\Sigma\times \mathbb{R}$ satisfies that $\GM_{\mu\nu}k^{\mu}k^{\nu} |_{\p}\geq 0$ for any null vector
$\vec{k}\in T_{\p}\Sigma\oplus \mathbb{R}$.
\end{defi}

Motivated by (\ref{killingform2}), we can define the Killing form
directly in terms of objects on the KID
\begin{defi}
The {\bf Killing form on a KID} is the 2-form $F_{\mu\nu}$
defined on $\left( T_{\p}\Sigma\oplus\mathbb{R}, \gl \right)$
given by
\begin{equation}\label{killingform}
F_{\mu\nu}= n_{\nu}\left(
\nablaSigma_{\mu}N+K_{\mu\alpha}Y^{\alpha} \right) -
n_{\mu}\left(
\nablaSigma_{\nu}N+K_{\nu\alpha}Y^{\alpha} \right) +
f_{\mu\nu},
\end{equation}
where $f_{\mu\nu}=
\nablaSigma_{[\mu}Y_{\nu]}$.
\end{defi}
In a spacetime setting it is well-known that for a non-trivial Killing vector
$\vec{\xi}$, the Killing form cannot vanish on a fixed point. Let us show that
the same happens in the KID setting.
\begin{lema}
\label{FixedPointF=0}
Let $\kid$ be a KID and $\p \in \Sigma$ a fixed point, i.e. $N |_{\p} = 0$ and
$\vec{Y} |_{\p} =0$. If $F_{\mu\nu} |_{\p} = 0$ then $N$ and $\vec{Y}$ vanish identically on
$\Sigma$.
\end{lema}

{\bf Proof.}
The aim is to obtain a suitable system of equations and show that, under the
circumstances of the lemma,  the solution must be identically zero.
Decomposing $
\nablaSigma_{i}Y_{j}$ in symmetric
and antisymmetric parts,
\begin{equation}\label{DY}
\nablaSigma_{i}Y_{j}=-NK_{ij}+f_{ij},
\end{equation}
and inserting into (\ref{kid2}) gives
\begin{equation}\label{DDN}
\nablaSigma_{i}
\nablaSigma_{ j}N=NQ_{i j}-Y^{l}\nablaSigma_{l}K_{i j}-K_{il}{f_{ j}}^{l}-K_{ jl}
{f_{i}}^{l},
\end{equation}
where $Q_{i j}=\RSigma_{i j}+\tr_{\,\Sigma}K K_{i j}-\tau_{i j} + \frac{1}{2}g_{i j}(\tr_{\,\Sigma} \tau-\rho)$.
In order to find an equation for $
\nablaSigma_{l}f_{i j}$, we take a derivative of
(\ref{kid1}) and write the three equations obtained by cyclic
permutation. Adding two of them and subtracting the third one, we
find,
\begin{equation}
\nablaSigma_l \nablaSigma_i Y_j = \RSigma_{klij} Y^k +
\nablaSigma_{j} ( N K_{li}) - \nablaSigma_i (N K_{lj}) - \nablaSigma_l (N K_{ij}),\nonumber
\end{equation}
after using the Ricci and first Bianchi identities.
Taking the antisymmetric part in $i,j$,
\begin{equation}\label{DF}
\nablaSigma_{l} f_{i j}=\RSigma_{k l i j}Y^{k}+\nablaSigma_{ j}NK_{li}-\nablaSigma_{i}NK_{l j} +
N\nablaSigma_{ j}K_{li}-N\nablaSigma_{i}K_{l j}.
\end{equation}
If $F_{\mu\nu} |_{\p}=0$, it follows that $f_{ij} |_{\p} =0$ and $\nablaSigma_i N |_{\p} =0$.
The equations given by (\ref{DY}), (\ref{DDN}) and (\ref{DF}) is a system of
PDE for the unknowns $N$, $Y_i$ and $f_{ij}$ written in normal form. It follows
(see e.g. \cite{Eisenhart}) that the vanishing of $N$, $\nablaSigma_iN$, $Y_i$ and $f_{ij}$
at one point implies its vanishing everywhere (recall that $\Sigma$ is connected). $\hspace*{1cm} \hfill \blacksquare$ \\

\subsection{Canonical Form of Null two-forms}
\label{canonical}

Let $F_{\mu\nu}$ be an arbitrary two-form on a spacetime $(M,g^{(4)})$.
It is well-known that the only two non-trivial scalars that can be constructed from
$F_{\mu\nu}$ are $I_{1}=F_{\mu\nu}F^{\mu\nu}$ and
$I_{2}= F^{\star}_{\mu\nu}F^{\mu\nu}$,
where $F^{\star}$ is the Hodge dual of $F$, defined by
$F^{\star}_{\mu\nu}=\frac12\eta^{(4)}_{\mu\nu\alpha\beta}F^{\alpha\beta}$,
with $\eta^{(4)}_{\mu\nu\alpha\beta}$ being the volume form of $(M,g^{(4)})$.
When both scalars vanish, the two-form is called \textit{null}.
Later on, we will encounter Killing forms which are null and we will exploit
the following well-known algebraic decomposition
which gives its {\bf canonical form}, see e.g. \cite{Israellibro} for a proof.
\begin{lema}
\label{lemmacanonicalform}
A null two-form $F_{\mu\nu}$ at a point $\p$ can be decomposed as
\begin{equation}\label{canonicalform}
F_{\mu\nu} |_{\p} =l_{\mu} w_{\nu}-l_{\nu}w_{\mu} |_{\p},
\end{equation}
where $\vec{l}\, |_{\p}$ is a null vector
and  $\vec{w} |_{\p}$ is spacelike and orthogonal to
$\vec{l}\, |_{\p}$.
\end{lema}

\setcounter{equation}{0}
\section{Staticity of a KID}
\label{staticityKID}


\subsection{Static KID}

To define a static KID we have to decompose the integrability equation $\xi_{[\mu}\nabla_{\nu}\xi_{\rho]}=0$
according to (\ref{killingdecomposition}).
By taking the normal-tangent-tangent part
(to $\Sigma$) and the completely tangential part (the other components
are identically zero by antisymmetry) we find
\begin{eqnarray}
&&N\nablaSigma_{[i}Y_{j]}+2Y_{[i}\nablaSigma_{j]}N+2Y_{[i}K_{j]l}Y^{l}=0,\label{static5}\\
&&Y_{[i}\nablaSigma_{j}Y_{k]}=0.\label{statictwo}
\end{eqnarray}
Since these expressions involve only objects on the KID, the following definition becomes
natural.
\begin{defi}
A KID $\kid$ satisfying (\ref{static5}) and (\ref{statictwo})
is called an {\bf integrable KID}.
\end{defi}

Multiplying equation (\ref{static5}) by $N$ and equation (\ref{statictwo}) by $Y^{k}$,
adding them up and using equation (\ref{kid1}),
we get the
following useful relation, valid everywhere on $\Sigma$,
\begin{equation}\label{static1}
\lambda \nablaSigma_{[i}Y_{j]}+Y_{[i}\nablaSigma_{j]}\lambda=0.
\end{equation}
If
$\lambda>0$ in some non-empty set of the KID, the Killing vector is
timelike in some
non-empty set of the spacetime. Hence
\begin{defi}
A {\bf static KID} is an integrable KID with $\lambda>0$ in some non-empty set.
\end{defi}

\subsection{Killing Form of a Static KID}\label{subsectionFonstaticKID}

In Subsection \ref{canonical} we introduced the invariant scalars $I_1$ and $I_2$
for any two-form in a spacetime. In this section we find their explicit
expressions for the Killing form of an integrable KID in the region $\{\lambda> 0 \}$.

Although not necessary, we will pass to the Killing development
(which is available in this case) since this simplifies
the proofs. We start with a lemma concerning the integrability of the Killing vector
in the Killing development.
\begin{lema}
\label{integrability}
The Killing vector field associated with the Killing development of an integrable KID is also integrable.
\end{lema}

{\bf Proof.} Let $\kid$ be an integrable KID. Suppose the Killing development
(\ref{killingdevelopment}) of a suitable open
set of $\Sigma$. Using $\vec{\xi} = \partial_t$ it follows
\begin{equation}\label{xidxi}
{\bm \xi}\wedge d{\bm \xi}=-\hat{\lambda}\partial_{i}\hat{Y}_{j}dt\wedge dx^{i}\wedge dx^{j} -
\hat{Y}_{i}\partial_{j}\hat{\lambda}dt\wedge dx^{i}\wedge dx^{j} +
\hat{Y}_{i}\partial_{j}\hat{Y}_{k}dx^{i}\wedge dx^{j}\wedge dx^{k},
\end{equation}
where $\hat{\lambda}$, $\hat{{\bf Y}}$ and $\hat{g}$ are defined in (\ref{killingdevelopment2}).
Integrability of $\vec{\xi}$ follows directly from
(\ref{statictwo}) and (\ref{static1}). $\hfill \blacksquare$ \\

The following lemma gives the explicit expressions for $I_{1}$ and $I_{2}$.
\begin{lema}
\label{lemmaI1}
The invariants of the Killing form in a static KID in the region
$\{ \lambda >0 \}$
read
\begin{equation}\label{I1}
I_{1}=-\frac{1}{2\lambda}\left( g^{ij}-\frac{Y^{i}Y^{j}}{N^{2}}
\right)\nablaSigma_{i}\lambda \nablaSigma_{j}\lambda,
\end{equation}
and
\begin{equation}\label{I2}
I_{2}=0.
\end{equation}
\end{lema}

{\bf Remark.} By continuity $I_{2}\big|_{\tbd\{\lambda>0\}}=0$. $\hfill \square$ \\

{\bf Proof.} Consider a static KID $\kid$ and let $\{ \lambda>0 \}_{0}$ be
a connected component of $\{\lambda>0\}$.
In $\{\lambda>0\}_{0}$ we have necessarily $N\neq 0$, so we can
construct the Killing development $(\{ \lambda>0 \}_{0},g^{(4)})$ and
introduce the so-called Ernst one-form, as
$\sigma_{\mu}=\nabla_{\mu} \lambda -i\omega_{\mu}$
where $\omega_{\mu}=\eta^{(4)}_{\mu\nu\alpha\beta}\xi^{\nu}\nabla^{\alpha}\xi^{\beta}$
is the twist of the Killing field
($\eta^{(4)}$ is the volume form of the Killing development).
The Ernst one-form satisfies the identity (see e.g. \cite{Mars})
$\sigma^{\mu}\sigma_{\mu}=-\lambda\left( F_{\mu\nu}+iF^{\star}_{\mu\nu} \right)
\left( F^{\mu\nu}+i{F^{\star}}^{\mu\nu} \right)$, which in the static case (i.e.
$\omega_{\mu}=0$) becomes
$\nabla_{\mu} \lambda \nabla^{\mu} \lambda=-2\lambda\left(F_{\mu\nu}F^{\mu\nu}
+iF_{\mu\nu}{F^{\star}}^{\mu\nu } \right)$
where the identity $F_{\mu\nu}F^{\mu\nu}=-F^{\star}_{\mu\nu}{F^{\star}}^{\mu\nu}$ has been used.
The imaginary part immediately gives (\ref{I2}). The real part gives
$I_{1}=-\frac{1}{2\lambda}|\nabla \lambda|_{\gM}^{2}$.
Taking coordinates $\{t,x ^{i}\}$ adapted to the Killing field $\partial_{t}$, it follows from (\ref{killingdevelopment2}) that
$|\nabla\lambda|_{\gM}^{2}={\gM}^{ij}\partial_{i}\lambda\partial_{j}\lambda$.
It is well-known (and easily checked) that the contravariant spatial components of $g^{(4)}$
are ${g^{(4)}}^{ij} = g^{ij}-\frac{Y^{i}Y^{j}}{N^2}$, where $g^{ij}$ is the inverse of $g_{ij}$
and (\ref{I1}) follows. $\hfill \blacksquare$ \\

This lemma allows us to prove the following result on the value of $I_1$ on the set ${\{\lambda>0\}}$.

\begin{lema}
$I_{1}|_{{\{\lambda>0\}}}\leq 0$ in a static KID.
\end{lema}

{\bf Proof.} 
Let $ \q\in \{ \lambda>0  \}\subset\Sigma$
and define the vector $\vec{\xi} \equiv N\vec{n}+\vec{Y}$ on the
vector space $(V_{\q}, \gl)$ introduced in Section \ref{subsectionkillingform}.
Since $\vec{\xi}$ is timelike at $\q$, we can
introduce its orthogonal projector
$h_{\mu\nu}=\gl_{\mu\nu}+\frac{\xi_{\mu}\xi_{\nu}}{\lambda}$ which is obviously
positive semi-definite. If we pull it back
onto $T_{\q} \Sigma$ we obtain a positive definite metric, called {\it orbit space metric},
\begin{equation}\label{quotientmetric}
h_{ij}=g_{ij}+\frac{Y_{i}Y_{j}}{\lambda}.
\end{equation}
It is immediate to check that the inverse of $h_{ij}$ is precisely
the term in brackets in (\ref{I1}). Consequently, $I_{1} |_{\q}  \leq 0$ follows.
$\hfill \blacksquare$ \\

{\bf Remark.} By continuity $I_{1}|_{\tbd \{\lambda>0\}}\leq 0$. $\hfill \square$

Furthermore, for the fixed points on the closure of $\{ \lambda > 0 \}$ we have the following result.
Notice that $\tbd\{\lambda>0 \} \subset
\overline{\{N \neq 0\}}$. Since the result involves points where $N$ vanishes, we cannot rely
on the Killing development for its proof and an argument directly on the initial data set
is needed.

\begin{lema}
\label{I1<0}
Let $\p \in \overline{\{\lambda >0 \}}$ be a fixed point of a static KID, then $I_1 |_{\p} < 0$.
\end{lema}

{\bf Proof.} From the previous lemma it follows that $I_{1}|_{p}\leq 0$.
It only remains to show that $I_1 |_{\p}$ cannot be zero. We argue by contradiction.
Assuming  that  $I_1 |_{\p} =0$ and using $I_2 |_{\p} =0$ by Lemma \ref{lemmaI1}, it follows
that $F_{\mu\nu}$ is null at $\p$. Lemma \ref{lemmacanonicalform} implies the
existence of a null vector $\vec{l}$ and a spacelike vector $\vec{w}$ on $V_{\p}$ such
that (\ref{canonicalform}) holds. Since $\vec{w}$ is defined up to
an arbitrary additive vector proportional to $\vec{l}$, we can choose
$\vec{w}$ normal to $\vec{n}$ without loss of generality. Decompose
$\vec{l}$ as $\vec{l}=a\left( \vec{x} + \vec{n} \right)$
with $x^{\mu}x_{\mu}=1$. We know from Lemma \ref{FixedPointF=0}
that $a \neq 0$ (otherwise $F_{\mu\nu} |_{\p} =0$ and $\{ \lambda > 0 \}$ would be empty).
Expression (\ref{killingform}) and the canonical form (\ref{canonicalform}) yield
\begin{equation*}
F_{\mu\nu} |_{\p}  = 2n_{ [ \nu}\nablaSigma_{\mu ] }N + \nablaSigma_{ [ \mu}Y_{\nu ] } |_{\p} = 2a \left( x_{ [ \mu}
w_{\nu ]}+n_{ [ \mu}w_{\nu ] } \right).
\end{equation*}
The purely tangential and normal-tangential components of this equation give, respectively
\begin{eqnarray}
  \nablaSigma_{i}Y_{j}\big|_{\p}=2a x_{[i}w_{j]}, \quad \nablaSigma_{i}N\big|_{\p}=-aw_{i}, \label{ab}
\end{eqnarray}
where $w_{i}$ is the projection of $w_{\mu}$ to $T_{\p} \Sigma$.
The Hessian of $\lambda$ at $\p$ is then
\begin{eqnarray*}
\nablaSigma_{i}\nablaSigma_{j}\lambda\big|_{\p}&=&2(\nablaSigma_{i}N\nablaSigma_{j}N-
\nablaSigma_{i}Y^{k}\nablaSigma_{j}Y_{k})\big|_{\p}\\
&=& -2a^2 w^{k}w_{k} x_{i}x_{j},
\end{eqnarray*}
where we have used $x^{i}x_{i}=1$ and $x^{i}w_{i}=0$ (which follows from $\vec{w}$ being orthogonal
to $\vec{l}$ ). This Hessian has therefore signature $\{-,0,0\}$.
The Gromoll-Meyer splitting Lemma (see Appendix \ref{ch:appendix2}) implies the existence
of an open neighbourhood $U_{\p}$ of $\p$ and coordinates $\{x,z^{A}\}$ in $U_{\p}$
such that $\p=(x=0,z^{A}=0)$ and $\lambda=-\hat{a}^{2}x^{2}+\zeta(z^{A})$ where
$\hat{a}>0$ and $\zeta$ is a smooth function satisfying
$\zeta\big|_{\p}=0$, $\nablaSigma_{i}\zeta\big|_{\p}=0$ and $\nablaSigma_{i}\nablaSigma_{j}\zeta\big|_{\p}=0$.
Since $\p\in\tbd\{\lambda>0\}$, there exists a curve
$\mu(s)=(x(s),z^{A}(s))$ in $U_{\p}\cap \{\lambda>0\}$, parametrized by $s\in (0,\epsilon)$ such that
$\mu(s)\underset{s\rightarrow 0}{\longrightarrow} \p$. Since $\lambda>0$ on the curve we have 
$-\hat{a}^{2}x^{2}(s)+\zeta(z^{A}(s))>0$, which implies
$\zeta(z^{A}(s))>0$. It follows that the curve $\gamma(s)\equiv\left(x(s)=\frac{1}{\hat{a}}\sqrt{\zeta(z^{A}(s))},z^{A}(s)\right)$
(also parametrized by $s$) 
belongs
to $\tbd \{\lambda>0\}$ and is composed by non-fixed points (because
$\nablaSigma_{i}\lambda\big|_{\gamma(s)}\neq 0$).
We can construct the Killing development (\ref{killingdevelopment})
near this curve, which is a static spacetime (see Lemma \ref{integrability}).
Applying Lemma \ref{lema:VC} by Vishveshwara and Carter it follows that $\gamma(s)$
(which belongs to $\tbd\{\lambda>0\}$ and has $N\neq 0$)
lies in an arc-connected component of a Killing prehorizon of the Killing development. Projecting equation
(\ref{kappa0}), valid on a Killing prehorizon, onto $\Sigma$, we get the relation
\begin{equation}\label{kappa}
\nablaSigma_{i}\lambda\big|_{\gamma(s)}
=2\kappa Y_{i}\big|_{\gamma(s)},
\end{equation}
where $\kappa$ is the surface gravity of the prehorizon. 
Therefore, $\kappa\big|_{\gamma(s)}\neq 0$.
Since $I_{1}=-2\kappa^{2}$ (see e.g. equation (12.5.14) in \cite{Wald}) and $\kappa$ remains
constant on $\gamma(s)$
(see Lemma \ref{lema:RW}),
it follows, by continuity of $I_{1}$, that $I_{1}\big|_{\p}=-2\kappa^{2}<0$.$\hfill \blacksquare$ \\


\subsection{Properties of $\tbd \{ \lambda>0 \}$ on a Static KID}

In this subsection we will show that, under suitable conditions, the boundary
of the region $\{\lambda> 0\}$ is a smooth surface.
Our first result on the smoothness of $\tbd\{ \lambda > 0 \}$ is the following.

\begin{lema}\label{surfaceNoFixed}
Let $\kid$ be a static KID
and assume that the set $\E=\tbd\{ \lambda>0 \} \cap \{ N \neq 0 \}$
is non-empty. Then $\E$ is a smooth submanifold of $\Sigma$.
\end{lema}
Recall that in this thesis, a submanifold is, by definition, injectively immersed,
but not necessarily embedded. Besides, it is worth to remark they are also not
necessarily arc-connected.

{\bf Proof.}
Since $N |_{\E}\ne0$,
we can construct the Killing development (\ref{killingdevelopment})
of a suitable neighbourhood of $\E\subset\Sigma$ satisfying $N \neq 0$ everywhere.
Moreover, by Lemma \ref{integrability}, $\vec{\xi} = \partial_t$ is integrable.
Applying Lemma \ref{lema:VC} by Vishveshwara and Carter, it follows
that the spacetime subset $\mathcal{N}_{\vec{\xi}}\equiv \tbd \{\lambda>0\}\cap \{\vec{\xi}\neq 0\}$
is a smooth null submanifold (in fact, a Killing prehorizon) of the Killing development and therefore transverse to $\Sigma$, which is spacelike.
Thus, $\E = \Sigma \cap  \mathcal{N}_{\vec{\xi}}$
is a smooth submanifold of $\Sigma$. $\hfill \blacksquare$ \\

This lemma states that the boundary of $\{ \lambda > 0 \}$ is smooth on the set of
non-fixed points. In fact, for the case of boundaries
having at least one fixed point, an explicit defining function for this surface
on the subset of non-fixed points can be given:

\begin{lema}\label{maldito}
Let $\kid$ be a static KID. If
an arc-connected component of $\tbd\{\lambda>0 \}$ contains at least one
fixed point, then $\nablaSigma_{i}\lambda\neq 0$ on all non-fixed
points in that arc-connected component.
\end{lema}

{\bf Proof.} Let $V$ be the set of non-fixed points in one of the arc-connected
components under consideration. This set is obviously open with at least one fixed point in its closure.
Constructing the Killing development as before, we know that $V$ belongs to a
Killing prehorizon $\mathcal{H}_{\vec{\xi}}$.
Projecting equation (\ref{kappa0}) onto $\Sigma$ we get
$\nablaSigma_{i}\lambda\big|_{\mathcal{H}_{\vec{\xi}\,}\cap\Sigma}=
2\kappa Y_{i}\big|_{\mathcal{H}_{\vec{\xi}\,}\cap\Sigma}$.
Since the surface gravity $\kappa$ is constant on each arc-connected component of $\mathcal{H}_{\vec{\xi}}$
and $I_{1}=-2\kappa^2$, Lemma \ref{I1<0} implies $\kappa\big|_{V} \neq 0$ and consequently
$\nablaSigma_{i}\lambda\big|_{V}\neq 0$.
$\hfill \blacksquare$ \\


Fixed points are more difficult to analyze. We first need a lemma
on the structure of $\nablaSigma_i N$ and $f_{ij}$ on  a fixed point.
\begin{lema}\label{lemmaFixed}
Let $\kid$ be a static KID and $\p \in \tbd\{ \lambda > 0 \}$ be a fixed point. Then
\begin{equation}
\nablaSigma_i N |_{\p}\ne 0 \nonumber
\end{equation}
and
\begin{equation}
\left. f_{ij} |_{\p} = \frac{b}{Q} \left ( \nablaSigma_i N X_j - \nablaSigma_j N X_i \right )\right |_{\p} \label{fijp}
\end{equation}
where $b$ is a constant, $X_i$ is unit and orthogonal to $\nablaSigma_i N |_{\p}$ and $Q = +\sqrt{\nablaSigma_i N {\nablaSigma}^i N}$.
\end{lema}

{\bf Proof.} From (\ref{killingform}),
\begin{equation}\label{FF}
I_{1}=F_{\mu\nu}F^{\mu\nu}=f_{ij}f^{ij}-2
\left( \nablaSigma_{i}N+K_{ij}Y^{j} \right)\left( {\nablaSigma}^{i}N+K^{ik}Y_{k} \right).
\end{equation}
Hence, $\nablaSigma_i N  |_{\p} \neq 0$ follows directly from $I_1 |_{\p} < 0$ (Lemma \ref{I1<0}).
For the second statement, let $u_i$ be unit and satisfy
$\nablaSigma_i N  = Q u_i$ in a suitable neighbourhood of $\p$.
Consider (\ref{static5}) in the region $N \neq 0$, which
gives
\begin{eqnarray}
f_{ij} = -2 N^{-1} Y_{[i}\left( \nablaSigma_{j]}N+K_{j]k}Y^{k} \right). \label{fYN}
\end{eqnarray}
Since $|\vec{Y}|/N$ stays bounded in the region $\{ \lambda > 0 \}$,
it follows that the second term tends to zero
at the fixed point $\p$. Thus,
let $\vec{X}_1$ and $\vec{X}_2$ be any pair of vector fields orthogonal to
$\vec{u}$. It follows by continuity that $f_{ij} X_1^i X_2^j |_{\p} =0$.
Hence for any orthonormal basis $\{ \vec{u}, \vec{X}, \vec{Z}\}$
at $\p$ it follows $f_{ij} X^i Z^j |_{\p} =0$ (because $\vec{X}$ and $\vec{Z}$ can be extended
to a neighbourhood of $\p$ while remaining orthogonal to $\vec{u}$). Consequently,
$f_{ij} |_{\p} = (b/Q)  ( \nablaSigma_i N  X_j - \nablaSigma_j N  X_i )
+ ( c/Q)  ( \nablaSigma_i N  Z_j - \nablaSigma_j N  Z_i  ) |_{\p}$
for some constants $b$ and $c$. A suitable rotation in the $\{\vec{X},\vec{Z}\}$ plane
allows us to set $c=0$ and (\ref{fijp}) follows. $\hfill \blacksquare$ \\


As we will see next, a consequence of this lemma is that an open subset of fixed points in
$\tbd \{\lambda>0\}$ is a smooth surface.
In fact, we will prove that this surface
is totally geodesic in $(\Sigma,g)$ and that the pull-back of the second fundamental form $K_{ij}$
vanishes there. This means from a spacetime perspective, i.e. when the initial data set is embedded
into a spacetime, that this open set of fixed points is  totally geodesic as a spacetime submanifold.
This is of course well-known in the spacetime setting from Boyer's results \cite{Boyer}, see
also \cite{Heuslerlibro}. In our initial data context, however, the result must be proven from scratch
as no Killing development is available at the fixed points.

\begin{proposition}
\label{Totally geodesic}
Let $\kid$ be a static KID
and assume that the set $\tbd\{ \lambda>0 \}$ is
non-empty.
If $\E\subset \tbd\{\lambda>0 \}$ is open and consists of fixed points, then
$\E$ is a smooth surface. Moreover,
the second fundamental form of $\E$ in $(\Sigma,g)$ vanishes and $K_{AB} \big |_{\E}=0$
\end{proposition}

{\bf Proof.} Consider a point $\p\in \E$. We know from Lemma \ref{lemmaFixed} that
$\nablaSigma_{i}N\big|_{\p}\neq 0$. This means that there exists an open neighbourhood $U_{\p}$
such that $\{N=\text{const}\}\cap U_{\p}$ defines a foliation by smooth and connected
surfaces, and moreover that $\nablaSigma_{i}N\neq 0$ everywhere on $U_{\p}$. Restricting
$U_{\p}$ if necessary we can assume that $\tbd\{\lambda>0\}\cap U_{\p}=\E\cap U_{\p}$
(because $\E$ is an open subset of $\tbd \{\lambda>0\}$). It is clear that
$\E\cap U_{\p}\subset\{N=0\}\cap U_{\p}$ (because $N$ vanishes on a fixed point). We only need to prove that these
two sets are in fact equal. Choose a continuous curve $\gamma:(-\epsilon,0)\rightarrow \{\lambda>0\}\cap U_{\p}$
satisfying $\text{lim}_{s\rightarrow 0}\gamma(s)=\p$. Assume that there is a
point $\q\in\{N=0\}\cap U_{\p}$ not lying in $\tbd \{ \lambda>0 \}$. This means that there is an
open neighbourhood $U_{\q}$ of
$\q$ (which can be taken fully contained in $U_{\p}$) which does not intersect $\{\lambda>0\}$. Take a point
$\r$ in $U_{\q}$ sufficiently close to $\q$ so that $N\big|_{\r}$ takes the same value as $N\big|_{\gamma(s_{0})}$
for some $s_{0}\in (-\epsilon,0)$ (this point $\r$ exists because $\nablaSigma_{i}N\big|_{\q}\neq 0$ and
$N\big|_{\q}=0$).
Since the surface $\{N=N\big|_{\r}\}\cap U_{\p}$ is connected and contains both $\r$ and $\gamma(s_{0})$, it
follows that there is a path in $U_{\p}$ with $N=N\big|_{\r}$ constant and connecting these two points.
This path must necessarily intersect $\tbd \{\lambda>0\}$ (recall that $\lambda\big|_{\gamma(s)}>0$ for all $s$).
But this contradicts the fact that
$\tbd\{\lambda>0\}\cap U_{\p}\subset \{N=0\}\cap U_{\p}$. Therefore,
$\E\cap U_{\p}=\{N=0\}\cap U_{\p}$, which proves that $\E$ is a smooth surface.

To prove the other statements,
let us introduce local coordinates $\{ u,x^{A} \}$
on $\Sigma$ adapted to $\E$ so that $\E\equiv\{ u=0 \}$ and let us prove that the linear
term in a Taylor expansion for $Y^i$ vanishes identically. Equivalently,
we want to show that $u^{j}\nablaSigma_{j}Y_{i} |_{\E}=0$ for
$\vec{u} = \partial_u$  (recall that on $\E$ we have
$Y_{i} |_{\E}=0$ and this covariant derivative coincides with the partial derivative).
Note that $\nablaSigma_{i}Y_{j} |_{\E}=f_{ij}$ (see (\ref{DY})), so that
$u^{i} u^j \nablaSigma_{i}Y_{j} |_{\E} =0 $ being the
contraction of a symmetric and an antisymmetric tensor. Moreover, for the tangential
vectors $e^i_A = \partial_A$ we find
$u^{j} e^i_A \nablaSigma_{i}Y_{j} |_{\E} = u^j \partial_A Y_j =0$ because $Y_j$ vanishes all along $\E$.
Consequently $u^{i}\partial_{i}Y_{j}|_{\E}=0$. Hence, the Taylor expansion reads
\begin{eqnarray}\label{expansions}
N&=&G(x^{A})u+O(u^{2}),\nonumber \\
Y_{i}&=&O(u^{2}).
\end{eqnarray}
Moreover, $G\ne0$ everywhere on $\E$  because substituting this Taylor expansion
in (\ref{I1}) and taking the limit $u \rightarrow 0$ gives
$I_{1} |_{\E}=-2g^{uu}G^{2}(x^{A})$ and we know
that $I_{1} |_{\E} \neq 0$ from Lemma \ref{I1<0}.

We can now prove that $\E$ is totally geodesic and that $K_{AB}=0$. For the first,
the Taylor expansion above gives
\begin{equation}\label{fij0}
f_{ij}|_{\E}=0
\end{equation}
and obviously
$N$ and $\vec{Y}$ also vanish on $\E$. Hence, from (\ref{DDN}),
\begin{equation}\label{DDN=0}
\nablaSigma_{i}\nablaSigma_{j}N |_{\E}=0.
\end{equation}
Since, by Lemma \ref{lemmaFixed}, $\nablaSigma_{i}N |_{\E}$ is proportional to the unit normal to $\E$
and non-zero, then $\nablaSigma_{i}\nablaSigma_{j}N |_{\E}=0$ is precisely the condition that
$\E$ is totally geodesic. In order to prove $K_{AB}|_{\E}=0$, we only need to
substitute the Taylor expansion (\ref{expansions}) in the $AB$ components of
(\ref{kid1}). After dividing by $u$ and taking the limit $u \rightarrow 0$,
$K_{AB} |_{\E}=0$ follows directly. $\hfill  \blacksquare$ \\

At this point, let us introduce a lemma on the constancy of $I_{1}$ on each arc-connected component of
$\tbd \{\lambda>0\}$.
\begin{lema}\label{lema:I1constant}
$I_{1}$ is constant on each arc-connected component of $\tbd \{\lambda>0\}$ in a static KID.
\end{lema}

{\bf Proof.}
For non-fixed points this is a consequence of the Vishveshwara-Carter Lemma (Lemma \ref{lema:VC})
and it has already been used several times before.
For an arc-connected open set $\E$ of fixed points, taking the derivative of equation (\ref{FF}) we
get
\[
\nablaSigma_{l}I_{1}=2f^{ij}\nablaSigma_{l}f_{ij}-4(\nablaSigma_{l}\nablaSigma_{i}N +
\nablaSigma_{l}K_{ij}Y^{j}+K_{ij}\nablaSigma_{l}Y^{j})({\nablaSigma}^{i}N+K^{ik}Y_{k}).
\]
Then, using the facts that $f_{ij}\big|_{\E}=0$ (equation (\ref{fij0})),
$\nablaSigma_{i}\nablaSigma_{j}N\big|_{\E}=0$
(equation (\ref{DDN=0})) and
$\nablaSigma_{i}Y_{j}=-NK_{ij}+f_{ij}$ (equation (\ref{DY})), it is immediate to obtain that
$\nablaSigma_{l}I_{1}\big|_{\E}=0$. Finally, continuity of $I_{1}$ leads to the result.
$\hfill\blacksquare$ \\

\begin{figure}
\begin{center}
\psfrag{Sb}{\color{red}{$S_{0}$}}
\psfrag{S}{$S$}
\psfrag{p}{$\p$}
\psfrag{Sigma}{$\Sigma$}
\includegraphics[width=9cm]{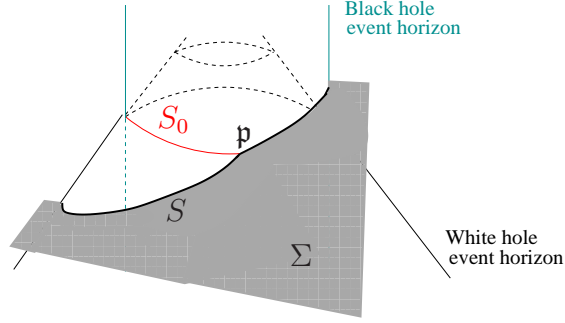}
\caption {An example of non-smooth boundary $\E=\tbd\{\lambda>0\}$ in an initial data set
$\Sigma$ of Kruskal spacetime with one dimension suppressed.
The region outside the cylinder and the cone corresponds to
one asymptotic region of the Kruskal spacetime. The initial data set $\Sigma$ intersects
the bifurcation surface $S_{0}$ (in red). The shaded region corresponds to the intersection of $\Sigma$
with the asymptotic region, and is in fact a connected component of the subset $\{\lambda > 0 \} \subset
\Sigma$. Its boundary is non-smooth at the point $\p$ lying on the bifurcation surface.}
\label{fig:figure1}
\end{center}

\end{figure}

We have already proved that both the open sets of fixed points and the open sets of non-fixed points
are smooth submanifolds.
Unfortunately, when $\tbd\{ \lambda >0 \}$ contains fixed points not lying on open sets, this boundary
is {\it not} a smooth submanifold in general.
Consider as an example
the Kruskal extension of the Schwarz\-schild
black hole and choose one of the asymptotic regions where the static Killing field is timelike in the domain
of outer communications.
Its boundary consists of one half of the black hole event horizon, one half
of the white hole event horizon and the bifurcation surface connecting both.
Take an initial data set $\Sigma$ that intersects the bifurcation surface
transversally and let us denote by $\ext$ the connected component
of the subset $\{ \lambda >0 \}$ within $\Sigma$ contained in the chosen asymptotic region.
The topological boundary $\tbd \ext$ is non-smooth because it has a
corner on the bifurcation surface where the black hole event horizon and the white hole event
horizon intersect (see example of Figure \ref{fig:figure1}).
We must therefore add some condition
on $\tbd\ext$  in order to
guarantee that this boundary does not intersect both a black and a white hole event horizon.
In terms of the Killing vector, this requires that $\vec{Y}$ points only to one side
of $\tbd\ext$. Lemma \ref{maldito} suggests that the
condition we need to impose is
$Y^{i} \nablaSigma_{i} \lambda \big|_{\tbd\ext}\ge0$
or $Y^{i} \nablaSigma_{i} \lambda \big|_{\tbd\ext}\le0$.
This condition is in fact sufficient
to show that $\tbd\ext$ is a smooth surface. Before giving the precise statement of
this result (Proposition \ref{C1} below) we need to prove a lemma on the structure of $\lambda$
near fixed points with $f_{ij}\neq 0$.
For this, the following definition will be useful.
\begin{defi}
A fixed point $\p\in \tbd\{\lambda>0\}$ is called {\bf transverse} if and only if $f_{ij} |_{\p}  \neq 0$
and {\bf non-transverse} if and only if $f_{ij} |_{\p} =0$
\end{defi}

\begin{lema}
\label{structurebneq0}
Let $\p\in\tbd \{\lambda>0\}$ be a transverse fixed point. Then,
there exists an open neighbourhood $U_{\p}$ of $\p$ and
coordinates $\{x,y,z\}$ on $U_{\p}$ such that $\lambda = \mu^2 x^2 - b^2 y^2$
for suitable constants $\mu>0$ and $b\neq 0$.
\end{lema}

{\bf Proof.} From Lemma \ref{lemmaFixed} we have $b\neq 0$. Squaring $f_{ij}$ we get $f_{il} f_{j}^{\,\,l} |_{\p} =
\left . b^2 \left ( \frac{\nablaSigma_i N \nablaSigma_j N}{Q_0^2} + X_i X_j \right ) \right |_{\p}$
and $f_{ij} f^{ij}|_{\p} = 2 b^2$, where $Q_0 = Q(\p)$. 
Being $\p$ a fixed point, both $\lambda$ and its gradient vanish at $\p$ and we have a critical point.
The Hessian of $\lambda$ at $\p$ is immediately computed to be
\begin{eqnarray}
\nablaSigma_i \nablaSigma_j \lambda |_{\p} &=& \left . 2 \nablaSigma_i N \nablaSigma_j N - 2 f_{il} f_{j}^{\,\,\,l} \right  |_{\p} \nonumber \\
&=&
\left . \frac{2 \left (Q_0^2 - b^2 \right )}{Q_0^2} \nablaSigma_i N \nablaSigma_j N - 2 b^2 X_i X_j \right |_{\p}.
\label{Hessianbneq0}
\end{eqnarray}
At a fixed point we have $I_1 |_{\p} = f_{ij} f^{ij} - 2 \nablaSigma_i N
{\nablaSigma}^i N |_{\p} = 2 (b^2 - Q_0^2) < 0$ (Lemma \ref{I1<0}). Let us define $\mu >0$ by
$\mu^2 = Q_0^2 - b^2$. The rank of the Hessian is therefore
two and the signature is $(+,-,0)$.
The Gromoll-Meyer splitting Lemma
(see Appendix \ref{ch:appendix2}) implies the existence of coordinates $\{x,y,z \}$  in a suitable
neighbourhood $U'_{\p}$ of $\p$ such that $\p = \{x=0,y=0,z=0\}$ and
$\lambda = \mu^2 x^2 - b^2 y^2 + h(z)$ on $U'_{\p}$. The function
$h(z)$ is smooth and
satisfies  $h(0) = h'(0)= h''(0)=0$, where prime stands for derivative with
respect to $z$. Moreover, evaluating the Hessian of $\lambda$ at $\p$ and comparing with
(\ref{Hessianbneq0}) we have $dx |_{\p} = Q_0^{-1} dN |_{\p} $ and $dy |_{\p} = \bm{X}$.
This implies $N = Q_0 x + O(2)$. Moreover, since $\nablaSigma_i Y_j |_{\p} = f_{ij} |_{\p} =
b (dx \otimes dy  - dy \otimes dx  )_{ij} |_{\p}$ we conclude $Y_x = - b y +
O(2)$,
$Y_y = b x + O(2)$, $Y_z = O(2)$. On the surface $\{z=0\}$,
the set of points
where $\lambda$ vanishes is given by the two lines $x=x_+ (y) \equiv b \mu^{-1} y $
and $x = x_{-} (y) \equiv - b \mu^{-1} y$.  Computing the gradient of $\lambda$
on these curves we find
\begin{eqnarray}
d \lambda |_{(x=x_{\pm}(y), z=0)}  = \pm 2 \mu b y dx - 2 b^2 y dy.
\label{dlambda1}
\end{eqnarray}
On the other hand, the Taylor expansion above for ${\bm Y}$
gives
\begin{eqnarray}
\bm{Y} |_{(x = x_{\pm}(y),z=0)} = -b y dx \pm  \frac{b^2}{\mu} y dy  + O(2).
\label{yd}
\end{eqnarray}
Let $\E$ be the arc-connected component of $\tbd\{\lambda>0\}$ containing $\p$.
On all non-fixed points in $\E$ we have $d \lambda = 2 \kappa \bm{Y}$, with
$\kappa^2 = -I_1/2$. Comparing (\ref{dlambda1}) with (\ref{yd}) yields
$\kappa= - \mu$ on the branch $x= x_{+} (y)$ and $\kappa = +\mu$
on the branch $x= x_{-}(y)$ (this is in agreement with $I_1 = -2 \kappa^2 =
- 2 \mu^2$ at every point in $\E$).  We already know 
that $\kappa$ must remain constant
on each arc-connected component of $\E \setminus F$, where $F = \{ \p \in \E, \p \,\,\,
\mbox{fixed point} \}$. Let us show that this  implies $h(z)=0$ on $U'_{\p}$. First,
we notice that the set of fixed points on $\E$ are precisely those
where $\lambda=0$ and $d \lambda=0$ (this is because in Lemma \ref{maldito}
we have shown that
$d \lambda \neq 0$ on every non-fixed point of any arc-connected
component of $\tbd\{\lambda >0\}$ containing at least one fixed point). From the
expression $\lambda = \mu^2 x^2 - b^2 y^2 + h(z)$, this implies that the fixed
points  in $U'_{\p}$ are those satisfying $\{ x=0, y=0, h(z)=0, h'(z)=0 \}$.
Assume that there is no neighbourhood $(-\epsilon,\epsilon)$ where $h$
vanishes identically. Then, there exists a sequence $z_n \rightarrow 0$
satisfying $h(z_n) \neq 0$. There must exist a subsequence (still
denoted by $\{z_n \}$) satisfying either $h(z_n)>0$, $\forall n \in \mathbb{N}$
or $h(z_n) <0$, $\forall n \in \mathbb{N}$. The two cases are similar, so we
only consider $h(z_n) = -a_n^2<0$. The set of points with
$\lambda=0$ in the surface $\{z=z_n \}$ are given by
$x = \pm \mu^{-1} \sqrt{b^2 y^2 + a_n^2}$. It follows that the points
$\{\lambda=0\} \cap \{ z=z_n\} $ in the quadrant $\{ x>0, y >0\}$ lie in the same
arc-connected component as the points $\{\lambda=0\} \cap \{ z=z_n \}$ lying
in the quadrant $\{x> 0, y<0\}$.
Since
$z_n$ converges to zero, it follows that the points $\{x=x_+(y),y>0,z=0\}$
lie in the same arc-connected component of $\E \setminus F$
than the points $\{x=x_-(y),y< 0,z=0\}$.
However, this is impossible because $\kappa$ (which is constant
on $\E\setminus F$) takes opposite values on the branch $x=x_{+}(y)$
and on the branch $x=x_{-}(y)$.
This gives a contradiction, and so
there must exist a neighbourhood $U_{\p}$ of $\p$ where $h(z)=0$. $\hfill \blacksquare$ \\

Now, we are ready to prove a smoothness result for $\tbd \{\lambda>0\}$.
\begin{proposition}\label{C1}
Let $\kid$ be a static KID
and consider a connected component $\{\lambda>0\}_{0}$ of $\{ \lambda>0 \}$. If $Y^i \nablaSigma_i \lambda \geq 0$
or $Y^i \nablaSigma_i \lambda \leq 0$ on an arc-connected component $\E$ of $\tbd \{\lambda>0\}_{0}$, then $\E$ is a smooth
submanifold (i.e. injectively immersed) of $\Sigma$.
\end{proposition}

{\bf Proof.}  If there are no fixed points in $\E$, the result
follows from Lemma \ref{surfaceNoFixed}. Let us therefore assume that
there is at least one fixed point $\p \in \E$.
The idea of the proof proceeds in three stages. The first stage will
consist in showing that
$Y^i \nablaSigma_i \lambda  \geq 0$ (or
$Y^i \nablaSigma_i \lambda \leq 0$) forces all fixed points in $\E$ to be non-transverse. The second one
consists in proving that,
in a neighbourhood of a non-transverse fixed point, $\E$ is a $C^1$ submanifold.
In the third and final stage we prove that $\E$ is, in fact, $C^{\infty}$.

{\it Stage 1.} 
We argue by contradiction. Assume the fixed point $\p$ is transverse.
Lemma \ref{structurebneq0} implies that either
$\{\lambda>0\}_{0} \cap U_{\p} = \{x > \frac{|b| |y|}{\mu} \}$ or $\{\lambda>0\}_{0} \cap U_{\p} = \{
x < - \frac{|b| |y|}{\mu} \}$. We treat the first case (the other is similar).
The boundary of $\{\lambda>0\}_{0} \cap U_{\p}$ is connected and given
by $x = x_+(y)$ for $y>0$ and
$x=x_{-}(y)$ for $y<0$. Using $d\lambda = 2 \kappa \bm{Y}$ on this boundary, it follows
$Y^i \nablaSigma_i \lambda = 2 \kappa Y_i Y^i$. But $\kappa$ has different signs on the
branch $x=x_{+}(y)$ and on the branch $x=x_{-}(y)$, so $Y^i \nablaSigma_i \lambda$ also
changes sign, against hypothesis. Hence $\p$ must be a
non-transverse fixed point.

{\it Stage 2.} Let us show that there exists a neighbourhood of
$\p$ where $\E$ is $C^1$. 
Being $\p$ non-transverse, we have $f_{ij} |_{\p} =0$
and, consequently, the Hessian of $\lambda$ reads
\begin{equation}\label{Hessianlambda}
\nablaSigma_i \nablaSigma_j \lambda |_{\p} = 2 \nablaSigma_i N \nablaSigma_j N |_{\p},
\end{equation}
which has signature $\{+,0,0\}$. Similarly as in Lemma \ref{I1<0}, the Gromoll-Meyer splitting Lemma (see Appendix \ref{ch:appendix2})
implies the existence of an open neighbourhood $U_{\p}$ of $\p$
and coordinates $\{x,z^A \}$ in $U_{\p}$ such that $\p = \{x=0, z^A =0\}$
and $\lambda = Q_0^2 x^2 - \zeta (z)$, where $\zeta$ is a smooth function
satisfying $\zeta|_{\p} =0$, $\nablaSigma_i \zeta |_{\p} =0$ and $\nablaSigma_{i}\nablaSigma_{j} \zeta |_{\p} =0$, and
$Q_{0}$ is a positive constant. Moreover,
evaluating the Hessian of $\lambda=Q^{2}_{0}x^{2}-\zeta(z)$ and comparing with (\ref{Hessianlambda})
gives $dx |_{\p} = Q_0^{-1} dN |_{\p}$.

Let us first show that there exists a neighbourhood $V_{\p}$ of $\p$ where $\zeta \geq 0$.
The surfaces $\{ N=0 \}$ and $\{ x=0 \}$ are tangent at $\p$. This implies that
there exists a neighbourhood $V_{\p}$ of $\p$ in $\Sigma$ such that the integral lines of $\partial_x$
are transverse to $\{N=0 \}$. Assume $\zeta(z) <0$ on any of these integral lines. If follows
that $\lambda = Q_0^2 x^2 - \zeta$ is positive everywhere on this line. But at the intersection with
$\{N=0\}$ we have $\lambda = N^2 - Y^i Y_i = - Y^i Y_i \leq 0$. This gives a contradiction and
hence $\zeta(z)\geq 0$ in $V_{\p}$ as claimed.

The set of points $\{ \lambda > 0 \} \cap V_{\p}$ is given by the union of two
disjoint connected sets namely $W_{+} \equiv \{ x > + \frac{\sqrt{\zeta}}{Q_{0}} \}$
and $W_{-} \equiv
\{ x < - \frac{\sqrt{\zeta}}{Q_{0}} \}$. On a connected component of $\{\lambda >0\}$ (in particular on $\{\lambda>0\}_{0}$)
we have that $N = \sqrt{\lambda + Y^i Y_i}$ must be either
everywhere positive or everywhere negative. On the other hand,
for $\delta >0$ small
enough $N |_{(x=\delta, z^A=0)}$ must have different sign
than $N |_{(x=-\delta, z^A=0)}$  (this is because $\partial_x N |_{\p}= dN
(\partial _x)|_{\p} = Q_0 dx (\partial_x) |_{\p} >0)$. It follows that either
$\{\lambda>0\}_{0} \cap V_{\p} = W_{+}$ (if $N>0$ in $\{\lambda>0\}_{0}$) or $\{\lambda>0\}_{0} \cap V_{\p} = W_{-}$ (if
$N<0$ in $\{\lambda>0\}_{0}$). Consequently, $\E$ is locally defined by
$x = \frac{\epsilon \sqrt{\zeta}}{Q_0}$, where $\epsilon$ is the sign of $N$
in $\{\lambda>0\}_{0}$. 
Now, we need to prove that $+ \sqrt{\zeta}$
is $C^1$. This requires studying the behavior of $\zeta$ at points where it
vanishes.

The set of fixed points $\p' \in V_{\p}$ is given by
$\{ x=0, \zeta(z)=0\}$ (this is a consequence of the fact that fixed
points in $\E$ are characterized by the equations $\lambda=0$ and $d\lambda=0$,
or equivalently $x=0$, $\zeta=0$, $d\zeta=0$. Since, for non-negative functions,
$\zeta=0$ implies $d\zeta=0$ the statement above follows).
The Hessian of $\lambda$ on any fixed point $\p' \subset V_{\p}$ reads
$\nablaSigma_i \nablaSigma_j \lambda |_{\p'} = 2 Q_0^2 (dx \otimes dx)_{ij} - \nablaSigma_{i}\nablaSigma_{j} \zeta |_{\p'}$.
Since $\p'$ must be a non-transverse fixed point, we have
$\nablaSigma_i Y_j |_{\p'} =f_{ij}|_{\p'}=0$ and hence
$\nablaSigma_{i}\nablaSigma_{j}\lambda |_{\p'}= 2\nablaSigma_{i}N\nablaSigma_{j}N |_{\p'}$ which has rank 1.
Consequently,
$\nablaSigma_{i}\nablaSigma_{j} \zeta |_{\p'} =0$. So, at all points where $\zeta$ vanishes
we not only have $d\zeta =0$ but also $\nablaSigma_{i}\nablaSigma_{j} \zeta =0$. We can now apply a theorem by
Glaeser (see Appendix \ref{ch:appendix2}) to conclude that
the positive square root $u \equiv \frac{+\sqrt{\zeta}}{Q_{0}}$ is $C^1$, as claimed.

{\it Stage 3.}
Finally, we will prove that $\E$ is, in fact, $C^{\infty}$ in a neighbourhood of $\p$
(we already know that $\E$ is smooth
at non-fixed points) 
This is equivalent to proving that the function $x=\epsilon u(z)$ is $C^{\infty}$.
Since $u=\frac{+\sqrt{\zeta}}{Q_{0}}$ and $\zeta\geq 0$, it follows that $u$ is smooth at any point
where $u>0$.
The proof will proceed in two steps. In the first step we will show that
$u$ is $C^{2}$ at points where $u$ vanishes and then, we will improve this to $C^{\infty}$.
Let us start with the $C^{2}$ statement.
At points where $u \neq 0$, we have $Y_{i} |_{(x=
\epsilon u(z),z^A)}= \frac{1}{2\kappa} \nablaSigma_i \lambda |_{(x=\epsilon u(z),z^A)}$. Hence $Y_i$ is non-zero and
orthogonal to $\E$ on such points.  Pulling back equation
$\nablaSigma_i Y_j + \nablaSigma_j Y_i + 2 N K_{ij}=0$ onto $\E\cap\{x\neq 0\}$, we get
\begin{equation}\label{kappaandK}
\kappa_{AB} + \epsilon \sigma K_{AB}=0,
\end{equation}
where $\sigma$
is the sign of $\kappa$ , $K_{AB}$ is the pull-back of $K_{ij}$
on the surface $\{x=\epsilon u(z)\}$ and $\kappa_{AB}$ is the second
fundamental form of this surface with respect to the unit normal pointing
inside  $\{\lambda >0 \}_{0}$.
By assumption $Y^i \nablaSigma_i \lambda$ has constant sign on
$\E$. This implies that $\sigma$ is either everywhere $+1$ or everywhere $-1$.
So, the graph $x= \epsilon u(z)$ satisfies the set of equations
$\kappa_{AB} + \epsilon \sigma K_{AB}=0$ on the open set $\{ z^A ; u(z) > 0 \}\subset \mathbb{R}^{2}$.
In the local coordinates $\{z^A\}$ these equations takes the form
\begin{eqnarray}
- \partial_{A} \partial_B u(z) + \chi_{AB}(u(z),\partial_C u(z), z )=0
\label{fulleq}
\end{eqnarray}
where $\chi$ is a smooth function of its arguments which satisfies
$\chi_{AB}(u=0,\partial_{C} u =0, z) = \epsilon\hat{\kappa}_{AB} (z) +  \sigma
\hat{K}_{AB} (z)$, where $\hat{\kappa}_{AB}$ is the second fundamental form
of the surface $\{ x=0 \}$ (with respect to the outer normal pointing towards $\{x>0\}$)
at the point with coordinates $\{z^{A}\}$ and
$\hat{K}_{AB}$ is the pull-back of $K_{ij}$ on this surface
at the same point. Take a fixed point $\p' \in \E$ not lying
within an open set of fixed points (if $\p'$ lies on an open set of fixed points we have $u\equiv 0$ on the open
set and the statement that $u$ is $C^{\infty}$ is trivial). It follows that $\p' \in \{x=0 \}$
and that the coordinates $z_0^A$ of $\p'$ satisfy $z_0^A \in \tbd\{ z^A ;  u(z)>0 \} \subset \mathbb{R}^2$.
By stage $2$ of the proof, the function
$u(z)$ is $C^1$ everywhere and its gradient vanishes wherever $u$ vanishes.
It follows that $u\big|_{z^A_0} = \partial_B u \big|_{z^A_0} = 0$. Being $u$ continuously
differentiable, it follows that the term $\chi_{AB}$ in (\ref{fulleq})
is $C^0$ as a function of $z^C$ and therefore admits a limit at $z^C_0$.
It follows that $\partial_{A} \partial_{B} u$ also has a well-defined limit at $z^C_0$,
and in fact this limit satisfies
\begin{eqnarray*}
\partial_A \partial_B u \big|_{z^C_0} = \hat{\kappa}_{AB} \big|_{z^C_0} + \epsilon \sigma \hat{K}_{AB}
\big|_{z^C_0}.
\end{eqnarray*}
This shows that $u$ is in fact $C^2$ everywhere. But taking the trace of
$\kappa_{AB}+\epsilon\sigma K_{AB}=0$, we get $p + \epsilon\sigma q =0$, where $p$ is the mean curvature of $\E$ and
$q$ is the trace of the pull-back of $K_{ij}$ on $\E$. This is an elliptic
equation in the coordinates $\{z^{A}\}$ (see e.g. \cite{AMS}), so $C^2$ solutions are smooth as a consequence of elliptic regularity
\cite{GilbargTrudinger}. Thus, the function $u (z)$ is
$C^{\infty}$. 
$\hfill \blacksquare$ \\

Knowing that this submanifold is differentiable, our next aim is to show that, under suitable circumstances
it has vanishing outer null expansion. This is the content of our next proposition.

\begin{proposition}
\label{is_a_MOTS}
Let $\kid$ be a static KID
and consider a connected component $\{\lambda>0\}_{0}$ of $\{ \lambda>0 \}$ with non-empty
topological boundary. Let $\E$ be an arc-connected component $\tbd \{\lambda>0\}_{0}$ and assume
\begin{itemize}
\item[(i)] $N Y^i \nablaSigma_i \lambda|_{\E} \geq 0$  if $\E$ contains at least one fixed
point.
\item[(ii)]
$N Y^i m_i |_{\E} \geq 0$  if $\E$ contains no fixed point,
where $\vec{m}$ is the unit normal pointing towards $\{\lambda>0\}_{0}$.
\end{itemize}
Then $\E$ is a smooth submanifold (i.e. injectively immersed) with $\theta^{+}=0$ provided the outer direction is defined as
the one pointing towards $\{\lambda>0\}_{0}$. Moreover, if $I_{1}\neq 0$ in $\E$, then
$\E$ is embedded.
\end{proposition}

{\bf Remark.} If the inequalities in (i) and (ii) are reversed, then $\E$ has $\theta^{-}=0$. $\hfill \square$ \\

{\bf Proof.}
Consider first the case when $\E$ has at least one fixed point.
Since, on $\E$, $N$ cannot change sign and vanishes only if $\vec{Y}$ also vanishes,
the hypothesis $N Y^i \nablaSigma_i \lambda |_{\E} \geq 0$
implies either $Y^i \nablaSigma_i \lambda |_{\E} \geq 0$ or
$Y^i \nablaSigma_i \lambda |_{\E} \leq 0$ and, therefore, Proposition \ref{C1}
shows that $\E$ is a smooth
submanifold. Let $\vec{m}$ be the unit normal pointing towards $\{\lambda>0\}_{0}$ and $p$ the
corresponding mean curvature.  We have to
show that $\theta^{+}=p + \gamma^{AB} K_{AB}$ (see equation (\ref{Hp})) vanishes.
Open sets of fixed points
are immediately covered by Proposition \ref{Totally geodesic} because
this set is then totally geodesic and $K_{AB}=0$, so that both null expansions
vanish.

On the subset $V\subset \E$ of non-fixed points we have
$Y_{i}\big|_{V}=\frac{1}{2\kappa}\nablaSigma_{i}\lambda\big|_{V}$ (see equation \ref{kappa})
and, therefore, $Y_{i}\big|_{V}=|N|\text{sign}(\kappa)m_{i}\big|_{V}$.
The condition $NY^{i}\nablaSigma_{i}\lambda\geq 0$ imposes
$\text{sign}(N)\text{sign}(\kappa)=1$ or, in the notation of the proof of Proposition \ref{C1},
$\epsilon\sigma=1$. Equation $p+q=0$ follows directly from
(\ref{kappaandK}) after taking the trace.

For the case $(ii)$, we know that $\E$ is smooth from Lemma \ref{surfaceNoFixed} and, hence,
$\vec{m}$ exists (this shows
in particular that hypothesis (ii) is well-defined).
Since $\E$ lies in a Killing prehorizon in the Killing development of the KID,
it follows that $\vec{\xi}$ is orthogonal to $\E$ and hence that $\vec{Y}$ is
normal to $\E$ in $\Sigma$. Since $\vec{Y}^2 = N^2$ on $\E$ it follows
$\vec{Y}|_{\E}=N\vec{m}|_{\E}$ and the same argument applies to conclude $\theta^{+}=0$.

To show that 
$\E$ is embedded if $I_{1}|_{\E}\neq 0$,
consider a point $\p\in \E$.
if $\p$ is a non-fixed point, we know that $\nablaSigma_{i}\lambda\big|_{\p}\neq 0$ and hence $\lambda$ is
a defining function for $\E$ in a neighbourhood of $\p$. This immediately implies that
$\E$ is embedded in a neighbourhood of $\p$.
When $\p$ is a fixed point, we have shown in the proof of Proposition
\ref{C1} that there exists an open neighbourhood $V_{\p}$ of $\p$ such that,
in suitable coordinates, $\overline{\{\lambda>0\}}\cap V_{\p}=\{x\geq u(z)\}$ or
$\overline{\{\lambda>0\}}\cap V_{\p}=\{x\leq -u(z)\}$ for a non-negative smooth
function $u(z)$. It is clear that the arc-connected component $\E$ is defined locally by
$x=u(z)$ or $x=-u(z)$ and hence it is embedded. $\hfill \blacksquare$ \\

\setcounter{equation}{0}
\section{The confinement result}
\label{mainresults}

Now, we are ready to state and prove our confinement result.
For simplicity, it will be formulated as a confinement result for outer trapped
surfaces instead of weakly outer trapped surfaces.
However, except for a singular situation,
it can be immediately extended to weakly outer trapped surfaces (see Remark 1 after the proof).

\begin{thr}\label{theorem2}
Consider a static KID $\kid$ satisfying the NEC and possessing a barrier $\Sb$ with interior
$\Omegab$ (see Definition \ref{defi:barrier}) which is outer untrapped and such that
such that $\lambda\big|_{\Sb}>0$.
Let $\{\lambda>0\}^{\text{ext}}$ be the connected component of
$\{\lambda>0\}$ containing $\Sb$. 
Assume that every arc-connected component of $\tbd \ext$
with $I_{1}=0$ is topologically closed and
\begin{enumerate}
\item $NY^{i}\nablaSigma_{i}\lambda\geq 0$ in each arc-connected component of $\tbd \ext$ containing
at least one fixed point.
\item $NY^{i}m_{i}\geq 0$ in each arc-connected component of $\tbd \ext$ which contains no fixed points, where
$\vec{m}$ is the unit normal pointing towards $\{\lambda>0\}^{\text{ext}}$.
\end{enumerate}
Consider any surface $S$ which is bounding with respect to $\Sb$. If $S$ is outer trapped then it does not intersect $\{\lambda>0\}^{\text{ext}}$.
\end{thr}

\begin{figure}
\begin{center}
\psfrag{S0}{\color{red}{$\E_{0}$}}
\psfrag{S}{\color{blue}{$S$}}
\psfrag{plambda}{}
\psfrag{pS}{$\bd \Sigma$}
\psfrag{Sigma}{$\Sigma$}
\psfrag{lambda}{$\{\lambda>0\}$}
\psfrag{Sb}{$S_{b}$}
\includegraphics[width=9cm]{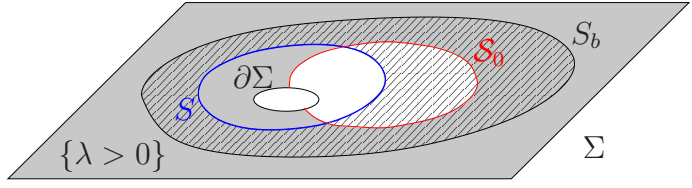}
\caption {Theorem \ref{theorem2} forbids the existence of an outer trapped surface $S$ like the one in the
figure (in blue). The striped area corresponds to the exterior of $S$ in $\Omegab$ and the shaded area corresponds
to the set $\{\lambda>0\}^{\text{ext}}$ whose boundary is $\E_{0}$ (in red). Note that $\E_{0}$ may intersect $\bd \Sigma$.}
\end{center}
\end{figure}


{\bf Proof.} We argue by contradiction. Let $S$ be an outer trapped surface which is
bounding with respect to
$\Sb$, satisfies the hypotheses
of the theorem and intersects $\{\lambda>0\}^{\text{ext}}$.
By definition of bounding, there exists a compact manifold
$\Sigmatilde$ whose boundary is the disjoint union of the
outer untrapped surface $\Sb$ and the outer trapped surface $S$. We work on $\Sigmatilde$
from now on. The Andersson and Metzger
Theorem \ref{thr:AM} implies that the topological
boundary of the weakly outer trapped region $\tbd T^{+}$ in $\Sigmatilde$ is a stable MOTS which
is bounding with
respect to $\Sb$. We first show that $\tbd T^+$ necessarily intersects $\{\lambda>0\}^{\text{ext}}$.
Indeed,
consider a point $\r \in S$ with $\lambda |_{\r} >0$ (this point exists by hypothesis) and
consider a path from $\r$ to $\Sb$ fully contained in
$\{\lambda>0\}^{\text{ext}}$ (this path exists because
$\{\lambda>0\}^{\text{ext}}$ is connected). Since $\r \in T^+$ it follows that
this path must intersect $\tbd T^+$
as claimed.
Furthermore, due to the maximum principle
for MOTS (see Proposition \ref{maximumprincipleforMOTS}), $\tbd T^{+}$ lies entirely in the exterior of $S$ in $\Omegab$
(here is where we use the hypothesis of $S$ being outer trapped instead of merely
being weakly  outer trapped).

Let us suppose for a moment that
$\tbd T^{+}\subset\overline{\{\lambda>0\}^{\text{ext}}}$.
Then the Killing vector $N\vec{n}+\vec{Y}$ is causal everywhere
on $\tbd T^{+}$, either future or past directed, and timelike somewhere on $\tbd T^{+}$.
Since $\tbd T^{+}$ intersects $\{\lambda>0\}^{\text{ext}}$, there must be non-fixed points on $\tbd T^{+}$.
If all points in $\tbd T^{+}$ are non-fixed, then we can construct the Killing development and
Theorem \ref{corollaryextended} can be applied at once giving a contradiction (note that $\tbd T^{+}$ is necessarily a locally outermost MOTS).
When $\tbd T^{+}$ has fixed points we cannot construct the Killing development everywhere. However, let
$V\subset\tbd T^{+}$ be a connected component of the set of non-fixed points in $\tbd T^+$
satisfying $V \cap \{\lambda >0 \} \neq \emptyset$
(this $V$ exists because $\lambda >0$ somewhere on $\tbd T^+$). Then,
the Killing development still exists in an open neighbourhood of $V$. In this portion we can repeat
the geometrical construction which allowed us to prove Theorem \ref{corollaryextended} and define
a surface $S'$ by moving $V$ a small, but finite amount $\tau$ along $\vec{\xi}$ to the past
and back to $\Sigma$ along the outer null geodesics.
Since $N$ and $\vec{Y}$ are smooth and approach zero at $\tbd V$
it follows that $S'$ and the set of fixed points in
$\tbd T^{+}$ join smoothly and
therefore define a closed surface $S''$. Clearly,
$S''$ is weakly outer trapped and lies, at least partially,
in the exterior of $\tbd T^{+}$, which is impossible.

Until now, we have essentially applied the ideas of Theorem \ref{corollaryextended}.
When $\tbd T^{+}\not\subset\overline{\{\lambda>0\}^{\text{ext}}}$ new methods are required. However, the
general strategy is still
to construct a weakly outer trapped surface outside $\tbd T^{+}$
in $\Sigmatilde$.

First of all, every arc-connected component $\E_{i}$ of $\tbd \ext$ with $I_1 \neq 0$
is embedded, as proven in Proposition \ref{is_a_MOTS}. For an  arc-connected component $\E_d$
with $I_1=0$ we note that, since no point on this set is a fixed point, it follows that
there exists an open neighbourhood $U$ of $\E_d$ containing no fixed points. Thus, the vector
field $\vec{Y}$ is nowhere zero on $U$. Staticity of the KID implies that $\bm{Y}$ is integrable
(see (\ref{statictwo})). It follows by the Fr\"obenius theorem that $U$ can be foliated by
maximal, injectively immersed submanifolds orthogonal to $\vec{Y}$. $\E_d$ is clearly
one of the leaves of this foliation because $\vec{Y}$ is orthogonal to $\E_d$ everywhere. By assumption,
$\E_d$ is topologically closed. Now, we can invoke a result on the theory of foliations that states
that any topologically closed leaf in a foliation is necessarily embedded (see e.g. Theorem 5 in
page 51 of \cite{Neto}).
Thus, each $\E_i$ is an embedded submanifold of $\Sigmatilde$.
Since we know that $\tbd T^+$ intersects $\ext$ and we are
assuming that $\tbd T^{+}\not\subset\overline{\{\lambda>0\}^{\text{ext}}}$, it follows that
at least one of the arc-connected components $\{\E_i\}$, say $\E_0$, must intersect both the interior
and the exterior of $\tbd T^+$ .
In Proposition \ref{is_a_MOTS} we have also shown that $\E_{0}$ has $\theta^{+}=0$ with respect to the direction
pointing towards $\{\lambda>0\}^{\text{ext}}$.

Thus, we have two intersecting surfaces $\tbd T^{+}$ and $\E_{0}$ which satisfy
$\theta^{+}=0$. Moreover, $\tbd T^{+}$ is a stable MOTS.
The idea is to use Lemma \ref{lemasmoothness} by Kriele and Hayward to construct a weakly outer trapped
surface $\hat{S}$ outside both $\tbd T^{+}$ and $\E_{0}$ and which is bounding with respect to $\Sb$.
However, Lemma \ref{lemasmoothness} can be applied directly only when both surfaces
$\tbd T^{+}$ and $\E_{0}$ intersect transversally in a curve and this need not happen for
$\E_{0}$ and $\tbd T^{+}$. To address this issue we use a technique developed by
Andersson and Metzger in their proof of Theorems 5.1 and 7.6 in \cite{AM}.

The idea is to use Sard Lemma
(see Appendix \ref{ch:appendix2}) in order to find a weakly outer trapped surface $\tilde{S}$
as close to $\tbd T^{+}$ as desired which does intersect $\E_{0}$ transversally. Then,
the Kriele and Hayward smoothing procedure applied to $\tilde{S}$ and $\E_{0}$
gives a weakly outer trapped surface penetrating $\Sigmatilde\setminus T^{+}$,
which is simply impossible.

So, it only remains to prove the existence of $\tilde{S}$.


Recall that $\tbd T^{+}$ is a stable MOTS. We will distinguish two cases.
If $\tbd T^{+}$
is strictly stable,
there exists a foliation
$\{\Gamma_{s}\}_{s\in \left(-\epsilon,0\right]}$ of
a one sided
tubular neighbourhood ${\cal W}$ of $\tbd T^+$ in $T^+$ such that
$\Gamma_{0}=\tbd T^+$
and all the surfaces $\{\Gamma_{s} \}_{s<0}$ have $\theta^{+}_{s}<0$.
To see this, simply choose a variation vector $\vec{\nu}$ such that $\vec{\nu}\big|_{\tbd T^+}=\psi\vec{m}$ where
$\psi$ is a positive principal eigenfunction of the stability operator $L_{\vec{m}}$ and $\vec{m}$
is the outer direction normal to $\tbd T^{+}$. Using $\delta_{\vec{\nu}}\theta^+=L_{\vec{m}}\psi=\lambda\psi>0$
it follows that the surfaces $\Gamma_{s}\equiv \varphi_{s}(\tbd T^+)$ generated by $\vec{\nu}$ are outer trapped
for $s\in(-\epsilon,0)$.
Next, define the mapping $\Phi: \E_{0}\cap  ({\cal W} \setminus \tbd T^+) \rightarrow
\left(-\epsilon,0\right)\subset\mathbb{R}$
which assigns to each point $\p\in \E_{0}\cap({\cal W} \setminus \tbd T^+)$
the corresponding
value of the parameter of the foliation $s\in\left(-\epsilon,0\right)$ on $\p$.
Sard Lemma (Lemma \ref{lema:Sard}) implies that
the set of regular values of the mapping $\Phi$ is dense in
$\left(-\epsilon,0\right)\subset\mathbb{R}$.
Select a regular value $s_{0}$ as close to $0$ as desired. Then, the surface $\tilde{S}\equiv \Gamma_{s_{0}}$
intersects transversally
$\E_{0}$, as required.

If $\tbd T^+$ is stable but {\it not strictly stable}, a foliation $\Gamma_s$
consisting on weakly outer trapped surfaces may not exist.
Nevertheless, following \cite{AM}, a suitable modification of
the interior of $\tbd T^+$ in $\Sigma$ solves this problem.
It is important to remark that, in this case, the contradiction which proves the theorem
is obtained by applying the Kriele and Hayward Lemma in the modified initial data set.
The modification is performed as follows.
Consider the same foliation $\Gamma_{s}$ as defined above
and replace the second fundamental
form $K$ on the hypersurface $\Sigma$ by the following.
\begin{equation}\label{Ktilde}
\tilde{K}=K-\frac 12 \phi(s) \gamma_{s},
\end{equation}
where $\phi : \mathbb{R}\rightarrow \mathbb{R}$ is a $C^{1,1}$
function such that $\phi(s)=0$ for $s\geq 0$ (so that the data remains unchanged outside
$\tbd T^{+}$) and $\gamma_s$ is the
projector to $\Gamma_s$.
Then, the outer null expansion of $\Gamma_s$ computed in the modified
initial data set $(\Sigma,g,\tilde{K})$
\[
{\tilde{\theta}^{+}}[\Gamma_{s}]={{\theta}^{+}}[\Gamma_{s}]-\phi(s),
\]
where ${{\theta}^{+}}[\Gamma_{s}]$ is the outer null expansion of $\Gamma_{s}$
in $(\Sigma,g,K)$.
Since $\tbd T^{+}$ was a stable but not strictly stable MOTS in $(\Sigma,g,K)$,
${\theta^+}[\Gamma_{s}]$ vanishes at least to second order at $s=0$.
On $s \leq 0$, define $\phi(s)=bs^2$ with $b$ a sufficient large constant. It follows that
for some $\epsilon >0$ we have
${\tilde{\theta}^{+}}[\Gamma_{s}]<0$ on all $\Gamma_s$ for $s \in (-\epsilon , 0)$.
Working with this foliation, Sard Lemma asserts that a weakly outer trapped surface
$\Gamma_{s_0}$ lying as close to $\tbd T^{+}$ as desired and intersecting $\E_0$ transversally
can be chosen in $(\Sigma,g,\tilde{K})$.

Furthermore, the surface $\E_{0}$ also has non-positive outer null expansion in the modified initial data,
at least for $s$ sufficiently close to zero.
Indeed, this outer null expansion $\tilde{\theta}^{+} [\E_0]$ reads
$\tilde{\theta}^{+} [\E_0]=p [\E_0]+\tr_{\E_{0}}\tilde{K}$.
By (\ref{Ktilde}), we have
$\tr_{\E_{0}}\tilde{K}\big|_{\r}=\tr_{\E_{0}}K\big|_{\r}-\frac12\phi(s_{\r})\tr_{\E_{0}}\gamma_{s_{\r}}$,
at any point $\r\in\E_{0}$, where $s_{\r}$ is the value of the leaf $\Gamma_{s}$ containing $\r$, i.e.
$\r \in \Gamma_{s_{\r}}$.
Since $\tr_{\E_{0}} \gamma_{s}\geq 0$ (because the pull-back of $\gamma_{s}$ is positive semi-definite)
we have $\tr_{\E_{0}}\tilde{K} = \tr_{\E_{0}} K$ for
$s \geq 0$ and  $\tr_{\E_{0}}\tilde{K} \leq  \tr_{\E_{0}} K$ for $s<0$ (small enough).
In any case $\tilde{\theta}^{+} (\E_0) \leq \theta^{+} (\E_0) =0$ and
we can apply the Kriele and Hayward Lemma to $\Gamma_{s_{0}}$ and $\E_{0}$ to construct a weakly outer trapped
surface which is bounding with respect to $\Sb$, lies in the topological
closure of the exterior of $\tbd T^+$ and penetrates this exterior somewhere.
Since the geometry outside $\tbd T^{+}$ has not been modified, this gives a contradiction. $\hfill\blacksquare$ \\

{\bf Remark 1.}
This theorem has been formulated for outer trapped surfaces instead of weakly outer trapped surfaces.
The reason is that in the proof we have used a foliation in the {\it inside} part of
a tubular neighbourhood of $\tbd T^{+}$.
If $S$ satisfies $\theta^+=0$, it is possible that $S=\bd \Sigma = \tbd T^+$
and then we would not have room to use this foliation. It follows that the hypothesis of the theorem can be relaxed to
$\theta^{+}\leq 0$ if one of the following conditions hold:
\begin{enumerate}
\item $S$ is not the outermost MOTS.
\item $S\cap \bd\Sigma=\emptyset$.
\item The KID $\kid$ can be isometrically embedded into another KID
$(\hat{\Sigma},\hat{g},\hat{K},\hat{N},\vec{\hat{Y}},\hat{\tau})$
with $\bd\Sigma\subset \text{int}(\hat{\Sigma})$
\end{enumerate}
In this case, Theorem \ref{theorem2} includes Miao's theorem in the particular case of asymptotically
flat time-symmetric vacuum static KID with minimal compact boundary.
This is because in the time-symmetric case all points with $\lambda=0$
are fixed points and hence there are no arc-connected components of
$\tbd \{ \lambda > 0 \}$ with $I_1=0$ and
$Y^{i}\nabla^{\Sigma}_{i}\lambda$ is identically zero on $\tbd \ext$. $\hfill \square$ \\


{\bf Remark 2.} In geometric terms, hypotheses $1$ and $2$ of the theorem exclude a priori the possibility
that $\tbd \ext$ intersects the
white hole Killing horizon at non-fixed points.  A similar theorem exists for initial
data sets which do not intersect the black hole Killing horizon (more precisely, such that
both inequalities in $1$ and $2$ are satisfied with the
reversed inequality signs). The conclusion of the theorem in this case is that
no bounding {\it past} outer trapped surface can intersect $\{\lambda>0\}^{\text{ext}}$
provided $\Sb$ is a {\it past} outer untrapped barrier (the proof of this statement can be
obtained by applying Theorem \ref{theorem2} to the static KID
$(\Sigma,g,-K;-N,\vec{Y}; \rho, -\vec{J}, \tau)$).

No version of this theorem, however,
covers the case when $\tbd \ext$ intersects both the black hole and the white hole Killing horizon. The reason is that, in this
setting, $\tbd \ext$ is, in general, not smooth and we cannot apply
the Andersson-Metzger theorem to $\Sigmatilde$. In the next chapter we will address this case in more detail. $\hfill \square$ \\

For the particular case of KID possessing an asymptotically flat end we have the following corollary,
which is an immediate consequence of Theorem \ref{theorem2}.

\begin{corollary}\label{corollarytheorem2}
Consider a static KID $\kid$
with a selected asymptotically flat end $\Sigma_{0}^{\infty}$ and satisfying the NEC.
Denote by $\{\lambda>0\}^{\text{ext}}$ the connected component of
$\{\lambda>0\}$ which contains the asymptotically flat end $\Sigma_{0}^{\infty}$.
Assume that every arc-connected component of $\tbd \ext$
with $I_{1}=0$ is closed and
\begin{enumerate}
\item $NY^{i}\nablaSigma_{i}\lambda\geq 0$ in each arc-connected component of $\tbd \{\lambda>0\}^{\text{ext}}$ containing
at least one fixed point.
\item $NY^{i}m_{i}\geq 0$ in each arc-connected component of $\tbd \{\lambda>0\}^{\text{ext}}$ which contains no fixed points, where
$\vec{m}$ is the unit normal pointing towards $\{\lambda>0\}^{\text{ext}}$.
\end{enumerate}
Then, any bounding (see Definition \ref{defi:bounding}) outer trapped surface $S$
in $\Sigma$ cannot
intersect $\{\lambda>0\}^{\text{ext}}$.
\end{corollary}


\chapter{Uniqueness of static spacetimes with weakly outer trapped surfaces}
\chaptermark{Uniqueness of static spacetimes with trapped surfaces}
\label{ch:Article4}

\setcounter{equation}{0}
\section{Introduction}\label{sectionintroduction}

In this chapter we will extend the classic static black hole uniqueness theorems to
asymptotically flat static KID containing weakly outer trapped surfaces. As emphasized in the previous chapter,
the first step for this extension
was given by Miao for the particular case of
asymptotically flat, time-symmetric, static and vacuum KID, with compact minimal boundary (Theorem \ref{thr:Miao}).
Indeed, our aim of extending the classic uniqueness theorems for static black holes to the quasi-local setting
can be reformulated
as generalizing Theorem \ref{thr:Miao} to non-vanishing matter (as long as the NEC is satisfied) and
arbitrary slices (not necessarily time-symmetric) containing weakly outer trapped surfaces.
In the previous chapter we obtained a generalization of this result as a confinement result.
In this chapter we address the extension of Miao's theorem as a uniqueness result.

As we already know, the most powerful method
to prove uniqueness of static black holes is the {\it doubling method} of
Bunting and Masood-ul-Alam. This method was described in some detail in Section
\ref{sc:UniquenessOfBlackHoles} where we gave a sketch of the proof of the
uniqueness theorem for static electro-vacuum black holes.
In the present chapter, our
strategy will be precisely to recover the framework of the doubling method from an arbitrary static
KID containing a weakly outer trapped surface. As it was discussed in Section \ref{sc:UniquenessOfBlackHoles},
this framework consists of an asymptotically flat spacelike hypersurface $\Sigma$ with
topological boundary $\tbd \Sigma$ which is a closed (i.e. compact and without boundary)
embedded topological manifold and such that
the static Killing field is causal on $\Sigma$ and null only on $\tbd \Sigma$.
As we pointed out in Section \ref{sc:UniquenessOfBlackHoles}, the existence of this topological manifold $\tbd \Sigma$ is ensured precisely by
the presence of a black hole.
Note that $\tbd \Sigma$ is not required to be smooth.

\begin{figure}[h]
\begin{center}
\psfrag{S}{\color{blue}{$S$}}
\psfrag{Sigmap}{{$\Sigma_{0}^{\infty}$}}
\psfrag{pS}{{$\bd \Sigma$}}
\psfrag{lambda}{\color{red}{$\tbd\ext$}}
\includegraphics[width=9cm]{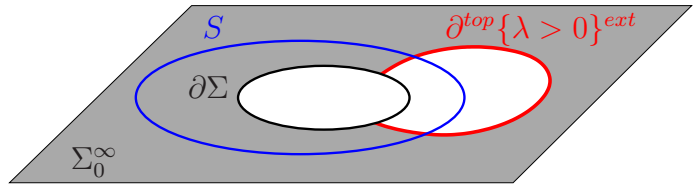}
\caption {The figure illustrates a situation where $\tbd \ext$ (in red) has non-empty manifold boundary (which lies in $\bd \Sigma$) and, therefore, is not closed.
Here, $S$ (in blue) represents a bounding MOTS and the grey region corresponds to $\ext$.
In a situation like this the doubling method cannot be applied.}
\label{fig:problem}
\end{center}
\end{figure}

Hence, our strategy to conclude uniqueness departing from a static KID $\kid$ with
an asymptotically flat end $\Sigma_{0}^{\infty}$ which contains a bounding MOTS $S$ will be therefore
to prove that the topological boundary $\tbd \ext$, where $\ext$ is the connected component
of $\{\lambda>0\}$ in $\Sigma$ which contains $\Sigma_{0}^{\infty}$, is a closed embedded topological
submanifold.
Since a priori MOTS have nothing to do with black holes, 
$\tbd \ext$ may fail to be closed
(see Figure \ref{fig:problem}) as required in the doubling method.
Consequently, throughout this chapter we will study under which conditions we can guarantee that
$\tbd \ext$ is closed. In fact, it turns out that
the confinement Theorem \ref{theorem2} and its Corollary
\ref{corollarytheorem2}
are already sufficient to conclude
that $\tbd \ext$ is a closed surface. This leads to our first
uniqueness result.

\begin{thr}\label{uniquenessthr0}
Consider a static KID $\kid$ with a selected asymptotically flat end
$\Sigma^{\infty}_0$ and satisfying the NEC. Assume that
$\Sigma$ possesses an outer trapped surface $S$ which is bounding. 
Denote by $\ext$ the connected component of
$\{\lambda>0\}$ which contains the asymptotically flat end $\Sigma_{0}^{\infty}$. If
\begin{enumerate}
\item Every arc-connected component of $\tbd \ext$ with $I_1=0$ is topologically closed.
\item $NY^{i}\nablaSigma_{i}\lambda\geq 0$ in each arc-connected component of $\tbd \ext$ containing
at least one fixed point.
\item $NY^{i}m_{i}\geq 0$ in each arc-connected component of $\tbd \ext$ which contains no fixed points, where
$\vec{m}$ is the unit normal pointing towards $\{\lambda>0\}^{\text{ext}}$.
\item The matter model is such that Bunting and Masood-ul-Alam doubling method gives uniqueness of
black holes.
\end{enumerate}
Then, $(\ext,g,K)$ is a slice of such a unique spacetime.
\end{thr}

{\bf Proof.}
Proposition \ref{is_a_MOTS} implies that $\tbd \ext$ is a smooth submanifold
with $\theta^+ =0$ with respect to the normal pointing towards $\ext$. We only need
to show that $\tbd \ext$ is closed (i.e. embedded, compact and without boundary)
in order to apply hypothesis 4 and conclude uniqueness.
By definition of bounding in the asymptotically flat setting
(see Definition \ref{defi:bounding})
we have a compact manifold  $\Sigmatilde$ with boundary $\partial \Sigmatilde = S \cup \Sb$,
where $\Sb = \{ r =r_0 \}$ is a sufficiently large coordinate sphere in $\Sigma^{\infty}_0$.
Take this sphere large enough so that $\{ r \geq r_0 \} \subset \ext$. We are
in a setting where all the hypothesis of
Theorem \ref{theorem2} hold. In the proof of this theorem we
have shown that $\tbd \ext$ is embedded and compact. Moreover,
$\tbd T^+$ lies in the interior $\mbox{int} (\Sigmatilde)$ and
does not intersect $\ext$. This, clearly prevents
$\tbd \ext$ from
reaching $S$, which in turn implies that $\tbd \ext$ has no boundary. $\hfill  \blacksquare$ \\

{\bf Remark.} This theorem applies in particular to static KID which are
asymptotically flat, without boundary and have
at least two asymptotic ends, as long as conditions 1 to 4 are fulfilled.
To see this, recall that
an asymptotically flat initial data is the union of a compact set
and a finite number of asymptotically flat ends. Select one of these ends
$\Sigma^{\infty}_0$  and define $S$ to be the union of coordinate spheres with sufficiently
large radius on all the other asymptotic ends. This surface is an
outer trapped surface which is bounding with respect to $\Sigma^{\infty}_0$
and we recover the hypotheses of Theorem  \ref{uniquenessthr0}. $\hfill \square$ \\

Theorem \ref{uniquenessthr0}
has been formulated for outer trapped surfaces instead of
weakly outer trapped surfaces for the same reason as in Theorem \ref{theorem2}.
Consequently, the hypotheses of this theorem can also be relaxed to
$\theta^{+}\leq 0$ if one of the following conditions hold: $S$ is not the outermost MOTS,
$S\cap\bd \Sigma=\emptyset$, or the KID can be extended.
Under these circumstances, this result already extends Miao's theorem as a uniqueness result.

Nevertheless, the theorem above requires several conditions on the boundary
$\tbd \ext$. Since $\tbd\ext$ is a fundamental object in the doubling procedure, it is rather
unsatisfactory to require conditions directly on this object. Out main aim in this chapter
is to obtain a uniqueness result which does not involve any a priori restriction on $\tbd \ext$.
As discussed in the previous chapter, $\tbd \ext$ is in general not a smooth submanifold
(see e.g. Figure \ref{fig:figure1})  and the
techniques of the previous chapter cannot be applied to conclude that $\tbd \ext$ is a closed embedded
topological submanifold.  The key difficulty lies in proving that $\tbd \ext$ is a manifold without boundary. In the previous theorem,
we used the non-penetration property of $\tbd T^+$ into $\ext$ in order to conclude that
$\tbd \ext$ must lie in the exterior of the bounding outer trapped surface $S$
(which implies that $\tbd \ext$ is a manifold without boundary).
In turn, this non-penetration property was strongly based
on the smoothness of $\tbd \ext$, which we do not have in general. The main problem is therefore: How can we
exclude the possibility that $\tbd \ext$ reaches $S$ in the general case? (see Figure \ref{fig:problem}).

To address this issue we need to understand better the structure
of $\tbd \ext$ (and, more generally, of $\tbd \{ \lambda >0\}$)
when conditions 2 and 3 are not satisfied. As we will discuss later,
this will force us to view KID as hypersurfaces embedded
in a spacetime, instead as abstract objects on their own, as we have done in the previous chapter.

To finish this introduction, let us give a briefly summary of the chapter. In Section \ref{sc:preliminaries5} we
define the concept of an {\it embedded static KID} and present some known results on the structure of the spacetime
in the neighbourhood of the fixed points of the isometry.
In Section \ref{sectionproposition} we will revisit the study of the properties of $\tbd\{\lambda>0\}$,
this time for embedded static KID. Finally, Section \ref{sectionuniqueness} is devoted to state and
prove the uniqueness theorem for asymptotically flat static spacetimes containing a bounding weakly outer trapped surface.

The results presented in this chapter have been summarized in
\cite{CMere4} and  will also be sent to publication \cite{CM4}.

\setcounter{equation}{0}
\section{Embedded static KID}
\label{sc:preliminaries5}

We begin this section with the definition of an embedded static KID. Recall
that,  according to our definitions, a spacetime has {\bf no boundary}.
\begin{defi}
An {\bf embedded static KID} $\kid$ is a static KID, possibly with boundary,
which is embedded in a spacetime $(M,\gM)$ with static Killing field $\vec{\xi}$ such that
$\vec{\xi}\, |_{\Sigma}=N\vec{n}+\vec{Y}$, where $\vec{n}$ is the unit future directed normal of
$\Sigma$ in $M$.
\end{defi}

{\bf Remark.} If a static KID has no boundary and
belongs to a matter model for which
the Cauchy problem is well-posed (e.g. vacuum, electro-vacuum, scalar field,
Yang-Mills field, $\sigma$-model, etc), it is clear that there exists a spacetime
which contains the initial data set as a spacelike hypersurface.
Whether this Cauchy development admits or not a Killing vector
$\vec{\xi}$ compatible with the Killing data has only been
answered in the affirmative for some special matter models, which include vacuum
and electro-vacuum \cite{Coll}. Even in these circumstances, it is at present
not known whether the spacetime thus
constructed is in fact {\it static} (i.e. such  that the Killing vector $\vec{\xi}$ is integrable).
This property is obvious near points where $N \neq 0$ (i.e. points where $\vec{\xi}$ is transverse
to $\Sigma$), but it is much less
clear near fixed points, specially those with $I_1 <0$. Indeed, these points
belong to a totally geodesic closed spacelike surface in the Cauchy development of the initial data set.
The points lying in the chronological future of this surface cannot be reached by integral
curves of the Killing vector starting on $\Sigma$. Proving that the Killing vector
is integrable on those points is an interesting and, apparently, not so trivial task.
In this thesis we do not explore this problem further and simply work with
the definition of embedded static KID stated above. $\hfill \square$ \\

In what follows, we will review some useful results concerning the structure of the spacetime
near fixed points of the static Killing $\vec{\xi}$.

\begin{proposition}\label{propositionRW}
Let $\kid$ be a static embedded KID  and let $(M,\gM)$ be the static spacetime where the KID is embedded.
Consider a fixed point $\p\in\tbd \{\lambda>0\} \subset \Sigma$ and let $S_0$ be the connected
spacelike surface of fixed points in $M$ containing $\p$ (which exists
by Theorem \ref{thr:Boyer}).
Then, there exists a neighbourhood ${\cal V}$ of $\p$ in $M$ and
coordinates $\{u,v,x^{A} \}$ on ${\cal V}$ such that $\{x^{A}\}$ are coordinates for
$S_0 \cap {\cal V}$ and the spacetime metric takes
the R\'acz-Wald-Walker form
\begin{equation}\label{RWmetric}
\gRW=2Gdudv+\gamma_{AB}dx^A dx^B,
\end{equation}
where $S_0 \cap {\cal V} = \{ u=v=0 \}$, $\partial_v$ is future directed and $G$ and $\gamma_{AB}$ are both positive definite
and depend smoothly on $\{w \equiv uv,x^A\}$.
\end{proposition}

{\bf Proof.} Theorem \ref{thr:Boyer} establishes that $\p$ belongs to a connected, spacelike, smooth surface
$S_{0}$ which lies in the closure of a non-degenerate Killing horizon.
Thus, we can use the R\'acz-Wald-Walker construction, see \cite{RW}, which shows that there exists
a neighbourhood ${\cal V}$ of $\p$ and coordinates $\{ u,v,x^A\}$ adapted
to $S_0 \cap {\cal V}$ such that the metric $\gM$ takes the form
\begin{equation}\label{RW}
\gM=2Gdudv+2vH_{A}dx^{A}du+\gamma_{AB}dx^{A}dx^{B},
\end{equation}
where $G$, $H_A$ and $\gamma_{AB}$ depend smoothly on $\{ w ,x^A \}$.
In these coordinates, the Killing vector $\vec{\xi}$ reads
\begin{equation}\label{killingRW}
\vec{\xi}=c^2\left( v\partial_{v}-u\partial_{u} \right),
\end{equation}
where $c$ is a (non-zero) constant and $\partial_v$ is future directed.
We only need to prove that staticity implies that $\{ u,v,x^A\}$ can be
chosen in such a way that $H_{A}=0$.
A straightforward computation shows that the integrability condition $\bm{\xi} \wedge d \bm{\xi} =0$
is equivalent to the following equations
\begin{eqnarray}
G \partial_{w} H_{A} - H_{A}\partial_{w}G&=&0, \label{RW1}\\
H_{[A}\partial_{B]}G + G\partial_{[A}H_{B]} &=& 0, \label{RW2} \\
H_{[A}\partial_{w}H_{B   ]}  & = & 0. \label{RW3}
\end{eqnarray}
Equation (\ref{RW1}) implies $H_{A}=f_{A}G$, where $f_{A}$ depend on $x^{C}$.
Inserting this in (\ref{RW2}), we get $\partial_{[A} f_{B]}=0$, which implies (after
restricting ${\cal V}$ if necessary) the existence of
a function $\zeta(x^{C})$ such that $f_{A}=\partial_{A}\zeta$. Equation (\ref{RW3})
is then identically satisfied.
Therefore, staticity is equivalent to
\begin{equation}\label{HG}
H_{A}(w,x^{C})=G(w,x^{C}) \partial_{A}\zeta(x^{C}).
\end{equation}

We look for a coordinate change $\{u,v,x^{C}\}\rightarrow \{u',v',x'^{C} \}$ which preserves the form of the
metric (\ref{RW})  and such that $H'_A=0$. It is immediate to check that an invertible change of the form
\[
\left\{ u=u(u'), v=v(v',{x'}^{C}),x^{A}={x'}^{A} \right\}
\]
preserves the form of the metric and transforms $H_A$ as
\begin{eqnarray}
v'H'_{A}&=&\frac{d u}{d u'}\left(\frac{\partial v}{\partial x'^{A}}G + vH_{A}\right),\label{RW5}
\end{eqnarray}
So, we need to impose $G \partial_A v  + vH_{A} = 0$, which in view
of (\ref{HG}), reduces to $\partial_A  v +v \partial_{A} \zeta=0$. Since
$v=v' e^{-\zeta}$ (with $v'$ independent of $x^A$) solves this equation, we conclude that the coordinate change
$$
\left\{ u=u', v=v' e^{-\zeta(x'^{C})}, x^{A}=x'^{A}\right\}
$$ brings the metric into
the form (\ref{RW}) (after dropping the primes). $\hfill \blacksquare$ \\

Now, let us consider an embedded static KID in a static spacetime with R\'acz-Wald-Walker metric $({\cal V},\gRW)$.
Since the vector $\partial_{v}$ is null on ${\cal V}$, it is transverse to $\Sigma \cap {\cal V}$ and, therefore,
the embedding of $\Sigma \cap {\cal V}$ can be written locally as
\begin{equation}\label{embedding}
\Sigma:(u,x^A)\rightarrow (u,v=\phi(u,x^A),x^A),
\end{equation}
where $\phi$ is a smooth function.
A simple computation using (\ref{killingRW}) leads to
\begin{eqnarray}
\left.\lambda\right|_{\Sigma \cap {\cal V}} &=& 2c^{4}\hat{G} u\phi, \label{lambdaRW}\\
\left. N \right|_{\Sigma \cap {\cal V}} &=& \left( \phi+u\partial_{u}\phi \right)\sqrt{\frac{c^4 \hat{G}}{2\partial_{u}\phi-\hat{G}
\partial_{A}\phi\partial^{A}\phi}}, \label{NRW}\\
\left. {\bf Y} \right|_{\Sigma \cap {\cal V}}&=& c^{2}\hat{G}\left(  \phi du - u d\phi \right) \label{YRW}.
\end{eqnarray}
where $\hat{G} \equiv G (w = u \phi, x^A)$ and indices $A,B,\dots$ are raised with the inverse of
$\hat{\gamma}_{AB} \equiv \gamma_{AB} (w = u \phi, x^A ) $.

Since $\Sigma$ is spacelike, the quantity ${2\partial_{u}\phi- \hat{G} \partial_{A}\phi\partial^{A}\phi}$ is positive.
In particular, this implies that
\begin{equation}\label{partialuphi}
\partial_{u}\phi>0,
\end{equation}
which will be used later.
For the sets $\{u=0\}$ and $\{\phi=0\}$ in $\Sigma \cap {\cal V}$
we have the following result.

\begin{lema}\label{lemau0v0smooth}
Consider an embedded static KID $\kid$ and use R\'acz-Wald-Walker coordinates
$\{u ,v, x^A \}$ in a spacetime neighbourhood ${\cal V}$ of a fixed point $\p \in \tbd \{ \lambda >  0\} \subset \Sigma$
such that the embedding of $\Sigma$ reads (\ref{embedding}).
Then the sets $\{u=0\}$ and $\{\phi=0\}$ in $\Sigma \cap {\cal V}$ are both smooth surfaces (not necessarily closed).
Moreover, a point
$\p\in \tbd\{ \lambda>0\}$ in $\Sigma \cap {\cal V}$ is a
non-fixed point if and only if $u\phi=0$ with either $u$ or $\phi$ non-zero.
\end{lema}
{\bf Proof:} The lemma follows directly from the fact that
both sets $\{ u=0 \}$ and $\{ \phi=0 \}$ in $\Sigma$ are
the intersections between $\Sigma$ and the null smooth embedded hypersurfaces
$\{u=0\}$ and $\{v=0\}$ in $({\cal V},\gRW)$, respectively.
The second statement of the lemma is a direct consequence of equations (\ref{killingRW}) and (\ref{lambdaRW}). $ \hspace*{1cm} \hfill \blacksquare$ \\

\setcounter{equation}{0}
\section{Properties of $\tbd\left\{\lambda>0\right\}$ on an embedded static KID}\label{sectionproposition}

In this section we will
explore in more detail the properties of the set $\tbd\left\{\lambda>0\right\}$ in $\Sigma$.
In particular, we will study the structure $\tbd\{\lambda>0\}$ in an embedded KID when no additional
hypothesis are made. First, we will briefly recall some results of the previous chapter which will be used below.
In Proposition \ref{Totally geodesic} we showed
that an open set of fixed points in
$\tbd \{\lambda>0\}$ in a static KID $\kid$ is a smooth and totally geodesic surface.
Moreover, Lemma \ref{surfaceNoFixed} and Proposition \ref{is_a_MOTS} imply that
every arc-connected component of the open set of non-fixed points in $\tbd\{\lambda>0\}\subset\Sigma$
is a smooth submanifold (not necessarily embedded)
of $\Sigma$ and has either $\theta^{+}=0$ or $\theta^{-}=0$.
The structure of those arc-connected components
of $\tbd \{\lambda>0\}$ having exclusively fixed points or exclusively non-fixed points
is therefore clear with no need of additional assumptions. 
However, for the case of arc-connected components having both types of points 
an additional assumption on the sign of $NY^{i}\nablaSigma_{i}\lambda$ was required to conclude smoothness (see
Propositions \ref{C1} and \ref{is_a_MOTS}).
This hypothesis was imposed in order to avoid the existence of {\it transverse} fixed points
in $\tbd \{\lambda>0\}$ (see stage 1 on the proof of Proposition \ref{C1}). Actually,
the existence of transverse points
is, by itself, not very problematic. Indeed, as we showed in Lemma \ref{structurebneq0},
the structure of $\tbd\{\lambda>0\}$ on a neighbourhood of transverse fixed points is well understood and
consists of two intersecting branches. The problematic situation happens when a sequence of
transverse fixed points tends to a non-transverse point $\p$.
In this case the intersecting branches can have a very complicated limiting behavior
at $\p$. If we consider the non-transverse limit
point $\p$, then we know from the previous chapter
(see stage 2 on the  proof of Proposition \ref{C1}) that locally near $\p$ there exists
coordinates such that $\lambda = Q_0^2x^2 - \zeta(z^A)$, with $\zeta$ a non-negative smooth
function. In order to understand the behavior of $\tbd \{\lambda > 0 \}$ we need to take
the square root of $\zeta$. Under the assumptions of Proposition
\ref{C1} we could show
that the {\it positive} square root is $C^1$. For general non-transverse points, this positive
square root is not $C^{1}$. In fact, is not clear at all whether there exists any $C^{1}$ square root
(even allowing this square root to change sign). The following example shows a function $\zeta$
which admits no $C^{1}$ square root. It is plausible that the equations that are satisfied
in a static KID forbid the existence of $\zeta$ functions with no $C^1$ square root. This is,
however, a difficult issue and we have not been able to resolve it. This is the reason why
we need to restrict ourselves to embedded static KID in this chapter. Assuming the existence
of a static spacetime where the KID is embedded, it follows that, irrespectively of the
structure of fixed points in $\Sigma$, a suitable square root of $\zeta$ always exists.

{\bf Example.} Non-negative functions do not have in general a $C^1$
square root. A simple example is given by the function $\rho = y^2 + z^2$
on $\mathbb{R}^2$. We know, however, that this type of example cannot occur for
the function $\zeta$ because the Hessian of $\zeta$  must vanish
at least on one point where $\zeta$ vanishes (and this is obviously not true for
$\rho$).

The following is an example of a non-negative function $\zeta$ for which the function
and its Hessian vanish at one point and which admits no $C^1$ square root.
Consider the function $\zeta(y,z)=z^2 y^{2}+z^{4}+f(y)$, where $f(y)$ is a smooth function
such that $f(y)=0$ for $y\geq 0$ and $f(y)>0$ for $y<0$. Recall that the set of fixed points
consists of the zeros of $\zeta$, and a fixed point is non-transverse if and only if
the Hessian of $\zeta$ vanishes (see the  proof of Proposition \ref{C1}). It follows
that the fixed points occur on the semi-line $\sigma\equiv \{y\geq 0,z=0\}$, with $(0,0)$ being non-transverse
and $(y>0,z=0)$ transverse. Consider the points $\p=(1,-1)$ and $\q=(1,1)$. First of all take a curve $\gamma$
joining them in such a way that it does not intersect $\sigma$. It is clear that
$\zeta$ remains positive along $\gamma$
and, therefore, its square root cannot change sign (if it is to be continuous). Now
consider the curve $\gamma'= \{y=1,-1\leq z\leq 1 \}$ joining $\p$ and $\q$ (which does intersect $\sigma$). Since
$\zeta\big|_{\gamma'}=z^{2}(1+z^{2})$, the only way to find a $C^{1}$ square root is by taking
$u=z\sqrt{1+z^{2}}$, which changes sign from $\p$ to $\q$. This is a contradiction to
the property above. So, we conclude that no $C^{1}$ square root of $\zeta$ exists.

Let us see that, in the spacetime setting, this behavior cannot occur.
Our first result of this section shows that the set $\tbd\{\lambda>0\}$ in an embedded KID is a union of compact, smooth surfaces which has
one of the two null expansions equal to zero. 

\begin{proposition}\label{proposition1}
Consider an embedded static KID $\kidtilde$, compact and possibly with boundary $\bd \Sigmatilde$.
Assume that every arc-connected component
of $\tbd \{ \lambda > 0 \}$ with $I_1 =0$ is topologically closed.
Then
\begin{equation}
\tbd \{\lambda>0\} = \underset{a}\cup S_a,
\end{equation}
where each $S_a$ is a smooth, compact, connected and orientable surface such that its boundary, if non-empty, satisfies
$\bd S_{a}\subset \bd\Sigmatilde$. Moreover, at least one of the two null expansions of $S_a$ vanishes everywhere.
\end{proposition}

{\bf Proof.}
Let $\{ \SS_{\alpha} \}$ be the collection of arc-connected components of $\tbd \{ \lambda > 0\}$.
We know that
the quantity $I_1$ is constant on each $\SS_{\alpha}$ (see Lemma \ref{lema:I1constant}). Consider an
arc-connected component $\SS_d$ of
$\tbd \{ \lambda > 0\}$ with $I_1 =0$. Since all
points in this component are non-fixed,
it follows  that $\SS_d$ is a smooth submanifold.
Using  the hypothesis 
that arc-connected components with $I_1=0$ are topologically closed it follows
that $\SS_d$ is, in fact, embedded. Choose $\vec{m}$ to be the unit normal satisfying
\begin{eqnarray}
\vec{Y} = N \vec{m}, \label{choice_m_degenerate}
\end{eqnarray}
 on $\SS_d$. This normal
is smooth (because neither $\vec{Y}$ nor $N$ vanish anywhere on $\SS_d$), which implies that $\SS_d$ is orientable.
Inserting $\vec{Y} = N \vec{m}$ into equation (\ref{kid1}) and taking the trace it follows
\begin{eqnarray}
p + q =0. \label{theta_plus_degenerate}
\end{eqnarray}

\begin{figure}[h]
\begin{center}
\psfrag{p}{$\q$}
\psfrag{I}{$I$}
\psfrag{II}{$II$}
\psfrag{III}{$III$}
\psfrag{IV}{$IV$}
\psfrag{l>}{$\lambda<0$}
\psfrag{l<}{$\lambda>0$}
\psfrag{S1}{$u=0$}
\psfrag{S2}{$\phi=0$}
\psfrag{S3}{$u=0$}
\psfrag{S4}{$\phi=0$}
\psfrag{>}{$N>0$}
\psfrag{<}{$N<0$}
\includegraphics[width=6cm]{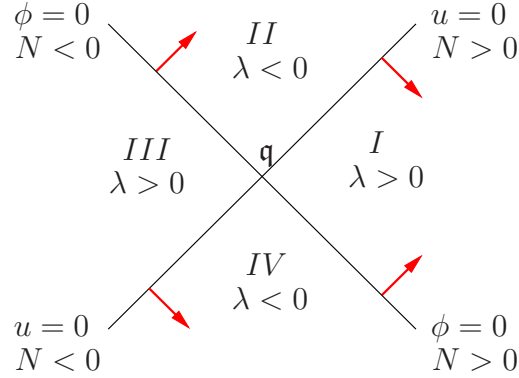}
\caption{In the R\'acz-Wald-Walker coordinate system we define four open regions by
$I=\{u>0\}\cap\{\phi> 0\}, II=\{u<0\}\cap\{\phi> 0\}, III=\{u<0\}\cap\{\phi< 0\}, IV=\{u>0\}\cap\{\phi< 0\}$.
The normal on its boundaries which satisfies (\ref{defm}) is depicted in red color. It is clear graphically that these normals
extend smoothly to the fixed points on the hypersurfaces $\{u=0\}$ and $\{ \phi=0 \}$, such as $\q$ in the figure.
This figure is, however, only schematic because one dimension has been
suppressed and fixed points need not be isolated in general.
A formal proof that $\vec{m}$ extends smoothly in all cases is given in the
text.}
\label{fig:cruz}
\end{center}
\end{figure}

Consider now a $\SS_{\alpha}$ with $I_1 \neq 0$. At non-fixed points we know that $\SS_{\alpha}$ is a smooth embedded surface
with $ \nablaSigma_i\lambda \neq 0$. On those points, define a unit normal $\vec{m}$ by the condition
\begin{eqnarray}
N \vec{m} (\lambda )  > 0 \label{defm}
\end{eqnarray}
We also know that $\nablaSigma_i \lambda = 2 \kappa Y_i$ where $I_1 = - 2 \kappa^2$. Let us see that $\SS_{\alpha} = \SS_{1,\alpha}
\cup \SS_{2,\alpha}$, where each $\SS_{1,\alpha}$ and $\SS_{2,\alpha}$ is a smooth, embedded, connected and orientable surface. To that aim,
define
\begin{eqnarray*}
\SS_{1,\alpha} & = &  \{ \p \in \SS_{\alpha} \mbox{ such that } \kappa \big|_{\p} > 0 \} \cup \{ \mbox{ fixed points in } \SS_{\alpha} \}, \\
\SS_{2,\alpha} & = &  \{ \p \in \SS_{\alpha} \mbox{ such that } \kappa \big|_{\p}< 0 \} \cup \{ \mbox{ fixed points in } \SS_{\alpha} \}.
\end{eqnarray*}
Notice that the fixed points are assigned to {\it both} sets. It is clear that at non-fixed points,
both $\SS_{1,\alpha}$ and $\SS_{2,\alpha}$ are smooth embedded surfaces. Let $\q$ be a fixed point in $\SS_{\alpha}$ and consider the
R\'acz-Wald-Walker coordinate system discussed in Proposition \ref{propositionRW}. The points in $\SS_{\alpha} \cap {\cal V}$
are characterized by $\{ u \phi =0 \}$ (due to (\ref{lambdaRW})). Inserting (\ref{lambdaRW})
and (\ref{YRW}) into $\nablaSigma_i \lambda = 2 \kappa Y_i$ yields, at any non-fixed point $\q' \in \SS_{\alpha} \cap {\cal V}$,
\begin{equation}
2c^2\left( \phi du+ud\phi \right) |_{\q'}= 2\kappa \left( \phi du -u d\phi \right) |_{\q'} \nonumber.
\end{equation}
Since $du\neq 0$ (because $u$ is a coordinate) and $d\phi\neq 0$ (see equation (\ref{partialuphi}))
we have
\begin{eqnarray}\label{sign(kappa)}
\kappa>0 & \text{on}\quad \{u=0,\phi\neq 0\}, \nonumber\\
\kappa<0 & \text{on}\quad \{u\neq 0,\phi=0\}.
\end{eqnarray}
Consequently, the non-fixed points in $\SS_{1,\alpha} \cap {\cal V}$ are defined by the condition $\{ u=0,\phi \neq 0 \}$
and the non-fixed points in $\SS_{2,\alpha} \cap {\cal V}$ are defined by the condition $\{ u \neq 0, \phi =0\}$. It is
then clear that $\SS_{1,\alpha} \cap {\cal V} = \{ u=0 \}$ and $\SS_{2,\alpha} \cap {\cal V} = \{ \phi = 0 \}$, which are smooth
embedded surfaces. It remains to see that the unit normal $\vec{m}$, which has been defined
only at non-fixed points via (\ref{defm}), extends to a well-defined normal to all of $\SS_{1,\alpha}$ and $\SS_{2,\alpha}$
(see Figure \ref{fig:cruz}).

This requires to
check that the condition (\ref{defm}), when evaluated on ${\cal V}$ defines a normal which extends smoothly to the fixed
points. Consider first the points $\{ u \neq 0, \phi =0 \}$. The unit normal to this surface is $\vec{m}  =
\epsilon | \nablaSigma \phi|^{-1}_{g} {\nablaSigma}\phi$
where $\epsilon = \pm 1$ and may, a priori, depend on the point. Since
\begin{eqnarray*}
\left. N \right|_{\{ u \neq 0 ,\phi =0\}} &=&  u\partial_{u}\phi \sqrt{\frac{c^4 \hat{G}}{2\partial_{u}\phi-\hat{G}
\partial_{A}\phi\partial^{A}\phi}}, \\
\left . \nablaSigma_i \lambda  \right|_{\{ u \neq 0 ,\phi =0\}} &=&  2 c^4 \hat{G} u \nablaSigma_i \phi,
\end{eqnarray*}
expression (\ref{defm}) implies
\begin{eqnarray*}
  0 < N \vec{m} (\lambda ) |_{ \{ u\neq0, \phi =0 \}} = 2 \epsilon c^4 \hat{G} u^2 \partial_u \phi |\nablaSigma \phi|_{g}
\sqrt{\frac{c^4 \hat{G}}{2\partial_{u}\phi-\hat{G} \partial_{A}\phi\partial^{A}\phi}}.
\end{eqnarray*}
Hence $\epsilon = 1$ at all points on $\{ u \neq 0, \phi =0 \}$. Thus the normal vector reads
$\vec{m} =  | \nablaSigma \phi|^{-1}_{g} {\nablaSigma} \phi $ at non-fixed points, and this field
clearly
extends smoothly to all points on $\SS_{1,\alpha} \cap {\cal V}$. This implies, in particular, that
$\SS_{1,\alpha}$ is orientable.

The argument for $\SS_{2,\alpha}$ is similar. Consider now the points
$\{ u = 0, \phi \neq 0 \}$. The unit vector
normal to this surface is $\vec{m}  = \epsilon^{\prime} | \nablaSigma u|^{-1}_{g} {\nablaSigma}u$
where
$\epsilon^{\prime} = \pm 1$. Using (\ref{lambdaRW}) and (\ref{NRW}) in (\ref{defm}) gives now
\begin{eqnarray*}
  0 < N \vec{m} (\lambda ) |_{ \{ u = 0, \phi \neq 0 \}} = 2 \epsilon^{\prime} c^4 \hat{G} \phi^2  |\nablaSigma u|_{g}
\sqrt{\frac{c^4 \hat{G}}{2\partial_{u}\phi-\hat{G} \partial_{A}\phi\partial^{A}\phi}},
\end{eqnarray*}
which implies $\epsilon^{\prime} = 1$
all points on $\{ u=0, \phi \neq 0\}$. The normal vector is  $\vec{m} =  |\nablaSigma u|^{-1}_{g} {\nablaSigma} u$
which again extends smoothly
to all points on $\SS_{2,\alpha} \cap {\cal V}$. As before, $\SS_{2,\alpha}$ is orientable.

Let us next check that $\SS_{1,\alpha}$ has $\theta^+ =0$ and $\SS_{2,\alpha}$ has $\theta^{-} =0$ (both with
respect to the normal $\vec{m}$ defined above). On open sets of fixed points this is a trivial
consequence of Proposition \ref{Totally geodesic} which implies both $p= q =0$. To discuss the  non-fixed points, we need
an expression for $\vec{Y}$ in terms of $\vec{m}$. Let $\vec{Y} = \epsilon^{\prime\prime} N \vec{m}$,
where $\epsilon^{\prime \prime} = \pm 1$. Using $\vec{Y} = \frac{1}{2\kappa} \nablaSigma \lambda$, we have
\begin{eqnarray*}
\frac{\epsilon^{\prime\prime}}{2 \kappa} | \nablaSigma \lambda|^2_g = \epsilon^{\prime\prime} \vec{Y} \left ( \lambda \right )
= N \vec{m} \left ( \lambda \right ) > 0
\end{eqnarray*}
Hence $\epsilon^{\prime\prime} = \mbox{sign}(\kappa)$ and
\begin{eqnarray}
\vec{Y} = \mbox{sign} (\kappa) N \vec{m}. \label{Y_in_terms_of_m}
\end{eqnarray}
Inserting this
into (\ref{kid1}) and taking the trace, it follows
\begin{eqnarray}
\mbox{sign} (\kappa ) p  + q =0 \label{theta_plus_non_degenerate}
\end{eqnarray}
This implies that $\theta^{+} = p + q =0$ at non-fixed points of $\SS_{1,\alpha}$ and
$\theta^{-} = -p  +q = 0$ at non-fixed points at $\SS_{2,\alpha}$. At fixed points not lying
on open sets, equations $\theta^+ =0$ (resp. $\theta^{-}=0$) follow by
continuity once we know that $\SS_{1,\alpha}$ (resp. $\SS_{2,\alpha}$) is smooth with a smooth unit normal.

The final step is to prove that $\SS_{1,\alpha}$ and $\SS_{2,\alpha}$ are topologically closed. Let us first
show that $\SS_{\alpha}$ is topologically closed. Consider a
sequence of points $\{\p_i \}$ in $\SS_{\alpha}$ converging to $\p$. It is clear that $\p \in \tbd \{\lambda > 0\}$, so we
only need to check that we have not moved to another arc-connected component. If $\p$ is a non-fixed
point, then $\{ \lambda =0 \}$ is a defining function for $\tbd \{ \lambda > 0 \}$ near $\p$
and the statement is
obvious. If $\p$ is a fixed point, we only need to use the R\'acz-Wald-Walker coordinate system near $\p$
to conclude that no change of arc-connected component can occur in the limit. To show that
each $\SS_{1,\alpha}$, $\SS_{2,\alpha}$ is topologically closed, assume now that $\p_i$ is a sequence on $\SS_{1,\alpha}$.
If the limit $\p$ is a fixed point, it belongs to $\SS_{1,\alpha}$ by definition. If the limit $\p$ is a non-fixed
point, we can take a subsequence $\{\p_i\}$ of non-fixed points. Since $\kappa$ remains constant
on the sequence, it takes the same value in the limit, which shows that $\p \in \SS_{1,\alpha}$, i.e.
$\SS_{1,\alpha}$ is topologically closed.

The surfaces $S_a$ in the statement of the theorem are the collection of $\{ \SS_{d} \}$
having $I_1 =0$ and the collection of pairs $\{ \SS_{1,\alpha}$, $\SS_{2,\alpha}\}$ for the
arc-connected components $\SS_{\alpha}$ with $I_1 \neq 0$. The statement that
$\bd S_a \subset \bd \Sigmatilde$ is obvious.
$\hfill \blacksquare$ \\

{\bf Remark 1.} In this proof we have tried to avoid using the existence of a spacetime
where $\kid$ is embedded as much as possible.
The only essential
information that we have used from the spacetime is that, near fixed
points, $\lambda$ can be written as the product of two smooth functions
with non-zero gradient, namely $u$ and $\phi$. This is the square root
of $\zeta$ that we mentioned above. To see this, simply note that
if a square root $h$ of $\zeta$ exists, then $\lambda = Q_0 x^2 - \zeta =
Q_0^2 x - h^2 = \left (Q_0 x - h  \right ) \left ( Q_0 x + h \right )$).
The functions $Q_0 x \pm h$ have non-zero gradient and are, essentially,
the functions $u$ and $\phi$ appearing the R\'acz-Wald-Walker coordinate system. $\hfill \square$ \\

{\bf Remark 2.} The assumption of every arc-connected component of
$\tbd \{\lambda>0\}$ with $I_{1}=0$ being topologically closed
is needed to ensure that these arc-connected components are embedded and compact.
From a spacetime perspective, this hypothesis
avoids the
existence of non-embedded degenerate Killing prehorizons which
would imply that, on an embedded KID, the arc-connected components of $\tbd \{\lambda>0\}$
which intersect these prehorizons could be non-embedded or non-compact
(see Figure \ref{fig:spiral} in Chapter \ref{ch:Preliminaires}).
Although it has not been proven, it may well be that non-embedded
Killing prehorizons cannot exist. A proof of this fact would allow us to drop
automatically this hypothesis in the theorem. $\hfill \square$ \\

We are now in  a situation where we can prove that $\tbd \ext =
\tbd T^+$ under suitable conditions on the trapped region and on the topology
of $\Sigmatilde$. This result is the crucial ingredient for our uniqueness
result later.
The strategy of the proof is, once again, to assume that $\tbd \ext \neq
\tbd T^+$ and to construct a bounding weakly outer trapped surface
outside $\tbd T^+$. This time, the surface we use to perform
the smoothing is more complicated than $\tbd \ext$, which
we used in the previous chapter. The newly constructed surface will
have vanishing
outer null expansion and will be closed and oriented. However, we
cannot guarantee a priori that it is bounding. To address
this issue we impose a topological condition on $\mbox{int} (
\Sigmatilde)$ which forces that all closed and orientable surfaces
separate the manifold into disconnected subsets. This topological
condition involves the first homology group $H_1 ( \mbox{int} (\Sigmatilde),
\mathbb{Z}_2)$ with coefficients in $\mathbb{Z}_2$ and imposes that
this homology group is trivial.  More precisely, the theorem that we will invoke is due to
Feighn \cite{Feighn} and reads as follows
\begin{thr}[Feighn, 1985]
\label{Feighn}
Let ${\cal N}$ and ${\cal M}$  be manifolds without
boundary of dimension $n$ and $n+1$ respectively.
Let $f: {\cal N} \rightarrow {\cal M}$ be a proper immersion
(an immersion is proper if
inverse images of compact sets are compact). If $H_1 ({\cal M},\mathbb{Z}_2
) = 0$ then ${\cal M} \setminus f ({\cal  N})$ is not connected. Moreover,
if two points $\p_1$ and $\p_2$ can be joined by an embedded curve
intersecting $f ({\cal N})$ transversally at just one point, then $\p_1$ and $\p_2$ belong to different
connected components of ${\cal M} \setminus f ({\cal N})$.
\end{thr}

The proof of this theorem requires that all embedded
closed curves in ${\cal M}$ are the boundary of an embedded
compact surface. This is a consequence of
$H_1 ({\cal M}, \mathbb{Z}_2)=0$ and this is the only
place where this topological condition enters
into the proof. This allows us to understand better what
topological restriction we are really imposing on
${\cal M}$, namely that every closed embedded curve
is the boundary of a compact surface.

Without entering into details of algebraic topology, we
just notice that $H_1 ({\cal M},\mathbb{Z}_2 )$
vanishes if $H_1 ({\cal M}, \mathbb{Z} )=0$ (see e.g.
Theorem 4.6 in \cite{Zomorodian}) and, in turn, this
is automatically satisfied in simply connected manifolds (see
e.g. Theorem 4.29 in \cite{Rotman}).

\begin{thr}\label{mainthr}
Consider an embedded static KID $\kidtilde$
compact, with boundary $\bd \Sigmatilde$ and satisfying the NEC.
Suppose that the boundary can be split into two non-empty disjoint components
$\bd \Sigmatilde= \bd^{-}\Sigmatilde\cup\bd^+\Sigmatilde$ (neither of which are necessarily connected).
Take $\bd^{+}\Sigmatilde$ as a barrier with interior $\Sigmatilde$ and assume
$\theta^{+}[ \bd^{-}\Sigmatilde ] \leq  0$
and $\theta^{+}[\bd^{+}\Sigmatilde]>0$
Let $T^{+}, T^-$ be, respectively, the weakly outer trapped
and the past weakly outer trapped regions of $\Sigmatilde$.
Assume also the following hypotheses:
\begin{enumerate}
\item Every arc-connected component of $\tbd \ext$ with $I_1 =0$ is topologically closed.
\item
$\left. \lambda \right|_{\bd^{+}\Sigmatilde}>0$.
\item
$H_1\left( \mbox{int} (\Sigmatilde),\mathbb{Z}_2 \right)=0$.
\item
$T^-$ is non-empty and $T^{-}\subset T^{+}$.
\end{enumerate}
Denote by
$\ext$ the connected component of $\{ \lambda>0 \}$
which contains $\bd^{+}\Sigmatilde$. Then
\[
\tbd\ext = \tbd T^{+},
\]
Therefore, $\tbd\ext$ is a non-empty stable MOTS which
is bounding with respect to $\bd^{+}\Sigmatilde$ and, moreover, it is the outermost bounding MOTS.
\end{thr}

{\bf Proof.} After replacing $\vec{\xi} \rightarrow - \vec{\xi}$ if necessary, we can assume
without loss of generality that $N >0$ on $\ext$.
From Theorem \ref{thr:AM}, we know that
the boundary of the weakly outer trapped region $T^{+}$ in $\Sigmatilde$ (which is non-empty because
$\theta^+[\bd^{-}\Sigmatilde] \leq 0$)
is a stable MOTS which is bounding with respect to $\bd^{+}\Sigmatilde$. $\tbd T^-$ is
also non-empty by assumption.

Since we are dealing with embedded KID, and
all spacetimes are boundaryless in this thesis, it follows that $\kid$
can be extended as  a smooth hypersurface in $(M,\gM)$\footnote{Simply consider $\bd \Sigmatilde$ as a surface
in $(M,\gM)$ and let $\vec{m}$ the be the spacetime normal to $\bd \Sigmatilde$ which
is tangent to $\Sigmatilde$. Take  a smooth hypersurface containing $\bd \Sigmatilde$ and tangent
to $\vec{m}$.  This hypersurface extends $\kid$. It is clear that
the extension can be selected as smooth as desired.}. Working on this extended KID allows us to assume
without loss of generality that
$\tbd T^+$ and $\tbd T^-$ lie in the {\it interior} of $\Sigmatilde$. This will be used when
invoking the Kriele and Hayward smoothing procedure below.

First of all, Theorem \ref{theorem1} implies that $\tbd\ext$ cannot lie completely in ${T^{+}}$ and
intersect the topological interior $\overset{\circ}{ T}{}^+$ (here is where we use the NEC).
Therefore, either $\tbd\ext$ intersects the exterior of $\tbd T^{+}$ or they both coincide.
We only need to exclude the first possibility. Suppose,
that $\tbd\ext$ penetrates into the exterior of $\tbd T^{+}$.
Let $\{\mathfrak{U}\}$ be the collection
of arc-connected components of $\tbd \{\lambda >0\}$ which have a non-empty intersection with $\tbd \ext$. In Proposition
\ref{proposition1} we have shown that
$\{\mathfrak{U}\}$  decomposes into a union of smooth surfaces $S_a$.
Define its unit normal $\vec{m}'$ as the smooth normal which points into $\ext$ at
points on $\tbd \ext$.
This normal exists because all $S_a$ are orientable.
By (\ref{defm}) and the fact that $N>0$ on $\ext$, we have that on the surfaces $S_{a}$ with $I_1 \neq 0$, the
normal $\vec{m}'$ coincides
with the normal $\vec{m}$ defined in the proof of Proposition \ref{proposition1}. On the surfaces $S_{a}$
with $I_1 =0$, this normal coincides with $\vec{m}$ provided $\vec{Y}$ points into $\ext$, see (\ref{choice_m_degenerate}).
Since, by assumption, $\tbd \ext$
penetrates into the exterior of $T^+$, it follows that there is at least
one $S_{a}$ with penetrates into the exterior of $T^+$. Let
$\{ S_{a'} \}$ be the subcollection of $\{ S_{a} \}$ consisting
on the surfaces which penetrate into the exterior of $\tbd T^+$.  A priori, none of the surfaces $S_{a'}$
need to satisfy $p +q=0$ with respect to the normal $\vec{m}'$. However, one of the following
two possibilities must occur:
\begin{enumerate}
\item There exists at least one surface, say $S_{0}$, in $\{S_{a'}\}$ containing a point
$\q \in \tbd \ext$ such that $\vec{Y} |_{\q}$
points inside $\ext$, or
\item All surfaces in $\{ S_{a'} \}$ have the property that, for any ${\q} \in S_{a'} \cap
\tbd \ext$ we have $\vec{Y} |_{\q}$ is either zero, or it points outside $\ext$.
\end{enumerate}
In case 1, we have that $S_0$ satisfies $p+q =0$ with respect to the normal
$\vec{m}'$. Indeed, we either have that $S_0$ satisfies $I_1=0$ or $I_1 \neq 0$.
If $I_1=0$ then, since $\vec{Y}$ points into $\ext$, we have that
$\vec{m}$ and $\vec{m'}$ coincide. Since $S_0$ satisfies $p +q=0$ with
respect  to $\vec{m}$ (see (\ref{theta_plus_degenerate})) the statement follows.
If $I_1 \neq 0$ then $\kappa>0$ on $S_0$ (from (\ref{Y_in_terms_of_m}) and the fact that
$\vec{m} = \vec{m}'$). Thus, $p+q =0$ follows from (\ref{theta_plus_non_degenerate}).

In case 2, all surfaces $\{ S_{a'}\}$ satisfy $\theta^{-} = - p +q =0$ with respect to $\vec{m}'$
and we cannot find a MOTS outside $\tbd T^+$. However, under assumption 3,
we have $T^{-} \subset T^+$ and hence each $S_{a'}$ penetrates into the exterior of $T^-$. We can therefore reduce
case 2 to case 1 by changing the time orientation (or simply replacing $\theta^+$ and
$T^+$ by $\theta^-$ and $T^-$ in the argument below).

\begin{figure}[h]
\begin{center}
\psfrag{S2}{\color{red}{$S_{1}$}}
\psfrag{S}{{$S_{0}$}}
\psfrag{lambda}{$\lambda>0$}
\psfrag{pS+}{$\bd^{+}\Sigmatilde$}
\psfrag{pS-}{$\bd^{-}\Sigmatilde$}
\psfrag{Sigma}{$\Sigma$}
\psfrag{T+}{\color{blue}{$\tbd T^{+}$}}
\includegraphics[width=9cm]{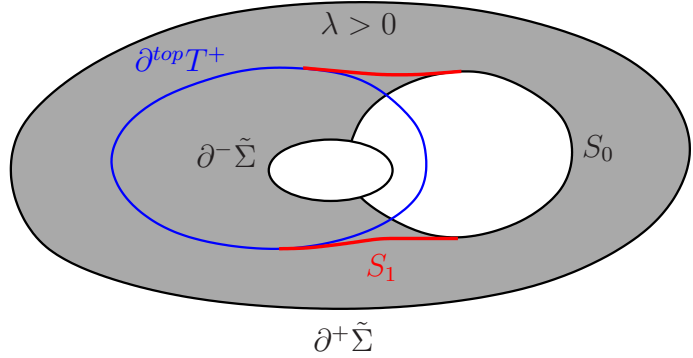}
\caption {The figure illustrates the situation when $S_{0}$ has boundary. The grey region represents the region with $\lambda>0$ in $\Sigmatilde$.
In this case we use the smoothing procedure of Kriele and Hayward to construct a smooth surface $S_{1}$ from
$S_{0}$ and $\tbd T^{+}$ (in blue). The red lines represent precisely the part of $S_{1}$ which comes from smoothing $S_{0}$ and $\tbd T^{+}$.}
\label{fig:mainthr}
\end{center}
\end{figure}

Let us therefore restrict ourselves to case 1. We know that $S_0$ either has no boundary, or the boundary
is contained in $\bd^- \Sigmatilde$. If $S_0$ has no boundary, simply rename this surface to $S_1$.
When $S_0$ has a non-empty boundary, it is clear that $S_0$ must intersect
$\tbd \T^+$. We can then use the smoothing procedure
by Kriele and Hayward (see Lemma \ref{lemasmoothness})
to construct a closed surface $S_1$ penetrating into the exterior
of $\tbd T^+$ and satisfying $\theta^+ \leq 0$ with respect to a normal
$\vec{m}''$ which coincides with $\vec{m}'$ outside the region where the smoothing is performed (see Figure \ref{fig:mainthr}).
As discussed in the previous chapter, when $S_0$
and $\tbd T^{+}$ do not intersect transversally we need to
apply the Sard Lemma to surfaces inside $\tbd T^+$.
If $\tbd T^+$ is only marginally stable, a suitable modification
of the initial data set inside $\tbd T^+$ is needed. The argument was discussed
in depth at the end of the proof of Theorem \ref{theorem2} and applies
here without modification.

So, in either case (i.e. irrespectively of whether $S_0$
has boundary of not), we have a closed surface $S_1$ penetrating
into the exterior of $\tbd T^+$ and satisfying $\theta^{+}\leq 0$ with respect to $\vec{m}''$. Here we apply  the topological
hypothesis  $3$ ($H_{1}(\mbox{int}(\Sigmatilde), \mathbb{Z}_{2})=0$).
Indeed $S_1$ is a closed manifold
embedded into $\mbox{int} (\Sigmatilde )$. Since $S_{1}$ is compact,
its embedding is obviously proper. Thus, the theorem by
Feighn \cite{Feighn} (Theorem \ref{Feighn}) implies that
$\mbox{int} (\Sigmatilde) \setminus S_{1}$ has at
least two connected components.
It is clear that one of the connected components $\Omega$ of
$\mbox{int} (\Sigmatilde)  \setminus S_1$ contains $\bd^+ \Sigmatilde$.
Moreover, by Feighn's theorem there is a tubular neighbourhood
of $S_1$ which intersects this connected component only to one side
of $S_1$. Consequently, $\overline{\Omega}$ is a compact manifold
with boundary $\bd \overline{\Omega} = S_1 \cap \partial^+ \Sigma$.
If follows that $S_1$ is bounding with respect to $\bd^+ \Sigmatilde$.
The choice of $\vec{m}''$ is such that $\vec{m}''$ points towards
$\bd^+ \Sigmatilde$. Consequently $S_1$ is a weakly outer trapped surface which is bounding
with respect
to $\bd^+ \Sigmatilde$ penetrating into the exterior of $\tbd T^+$,
which is impossible.
$\hfill\blacksquare$ \\

\begin{figure}
\begin{center}
\psfrag{ext}{{$\tbd \ext$}}
\psfrag{T+}{\color{blue}{$\tbd T^{+}$}}
\psfrag{T-}{{$\tbd T^{-}$}}
\psfrag{theta+}{\color{red}{{$\theta^{+}=0$}}}
\psfrag{theta-}{\color{red}{{$\theta^{-}=0$}}}
\psfrag{pSigma}{{$\partial^{+}\Sigma$}}
\psfrag{Sigma}{$\Sigma$}
\includegraphics[width=12cm]{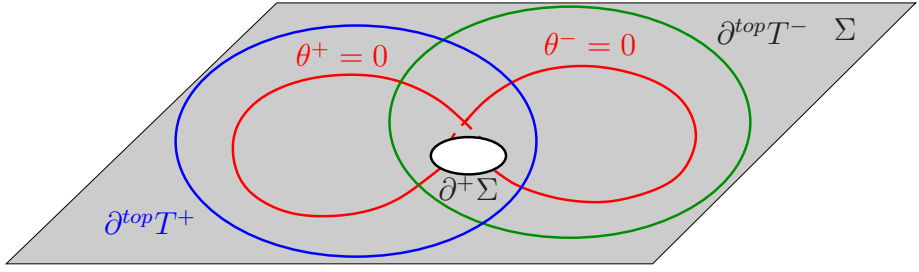}
\caption{The figure illustrates a hypothetical situation where $T^{+}\subset T^{-}$ does not hold and the conclusions of the
Theorem \ref{mainthr} would not be true. The red continuous line represents the set $\tbd \ext$ which
is composed by a smooth surface with $\theta^{+}=0$, lying inside of $\tbd T^{+}$ (in blue) and partly outside of $\tbd T^{-}$ (in green),
and a smooth surface with $\theta^{-}=0$, which lies partly outside of $\tbd T^{+}$ and inside of $\tbd T^{-}$.}
\label{figT+subsetT-}
\end{center}
\end{figure}

{\bf Remark 1.} If the hypothesis $T^{-}\subset T^{+}$ is not assumed, then the possibility $2$
in the proof of the Theorem would not lead to a contradiction (at least with our method
of proof). To understand this better, without the
assumption  $T^{-}\subset T^{+}$
it may happen  a priori that all the surfaces $S_{a'}$
(which have  $\theta^{-}=0$  and penetrate
in the exterior of $\tbd T^{+}$) are fully contained in $T^{-}$.
A situation like this is illustrated in Figure \ref{figT+subsetT-}, where
$\tbd T^{-}$ intersects $\tbd T^{+}$. It would be interesting to either prove this
theorem without the assumption $T^- \subset T^+$ or else find a counterexample
of the statement $\tbd \ext = \tbd T^+$ when assumption 4 is dropped. The problem, however,
appears to be difficult. $\hfill \square$ \\

\setcounter{equation}{0}
\section{The uniqueness result}\label{sectionuniqueness}

Finally, we are ready to state and prove the uniqueness result for static spacetimes containing
trapped surfaces.

\begin{thr}\label{uniquenessthr}
Let $\kid$ be an embedded static KID with a selected asymptotically flat end $\Sigma_{0}^{\infty}$ and
satisfying the NEC. Assume that
$\Sigma$ possesses a weakly  outer trapped surface $S$ which is bounding. 
Assume the following:
\begin{enumerate}
\item Every arc-connected component of $\tbd \ext$ with $I_1 =0$ is topologically closed.
\item
$T^-$ is non-empty and $T^{-}\subset T^{+}$.
\item
$H_{1}\left(\Sigma,\mathbb{Z}_{2} \right)=0$.
\item
The matter model is such that Bunting and Masood-ul-Alam doubling method for time-symmetric
initial data sets gives uniqueness of black holes.
\end{enumerate}
Then $\left( \Sigma \setminus {T}^{+},g,K \right)$ is a slice of such a unique spacetime.
\end{thr}

{\bf Proof.} Take a coordinate sphere $S_b \equiv \{r = r_0 \}$ in the asymptotically flat end
$\Sigma^{\infty}_0$ with $r_0$ large enough so that $\lambda >0$
on $\{ r \geq r_0 \} \subset \Sigma^{\infty}_0$ and  all
the surfaces $\{ r = r_1 \}$ with $r_1 \geq r_0$
are outer untrapped
with respect to the unit normal pointing towards increasing $r$. $S_b$ is a barrier with
interior $\Omegab=\Sigma\setminus\{r>r_{0}\}$.

%

Take $\Sigmatilde$ to be the topological closure of
the exterior of $S$ in $\Omega_b$. Then define
$\bd^- \Sigmatilde = S$ and
$\bd^{+}\Sigmatilde=S_{b}$. Let $\ext$ be the connected component
of $\{\lambda >0\} \subset \Sigmatilde$ containing $\Sb$.
All the hypothesis of Theorem \ref{mainthr} are satisfied and we can
conclude $\tbd \ext = \tbd T^+$.
This implies that the manifold $\Sigma \setminus {T}^{+}$ is an asymptotically flat spacelike hypersurface
with topological boundary $\tbd (\Sigma \setminus {T}^{+})$ which is compact and embedded (moreover, it is smooth)
such that the static Killing vector is timelike on $\Sigma \setminus {T}^{+}$ and
null on $\tbd(\Sigma \setminus{T}^{+})$. Under these assumptions,
the doubling method of Bunting and Masood-ul-Alam \cite{BMuA} can be applied. Hence, hypothesis $4$ gives uniqueness.
$\hfill \blacksquare$ \\

{\bf Remark 1.} In contrast to Theorems \ref{theorem2} and \ref{uniquenessthr0},
this result has been formulated for weakly outer trapped surfaces instead
of outer trapped surfaces. As mentioned in the proof of
Theorem \ref{mainthr} this is because, $\kid$ being an embedded static KID,
it can be extended smoothly as a hypersurface  in the spacetime.
It is clear however, that we are hiding the possible difficulties in the
definition of {\it embedded static KID}. Consider, for instance, a {\it static KID} with boundary
and assume that the KID is vacuum. The Cauchy problem is of course well-posed
for vacuum initial data. However, since $\Sigma$ has boundary, the spacetime
constructed by the Cauchy development also has boundary and we cannot
a priori guarantee that the static KID is an embedded static KID (this would require
extending the spacetime, which is as difficult -- or more -- than extending
the initial data).

Consequently, Theorem \ref{uniquenessthr} includes Miao's theorem
in vacuum as a particular case only for vacuum static KID for which either (i)
$S$ is not the outermost MOTS, (ii) $S \cap \bd \Sigma = \emptyset$ or (iii)
the KID can be extended as a vacuum static KID. Despite this subtlety, we emphasize that all
the other conditions of the theorem are fulfilled for asymptotically
flat, time-symmetric vacuum KID with a compact minimal boundary. Indeed,
condition 4 is obviously satisfied for vacuum.
 Moreover, the property of time-symmetry
implies that all points with $\lambda =0$ are fixed points and hence no
arc-connected component of $\tbd \{ \lambda > 0 \}$ with $I_1 =0$ exists. Thus,
condition 1 is automatically satisfied. Time-symmetry also implies
$T^- = T^+$ and condition 2 is trivial.
Finally, the region outside the outermost minimal surface
in a Riemannian manifold with non-negative Ricci scalar is $\mathbb{R}^3$ minus a finite
number of closed balls (see e.g. \cite{HI}). This manifold is simply connected and hence satisfies
condition 3. $\hfill \square$ \\

{\bf Remark 2.} Condition 4 in the theorem could be replaced by a statement of the form
\begin{itemize}
\item[4'.] The matter model is such that static black hole initial data implies uniqueness, where
a {\it black hole static initial data} is an asymptotically flat static KID possibly with boundary
with an asymptotically flat end $\Sigma^{\infty}_0$
such that $\tbd \ext$ (defined
as the connected component of $\{\lambda > 0 \}$ containing
the asymptotic region in $\Sigma^{\infty}_0$)
is a topological manifold without boundary and compact.
\end{itemize}
The Bunting and Masood-ul-Alam method is, at present, the most powerful method to prove uniqueness
under the circumstances of 4'. However, if a new method is invented, Theorem
\ref{uniquenessthr} would still give uniqueness. $\hfill \square$ \\

{\bf Remark 3.} A comment on the condition $T^- \subset T^+$
is in order. First of all, in the  static regime, $T^{+}$ and $T^{-}$ are expected
to be the intersections of
both the black and the white hole with $\Sigmatilde$. Therefore,
the hypothesis $T^{-}\subset T^{+}$ could be understood as the requirement
that the first intersection, as coming from $\bd^{+}\Sigmatilde$, of $\Sigmatilde$ with
an event horizon occurs with the black hole event horizon. Therefore, this hypothesis is
similar to the hypotheses on $\tbd \ext$ made in Theorem \ref{is_a_MOTS}. However, there is
a fundamental difference between them: The hypothesis $T^{-}\subset T^{+}$ is an hypothesis on
the weakly outer trapped regions which, a priori, have nothing to do with the location and properties of
$\tbd \ext$. In a physical sense,
the existence of past weakly outer trapped surfaces in the
spacetime reveals the presence of a white hole region. Moreover, given
a (3+1) decomposition of a spacetime satisfying the NEC,
the Raychaudhuri equation implies
that $T^-$ shrinks to the future while $T^+$ grows to the future (see \cite{AMMS}) (``grow'' and
``shrink'' is with respect to any timelike congruence in the spacetime). It is plausible that
by letting the initial data evolve sufficiently long, only the black hole event horizon is
intersected by $\Sigma$. The uniqueness theorem \ref{uniquenessthr} could be applied to this
evolved initial data. Although this requires much less global
assumptions than for the theorem that ensures
that  no MOTS can penetrate into the domain of outer communications, it still requires
some control on the evolution of the initial data.
In any case,  we believe that the condition $T^- \subset T^+$ is probably not necessary for the validity
of the theorem. It is an interesting open problem to analyze this issue further. $\hfill \square$ \\

We conclude with a trivial corollary of Theorem \ref{uniquenessthr}, which is nevertheless interesting.

\begin{corollary}
Let $(\Sigma,g,K=0;N,\vec{Y}=0;\rho,\vec{J} =0,\tau_{ij};\vec{E})$ be a ti\-me-sy\-mme\-tric
electrovacuum embedded static KID, i.e
a static KID with an electric field $\vec{E}$ satisfying
\begin{eqnarray*}
\nablaSigma_i E^i = 0, \quad \rho = |\vec{E}|^2_{g}, \quad
\tau_{ij} = | \vec{E} |^2 g_{ij} - 2 E_i E_j.
\end{eqnarray*}
Let $\Sigma = \mathcal{K} \cup \Sigma^{\infty}_0$ where $\mathcal{K}$ is a compact and $\Sigma^{\infty}_0$
is an asymptotically flat end and assume that $\bd \Sigma \neq \emptyset$ with
mean curvature with respect to the normal which points inside $\Sigma$ satisfying $p \leq 0$. Then $(\Sigma\setminus T^+,g, K=0; N, \vec{Y}=0, \rho, \vec{J} =0, \tau_{ij}, \vec{E})$
can be isometrically embedded in the Reissner-Nordstr\"om spacetime with $M > |Q|$, where
$M$ is the ADM mass of $(\Sigma,g)$ and $Q$ is the total electric charge of $\vec{E}$, defined
as $Q = \frac{1}{4 \pi}  \int_{S_{r_0}} E^i m_i \eta_{S_{r_0}}$ where $S_{r_0} \subset \Sigma^{\infty}_{0}$
is the coordinate sphere $\{ r =  r_0 \}$ and $\vec{m}$ its unit normal pointing towards infinity.
\end{corollary}

{\bf Remark.} The standard Majumdar-Papapetrou spacetime cannot occur because it possesses degenerate
Killing horizons which are excluded in the hypotheses of the corollary
(recall that, by Proposition \ref{prop:Chrusciel}, degenerate Killing horizons implies
cylindrical ends in time-symmetric slices). $\hfill \square$

\newpage
\thispagestyle{empty}
\mbox{}
\newpage

\chapter{A counterexample of a recent proposal on the Penrose inequality}
\chaptermark{Counterexample of a recent proposal on the Penrose inequality}
\label{ch:Article3}

\setcounter{equation}{0}
\section{Introduction}

In this chapter we will give a counter-example of the Penrose inequality proposed by Bray and Khuri
in \cite{BK}.

As discussed in Chapter \ref{ch:Preliminaires}, in a consistent attempt \cite{BK} to prove the standard Penrose inequality (equation (\ref{penrose1}))
in the general case (i.e. non-time-symmetric),
Bray and Khuri were led to conjecture a new version of the Penrose inequality
in terms of the outermost generalized apparent horizon (see Definition \ref{defi:GAH}) as follows.
\begin{equation}\label{penroseBK2}
M_{\scriptscriptstyle ADM}\geq \sqrt{\frac{|S_{{out}}|}{16\pi}},
\end{equation}
where $M_{\scriptscriptstyle ADM}$
is the ADM mass of a spacelike hypersurface $\Sigma$,
which contains an asymptotically flat end $\Sigma_{0}^{\infty}$, and $|S_{{out}}|$ denotes the area of the
outermost bounding generalized apparent horizon $S_{{out}}$ in $\Sigma$.
As we already remarked in Section \ref{sc:PenroseInequality}, this inequality has several convenient properties
such as the invariancy under time reversals, no need of taking the minimal area enclosure of $S_{{out}}$, and the facts that
it is stronger than (\ref{penrose1}) and covers a larger number of slices of Kruskal with equality than
(\ref{penrose1}). Furthermore, it also has good analytical properties which potentially can lead to
its proof in the general case.
Indeed, Bray and Khuri proved
that if a certain system of PDE admits solutions with the right
boundary behavior, then (\ref{penroseBK2}) follows.

Nevertheless, as we also pointed out in Section \ref{sc:PenroseInequality}, inequality (\ref{penroseBK2})
is not directly supported by cosmic censorship. In fact, it is not difficult to obtain particular situations where
$S_{{out}}$ lies, at least partially, outside the event horizon, as for example for a slice $\Sigma$ in the Kruskal spacetime for which
$\tbd T^{+}$ and $\tbd T^{-}$ intersect transversally. In this case, Eichmair's theorem
(Theorem \ref{thr:Eichmair}) implies that there exists a $C^{2,\alpha}$ outermost generalized apparent
horizon lying, at least partially, in the domain of outer communications of the Kruskal spacetime.

Thus, it becomes natural to study the outermost generalized apparent horizon in slices of this
type in order to check whether (\ref{penroseBK2}) holds or not.
Surprisingly, the result we will find is that there are examples for which
inequality (\ref{penroseBK2}) turns out to be violated. More precisely,
\begin{thr}\label{theorem}
In the Kruskal spacetime with mass $M_{Kr}>0$, there exist asymptotically flat, spacelike
hypersurfaces with an
outermost generalized apparent horizon $S_{out}$ satisfying $|S_{out}|>16\pi M_{Kr}^{2}$.
\end{thr}

For the systems of PDE proposed in \cite{BK}, this means that
a general existence theory cannot be expected with boundary conditions compatible with
generalized apparent horizons. However, simpler boundary conditions (e.g.
compatible with
future and past apparent horizons) are not ruled out. This may in fact
simplify the analysis of these equations.

The results on this chapter have been published in \cite{CM3}, \cite{CMere3}.

\setcounter{equation}{0}
\section{Construction of the counterexample}

Let us
consider the Kruskal spacetime of mass $M_{Kr}>0$ with metric
\[
ds^2= \frac{32M_{Kr}^{3}}{r}e^{-r/2M_{Kr}}d\hat{u}d\hat{v}+ r^2 \left( d\theta^2 + \sin^2{\theta} d\phi^2 \right),
\]
where $r(\hat{u}\hat{v})$ solves the implicit equation
\begin{equation}\label{uvr}
\hat{u}\hat{v}= \frac{r-2M_{Kr}}{2M_{Kr}}e^{r/2M_{Kr}}.
\end{equation}
In this metric
$\partial_{\hat{v}}$ is future directed and $\partial_{\hat{u}}$ is past directed. The
region $\{\hat{u}>0,\hat{v}>0\}$ defines the domain
of outer communications and
$\{\hat{u}=0\}$, $\{\hat{v}=0\}$ define, respectively, the black hole and white hole
event horizons.
Consider the one-parameter family of axially-symmetric embedded hypersurfaces
$\Sigma_{\epsilon} = \mathbb{R} \times \mathbb{S}^2$, with intrinsic coordinates $\ycoor \in \mathbb{R}$, $x \in [-1,1]$,
$\phi \in [0,2 \pi]$, defined by the embedding
\begin{eqnarray*}
\Sigma_{\epsilon}\equiv \left\{ \hat{u}=\ycoor- \epsilon  x, \hat{v}=\ycoor+ \epsilon x,
\cos \theta = x , \phi = \phi\right\}.
\end{eqnarray*}
Inserting this embedding functions into equation (\ref{uvr}) we get
\begin{equation}\label{uvrembedded}
\ycoor^{2}-\epsilon^{2}x^{2}=\frac{r-2M_{Kr}}{2M_{Kr}}e^{r/2M_{Kr}},
\end{equation}
from which it is immediate to
show that, for $|\epsilon|<1$, $\Sigma_{\epsilon}$
does not touch the Kruskal singularity ($r=0$)
for any value of $\{\ycoor,x\}$ in their coordinate range.
It is also immediate to check that the hypersurfaces $\Sigma_{\epsilon}$ are smooth everywhere, included the north and
south poles defined by $|x|=1$.
It is straightforward to prove that the induced metric $g_{\epsilon}$ on $\Sigma_{\epsilon}$
is positive
definite and
satisfies (for $\epsilon$ is small enough) $g_{\epsilon}=dr^{2}+r^{2}\left( \frac{dx^2}{1-x^{2}} +(1-x^{2})d{\phi}^{2}\right)+O^{(2)}(\frac{1}{r})$,
where $r$ is defined in (\ref{uvrembedded}).
Consequently, the hypersurfaces $\Sigma_{\epsilon}$ are spacelike and asymptotically flat.
Let us select ${\Sigma_{\epsilon}}_{0}^{\infty}$ to be the asymptotically
flat end of the region $\{\hat{u}>0,\hat{v}>0\}$.

The discrete isometry of the Kruskal
spacetime defined by
$\left\{\hat{u},\hat{v}\right\}\rightarrow
\left\{\hat{v},\hat{u}\right\}$ implies
that under reflection with respect to the equatorial plane, i.e. $(\ycoor,x,\phi) \rightarrow (\ycoor,-x,\phi)$,
the induced metric of $\Sigma_{\epsilon}$ remains invariant, while the second fundamental
form of $\Sigma_{\epsilon}$ changes sign. The latter is due to the fact that $\Sigma_{\epsilon}$ is
defined by $\hat{u}-\hat{v}+2\epsilon x=0$ and hence the future directed unit normal to
$\Sigma_{\epsilon}$
is proportional (with metric coefficients which only depend on $uv$ and $x^{2}$)
to $d\hat{u}-d\hat{v}+2\epsilon dx$ and, therefore, it changes sign under a reflection
$(\ycoor,x,\phi) \rightarrow (\ycoor,-x,\phi)$ and a simultaneous spacetime isometry
$\left\{\hat{u},\hat{v}\right\}\rightarrow
\left\{\hat{v},\hat{u}\right\}$ (notice that this isometry reverses the time orientation).
Let us denote by $\Sigma^{+}_{\epsilon}$ the intersection of $\Sigma_{\epsilon}$ with the
domain of outer communications $\{\hat{u}>0,\hat{v}>0\}$, which
is given by $\{\ycoor - | \epsilon x |>0\}$. For
$\epsilon  \neq 0$, $\tbd \Sigma^{+}_{\epsilon}$ is composed by a portion of the
black hole event horizon and a portion of the white hole event horizon. Moreover, $\tbd \T^{+}$
is given by $\{\ycoor - \epsilon x=0\}$, while $\tbd \T^{-}$
is $\{\ycoor + \epsilon x=  0 \}$ so that these surfaces intersect transversally on the circumference
$\{\ycoor=0, x= 0\}$ provided $\epsilon \neq 0$\footnote{A graphic example of this type of hypersurface was
already given in Figure \ref{fig:figure1} (where one spatial dimension was suppressed), where the portions of
$S$ intersecting the black hole event horizon and the white hole event horizon represent part of the sets $\tbd T^+$ and $\tbd T^-$, respectively.
The tip of $S$ at
the intersection with the bifurcation surface $S_{0}$ corresponds to the circumference $\{\ycoor=0, x=0\}$.}.
By Eichmair's theorem (Theorem \ref{thr:Eichmair}),
there exists
a $C^{2,\alpha}$ outermost generalized apparent horizon $S_{{out}}$ which is bounding and contains both $\tbd \T^{+}$ and
$\tbd \T^{-}$. Uniqueness implies that this surface must be axially symmetric and have equatorial symmetry.
In what follows we will estimate the area of $S_{{out}}$ from below . To that aim we will proceed in two
steps. Firstly, we will prove that an axial and equatorially symmetric generalized
apparent horizon $\hat{S}_{\epsilon}$ of spherical topology and lying in a sufficiently small neighbourhood of $\{\ycoor =0\}$ exists (provided $\epsilon$ is small enough) and determine its
embedding function. In the second step we will compute its area and prove that it is
smaller or equal than the area of the outermost generalized apparent horizon $S_{{out}}$.

\subsection{Existence and embedding function}

This subsection is devoted to prove the existence of $\hat{S}_{\epsilon}$ and to calculate its embedding function
up to first order in $\epsilon$.
For that, we will consider surfaces $S_{\epsilon}$ of spherical topology
defined by embedding functions
$\{ \ycoor = \y (x, \epsilon), x=x, \phi = \phi \}$ in $\Sigma_{\epsilon}$ and satisfying $\y(-x,\epsilon)=\y(x,\epsilon)$.
Since the outermost generalized apparent horizon is known to be $C^{2,\alpha}$ it is natural to
consider the spaces of functions
\[
U^{m,\alpha} \equiv
\left\{ \y\in C^{m,\alpha}(\mathbb{S}^2):\partial_{\phi}\y=0,\y(-x)= \y(x)\right\} ,
\]
i.e. the spaces of $m$-times differentiable functions on the unit sphere, with H\"older continuous
$m$-th derivatives with exponent $\alpha \in (0,1]$
and invariant under the axial Killing vector
on $\mathbb{S}^2$ and under reflection about the equatorial plane. Each space $U^{m,\alpha}$ is a closed
subset of the Banach space $C^{m,\alpha}(\mathbb{S}^2)$ and hence a Banach space itself.
Let $I \subset \mathbb{R}$ be the closed interval where $\epsilon$ takes values.
The expression that defines a generalized apparent horizon is $p-|q|=0$, where
$p$ is the mean curvature of the corresponding surface $S_{\epsilon}$ in $\Sigma_{\epsilon}$ with respect to the direction pointing into ${\Sigma_{\epsilon}}_{0}^{\infty}$ and $q$ is the trace on $S_{\epsilon}$ of the pull-back of the second fundamental
form $K$ of $\Sigma_{\epsilon}$.
For each function
$\y \in U^{2,\alpha}$
the expression $p - |q|$
defines a non-linear map
$\f:U^{2,\alpha}\times I \rightarrow U^{0,\alpha}$.
Thus, we are looking for
solutions $\y\in U^{2,\alpha}$ of the equation $\f=0$.

We know that when $\epsilon=0$,
the hypersurface $\Sigma_{\epsilon}$
is totally geodesic, which implies $q=0$ for any surface on it.
Consequently, all
generalized apparent horizons on $\Sigma_{\epsilon=0}$ satisfy $p=0$ and are, in fact,
minimal surfaces. The only closed minimal surface in $\Sigma_{\epsilon =0}$ is
the bifurcation surface $S_{0}=\left\{ \hat{u}=0, \hat{v}=0 \right\}$. Thus, the equation
$\f(\y,\epsilon)=0$ has $\y=0$ as the unique solution when $\epsilon=0$.
It becomes natural to use the implicit function theorem for Banach spaces to show that
there exists a unique solution $\y\in U^{2,\alpha}$ of $\f=0$ in a neighbourhood of $\y=0$ for $\epsilon$ small enough. To apply the implicit function theorem it will be necessary to know the explicit form of the
linearization of the differential equation $\f(y,\epsilon)=0$.
The following lemma gives precisely the explicit form of $\f$ up to first order in $\epsilon$.

\begin{lema}\label{lema:explicitf}
Let $\Sigma_{\epsilon}$ be the one-parameter family of axially-symmetric hypersurfaces embedded in the Kruskal spacetime with mass $M_{Kr}>0$,
with intrinsic coordinates $\ycoor \in \mathbb{R}$, $x \in [-1,1]$,
$\phi \in [0,2 \pi]$, defined by
\begin{eqnarray*}
\Sigma_{\epsilon}\equiv \big\{ \hat{u}=\ycoor- \epsilon  x, \hat{v}=\ycoor+ \epsilon x,
\cos \theta = x , \phi = \phi\big\}.
\end{eqnarray*}
Consider the surfaces $S_{\epsilon}\subset \Sigma_{\epsilon}$
defined by $\{\ycoor=y(x),x,\phi\}$ where the embedding function has the form $y=\epsilon Y$, with $Y\in U^{m,\alpha}(\mathbb{S}^2)$.
Then,
$p$ and $q$ satisfy
\begin{eqnarray}
p(y=\epsilon Y,\epsilon)&=&\frac{1}{M_{Kr}\sqrt{e}}\, L[Y(x)]\epsilon+O(\epsilon^{2}),\label{explicitp} \\
q(y=\epsilon Y,\epsilon)&=&-\frac{1}{M_{Kr}\sqrt{e}}\, 3x\epsilon+O(\epsilon^{2}), \label{explicitq}
\end{eqnarray}
where $L[z(x)]\equiv -(1-x^{2})\ddot{z}(x)+2x\dot{z}(x)+z(x)$ and where the dot denotes derivative with respect to
$x$.
\end{lema}
{\bf Proof.} The proof is by direct computation. 
Let us define $H=\frac{32M_{Kr}^{3}}{r}e^{-r/2M_{Kr}}$, $Q=r^{2}$ and $x=\cos{\theta}$, so that
the Kruskal metric takes the form
\[
\gM=Hd\hat{u}d\hat{v}+\frac{Q}{1-x^{2}}dx^{2}+(1-x^{2})Q d\phi^{2}.
\]
The induced metric $g_{\epsilon}$ on $\Sigma_{\epsilon}$ is
\begin{equation}\label{explicitg}
g_{\epsilon}=\hat{H}d\ycoor^{2}+\left( \frac{\hat{Q}}{1-x^{2}}-\epsilon^{2}\hat{H}\right)dx^{2}+(1-x^{2})\hat{Q}d\phi^{2},
\end{equation}
where $\hat{H}$, $\hat{Q}$ are obtained from $H$, $Q$ by expressing $r$ in terms of $(\ycoor,x)$ according to
(\ref{uvrembedded}).
The induced metric $\gamma_{\epsilon}$ on $S_{\epsilon}$ satisfies
\begin{equation}\label{explicitgamma}
\gamma_{\epsilon}=\left[ \frac{\tilde{Q}}{1-x^{2}}+\epsilon^{2}\left(\dot{Y}^{2}(x)-1\right) \tilde{H} \right]dx^{2}+(1-x^{2})\tilde{Q}d\phi^{2},
\end{equation}
where $\tilde{H}$, $\tilde{Q}$ are obtained from $\hat{H}$ and $\hat{Q}$ by inserting
$\ycoor=\epsilon Y(x)$.
Firstly, let us deal with the computation of $p=-m_{i}\gamma_{\epsilon}^{AB}\nabla^{\Sigma_{\epsilon}}_{\vec{e}_{A}}e_{B}^{i}$,
where $\bold{m}$ is the unit vector tangent to $\Sigma_{\epsilon}$ normal to $S_{\epsilon}$ which points to the
asymptotically flat end in $\{\hat{u}>0,\hat{v}>0\}$ and $\{\vec{e}_{A}\}$ is a basis for $T S_{\epsilon}$.
In our coordinates
\begin{eqnarray*}
&&\vec{e}_{x}=\partial_{x}+\epsilon\dot{Y}(x)\partial_{\ycoor},\\
&&\vec{e}_{\phi}=\partial_{\phi}.
\end{eqnarray*}
The unit normal is therefore
\begin{equation}\label{explicitm}
\bold{m}=\sqrt{\frac{\tilde{H}\left(\tilde{Q}-\epsilon^{2}(1-x^2)\tilde{H}\right)}{\tilde{Q}+\epsilon^{2}(1-x^2)
(\dot{Y}^{2}-1)\tilde{H}}}
\,\left(d\ycoor-\epsilon\dot{Y}(x)dx\right).
\end{equation}
Since $\gamma_{\epsilon}$ is diagonal, we only need to calculate
$\nabla^{\Sigma_{\epsilon}}_{\vec{e}_{x}}e_{x}^{\ycoor}$,
$\nabla^{\Sigma_{\epsilon}}_{\vec{e}_{\phi}}e_{\phi}^{\ycoor}$,
$\nabla^{\Sigma_{\epsilon}}_{\vec{e}_{x}}e_{x}^{x}$ and $\nabla^{\Sigma_{\epsilon}}_{\vec{e}_{\phi}}e_{\phi}^{x}$
up to first order. The results are the following.
\begin{eqnarray}
\nabla^{\Sigma_{\epsilon}}_{\vec{e}_{x}}e_{x}^{\ycoor}&=&
    -
    \frac{\partial_{\ycoor}\hat{Q}}{2(1-x^2)\tilde{H}}+
    \epsilon\left(\ddot{Y}+\dot{Y}\partial_{x}\ln{\hat{H}}\right)+
    O(\epsilon^{2}),
    \label{explicitDxxy}\\
\nabla^{\Sigma_{\epsilon}}_{\vec{e}_{x}}e_{x}^{x}&=&
    \frac{2x+(1-x^{2})\partial_{x}\ln{\hat{Q}}}{2(1-x^{2})}+
    \epsilon\dot{Y}\partial_{\ycoor}\ln{\hat{Q}}  +O(\epsilon^{2}) ,
    \label{explicitDxxx}\\
\nabla^{\Sigma_{\epsilon}}_{\vec{e}_{\phi}}e_{\phi}^{\ycoor}&=&
    -\frac{(1-x^{2})\partial_{\ycoor}\hat{Q}}{2\tilde{H}},
    \label{explicitDphiphiy}\\
\nabla^{\Sigma_{\epsilon}}_{\vec{e}_{\phi}}e_{\phi}^{x}&=&
    \frac{(1-x^{2})\left( 2x-(1-x^{2})\partial_{x}\ln{\hat{Q}}\right)}{2}+O(\epsilon^{2}),
    \label{explicitDphiphix}
\end{eqnarray}
where $\partial_{\ycoor}\hat{Q}$ means taking derivative with respect to
$\ycoor$ of $\hat{Q}$ and afterwards, substituting $\ycoor=\epsilon Y(x)$
(and similarly for the other derivatives).

In order to compute the derivatives of $\hat{H}$ and $\hat{Q}$, we need to
calculate the derivatives $\partial_{\ycoor}r(\ycoor,x)$ and $\partial_{x}r(\ycoor,x)$.
This can be done by taking derivatives of (\ref{uvrembedded})
with respect to $x$ and $\ycoor$, which gives,
\begin{eqnarray*}
&&\partial_{\ycoor}r=\epsilon\frac{8M_{Kr}^{2}}{r}e^{-r/2M_{Kr}}  Y, \label{Dyr}\\
&&\partial_{x}r=-\epsilon^{2}\frac{8M_{Kr}^{2}}{r}e^{-r/2M_{Kr}} x. \label{Dxr}
\end{eqnarray*}
At $\epsilon=0$ we have $y=0$ and equation (\ref{uvrembedded}) gives $r\big|_{S_{\epsilon=0}}=2M_{Kr}$.
Then $r\big|_{S_{\epsilon}}=2M_{Kr}+O(\epsilon)$
This allows us to compute the derivatives of $\hat{H}$ and $\hat{Q}$ up to first order in $\epsilon$. The result is
\begin{eqnarray*}
\partial_{x}\hat{H}&=&O(\epsilon^{2}),\label{DxH}\\
\partial_{\ycoor}\hat{Q}&=&\epsilon\frac{16M_{Kr}^{2}}{e} Y +O(\epsilon^2),\label{DyQ}\\
\partial_{x}\hat{Q}&=&O(\epsilon^{2}).\label{DxQ}
\end{eqnarray*}
Inserting these equations into
(\ref{explicitgamma}), (\ref{explicitm}), (\ref{explicitDxxy}), (\ref{explicitDxxx}),
(\ref{explicitDphiphiy}) and (\ref{explicitDphiphix}), and putting all these results together,
we finally obtain that
$p=-m_{i}\gamma_{\epsilon}^{AB}\nabla^{\Sigma_{\epsilon}}_{\vec{e}_{A}}e^{i}_{B}$ satisfies (\ref{explicitp}).

Next, we will study $q=\gamma_{\epsilon}^{AB}e^{i}_{A}e^{j}_{B}K_{ij}$, where $K$ is the second fundamental form
of $\Sigma_{\epsilon}$ with respect to the future directed unit normal.
Since, $\gamma_{\epsilon}$ is diagonal, we just have to compute
$e^{i}_{x}e^{j}_{x}K_{ij}=\dot{y}^{2}K_{yy}+2\dot{y}K_{xy}+K_{xx}$ and $e^{i}_{\phi}e^{j}_{\phi}K_{ij}=K_{\phi\phi}$
up to first order.
To that aim, it is convenient to take coordinates
$\{T=\frac12(\hat{v}-\hat{u}),\ycoor=\frac12(\hat{v}+\hat{u}),x,\phi\}$ in the Kruskal spacetime for which the
metric $\gM$ is diagonal. In these coordinates $\Sigma_{\epsilon}$ is defined by
$\{T=\epsilon x,\ycoor,x,\phi\}$ and the future directed unit normal to $\Sigma_{\epsilon}$ reads
\[
\bold{n}=\sqrt{\frac{\hat{H}\hat{Q}}{\hat{Q}-\epsilon^{2}(1-x^{2})\hat{H}}}\left( -dT+\epsilon dx \right).
\]
The computation of the second fundamental form is straightforward and gives
\begin{eqnarray}
\dot{y}^{2}K_{yy}&=&O(\epsilon^{2}),\label{Kyy}\\
2\dot{y}K_{xy}&=&O(\epsilon^{2}),\label{Kxy}\\
K_{xx}&=&\sqrt{\tilde{H}}
    \left[ \frac{\partial_{\,T}Q'}{2(1-x^{2})\tilde{H}} -
    \epsilon\frac{x}{1-x^2}\right]+O(\epsilon^{2})
    \label{Kxx}\\
K_{\phi\phi}&=&\sqrt{\tilde{H}}
    \left[ \frac{(1-x^{2})\partial_{\,T}Q'
    }{2\tilde{H}} -\epsilon (1-x^{2})x\right]+O(\epsilon^{2})\label{Kphiphi},
\end{eqnarray}
where we have denoted by $Q'$ the function obtained from $Q$ by expressing $r$ in
terms of $(T,\ycoor)$ according to $\hat{u}\hat{v}=\ycoor^{2}-T^{2}=\frac{r-2M_{Kr}}{2M_{Kr}}e^{r/2M_{Kr}}$.
This expression also allows us to compute $\partial_{\,T}Q'$ which, on $\Sigma_{\epsilon}$
(where $T=\epsilon x$) and using $r=2M_{Kr}+O(\epsilon)$, takes the form
\[
\partial_{\,T}Q'=-\epsilon \frac{16M_{Kr}^{2}}{e} x+O(\epsilon^2).
\]
Inserting this into (\ref{Kxx}) and (\ref{Kphiphi}), and using (\ref{explicitgamma}), it is a matter of simple computation to
show that $q=\gamma_{\epsilon}^{AB}e^{i}_{A}e^{j}_{B}K_{ij}$ satisfies (\ref{explicitq}).
$\hfill \blacksquare$ \\

From this lemma we conclude that $f(y=\epsilon Y, \epsilon)\equiv p(y=\epsilon Y, \epsilon)-|q(y=\epsilon Y, \epsilon)|$ reads
\begin{equation}\label{explicitf}
\f(y= \epsilon Y,\epsilon)=\frac{1}{M_{Kr}\sqrt{e}}\,(L[Y(x)]-3|x|)\epsilon+O(\epsilon^{2}).
\end{equation}

The implicit function theorem requires the operator $\f$ to have a continuous Fr\'echet derivative and the
partial derivative $\left.D_{\y}\f\right|_{(\y=0,\epsilon=0)}$ to be an isomorphism
(see Appendix \ref{ch:appendix2}).
The problem is not trivial in our case because the appearance of $|x|$ makes the
Fr\'echet derivative of $\f$ potentially
discontinuous\footnote{We thank M. Khuri for pointing out this issue.}.
However, the problem can be solved considering a suitable modification of $\f$,
as we discuss in detail next.

\begin{proposition}\label{proposition}
There exists a neighborhood $\tilde{I}\subset I$ of $\epsilon=0$
such that $\f(\y,\epsilon)=0$ admits a solution
$\y(x,\epsilon) \in U^{2,\alpha}(\mathbb{S}^2)$  for all $\epsilon \in \tilde{I}$.
Moreover, $\y(x,\epsilon)$ is $C^1$ in $\epsilon$
and satisfies $\y(x,\epsilon=0)=0$.
\end{proposition}

{\bf Proof.} Firstly, let us consider surfaces $S_{\epsilon}$ in $\Sigma_{\epsilon}$ defined by
$\left\{ \ycoor=y(x,\epsilon),x,\phi \right\}$ such that the embedding function has the form
$\y = \epsilon \Y$, where $\Y \in U^{2,\alpha}$. 
Since we are considering surfaces with axial symmetry, neither $p$ nor $q$ depend on $\phi$.
Let $\eta^{\mu}$ denote the spacetime coordinates, $z^{i}$ the coordinates on $\Sigma_{\epsilon}$,
$x^{A}$ the coordinates on $S_{\epsilon}$, $\eta^{\mu}(z^{i})$ the embedding functions of $\Sigma$ in $M$
(which depend smoothly on $z^{i}$), and
$z^{i}(x^{A})$ the embedding functions of $S$ in $\Sigma$ (which depend smoothly on $x^A$).
Thus, by definition, we have
\[
p(x,\epsilon)=-\gamma^{AB}m_{i}
    \left[ \frac{\partial^{2}z^{i}}{\partial x^{A}\partial x^{B}}
    +{\Gamma^{\Sigma_{\epsilon}}}_{jk}^{i}(z(x))\frac{\partial z^{j}}{\partial x^{A}}\frac{\partial z^{k}}{\partial x^{B}}\right],
\]
where ${\Gamma^{\Sigma_{\epsilon}}}_{jk}^{i}$ are the Christoffel symbols of $\Sigma_{\epsilon}$.
In this expression all terms depend smoothly on
$(\dot y(x),y(x),x,\epsilon)$, except $\frac{\partial^{2}z^{i}}{\partial x^{A}\partial x^{B}}$
which also depends on $\ddot y(x)$. Therefore, $p$ can be viewed as a smooth
function of $(\ddot{y}(x),\dot{y}(x),y(x),x,\epsilon)$. Similarly, by definition,
\[
q(x,\epsilon)=\left . -\gamma^{AB}n_{\mu}e^{i}_{A}
    e_{B}^{j}
    \left[ \frac{\partial^{2}\eta^{\mu}}{\partial z^{i}\partial z^{j}}
    +{\Gamma}_{\nu\beta}^{\mu}(\eta(z)) \frac{\partial \eta^{\nu}}{\partial z^{i}}\frac{\partial \eta^{\beta}}{\partial z^{j}} \right]\right |_{z^i = z^i (x^A)},
\]
where all terms depend smoothly on $(\dot{y}(x),y(x),x,\epsilon)$
Therefore, setting $y=\epsilon Y$ and since both $p$ and $q$ are $O(\epsilon)$ (see equations (\ref{explicitp})
and (\ref{explicitq})), we can write
\[
p = \epsilon \Pcal (\Y(x),\dot{\Y}(x),\ddot{\Y}(x),x, \epsilon)
\]
and
\[
q = \epsilon \Qcal (\Y(x),\dot{\Y}(x),x, \epsilon),
\]
where
$\Pcal : \mathbb{R}^3 \times[-1,1] \times I \rightarrow \mathbb{R}$ and $\Qcal : \mathbb{R}^2 \times[-1,1] \times I \rightarrow \mathbb{R}$ are smooth functions.
Moreover,
the function $\Qcal$ has the symmetry $\Qcal\left(x_1,x_2,x_3,x_4\right) = -
\Qcal\left(x_1,-x_2,-x_3,x_4\right)$, which reflects the fact that the extrinsic curvature
of $\Sigma_{\epsilon}$ changes sign under a transformation $x \rightarrow -x$ 
and the symmetry $Y(-x)=Y(x)$.
Let us write
$\P(Y,\epsilon)(x) \equiv \Pcal(\Y(x),\dot{\Y}(x),\ddot{\Y}(x),x,\epsilon)$ and
similarly $\Q(Y,\epsilon)(x) \equiv \Qcal (\Y(x),\dot{\Y}(x),x,\epsilon)$.

Now, instead of $\f$, let us consider
the functional $\F : U^{2,\alpha} \times I \rightarrow U^{0,\alpha}$
defined by $\F (\Y,\epsilon)= \P (\Y,\epsilon) -
|\Q (\Y,\epsilon)|$. This functional has the property that, for
$\epsilon>0$, the solutions of $\F(\Y,\epsilon)=0$ correspond exactly
to the solutions of $\f(\y,\epsilon)=0$
via the relation $\y = \epsilon \Y$.
Moreover, the functional $\F$ is well-defined for all $\epsilon \in I$, in
particular at $\epsilon=0$.
Therefore, by proving that $\F=0$ admits solutions
in a neighbourhood of $\epsilon=0$, we will conclude that $\f=0$
admits solutions for $\epsilon > 0$ and the solutions will in fact belong to a
neighbourhood of $\y=0$ since $\y = \epsilon \Y$.

In order to show that $\F$ admits solutions we will use the implicit function
theorem. Equation (\ref{explicitf}) yields
\begin{equation}\label{explicitF}
\F (\Y,\epsilon=0)(x) = c
\left ( L[\Y(x)] -  3|x| \right)
\end{equation}
where $c$ is the constant $1/(M_{Kr}\sqrt{e})$ and
$L[\Y] \equiv -(1-x^2)\ddot{\Y} +2x\dot{\Y}+\Y$. As it is well-known the eigenvalue problem
$(1-x^{2})\ddot{z}(x)-2x\dot{z}(x)+\lambda z(x)=0$
has non-trivial smooth solutions on $[-1,1]$ (the Legendre polynomials)
if and only if $\lambda=l(l+1)$, with $l\in\mathbb{N}\cup\{0\}$.
Thus, the kernel
of $L[Y]$ (for which $\lambda=1$) is $Y=0$. We conclude that $L$ is an isomorphism
between $U^{2,\alpha}$ and $U^{0,\alpha}$. Let $\Ysol \in U^{2,\alpha}$ be the unique
solution of the equation $L[\Y]=3|x|$. For later use, we note that $\Q(\Ysol,\epsilon=0) =
-3 c x$ (see equation (\ref{explicitq})).
This vanishes {\it only} at $x=0$. This is the key property that allows
us to prove that $\F$ is $C^1(U^{2,\alpha}\times I)$.

The $C^1(U^{2,\alpha}\times I)$ property of the functional $\P (\Y,\epsilon)$ is immediate from
Theorem \ref{thr:GT} in the Appendix \ref{ch:appendix2}.
More subtle is to show that $|\Q|$ is $C^1(U^{2,\alpha}\times I)$ in a
suitable neighbourhood of $(\Ysol,\epsilon=0)$. Let $r_0 >0$ and
define
\begin{equation}\label{Nur0}
{\cal V}_{r_0} = \{ (\Y,\epsilon)\in U^{2,\alpha}\times I:
\| (\Y-\Ysol,\epsilon) \|_{U^{2,\alpha}\times I} \leq r_0 \}.
\end{equation}
First of all we need
to show that $|\Q|$ is Fr\'echet-differentiable on ${\cal V}_{r_0}$, i.e.
that for all $(\Y,\epsilon) \in {\cal V}_{r_0}$
there exists a bounded linear map $D_{\Y,\epsilon}|\Q|:U^{2,\alpha}\times
I \rightarrow U^{0,\alpha}$ such
that, for all $(Z,\delta) \in U^{2,\alpha} \times I$,
$|\Q(\Y+Z,\epsilon+\delta)|-|\Q(\Y,\epsilon)| =D_{\Y,\epsilon}|\Q| (Z,\delta)+R_{\Y,\epsilon}(Z,\delta)$
where $\| R_{\Y,\epsilon}(Z,\delta)\|_{U^{0,\alpha}}=o(\|(Z,\delta)\|_{U^{2,\alpha}\times I})$.
The key observation is that, by choosing $r_0$ small enough in Definition \ref{Nur0}, we have
\begin{eqnarray}
|\Q(\Y,\epsilon)(x)|=-\sigma(x)\Q(\Y,\epsilon)(x) \quad \forall (Y,\epsilon) \in {\cal V}_{r_0}, \label{key}
\end{eqnarray}
where $\sigma(x)$ is the {\it sign}
function, (i.e. $\sigma(x)=+1$ for $x \geq 0$ and $\sigma(x)=-1$ for $x<0$).
To show this we need to distinguish two cases: when $x$ lies in a sufficiently small neighbourhood $(-\varepsilon,\varepsilon)$ of
$0$ and when $x$ lies outside this neighbourhood. Consider first the latter case. As already mentioned, we have
$\Q(\Ysol,\epsilon=0)=-3cx$ which is negative for $x>0$ and positive for $x<0$.
Taking $r_0$ small enough, and using that $\Qcal$ is a smooth function of its arguments it follows that the inequalities
$\Q(\Ysol,\epsilon)<0$ for $x\geq \varepsilon$ and $\Q(\Ysol,\epsilon)>0$ for $x \leq - \varepsilon$ still hold
for any $(\Y,\epsilon) \in {\cal V}_{r_0}$. For the points $x\in (-\varepsilon,\varepsilon)$, the function $\Q (\Y,\epsilon)(x)$
is odd in $x$, so it passes through zero at $x=0$. Hence,
the relation (\ref{key})
holds in $(-\varepsilon,\varepsilon)$ provided we can prove that $\Q(\Y,\epsilon)$ is strictly decreasing at $x=0$.
But this follows immediately from the fact that $\frac{d \Q (\Ysol,\epsilon=0)}{dx} |_{x=0} = -3 c$ and
$\Qcal$ is a smooth function of its arguments.

From its definition, it follows that $\Q (\Y,\epsilon)(x)$ is $C^{1,\alpha}$ (note that only first derivatives of
$\Y$ enter in $\mathcal{Q}$) and
that the functional $\Q(\Y,\epsilon)$ has
Fr\'echet derivative (see Theorem \ref{thr:GT} in Appendix \ref{ch:appendix2})
\begin{equation}
D_{\Y,\epsilon}\Q(Z,\delta)(x)=A_{\Y,\epsilon}(x)Z(x)+B_{\Y,\epsilon}(x)\dot{Z}(x)+
C_{\Y,\epsilon}(x)\delta, \nonumber
\end{equation}
where $A_{\Y,\epsilon}(x) \equiv \partial_{1} \Qcal |_{(\Y(x),\dot{\Y}(x),x,\epsilon)}$,
$B_{\Y,\epsilon}(x)  \equiv \partial_{2} \Qcal |_{(\Y(x),\dot{\Y}(x),x,\epsilon)}$ and
$C_{\Y,\epsilon}(x)  \equiv  \partial_{4} \Qcal |_{(\Y(x),\dot{\Y}(x),x,\epsilon)}$. We note that these three
functions are $C^{1,\alpha}$ and that $A_{\Y,\epsilon}$, $C_{\Y,\epsilon}$ are odd, while
$B_{\Y,\epsilon}$ is even (as a consequence of the symmetries of $\Qcal$). Defining the linear map
\begin{equation}
D_{\Y,\epsilon}|\Q| (Z,\delta) \equiv -\sigma( A_{\Y,\epsilon}Z+B_{\Y,\epsilon}\dot{Z}+C_{\Y,\epsilon}\delta ),\nonumber
\end{equation}
it follows from (\ref{key}) that
\[
|\Q(\Y+Z,\epsilon+\delta)|-|\Q(\Y,\epsilon)|= D_{\Y,\epsilon} |\Q| (Z,\delta) + R_{\Y,\epsilon}
(Z,\delta),
\]
with $\| R(Z,\delta)\|_{U^{0,\alpha}}=o(\|(Z,\delta)\|_{U^{2,\alpha}\times I})$.
In order to conclude that $D_{\Y,\epsilon} |\Q|$ is the derivative of $|\Q(\Y,\epsilon)|$,
we only need to check that, it is (i) well-defined (i.e. that its image belongs to $U^{0,\alpha}$) and (ii)
that it is bounded, i.e. that $\| D_{\Y,\epsilon} |\Q| (Z,\delta)
\|_{U^{0,\alpha}}
< C \|(Z,\delta) \|_{U^{2,\alpha}\times I}$
for some constant $C$.

To show (i), let us concentrate on the most difficult term which is
$-\sigma B_{\Y,\epsilon} \dot{Z}$ (because $B_{\Y,\epsilon}(x)$ is even and need not vanish at $x=0$).
Since $\dot{Z}$ is an odd function, $-\sigma B_{\Y,\epsilon} \dot{Z}$ is continuous.
To show it is also H\"older continuous, we only need to consider points $x_1 = -a$ and $x_2 = b$
with $0 < a < b$ (if $x_1\cdot x_2\geq 0$, the {\it sign} function remains constant, so
$-\sigma B_{\Y,\epsilon} \dot{Z}$ is in fact $C^{1,\alpha}$). Calling $w(x) \equiv
-\sigma(x) B_{Y,\epsilon} (x) \dot{Z}(x)$
and using that $w(x)$ is even, we find
\begin{eqnarray}
&&|w(x_2) - w(x_1)| = |w(b) - w(-a)| = |w(b) - w(a) | = \nonumber \\
&&\left |\left .
\frac{d (B_{Y,\epsilon} \dot{Z})}{dx} \right |_{x=\zeta} \right |
|b-a|
= \left | \left . \frac{d (B_{Y,\epsilon} \dot{Z})}{dx} \right |_{x=\zeta}
\right |
|b-a|^{1-\alpha}|b-a|^{\alpha}\leq \nonumber\\
&&\left | \left . \frac{d (B_{Y,\epsilon} \dot{Z})}{dx} \right |_{x=\zeta}
\right |
|b-a|^{1-\alpha}
|x_2 - x_1|^{\alpha}\leq
\left | \left .
\frac{d (B_{Y,\epsilon} \dot{Z})}{dx} \right |_{x=\zeta} \right |
|x_2 - x_1|^{\alpha}\nonumber \\
&&\leq \sup_{x}\left | \left .
\frac{d (B_{Y,\epsilon} \dot{Z})}{dx}\right. \right ||x_2 - x_1|^{\alpha}
\label{estim}
\end{eqnarray}
where the mean value theorem has been applied in the third equality and $\zeta \in (a,b)$. We also
have used that $|b-a|^{\alpha} \leq |b+a|^{\alpha} = |x_2 - x_1|^{\alpha}$
and $|b-a| < 1$. This proves that $- \sigma \B \dot{Z}$ is H\"older continuous with exponent $\alpha$.
The remaining terms $-\sigma(x)A_{\Y,\epsilon}(x)Z(x)$ and $-\sigma(x)C_{\Y,\epsilon}(x)\delta$
are obviously continuous
because they vanish at $x=0$. To show H\"older continuity the same argument that for
$-\sigma(x) B_{Y,\epsilon} (x) \dot{Z}$ works.

To check (ii), we have to find and upper bound for the norm
$\| w(x)
\|_{U^{0,\alpha}}$.
\begin{eqnarray*}
&&\| w(x)
\|_{U^{0,\alpha}}=\sup_{x}|w(x)|+\sup_{x_1\neq x_2}
\frac{|w(x_{2})-w(x_{1})|}{|x_{2}-x_{1}|^{\alpha}}\\
&&\leq \sup_{x}|B_{Y,\epsilon}(x)|\sup_{x}|\dot{Z}(x)|+\sup_{x}\left| \frac{d(B_{Y,\epsilon}\dot{Z})}{dx}\right|
\\
&&\leq
\sup_{x}|B_{Y,\epsilon}(x)|\sup_{x}|\dot{Z}(x)|+\sup_{x}| \dot{B}_{Y,\epsilon}(x)|\sup_{x}|\dot{Z}(x)|+\sup_{x}
|B_{Y,\epsilon}(x)|\sup_{x}|\ddot{Z}(x)|\\
&&\leq ( 2\sup_{x}|B_{Y,\epsilon}(x)|+\sup_{x}|\dot{B}_{Y,\epsilon}(x)| ) \|(Z,\delta)\|_{U^{2,\alpha}\times I},
\end{eqnarray*}
where, in the first inequality, (\ref{estim}) has been used. Since $B_{Y,\epsilon}(x)$ is $C^{1,\alpha}$, then
$( 2\sup_{x}|B_{Y,\epsilon}(x)|+\sup_{x}|\dot{B}_{Y,\epsilon}(x)| )$ is bounded
in the compact set $[-1,1]$ and, therefore, there exists a constant $C$ such that
$\| -\sigma B_{Y,\epsilon} \dot{Z}
\|_{U^{0,\alpha}}
< C \|(Z,\delta) \|_{U^{2,\alpha}\times I}$. A similar argument applies to $- \sigma A_{\Y,\epsilon} Z$
and $- \sigma C_{\Y,\epsilon} \delta$ and we conclude that
$D_{\Y,\epsilon} |\Q|$ is indeed a continuous operator.

In order to apply the implicit function theorem, it is furthermore necessary
that $|\Q|\in C^1(U^{2,\alpha}\times I)$ (i.e. that
$D_{Y,\epsilon} |\Q|$ depends continuously on $(\Y,\epsilon)$).
This means that given any convergent sequence $(\Y_n,\epsilon_{n})\in
{\cal V}_{r_0}$, the corresponding operators $D_{\Y_n,\epsilon_n} |\Q|$ also
converge. Denoting by $(\Y,\epsilon) \in {\cal V}_{r_0}$
the limit of the sequence, we need to
prove that
\[\| D_{\Y_n,\epsilon_n} |\Q| - D_{\Y,\epsilon} |\Q|
\|_{\pounds (U^{2,\alpha} \times I, U^{0,\alpha})} \rightarrow 0,
\]
where, for any linear operator $\mathscr{L}:U^{2,\alpha} \times I\rightarrow U^{0,\alpha}$, the operator norm is
\[
\| \mathscr{L}
\|_{\pounds (U^{2,\alpha} \times I, U^{0,\alpha})}\equiv \sup_{\scriptscriptstyle{(Z,\delta)\neq (0,0)}} 
\frac{\| \mathscr{L}(Z,\delta)
\|_{U^{0,\alpha}}}
{\|(Z,\delta)\|_{U^{2,\alpha}\times I}}.
\]
For that it suffices to
find a constant $K$ (which may depend on $(\Y,\epsilon)$), such that
\begin{eqnarray}
&&\hspace{-18mm}\| ( D_{\Y_n,\epsilon_n} |\Q| -
D_{\Y,\epsilon} |\Q| ) (Z,\delta) \|_{U^{0,\alpha}}\nonumber \\
&&\qquad\qquad \leq
K \|(Z,\delta)\|_{U^{2,\alpha}\times I} \|( \Y_n-\Y,\epsilon_{n}-\epsilon)\|_{U^{2,\alpha} \times I}
\label{convergence}
\end{eqnarray}
for all $(Z,\delta)\in U^{2,\alpha} \times I$.
Indeed, if (\ref{convergence}) holds then the right-hand side tends to zero when
$(Y_{n},\epsilon_{n})\rightarrow (Y,\epsilon)$
Again, the most difficult case
involves $\sigma (B_{Y,\epsilon} - B_{Y_n,\epsilon_n}) \dot{Z}$, so let us concentrate on this term
(the same argument works for the remaining terms in $D_{\Y_n,\epsilon_n} |\Q| - D_{\Y,\epsilon}|\Q|$).

With the definition $z \equiv \sigma (B_{\Y,\epsilon} - B_{\Y_n,\epsilon_n}) \dot{Z}$, we have
\[
\sup_x |
z(x)| \leq \sup_{x}|{B}_{\Y,\epsilon}(x) - {B}_{\Y_n,\epsilon_n}(x)|\sup_{x}|\dot{Z}(x)|.
\]
To bound the $C^0$-norm of $z$
in terms of $\|(Z,\delta)\|_{U^{2,\alpha}\times I} \|( \Y_n-\Y,\epsilon_{n}-\epsilon)\|_{U^{2,\alpha} \times I}$,
we have to
use the mean value theorem on the function
${\cal B} \equiv \partial_2 \Qcal$ (recall that $B_{Y,\epsilon}(x) = {\cal B} |_{(Y(x),\dot{Y}(x),x,
\epsilon)}$).
By the definition of $\mathcal{V}_{r_{0}}$ (see (\ref{Nur0})) any element $(Y,\epsilon)\in \mathcal{V}_{r_{0}}$
satisfies that $|Y-Y_{1}|(x)\leq r_{0}$ and
$|\dot{Y}-\dot{Y}_{1}|(x)\leq r_{0}$ $\forall x\in[-1,1]$.
This implies that there is a compact set $\mathbb{K}\subset \mathbb{R}^{4}$
depending only on $r_{0}$ and $Y_{1}$ such that
$(\Y(x), \dot{\Y}(x),x,\epsilon) \in \mathbb{K}$, for all $x \in [-1,1]$ and $(Y,\epsilon)\in{\cal V}_{r_{0}}$.
When applying the mean value theorem to
the derivatives $\partial_{1}{\cal B}$, $\partial_{2}{\cal B}$ and
$\partial_{4}{\cal B}$ all mean value points will therefore belong to $\mathbb{K}$. Taking the supremum
of these derivatives in $\mathbb{K}$, we get the following bound.
\begin{eqnarray}
\hspace{-6mm}
\sup_x |
z(x)|\leq
\sup_{\mathbb{K}}\left( |\partial_{1}{\cal B}|+|\partial_{2}{\cal B}|+|\partial_{4}{\cal B}| \right)
\sup_x | \dot{Z} | \| (Y_n  - Y,\epsilon_{n}-\epsilon) \|_{U^{2,\alpha}\times I}. \label{ineq1}
\end{eqnarray}
Since ${\cal B}$ is smooth, (\ref{ineq1})
is already of the form (\ref{convergence}).

It only remains to bound the H\"older norm
of $z$ in a similar way.
As before,
this is done
by distinguishing two cases, namely when $x_1 \cdot x_2 \geq 0$ and when
$x_1 \cdot x_2 <0$. If $x_1 \cdot x_2 \geq 0$ then $\sigma(x)$ is a constant function and therefore,
to obtaining an inequality of the form
\[
\sup_{x_1 \neq x_2} \frac{| z(x_2) - z(x_1)|}{|x_2 - x_1|^{\alpha}} \leq K_1 \|(Z,\delta)\|_{U^{2,\alpha}\times I} \|(\Y_n-\Y,\epsilon_{n}-\epsilon)\|_{U^{2,\alpha} \times I}
\]
is standard (and a consequence of Theorem \ref{thr:GT}).
When $x_1 \cdot x_2 <0$, we exploit the parity of the functions
as in (\ref{estim}) to get
\[
| z(x_2) - z(x_1) | \leq
\left | \left .  \frac{d ((B_{\Y_n,\epsilon_n} - B_{\Y,\epsilon}) \dot{Z})}{dx}
\right |_{x=\zeta}  \right |
|x_2 - x_1|^{\alpha},
\]
where $\zeta \in (a,b)$ and we are assuming $x_1 = -a, x_2 = b, 0 < a < b$
without loss of generality. Since the sign function $\sigma(x)$ has already
disappeared, a bound for the right hand side
in terms of $K_2
\|(Z,\delta)\|_{U^{2,\alpha}\times I} \|(\Y_n-\Y,\epsilon_{n}-\epsilon)\|_{U^{2,\alpha} \times I} |x_2 - x_1|^{\alpha}$
is guaranteed by Theorem \ref{thr:GT}. This, combined with (\ref{ineq1}) gives (\ref{convergence}) and hence
continuity of the derivative of $D_{\Y,\epsilon} |\Q|$ with respect to
$(\Y,\epsilon) \in {\cal V}_{r_0}$.

The final requirement to apply the implicit function theorem to $\F = \P  - |\Q|$  is to
check that $D_{\Y} \F |_{(\Ysol,\epsilon=0)}$ is an isomorphism between $U^{2,\alpha}$ and $U^{0,\alpha}$.
This is immediate from equation (\ref{explicitF}) that implies
\[
D_\Y \F |_{(\Ysol,\epsilon=0)} (Z) = \F(Y_{1}+Z,\epsilon=0)-\F(Y_{1},\epsilon=0)= c L (Z),
\]
and we have already shown that $L$ is an isomorphism.

Thus, the implicit function theorem
can be used to conclude that there exists an open
neighbourhood $\tilde {I} \subset I$ of $\epsilon=0$ and a $C^1$ map
$\tilde{Y}: \tilde{I} \rightarrow U^{2,\alpha}$
such that $\tilde{Y} (\epsilon=0) = \Ysol$ and $\y = \epsilon
\tilde{Y}(\epsilon)$ defines a $C^{2,\alpha}$ generalized
apparent horizon embedded in $\Sigma_{\epsilon}$.

$\hfill \blacksquare$ \\

\begin{figure}
\begin{center}
\includegraphics[width=12cm]{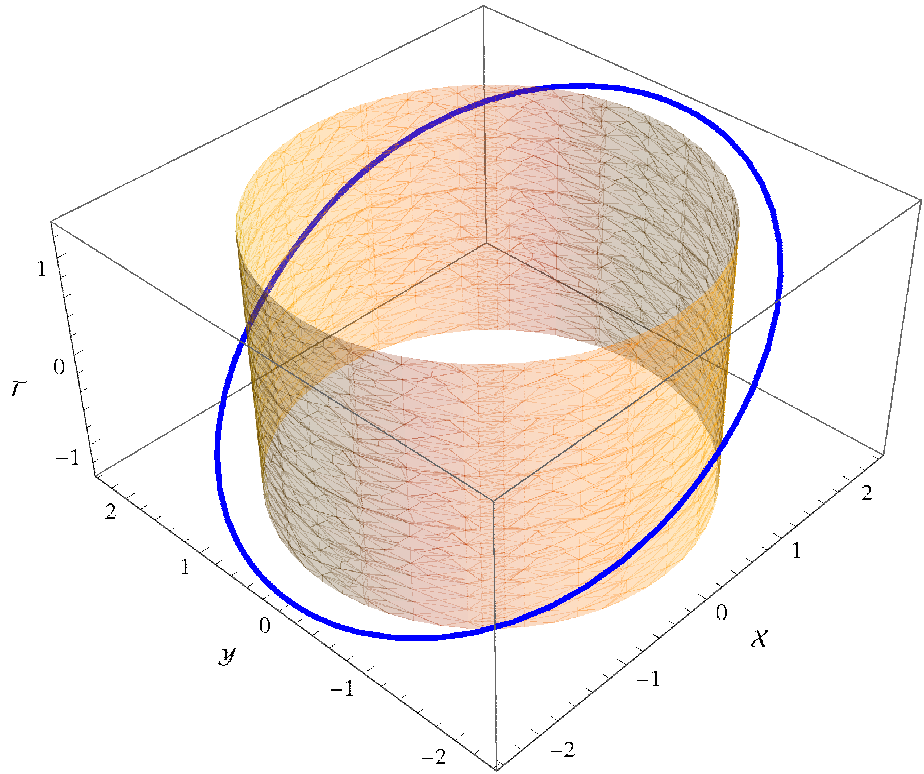}
\psfrag{x}{$\overline{x}$}
\psfrag{y}{$\overline{y}$}
\psfrag{t}{$\overline{t}$}
\caption{Parametic plot of the solution $Y_{1}(\cos{\theta})$ (in blue) in coordinates $\mathcal{T}\equiv M_{Kr}\ln{\frac{\hat{v}}{\hat{v}}}$,
$\mathcal{X}=r \cos{\theta}$ and
$\mathcal{Y}=r \sin{\theta}$ where $\theta$ has been allowed to vary between $0$ and $2\pi$, $M_{Kr}=1$ and $\epsilon=0.5$.
The figure also shows the set $\{r=2M\}$ (in gold) in these coordinates. Note that the solution lies entirely outside the region $\{r\leq 2M\}$ (i.e. the region inside the cylinder).}
\label{fig:GAHKr}
\end{center}
\end{figure}

We will denote by $\hat{S}_{\epsilon}$ the surface defined by this solution.
The proposition above implies that we can expand $\y(x, \epsilon)=
Y_{1}(x)\epsilon + o(\epsilon)$.
From
(\ref{explicitf}) it follows
that $Y_{1}$ satisfies the linear equation $L[Y_{1}(x)]= 3 |x|$.
Decomposing $Y_{1}(x)$ into Legendre polynomials $P_{l}(x)$,
as $Y_{1}(x)=\sum_{l=0}^{\infty} a_{l}P_{l}(x)$, where convergence is in $L^{2}[-1,1]$, this equation reads
\[
L[Y_{1}(x)]=\sum_{l=0}^{\infty} a_{l}L[P_{l}(x)]=3|x|.
\]
The Legendre equation, $-(1-x^{2})\ddot{P}_{l}(x)+2x\dot{P}_{l}(x)-l(l+1)P_{l}(x)=0$,
implies that $L[P_{l}(x)]=(l(l+1)+1)P_{l}(x)$.
We can also decompose $|x|$ in terms of Legendre polynomials. This computation can be found in
\cite{Bravo} and gives
\begin{equation}
|x|=\frac12 + \sum_{l=1}^{\infty} b_{2l}P_{2l}(x),\nonumber
\end{equation}
where
\begin{equation}
b_{2l}=\frac{(4l+1)(-1)^{l+1}}{2^{2l}} \frac{(2l-2)!}{(l-1)!(l+1)!}, \qquad l\geq 1.\nonumber
\end{equation}
It follows that the unique solution to the equation $L[Y_{1}(x)]=3|x|$ is
\begin{eqnarray}\label{Y1}
\hspace{-1cm} Y_1(x)  =   \frac32 + \sum_{l=1}^{\infty}a_{2l}P_{2l}(x),
\end{eqnarray}
with
\begin{eqnarray}\label{a2n}
a_{2l}   =  \frac{3(4l+1) (-1)^{l+1}}{\left[ 2l(2l+1)+1 \right]2^{2l}}\frac{(2l-2)!}{(l-1)!(l+1)!}, \qquad l\geq 1
\end{eqnarray}
(see Figure \ref{fig:GAHKr}).

\subsection{Area of the outermost generalized trapped horizon}

In this subsection
we will compute the area of $\hat{S}_{\epsilon}$, to second order in
$\epsilon$, and we will obtain that it is greater than $16\pi M_{Kr}^{2}$.
Then, we will prove that any generalized apparent horizon enclosing
$\hat{S}_{\epsilon}$ has greater or equal area than $\hat{S}_{\epsilon}$ which
will complete the proof of Theorem \ref{theorem}.



Integrating the volume element of $\hat{S}_{\epsilon}$, it is straightforward to get
\begin{eqnarray}\label{area0}
\hspace{-1cm}|\hat{S}_{\epsilon}|&=&\int_{-1}^{1}\int_{0}^{2\pi} r^{2}\sqrt{1+\epsilon^{2} \frac{32M_{Kr}^{3}}{r^{3}}e^{-r/2M_{Kr}}
(1-x^2)(\dot{Y}_{1}^{2}-1)+O(\epsilon^{3})}d\phi dx\nonumber \\
\hspace{-1cm} &=&\int_{-1}^{1}\int_{0}^{2\pi}\left[ r^{2}+\epsilon^{2}\frac{16M_{Kr}^{3}}{r}e^{-r/2M_{Kr}}(1-x^2)(\dot{Y}_{1}^{2}-1)+
O(\epsilon^{3}) \right]d\phi dx,\nonumber
\end{eqnarray}
where $r$ still depends on $\epsilon$.
Let us expand $r=r_{0}+r_{1}\epsilon+r_{2}\epsilon^{2}+O(\epsilon^{3})$.
Using equation (\ref{uvrembedded}) and expanding the exponential therein, it follows
\begin{equation}\label{r(epsilon)}
r=2M_{Kr}+\frac{2M_{Kr}}{e}(Y_{1}^{2}-x^{2})\epsilon^{2}+O(\epsilon^{3}).
\end{equation}
Then, after inserting (\ref{Y1}), (\ref{a2n}) and (\ref{r(epsilon)})
into the integral and using the orthogonality properties of the Legendre polynomials, we find
\begin{equation*}
|\hat{S}_{\epsilon}| = 16 \pi M_{Kr}^2 +  \frac{8  \pi M_{Kr}^2 \epsilon^2}{e} \left (
5 + 4 \sum_{l=1}^{\infty}  \frac{2l(2l+1)+1}{4l+1}a_{2l}^2 \right) + O (\epsilon^3).
\end{equation*}
Since the second term is strictly positive, it follows that $|\hat{S}_{\epsilon}| > 16 \pi M_{Kr}^2$.
This is not yet a counterexample of (\ref{penroseBK2}) because $\hat{S}_{\epsilon}$ is not known to
be the outermost generalized apparent horizon.
Before turning into this point, however,
let us give an alternative argument to show that the area increases. This will shed some light into
the underlying reason why the area of $\hat{S}_{\epsilon}$ is larger than $16 \pi M_{Kr}^2$.

To that aim, let us now use coordinates $\{\hat{u},x,\phi\}$ in $\Sigma_{\epsilon}$.
Then, the embedding of $\Sigma_{\epsilon}$ becomes
$\Sigma_{\epsilon}\equiv\left\{ \hat{u},\hat{v}=\hat{u}+2 \epsilon x,x,\phi \right\}$, and the
corresponding embedding
in $\Sigma_{\epsilon}$ for the surfaces $\hat{S}_{\epsilon}$
is $\hat{S}_{\epsilon}=\left\{ \hat{u} = u(x,\epsilon),x,\phi \right\}$. Again,
$u$ admits an expansion $u = U_1(x)  \epsilon + o(\epsilon)$. The
relationship between $U_1$ and $Y_1$ is simply $Y_{1}=U_{1}+x$. It follows
that $U_1$ satisfies $L[U_{1}(x)]= 3(|x|- x)$.
Similarly, if we take $\left\{ \hat{v},x,\phi \right\}$ as coordinates for $\Sigma_{\epsilon}$,
then  the embedding of $\hat{S}_{\epsilon}$ reads $\hat{v} = V_1 (x) \epsilon + o (\epsilon)$, with
$V_1$ satisfying $Y_{1}=V_{1}-x$ and therefore
$L[V_1(x)]= 3 (|x|+ x)$.
Thus,  $L[U_{1}(x)]\geq 0$ and $L[V_1(x)] \geq 0 $ and neither of them is identically zero.
Since $L$ is an elliptic operator with positive zero order term,
we can use the maximum principle 
to conclude that $U_1(x)>0$ and $V_{1}(x)>0$ everywhere. Geometrically, this means
that $\hat{S}_{\epsilon}$ lies fully in $\Sigma^{+}_{\epsilon}$
for $\epsilon$ small enough (c.f. Figure \ref{fig:GAHKr}). In fact, the maximum principle applied to $L[Y_1] = 3 |x|$
also implies $Y_1 > 0$. This will be used below.

We can now view $\hat{S}_{\epsilon}$ as a first order spacetime variation of
the bifurcation surface $\hat{S}_{\epsilon=0}$. The variation vector $\partial_{\epsilon}$ is defined
as the tangent vector to the curve generated when a point with fixed coordinates $\{x,\phi\}$ in
$\hat{S}_{\epsilon}$ moves
as $\epsilon$ varies. This vector satisfies
$\partial_{\epsilon}=U_{1}\partial_{\hat{u}}+V_{1}\partial_{\hat{v}}+O(\epsilon)$ and
is spacelike everywhere on the unperturbed
surface $\hat{S}_{\epsilon=0}$. If we do a Taylor expansion of $|\hat{S}_{\epsilon}|$ around $\epsilon=0$,
we see that the zero order term is $|\hat{S}_{\epsilon=0}| = 16 \pi M_{Kr}^2$, as this is the
area of the bifurcation surface.
The bifurcation surface is totally geodesic so that, in particular, its
mean curvature vector vanishes.
Consequently, the linear term in the expansion is identically zero as a consequence
of the first variation of area (\ref{firstvariation}). For any
$\epsilon \geq 0$ we have
\begin{eqnarray}\label{firstvariationarea}
&&\hspace*{-1.6cm}\frac{d |\hat{S}_{\epsilon}|}{d \epsilon} = \int_{\hat{S}_{\epsilon}} ( \vec{H}_{\hat{S}_{\epsilon}}, \partial_{\epsilon} )
\bm{\eta_{\hat{S}_{\epsilon}}}\nonumber\\
&&\hspace*{-6mm} = \int_{\hat{S}_{\epsilon}} \left( -\frac 12 \left[(p+q)\vec{l}_{-}+(-p+q)\vec{l}_{+}\right], U_{1}\partial_{\hat{u}}
+V_{1}\partial_{\hat{v}}+O(\epsilon)\right)\bm{\eta_{\hat{S}_{\epsilon}}}
\end{eqnarray}
where $\vec{H}_{\hat{S}_{\epsilon}}$ is the spacetime mean curvature vector of $\hat{S}_{\epsilon}$, $( \, , \, )$ denotes
the scalar product with the spacetime metric, and $\vec{l}_{+}$ and $\vec{l}_{-}$
are the outer and the inner null vectors which are future directed and satisfy $(\vec{l}_{+},\vec{l}_{-})=-2$.
Since on $\hat{S}_{\epsilon=0}$ the vectors $\partial_{\hat{v}}$ and $-\partial_{\hat{u}}$ are proportional to
$\vec{l}_{+}$ and $\vec{l}_{-}$, we have
\begin{eqnarray*}
\vec{l}_{+}\big|_{\hat{S}_{\epsilon}}&=&\sqrt{\frac{e}{8M_{Kr}^{2}}}\partial_{\hat{v}}+O(\epsilon),\\
\vec{l}_{-}\big|_{\hat{S}_{\epsilon}}&=&\sqrt{\frac{e}{8M_{Kr}^{2}}}(-\partial_{\hat{u}})+O(\epsilon),
\end{eqnarray*}
where the factor $\sqrt{\frac{e}{8M_{Kr}^{2}}}$ is due to the normalization $(l_{+},l_{-})=-2$.
Besides, $\bm{{\eta}_{\hat{S}_{\epsilon}}}=4M_{Kr}^{2}dx\wedge d\phi+O(\epsilon)$.
Then, inserting these expressions into the first variation integral
(\ref{firstvariationarea})
and taking the derivative
with respect to $\epsilon$ at $\epsilon=0$, we obtain 
\begin{eqnarray*}
\left. \frac{d^2 |\hat{S}_{\epsilon}|}{d\epsilon^2} \right|_{\epsilon=0}= \frac{16\sqrt{2}\pi M_{Kr}^2}{e}\int _{-1}^1 \left[
\frac{}{} U_1(x)L[V_1(x)]+V_1(x)L[U_1(x)]\right] dx,
\end{eqnarray*}
where (\ref{explicitp}), (\ref{explicitq})
and the relations $Y_{1}=U_{1}+x$ and $Y_1=V_{1}-x$ has been used.
Since $U_1$ and $V_1$ are strictly positive and
$L[U_{1}(x)]$, $L[V_1(x)]$ are non-negative and not identically zero, it follows
$\left. \frac{d^2 |\hat{S}_{\epsilon}|}{d\epsilon^2} \right|_{\epsilon=0}>0$ and hence that the area of $\hat{S}_{\epsilon}$
is larger than $16 \pi M_{Kr}^2$ for small $\epsilon$.

We have obtained that the second order variation of area
turns out to be strictly positive along the direction joining the bifurcation surface with $\hat{S}_{\epsilon}$,
which is tied to the fact that $L[U_1]$ and $L[V_1]$ have a sign.
The right hand sides of these operators are (except for a constant) the linearization of $|q| \pm q$
and these objects are obviously non-negative
in all cases. We conclude, therefore, that the fact that the area of $\hat{S}_{\epsilon}$ is larger than $16\pi M_{Kr}^2$
is closely related to the defining equation $p = |q|$.
It follows that the increase of area is a robust property
which does not depend strongly on the choice of hypersurfaces $\Sigma_{\epsilon}$ that we have made.
In fact, had we chosen hypersurfaces
$\Sigma_{\epsilon}\equiv \left\{ u=y- \epsilon \beta(x) , v=y+ \epsilon \beta(x)
, \cos \theta = x, \phi = \phi \right\}$,
the corresponding equations would have been
$L[U_{1}(x)]=|L[\beta(x)]| - L[\beta(x)] $ and
$L[V_{1}(x)]=|L[\beta(x)]| + L[\beta(x)]$. The same conclusions would follow
provided the right hand sides are not identically zero.

Having shown that $|\hat{S}_{\epsilon}| > 16 \pi M_{Kr}^2$ for $\epsilon \neq 0$
small enough, the next step is to analyze whether $|\hat{S}_{\epsilon}|$ is
a lower bound for the area of
the
outermost generalized apparent horizon. Indeed, in order to have a counterexample
of (\ref{penroseBK2}) we only need
to make sure that no generalized apparent horizon with less area than
 $\hat{S}_{\epsilon}$
and enclosing $\hat{S}_{\epsilon}$ exists in $\Sigma_{\epsilon}$.

We will argue by contradiction.
Let ${S}_{\epsilon}'$ be a generalized
apparent horizon enclosing $\hat{S}_{\epsilon}$ and
with $|{S}_{\epsilon}' | < |\hat{S}_{\epsilon}|$. In these circumstances, $\hat{S}_{\epsilon}$ cannot be area outer minimizing.
Thus, its minimal area enclosure $\hat{S}_{\epsilon}'$ does not coincide with it. Now, two
possibilities arise: (i) either $\hat{S}_{\epsilon}'$ lies completely outside $\hat{S}_{\epsilon}$, or (ii)
it coincides with $\hat{S}_{\epsilon}$ on a closed subset $\mathcal{K}$, while the complement
$\hat{S}_{\epsilon}' \setminus \mathcal{K}$ (which is non-empty) has vanishing mean
curvature $p$ everywhere.

To exclude case (i), consider the foliation of $\Sigma_{\epsilon}$ defined
by the surfaces $\{\ycoor=y_0,x,\phi\}$, where $y_0$ is a constant.
We then compute the mean curvature
$p_{y_{0}}$ of these surfaces.
The induced metric is
\[
\gamma^{y_{0}}_{AB}=\left( \frac{r^{2}}{1-x^2}-\epsilon^{2}\frac{32M_{Kr}^{3}}{r}e^{-r/2M_{Kr}}\right) dx^{2}
+(1-x^{2})r^{2}d\phi^{2}.
\]
The tangent vectors and the unit normal one-form are
\begin{eqnarray*}
\vec{e}_{x}=\partial_{x}, \quad
\vec{e}_{\phi}=\partial_{\phi}, \quad
\bold{m}=A d\ycoor,
\end{eqnarray*}
where $A=\sqrt{\frac{32M_{Kr}^{3}}{r}e^{-r/2M_{Kr}}}$ is the normalization factor.
Since $\gamma^{\ycoor_{0}}$ is diagonal we just need the following derivatives
\begin{eqnarray*}
\nabla^{\Sigma_{\epsilon}}_{\vec{e}_{x}}e_{x}^{\ycoor}&=&-
    \frac{r^{3}+8\epsilon^{2}M_{Kr}^{2}(2M_{Kr}+r)(1-x^{2})e^{-r/2M_{Kr}}}{4M_{Kr}(1-x^{2})r^2}y_{0}\\
\nabla^{\Sigma_{\epsilon}}_{\vec{e}_{\phi}}e_{\phi}^{\ycoor}&=&-\frac{(1-x^{2})r}{4M_{Kr}}y_{0}.
\end{eqnarray*}
Inserting all these expressions in
$p_{y_{0}}=-m_{i}\gamma^{AB}\nabla^{\Sigma_{\epsilon}}_{\vec{e}_{A}}e_{B}^{i}$
we obtain
\begin{eqnarray*}
p_{y_{0}}=A\left(
        \frac{r^{3}+8\epsilon^{2} M_{Kr}^2(2M_{Kr}+r)(1-x^2)e^{-r/2M_{Kr}}}{4M_{Kr}r\left(r^{3}-32\epsilon^{2}M_{Kr}^{3}(1-x^{2})e^{-r/2M_{Kr}}\right)}+
    \frac{1}{4M_{Kr}r} \right)
    y_{0}.
\end{eqnarray*}
Thus, taking $-1<\epsilon<1$ small enough so that
\begin{equation*}\label{epsilon}
\epsilon^{2}<\frac{r^{3}_{\text{min}}e^{r_{\text{min}}/2M_{Kr}}}{32M_{Kr}^{3}},
\end{equation*}
where $r_{\text{min}}$ is the minimum value of $r$ in $\Sigma_{\epsilon}$ (recall that $r_{\text{min}}>0$ provided
$|\epsilon|<1$), we can assert that $p_{y_{0}}>0$ for all
$y_{0}>0$.

\begin{figure}
\begin{center}
\psfrag{S}{$\hat{S}_{\epsilon}$}
\psfrag{S2}{\color{red}{$\hat{S}_{\epsilon}'$}}
\psfrag{y}{\color{blue}{$\{\ycoor=y_{\epsilon}\}$}}
\includegraphics[width=5cm]{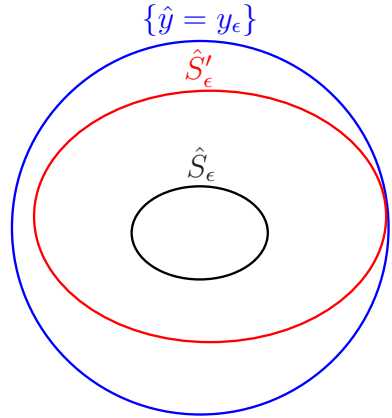}
\caption {If the minimal area enclosure $\hat{S}_{\epsilon}'$ (in red) lies completely outside $\hat{S}_{\epsilon}$ then $\hat{S}_{\epsilon}'$,
which is a minimal surface, must touch tangentially from the inside
a surface $\{\ycoor=y_{\epsilon}\}$ (in blue) which has $p_{y_{\epsilon}}>0$.}
\label{fig:BK1}
\end{center}

\end{figure}

We noted above that $Y_1(x) >0$
everywhere. Thus, for small enough positive $\epsilon$, the function $y(x,
\epsilon)$  is also strictly
positive. Since $\hat{S}_{\epsilon}'$ lies fully outside $\hat{S}_{\epsilon}$, the coordinate function $\ycoor$ restricted to
$\hat{S}_{\epsilon}'$ achieves a positive maximum $y_{\epsilon}$ somewhere. At this point, the two
surfaces $\hat{S}_{\epsilon}'$ and $\{\ycoor = y_{\epsilon}\}$ meet tangentially, with $\hat{S}_{\epsilon}'$ lying fully inside
$\{ \ycoor = y_{\epsilon} \}$ (see Figure \ref{fig:BK1}).
This is a contradiction to the maximum principle for minimal
surfaces (see Proposition \ref{maximumprincipleforMOTS} with $K=0$ in Appendix \ref{ch:appendix2}).

\begin{figure}
\begin{center}
\psfrag{S}{$\hat{S}_{\epsilon}$}
\psfrag{S2}{\color{red}{$\hat{S}_{\epsilon}'$}}
\psfrag{y+}{\color{blue}{$\{\ycoor=y_{\max}\}$}}
\psfrag{y-}{\color{blue}{$\{\ycoor=y_{\min}\}$}}
\includegraphics[width=6cm]{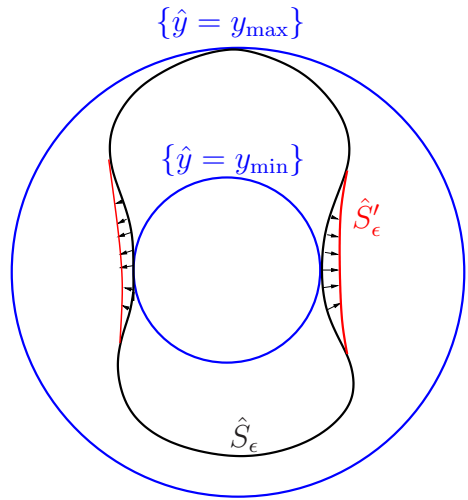}
\caption {In the case (ii), the minimal area enclosure $\hat{S}_{\epsilon}'$ coincides with $\hat{S}_{\epsilon}$ in a compact set. The coordinate function
$\ycoor$ restricted to $\hat{S}_{\epsilon}'$ cannot achieve a local maximum in the set where $\hat{S}_{\epsilon}'$ and $\hat{S}_{\epsilon}$ do not coincide (in red). Then,
this set can be viewed as an outward variation of order $\epsilon$ of the corresponding points in $\hat{S}_{\epsilon}$.}
\label{fig:BK2}
\end{center}
\end{figure}

It only remains to deal with case (ii). The same argument above shows that the coordinate function
$\ycoor$ restricted to $\hat{S}_{\epsilon}' \setminus \mathcal{K}$ cannot reach a local maximum. It follows that
the range of variation of $\ycoor$ restricted to $\hat{S}_{\epsilon}'$ is contained in the range of variation of
$\ycoor$ restricted to $\hat{S}_{\epsilon}$ (see Figure \ref{fig:BK2}).

Since $\max_{\hat{S}_{\epsilon}} \ycoor -
\min_{\hat{S}_{\epsilon}} \ycoor = O (\epsilon)$,
it follows that we can regard $\hat{S}_{\epsilon}'$ as an outward variation of $\hat{S}_{\epsilon}$ of order $\epsilon$ when
$\epsilon$ is taken small enough. The corresponding
variation vector field $\vec{\nu}$ can be taken orthogonal
to $\hat{S}_{\epsilon}$ without loss of generality, i.e.
$\vec{\nu}=\nu \vec{m}$, where $\vec{m}$ is the outward unit normal to $\hat{S}_{\epsilon}$.
The function $\nu$ vanishes on $\mathcal{K}$ and is positive in its complement
$U \equiv \hat{S}_{\epsilon} \setminus \mathcal{K}$. Expanding to second
order and using the first and second variation of area
(see e.g. \cite{Chavel}) gives
\begin{eqnarray*}
|\hat{S}_{\epsilon}' |&=&  |\hat{S}_{\epsilon}|  +
\epsilon \int_{U}   p_{\hat{S}_{\epsilon}}\nu
\bm{\eta_{\hat{S}_{\epsilon}}}  \\
&&  +  \frac{\epsilon^2}{2}
\int_{U}
\left ( |\nabla_{\hat{S}_{\epsilon}}\nu|^2 +
\frac{\nu^2}{2}
\left( R^{\hat{S}_{\epsilon}}-R^{\Sigma_{\epsilon}}-|\kappa_{\hat{S}_{\epsilon}}|^{2} + p_{\hat{S}_{\epsilon}}^2
\right )  + p_{\hat{S}_{\epsilon}} \frac{d\nu}{d\epsilon}  \right )
\bm{\eta_{\hat{S}_{\epsilon}}} + O (\epsilon^3),
\end{eqnarray*}
where $\nabla_{\hat{S}_{\epsilon}}$, $R^{\hat{S}_{\epsilon}}$ and $\kappa_{\hat{S}_{\epsilon}}$ are, respectively,
the gradient, scalar curvature and second fundamental form of $\hat{S}_{\epsilon}$, and $R^{\Sigma_{\epsilon}}$
is the scalar curvature of $\Sigma_{\epsilon}$. Now, the
mean curvature $p_{\hat{S}_{\epsilon}}$ of $\hat{S}_{\epsilon}$ reads $p_{\hat{S}_{\epsilon}} =\frac{3 \epsilon}{M_{Kr} \sqrt{e}}  |x| + o (\epsilon)$ (see equation (\ref{explicitp}))
and both $R^{\Sigma_{\epsilon}} $ and $\kappa_{\hat{S}_{\epsilon}}$ are of order $\epsilon$ (because $\Sigma_{\epsilon=0}$
has vanishing scalar curvature and $\hat{S}_{\epsilon=0}$ is totally geodesic).
Moreover,
$R^{\hat{S}_{\epsilon}} = 1/(2M_{Kr}^2) + O(\epsilon)$. Thus,
\begin{eqnarray*}
|\hat{S}_{\epsilon}' |= |\hat{S}_{\epsilon}| + \epsilon^2 \left\{ \int_{U}
\left [ \frac{3|x|\nu}{M_{Kr}\sqrt{e}}+ \left(\frac{|\nabla_{\hat{S}_{\epsilon}}\nu|^2}{2}+\frac{\nu^2}{8M_{Kr}^2} \right) \right ] \bm{\eta_{\hat{S}_{\epsilon}}} \right\}+ O(\epsilon^3).
\end{eqnarray*}
It follows that, for small enough $\epsilon$, the area of $\hat{S}_{\epsilon}'$ is
larger than $\hat{S}_{\epsilon}$ contrarily to our assumption. This proves
Theorem \ref{theorem} and, therefore, the existence of
counterexamples to the  version (\ref{penroseBK2}) of the
Penrose inequality.

It is important to remark that the existence of this counterexample does not
invalidate the approach suggested by Bray and Khuri 
to study the general Penrose inequality. It means, however, that the emphasis should not be put on
generalized apparent horizons. It may be that the approach can serve to prove the standard
version (\ref{penrose1}) as recently discussed in \cite{BK2}.

\newpage
\thispagestyle{empty}
\mbox{}
\newpage

\chapter{Conclusions}

In this thesis we have studied some questions within the framework of the theory of General Relativity.
In particular, we have concentrated on some of the properties of marginally outer trapped surfaces (MOTS)
and weakly outer trapped surfaces
in spacetimes with symmetries, specially static isometries, and its application
to the uniqueness theorems of black holes and the Penrose inequality.
We can summarize the main results of this thesis in the following list.\\

\begin{enumerate}
\item We have obtained a general expression for the first variation of the outer null expansion $\theta^+$
of a surface $S$ along an arbitrary vector field $\vec{\xi}$ in terms of the deformation tensor of the spacetime metric
associated with the vector $\vec{\xi}$. This expression has been particularized when $S$ is a MOTS.

\item Starting from a geometrical idea that generates a family of surfaces by moving first along $\vec{\xi}$ and then along
null geodesics, 
we have used the theory of linear elliptic second order operators to obtain restrictions on any
vector field on stable and strictly stable MOTS. Using the expression mentioned in the previous point,
these results have been particularized
to generators of symmetries of physical interest, such as Killing vectors, homotheties and
conformal Killing vectors. As an application we have shown that there exists no stable MOTS
in any spacelike hypersurface of a large class of Friedmann-Lema\^itre-Robertson-Walker cosmological models,
which includes all classic models of matter and radiation dominated eras and those models with accelerated
expansion which satisfy the null energy condition (NEC).

\item For the situations when the elliptic theory is not useful,
we have exploited
the geometrical idea mentioned before to obtain similar restrictions for Killing vectors and homotheties
on outermost and locally outermost MOTS. As a consequence of these results, 
we have shown that, on a spacelike hypersurface possessing an untrapped barrier $S_{b}$,
a Killing vector or a homothety $\vec{\xi}$ cannot
be timelike anywhere on a bounding weakly outer trapped surface
whose exterior lies in the region where $\vec{\xi}$ is timelike, 
provided the NEC holds in the spacetime.

For the more general cases when the elliptic theory simply cannot be applied, a suitable variation of the geometrical idea
has allowed us to obtain weaker restrictions on
any vector field $\vec{\xi}$ on locally outermost MOTS.
This results have also been particularized to
Killing vectors, homotheties and conformal Killing vectors.

\item Analyzing the Killing form in a static Killing initial data (KID) $\kid$
we have shown, at the initial data level, that the topological
boundary of each connected component $\{\lambda>0\}_{0}$ of the region where the Killing vector is timelike is
a smooth injectively immersed submanifold with $\theta^{+}=0$ with respect to the outer normal which points into $\{\lambda>0\}_{0}$, provided
\begin{itemize}
\item[(i)] $N Y^i \nablaSigma_i \lambda|_{\tbd\{\lambda>0\}_{0}} \geq 0$  if $\tbd\{\lambda>0\}_{0}$ contains at least one fixed
point.
\item[(ii)]
$N Y^i m_i |_{\tbd\{\lambda>0\}_{0}} \geq 0$  if $\tbd\{\lambda>0\}_{0}$ contains no fixed point,
where $\vec{m}$ is the unit normal pointing towards $\{\lambda>0\}_{0}$.
\end{itemize}
There are examples in the Kruskal spacetime where these conditions do not hold and $\tbd \{\lambda>0\}_{0}$ fails to be smooth and has $\theta^{+}\neq 0$.

\item
Under the same hypotheses as before
we have
proven a confinement result for MOTS in arbitrary spacetimes satisfying the NEC and for arbitrary spacelike hypersurfaces, not
necessarily time-symmetric. The hypersurfaces need not be asymptotically flat either and are only required to have an outer untrapped barrier
$S_b$.
This result, which also have been proved at the initial data level, asserts that no bounding weakly outer trapped surface can intersect $\ext$, where $\ext$ denotes the connected component
of $\{\lambda>0\}$ which contains $\Sb$.
A condition which ensures that all arc-connected components of $\tbd \{\lambda>0\}$ are topologically closed is required.
This condition is automatically fulfilled in spacetimes containing no non-embedded Killing prehorizons.

\item
We have proven that the set $\tbd \{\lambda>0\}$ in an embedded static KID is a union of smooth 
injectively immersed surfaces with at least one of the two null expansions equal to zero (provided
the topological  condition mentioned in the previous point is satisfied).

\item Using the previous result,
we have shown that, in a static embedded KID which satisfies the NEC and possesses an outer untrapped
barrier $S_{b}$ and a bounding weakly outer trapped surface, the set $\tbd \ext$ is the outermost bounding
MOTS provided that every arc-connected component of $\tbd \ext$ is topologically closed, the past weakly outer trapped region
$T^{-}$ is contained in the weakly outer trapped region $T^{+}$ and a topological condition which ensures that all
closed orientable surfaces separate the manifold.

\item With the previous result at hand, we have obtained
a uniqueness theorem for embedded static KID containing an asymptotically flat end which
satisfy the NEC and possess a bounding weakly outer trapped surface.
The matter model is arbitrary as long as it admits a static black hole uniqueness proof with the Bunting and Masood-ul-Alam doubling method.
This result extends a previous theorem by Miao valid on vacuum and time-symmetric slices,
and allows to conclude that, at least regarding uniqueness of black holes,
event horizons and MOTS do coincide in static spacetimes.
This result requires the same hypotheses as the result in the previous point.
As we have mentioned before, the condition on the arc-connected components of $\tbd \ext$
is closely related with the non-existence of non-embedded Killing prehorizons and
can be removed if a result on the non-existence of these type of
prehorizons is found. The condition $T^{-}\subset T^{+}$ is needed for out argument to work.
Trying to drop this hypotheses is a logical next step, but it would require a different method of proof.

\item Finally, we have proved that there exist slices in the Kruskal spacetime where the outermost
generalized apparent horizon has area greater than $16\pi M_{Kr}^{2}$, where $M_{Kr}$ is the mass of the Kruskal spacetime. This gives a counterexample of a Penrose inequality recently proposed by Bray and Khuri
(in terms of the area of the outermost apparent horizon) in
order to address the general proof of the standard Penrose inequality.
The existence of this counterexample
does not invalidate the approach of these authors but indicate that the emphasis must not be on
generalized apparent horizons.

\end{enumerate}

\newpage
\thispagestyle{empty}
\mbox{}
\newpage

\newcounter{ape}
\setcounter{ape}{1}
\renewcommand{\theequation}{\Alph{ape}.\arabic{equation}}
\renewcommand{\thechapter}{\Alph{ape}}

\newtheorem{thra}{Theorem}[chapter]
\newtheorem{defia}[thra]{Definition}
\newtheorem{lemaa}[thra]{Lemma}
\newtheorem{corollarya}[thra]{Corollary}
\newtheorem{propositiona}[thra]{Proposition}
\appendix
\addappheadtotoc


\chapter{Differential manifolds}
\label{ch:appendix1}

In this Appendix, we will give a definition of a differentiable manifold which allows us to consider manifolds with and without boundary at the same time. We follow \cite{Hirsch}.\\

Consider the vector space $\mathbb{R}^{n}$ and let $\bf \omega_{\alpha}$ be a one-form defined on this vector space (the
index $\alpha$ is simply a label at this point).
Let us define the set $H_{\alpha}=\{ \vec{r}\in\mathbb{R}^{n}: {\bf \omega}_{\alpha}(\vec{r}\,)\geq 0 \}$,
which is either a half plane if $\omega_{\alpha}\neq 0$ or the whole space if $\omega_{\alpha}=0$.
The concept of differentiable manifold may be defined as follows.
\begin{defia}\label{defi:manifold}
A {\bf differentiable manifold} is a topological space $M$ together with a collection of open sets $U_{\alpha}\subset M$ such
that:
\begin{enumerate}
\item The collection $\{U_{\alpha}\}$ is an open cover of $M$, i.e. $M=\underset{\alpha}{\bigcup}U_{\alpha}$. 
\item For each $\alpha$ there is a bijective map $\varphi_{\alpha}: U_{\alpha}\rightarrow V_{\alpha}$, where $V_{\alpha}$ is an open subset
of $H_{\alpha}$ with the induced topology of $\mathbb{R}^{n}$.
Every set $(U_{\alpha},\varphi_{\alpha})$ is called a {\it chart} or a {\it local coordinate system}. The collection $\{ (U_{\alpha},\varphi_{\alpha}) \}$ is called an {\it atlas}.
\item Consider two sets $U_{\alpha}$ and $U_{\beta}$ which overlap, i.e. $U_{\alpha}\cap U_{\beta}\neq \emptyset$, and consider the map
    $\varphi_{\beta}\circ\varphi_{\alpha}^{-1}: \varphi_{\alpha}(U_{\alpha}\cap U_{\beta})\rightarrow \varphi_{\beta}(U_{\alpha}\cap U_{\beta})$. Then, there exists a map $\varphi_{\alpha\beta}: W_{\alpha}\rightarrow W_{\beta}$, where $W_{\alpha}$ and $W_{\beta}$ are open subsets of $\mathbb{R}^{n}$ which, respectively, contain
    $\varphi_{\alpha}(U_{\alpha}\cap U_{\beta})$ and $\varphi_{\beta}(U_{\alpha}\cap U_{\beta})$
    such that $\varphi_{\alpha\beta}$ is a differentiable bijection, with differentiable inverse and satisfying
    $\left.\varphi_{\alpha\beta}\right|_{\varphi_{\alpha}(U_{\alpha}\cap U_{\beta})}=\varphi_{\beta}\circ \varphi_{\alpha}^{-1}$.
\end{enumerate}
\end{defia}

{\bf Remark.} Since no confusion arises, we will denote a differential manifold $(M,\{U_{\alpha}\})$
simply by $M$. Note that manifolds need not be connected according to this definition. $\hfill \square$

\begin{defia}\label{defi:smoothmanifold}
A differentiable manifold $M$ is {\bf of class $C^{k}$} if the mappings $\varphi_{\alpha\beta}$ and their inverses are $C^{k}$.\\
A differentiable manifold $M$ is {\bf smooth} (or $C^{\infty}$) if it is $C^{k}$ for all $k\in\mathbb{N}$.
\end{defia}

\begin{defia}\label{defi:manifoldwithboundary}
$M$ is a {\bf differentiable manifold with boundary} if for at least one chart $U_{\alpha}$, we have 
$\omega_{\alpha}\neq 0$.
In this case, the {\bf boundary} of $M$ is defined as $\bd M=\underset{\alpha ,\omega_{\alpha}\neq 0}{\bigcup} \{\p\in U_{\alpha} \text{ such that } \omega_{\alpha}\left( \varphi_{\alpha}(\p) \right)=0\}$
\end{defia}

{\bf Remark.} Along this thesis the sign $\bd$ will denote the boundary of a manifold while the sign $\tbd$ will refer to the
{\it topological}
boundary of any subset of a topological space 
(both concepts are in general completely different). $\hfill \square$

\begin{defia}
$M$ is a {\bf differentiable manifold without boundary} if
$\omega_{\alpha}=0$ for all $\alpha$.
\end{defia}

It can be proven that $\bd M$ is a differentiable manifold without boundary.

\begin{defia}
The interior $\text{int}({M})$ of a manifold $M$ is defined as $\text{int}({M})=M\setminus \bd M$.
\end{defia}

We will denote by $\overline{U}$ the topological closure of a set $U$ and by $\overset{\circ}{U}$ its topological
interior.


\begin{defia}\label{defi:orientablemanifold}
A differentiable manifold, with or without boundary, is {\bf orientable} if there exists an atlas such that for any two charts $(U_{\alpha},\varphi_{\alpha})$
and $(U_{\beta},\varphi_{\beta})$ which overlap, i.e. $U_{\alpha}\cap U_{\beta}\neq 0$, the Jacobian of
$\left.\varphi_{\alpha\beta}\right|_{U_{\alpha}\cap U_{\beta}}$
on $U_{\alpha}\cap U_{\beta}$ is positive. Such an atlas will be called {\bf oriented atlas}\\
A differentiable manifold with an oriented atlas is said to be {\bf oriented}.
\end{defia}

\begin{defia}
Consider an oriented manifold $M$ endowed with a metric $\gN$.
The {\bf volume element} ${\bf \eta}^{(n)}$ of $(M,\gN)$ is the $n$-form
$\eta^{(n)}_{\alpha_{1}...\alpha_{n}}=\sqrt{|\text{det } \gN|} \epsilon_{\alpha_{1}...\alpha_{n}}$ in
any coordinate chart of the oriented atlas. Here, $\epsilon_{\alpha_{1}...\alpha_{n}}$ is
the totally antisymmetric symbol and $\text{det } \gN$ is the determinant of $\gN$ in this chart.
\end{defia}

All manifolds in thesis are assumed to be Hausdorff and paracompact.
These concepts are defined as follows.

\begin{defia}
A topological space $M$ is {\bf Hausdorff} if for each pair of points $\p,\q$ with $\p\neq \q$, there exist two disjoint
open sets $U_{\p}$
and $U_{\q}$ such that $\p\in U_{\p}$ and $\q\in U_{\q}$. 
\end{defia}

\begin{defia}
Let $M$ be a topological space and let $\{U_{\alpha}\}$ be an open cover of $M$. An open cover $ \{ V_{\beta}\}$ is said to be a
{\it refinement} of $\{U_{\alpha}\}$ if for each $V_{\beta}$ there exists an $U_{\alpha}$ such that $V_{\beta}\subset U_{\alpha}$.
The cover $\{V_{\beta }\}$ is said to be {\it locally finite} if each $\p\in M$ has an open neighbourhood $W$ such that only
finitely many $V_{\beta}$ satisfy $W\cap V_{\beta}\neq \emptyset$.\\
The topological space $M$ is said to be {\bf paracompact} if every open cover $\{ U_{\alpha} \}$ of $M$ has a locally finite refinement $\{ V_{\beta} \}$.
\end{defia}

\newpage
\thispagestyle{empty}
\mbox{}
\newpage

\addtocounter{ape}{1}
\chapter{Elements of mathematical analysis}
\label{ch:appendix2}

This Appendix is devoted to introducing some elements of mathematical analysis which are used
throughout this thesis.

Firstly, recall that a Banach space is a normed vector space which is complete.
Let ${\cal X}$, ${\cal Y}$ be Banach spaces
with respective norms $|| \cdot ||_{\cal X}$ and $|| \cdot ||_{\cal Y}$.
Let  $U_{\cal X} \subset {\cal X}$,
$U_{\cal Y} \subset {\cal Y}$ be open sets. A function
$f : U_{\cal X} \rightarrow U_{\cal Y}$ is said to be Fr\'echet-differentiable
at $x \in U_{\cal X}$ if there exists
a linear bounded map $D_x f: {\cal X} \rightarrow {\cal Y}$ such that
\begin{eqnarray*}
\lim_{ h \rightarrow 0}
\frac{ ||f (x+h) - f(x) - D_xf (h) ||_{\cal Y}}{||h||_{\cal X}}=0.
\end{eqnarray*}
$f$ is said to be  $C^1$ if it is differentiable at every
point $x \in U_{\cal X}$ and  the map $Df: U_{\cal X} \rightarrow
L ({\cal X},{\cal Y})$ defined by $Df (x) = D_x f$ is continuous. Here $L({\cal X},{\cal Y})$
is the Banach space of linear bounded maps between ${\cal X}$ and
${\cal Y}$ with the operator norm.

A key tool in analysis is the {\it implicit function theorem}.
\begin{thra}[Implicit function theorem (e.g. \cite{ChoquetBruhat})]\label{thr:implicitfunction}
Let ${\cal X}$, ${\cal Y}$, ${\cal Z}$ be Banach spaces and
$U_{\cal X}$, $U_{\cal Y}$, $U_{\cal Z}$ respective open sets with $0 \in U_{\cal Z}$.
Let $f: U_{\cal X} \times U_{\cal Y} \rightarrow  U_{\cal Z}$ be $C^1$
with Fr\'echet-derivative $D_{(x,y)} f$.

Let  $x_0 \in U_{\cal X}$, $y_0 \in {\cal Y}$ satisfy $f(x_0,y_0)=0$ and assume
that the linear map
\begin{eqnarray*}
D_y f |_{(x_0,y_0)} : {\cal Y} & \rightarrow & {\cal Z}, \\
 \hat{y}  & \rightarrow & D_{(x_0,y_0)} f (0,\hat{y})
\end{eqnarray*}
is invertible, bounded and with bounded inverse. Then there exist open neighbourhoods
$x_0 \in {U}_{x_0} \subset U_{\cal X}$ and
$y_0 \in {U}_{y_0} \subset U_{\cal Y}$  and a $C^1$ map $g : U_{x_0}  \rightarrow U_{y_0}$
such that $f (x,g(x))=0$ and, moreover, $f(x,y)=0$ with $(x,y) \in U_{x_0} \times U_{y_0}$ implies $y =
g(x)$.
\end{thra}

In the context of partial differential equations, one important class of Banach spaces
are the H\"older spaces.

Let $\Omega \subset \mathbb{R}^n$ be a domain and $f : \overline{\Omega} \rightarrow \mathbb{R}$. Let $\beta =
(\beta_1, \cdots, \beta_n)$ be multi-index (i.e. $\beta_i \in \mathbb{N} \cup \{ 0 \}$ for all
$i \in \{ 1, \cdots n\}$) and define $|\beta| = \sum_{i=1}^{n} \beta_i$ .
Denote by $D^{\beta} f$ the partial derivative
$D^{\beta} f = \partial_{x_1^{\beta_1}} \cdots \partial_{x_n^{\beta_n}} f$ when this exists.
For $k \in  \mathbb{N} \cup \{0 \}$ we
denote by $C^k (\overline{\Omega})$  the set of functions $f$ with continuous derivatives $D^{\beta} f$
for all $\beta$ with $|\beta|\leq k$.

Let $0 < \alpha \leq  1$. The function $f$ is {\it H\"older continuous with exponent $\alpha$}
if
\begin{eqnarray*}
[f]_{\alpha} \equiv \sup_{\underset{x \neq y}{x,y\in \overline{\Omega}}}
\frac{|f(x) - f(y)|}{|x-y|^{\alpha}}
\end{eqnarray*}
is finite. When $\alpha =1$, the function is called Lipschitz continuous.

\begin{defia}
For $0 < \alpha \leq 1$ and $k\in \mathbb{N}\cup\{0\} $ the H\"older space
$C^{k,\alpha} (\overline{\Omega})$ is the Banach space of all functions $u \in C^k (\overline{\Omega})$
for which the norm
\begin{eqnarray*}
[f]_{k,\alpha} = \sum_{|\beta|=0}^{k} \sup_{\overline{\Omega}} |D^{\beta} f | + \max_{|\beta| = k} [D^{\beta} f]_{\alpha}
\end{eqnarray*}
is finite.
\end{defia}

The definition extends to Riemannian manifolds if we replace $|x - y |$ by  the distance
function $d(x,y)$ between two points.

The following result appearing in \cite{GilbargTrudinger} (pages 448-449 and problem
17.2) is useful when we apply the implicit function theorem in Chapter \ref{ch:Article3}.
\begin{thra}\label{thr:GT}
Let $\psi\in C^{2,\alpha}(\overline{\Omega})$ with $\Omega\subset \mathbb{R}$ a domain and
consider the maps
\[
F:C^{2,\alpha}(\overline{\Omega})\longrightarrow C^{0,\alpha}(\overline{\Omega})
\]
and
\[
\mathcal{F}:\Gamma=\overline{\Omega_{2}}\times \overline{\Omega}\longrightarrow \mathbb{R},
\]
where $\Omega_2\subset \mathbb{R}^{3}$ is a domain, which are related by
\[
F(\psi)(x)=\mathcal{F}(\ddot{\psi}(x),\dot{\psi}(x),\psi(x),x).
\]
Assume that $\mathcal{F}\in C^{2,\alpha}({\Gamma})$.
Then $F$ has continuous Fr\'echet derivative given by
\begin{eqnarray*}
D_{\psi}F(\varphi)&=&\partial_{1}\mathcal{F}\big|_{(\ddot{\psi}(x),\dot{\psi}(x),\psi(x),x)}\ddot{\varphi}(x)
+\partial_{2}\mathcal{F}\big|_{(\ddot{\psi}(x),\dot{\psi}(x),\psi(x),x)}\dot{\varphi}(x)\\
&&\quad +\partial_{3}\mathcal{F}\big|_{(\ddot{\psi}(x),\dot{\psi}(x),\psi(x),x)}\varphi(x).
\end{eqnarray*}
\end{thra}

Consider a manifold $S$ with metric $g$ and let $\nabla$ be the corresponding
covariant derivative. Let $a^{ij}$ be a symmetric tensor field , $b^i$ a
vector field and $c$ a scalar. Consider
a linear second order differential operator $L$ on the form
\begin{equation}\label{ellipticoperator}
L\psi=-a^{ij}(x) \nabla_{i} \nabla_{j}\psi+b^{i}(x) \nabla_{i}\psi+c(x)\psi,
\end{equation}

\begin{defia}
L is {\bf elliptic} at a point $x\in S$ if the matrix $[a^{ij}](x)$ is positive definite.
\end{defia}

Assume that $S$ is orientable and
denote by $<,>_{L^{2}}$ the $L^2$ inner product of two functions $\psi,\phi : S \rightarrow \mathbb{R}$ defined by
$<\psi,\phi>_{L^{2}}\equiv \int_{S}\psi\phi \bm{\eta}_S$, where $\bm{\eta}_S$ is the (metric) volume form on $S$.
Given a second order linear differential operator, the formal adjoint
$L^{\dagger}$ is the linear second order differential operator which satisfies
\[
<\psi,L^{\dagger}\phi>_{L^{2}}=<\phi,L\psi>_{L^{2}}.
\]
for all pairs of smooth functions with compact support.
A linear operator $L$ is {\it formally self-adjoint} with respect to the product $L^{2}$ if $L^{\dagger}=L$.

When acting on the H\"older space $C^{2,\alpha}(S)$ for $0 < \alpha <1$, the linear second order operator $L$
becomes a bounded linear operator $L: C^{2,\alpha} (S)  \rightarrow C^{0,\alpha}(S)$.
The formal adjoint is also a map $L^{\dagger}: C^{2,\alpha} (S)  \rightarrow C^{0,\alpha}(S)$. An {\it eigenvalue}
of $L$ is a number
$\mu \in  \mathbb{C}$ for which
there exist functions $u,v \in C^{2,\alpha} (S)$ such that $L [u]+ i L[v] = \mu \left ( u+ i v\ \right )$.
The complex function $u + i v$ is called an {\it eigenfunction}.

The following lemma concerns the existence and uniqueness of the {\it principal eigenvalue}
(i.e. the eigenvalue with smallest real part) of $L$ and $L^{\dagger}$. This result
is an adaptation of a standard result of elliptic theory to the case of compact connected manifolds without
boundary (see Appendix B of \cite{AMS}).
\begin{lemaa}
\label{PrincipleEigenvalue}
Let $L$ be a linear second order elliptic operator on a compact manifold $S$. Then
\begin{enumerate}
\item There is a real eigenvalue $\auto$, called the principal eigenvalue, such that for
     any other eigenvalue $\mu$ the inequality $\text{Re}(\mu)\geq \auto$ holds. The corresponding
     eigenfunction $\phi$, $L\phi=\auto\phi$ is unique up to a multiplicative constant and can
     be chosen to be real and everywhere positive.
\item The formal adjoint $L^{\dagger}$ (with respect to the $L^{2}$ inner product) has the same principal
     eigenvalue $\auto$ as $L$.
\end{enumerate}
\end{lemaa}

For formally self-adjoint operators, the principal eigenvalue $\auto$ satisfies
\begin{equation}\label{Rayleigh-Ritz}
\auto= \underset{\underset{\scriptscriptstyle{\psi\neq 0}}{\scriptscriptstyle{\psi\in C^{2,\alpha}(S^{2})}}}{\mbox{inf}} \frac{<\psi,L\psi>_{L^{2}}}{<\psi,\psi>_{L^{2}}},
\end{equation}
where the quotient $\frac{<\psi,L\psi>_{L^{2}}}{<\psi,\psi>_{L^{2}}}$ is called the Rayleigh-Ritz ratio of the function $\psi$. This formula,
which reflects the connection between the eigenvalue problems and the variational problems,
is also useful to obtain upper bounds for $\auto$.

An important tool in the analysis of the properties of the elliptic operator $L$ is the maximum principle.
The standard formulations of the maximum principle for elliptic operators requires
that the coefficient $c$ in (\ref{ellipticoperator}) is non-negative (see e.g. Section 3 of \cite{GilbargTrudinger}).
The following formulation of the maximum principle, which is more suitable for our purposes,
requires non-negativity of the principal eigenvalue. Its proof can be found in Section 4 of \cite{AMS}.

\begin{lemaa}\label{lemmaelliptic}
Consider a linear second order elliptic operator $L$ on a
compact manifold $\S$ with principal eigenvalue $\auto\geq 0$
and principal eigenfunction $\phi$ and let $\psi$ be a smooth function
satisfying $L \psi \geq 0$ ($L \psi \leq 0$).
\begin{enumerate}
\item If $\auto=0$, then $L \psi \equiv 0$ and $\psi=C\phi$ for some constant
$C$.
\item If $\auto>0$ and $L \psi \not\equiv 0$, then $\psi>0$ ($\psi<0$) all over
$\S$.
\item If $\auto>0$ and $L \psi \equiv 0$, then $\psi\equiv 0$.
\end{enumerate}
\end{lemaa}

For surfaces $S$ embedded in an initial data set $\id$,
the outer null expansion $\theta^{+}$ (also the inner
null expansion $\theta^{-}$)
is a {\it quasilinear} second order elliptic operator\footnote{A quasilinear second order elliptic operator $Q$ has the form
$Q\psi=-a^{ij}(x,\psi,\nabla\psi)\nabla_{i} \nabla_{j}\psi+b(x,\psi,\nabla\psi)$,
with the matrix $[a^{ij}]$ being positive definite.}
acting on the embedding functions of $S$. In this case, there also exists
a maximum principle which is useful (see e.g. \cite{AM}).
\begin{propositiona}\label{maximumprincipleforMOTS}
Let $\id$ be an initial data set and let $S_1$ and $S_2$ be two connected
$C^2$-surfaces touching at one point $\p$, such that the outer normals of
$S_{1}$ and $S_{2}$ agree at $\p$. Assume furthermore that $S_2$ lies to the outside of $S_1$,
that is in direction of its outer normal near $\p$, and that
\[
\sup_{S_1}\theta^{+}[S_1]\leq \inf_{S_2}\theta^{+}[S_2].
\]
Then $S_1=S_2$.
\end{propositiona}
In particular, if two MOTS touch at one point and the outer normals
agree there then the two surfaces must coincide.
This maximum principle can be viewed as an extension of the maximum principle for minimal surfaces which asserts precisely
that two minimal surfaces touching at one point are the same surface (see e.g. \cite{DHT}).

We discuss next the Sard Lemma, which is needed at several places in the
main text.
First
we define regular and critical value for a smooth map.

Let $f : {\cal N} \rightarrow {\cal M}$ be a smooth map.
A point $\p \in {\cal N}$ is a {\bf regular point}
if
$D_{\p} f: T_{\p} {\cal N} \rightarrow T_{f(\p)} {\cal M}$ has maximum rank (i.e.
$\text{rank}(D_{\p} f)=\mbox{min} (n,m)$, where $n$ is the dimension of ${\cal N}$
and $m$ is the dimension of ${\cal M}$). A {\bf critical point} $\p \in {\cal M}$
is a point which is not regular.
A point $\q \in {\cal M}$ is a {\bf regular value}
if $f^{-1} (\q)$ is either empty or  all $\p \in f^{-1} (\q)$
are regular points.  A point $\q \in {\cal M}$
is a {\bf critical value} if it is not a regular value.

We quote Theorem 1.2.2 in \cite{Artino}
\begin{thra}[Sard]\label{lema:Sard}
Let ${\cal N}$ and ${\cal M}$ be paracompact manifolds,
then the set of critical values of a smooth map $f :
{\cal N} \rightarrow {\cal M}$ has measure zero in ${\cal M}$.
\end{thra}

This theorem is equivalent to saying that the set
of regular values of $f: {\cal N} \rightarrow {\cal M}$ is dense in ${\cal M}$.

For maps $f : {\cal N} \rightarrow \mathbb{R}$ the definition above states
that $\p\in\mathcal{N}$ is a critical point if and only if $df |_{\p} =0$.
Let $\p \in {\cal N}$
be a critical point and $H_{\p}$ the Hessian at $\p$ (i.e.
$H_{\p} (\vec{X}, \vec{Y}) = \vec{X} (\vec{Y} (f)) |_{\p}$).
For any isolated critical point $\p \in {\cal N}$ with non-degenerate Hessian, the Morse
Lemma (see e.g. Theorem 7.16 in \cite{Morse}) asserts that there exists
neighbourhood $U_{\p}$ of $\p$ and coordinates $\{ x_1,\cdots, x_n\}$ on $U_{\p}$ such that $\p =
(0, \cdots 0)$ and
$f$ takes the form
$f (x) = f(\p) - (x_1 )^2 - \cdots - (x_q)^2 + (x_{q+1})^2 + \cdots (x_{n})^2$ where
the signature of $H_{\p}$ is $n - q$. For arbitrary critical points this Lemma has been
generalized by Gromoll and Meyer \cite{GromollMeyer}. The generalization allows for Hilbert manifolds
of infinite dimensions. In the finite dimensional case Lemma 1 in \cite{GromollMeyer} can be rewritten in the following form.

\begin{lemaa}[Gromoll-Meyer splitting Lemma, 1969]
Let ${\cal N}$ be a manifold of dimension $n$ and $f: {\cal N} \rightarrow \mathbb{R}$ a smooth map.
Let  $\p$ be a critical point (not
necessarily isolated) and $H_{\p}$ the Hessian of $f$ at $\p$. Assume that the signature of $H_{\p}$ is
$\{ \underbrace{+, \cdots, +}_{q},
\underbrace{-, \cdots, -}_{r},
\underbrace{0, \cdots, 0}_{n-q-r} \}$

Then, there exists an open neighbourhood $U_{\p}$ of $\p$ and
coordinates $\{ x_1 , \cdots, x_{n} \}$ such that $
\p = \{ 0, \cdots 0 \}$ and $f$ takes the form
\begin{eqnarray*}
f (x) = f(\p) + (x_1 )^2 + \cdots + (x_q)^2 - (x_{q+1})^2 - \cdots
(x_{q+r})^2 + h (x_{q+r+1},\cdots,x_n )
\end{eqnarray*}
where $h$ is smooth and this function, its gradient and its Hessian vanishes
at  $(x_{q+r+1}=0,\cdots,x_n=0 )$.
\end{lemaa}

Finally, the following result by Glaeser \cite{Glaeser}
is needed in Chapter \ref{ch:Article1} (proof of Proposition \ref{C1}) when
dealing with positive square roots of non-negative functions.

\begin{thra}[Glaeser, 1963 \cite{Glaeser}]
\label{Glaser}
Let $U$ be an open subset of $\mathbb{R}^n$ and
$f: U  \rightarrow  \mathbb{R}$ be $C^2$ and
satisfy $f \geq 0$ everywhere. If the Hessian of $f$ vanishes
everywhere on the set $F = \{ \p \in  U, \mbox{such that } f(\p)=0 \}$, then
$g = + \sqrt{f}$ is $C^1$ on $U$.
\end{thra}

\label{sc:holderspaces}

\end{document}